\documentclass[fleqn,usenatbib]{mnras}

\usepackage{newtxtext,newtxmath}

\usepackage[T1]{fontenc}

\DeclareRobustCommand{\VAN}[3]{#2}
\let\VANthebibliography\thebibliography
\def\thebibliography{\DeclareRobustCommand{\VAN}[3]{##3}\VANthebibliography}

\usepackage{graphicx}	
\usepackage{amsmath}	
\usepackage{multirow}
\usepackage{booktabs}

\newcommand{\LCDM}{\ensuremath{\Lambda\textrm{CDM}}}
\newcommand{\wCDM}{\ensuremath{w\textrm{CDM}}}
\newcommand{\OmegaM}{\ensuremath{\Omega_{\mathrm{m}}}}

\newcommand{\Omegab}{\ensuremath{\Omega_{\mathrm{b}}}}
\newcommand{\Omegak}{\ensuremath{\Omega_{\mathrm{k}}}}
\newcommand{\Hnow}{\ensuremath{H_{0}}}
\newcommand{\sigmaeight}{\ensuremath{\sigma_{8}}}
\newcommand{\seight}{\ensuremath{S_{8}}}
\newcommand{\ns}{\ensuremath{n_{\mathrm{s}}}}
\newcommand{\w}{\ensuremath{w}}

\newcommand{\Msunh}{\ensuremath{h^{-1}\mathrm{M}_{\odot}}}
\newcommand{\Mpc}{\ensuremath{\mathrm{Mpc}}}
\newcommand{\Mpch}{\ensuremath{h^{-1}\mathrm{Mpc}}}

\newcommand{\Rfiveoo}{\ensuremath{R_{500\mathrm{c}}}}
\newcommand{\Mfiveoo}{\ensuremath{M_{500\mathrm{c}}}}

\newcommand{\redshift}{\ensuremath{z}}
\newcommand{\mass}{\ensuremath{M}}
\newcommand{\dif}{\ensuremath{\mathrm{d}}}

\newcommand{\rhocrit}{\ensuremath{\rho_{\mathrm{c}}}}

\newcommand{\PLANCK}{\emph{Planck}}

\newcommand{\eROSITA}{\emph{eROSITA}}
\newcommand{\eFEDS}{eFEDS}
\newcommand{\ROSAT}{\emph{ROSAT}}
\newcommand{\XXL}{\emph{XXL}}

\newcommand{\fcont}{\ensuremath{f_{\mathrm{cont}}}}
\newcommand{\rich}{\ensuremath{\lambda}}
\newcommand{\Ldet}{\ensuremath{\mathcal{L}_{\mathrm{det}}}}
\newcommand{\ext}{\ensuremath{\mathtt{EXT}}}
\newcommand{\Lext}{\ensuremath{\mathcal{L}_{\mathrm{ext}}}}
\newcommand{\rate}{\ensuremath{\eta}}
\newcommand{\Comp}{\ensuremath{\mathcal{C}}}
\newcommand{\texp}{\ensuremath{t_{\mathrm{exp}}}}
\newcommand{\snr}{\ensuremath{\mathtt{SNR}_{\eta}}}
\newcommand{\wl}{\ensuremath{\mathrm{WL}}}
\newcommand{\zcl}{\ensuremath{z_{\mathrm{cl}}}}

\newcommand{\gshear}{\ensuremath{g_{+}}}

\newcommand{\bwl}{\ensuremath{b_{\mathrm{WL}}}}
\newcommand{\Mwl}{\ensuremath{M_{\mathrm{WL}}}}

\newcommand{\fmis}{\ensuremath{f_{\mathrm{mis}}}}
\newcommand{\sigmamis}{\ensuremath{\sigma_{\mathrm{mis}}}}
\newcommand{\Xlabel}{\ensuremath{\mathcal{X}}}
\newcommand{\sobs}{\ensuremath{\mathcal{S}}}
\newcommand{\fobs}{\ensuremath{\mathcal{O}}}
\newcommand{\mpiv}{\ensuremath{M_{\mathrm{piv}}}}
\newcommand{\zpiv}{\ensuremath{z_{\mathrm{piv}}}}
\newcommand{\Ez}{\ensuremath{E(z)}}
\newcommand{\Ezpiv}{\ensuremath{E(z_{\mathrm{piv}})}}
\newcommand{\Arate}{\ensuremath{A_{\eta}}}
\newcommand{\Brate}{\ensuremath{B_{\eta}}}
\newcommand{\deltarate}{\ensuremath{\delta_{\eta}}}
\newcommand{\gammarate}{\ensuremath{\gamma_{\eta}}}
\newcommand{\sigmarate}{\ensuremath{\sigma_{\eta}}}

\newcommand{\brate}{\ensuremath{b_{\eta}}}
\newcommand{\Bgrp}{\ensuremath{B_{\mathrm{grp}}}}
\newcommand{\mgrp}{\ensuremath{M_{\mathrm{grp}}}}
\newcommand{\Af}{\ensuremath{A_{\mathrm{f}}}}
\newcommand{\Bf}{\ensuremath{B_{\mathrm{f}}}}
\newcommand{\deltaf}{\ensuremath{\delta_{\mathrm{f}}}}
\newcommand{\gammaf}{\ensuremath{\gamma_{\mathrm{f}}}}

\newcommand{\Awl}{\ensuremath{A_{\mathrm{WL}}}}
\newcommand{\Bwl}{\ensuremath{B_{\mathrm{WL}}}}
\newcommand{\deltawl}{\ensuremath{\delta_{\mathrm{WL}}}}
\newcommand{\gammawl}{\ensuremath{\gamma_{\mathrm{WL}}}}
\newcommand{\sigmawl}{\ensuremath{\sigma_{\mathrm{WL}}}}
\newcommand{\Arich}{\ensuremath{A_{\lambda}}}
\newcommand{\Brich}{\ensuremath{B_{\lambda}}}
\newcommand{\deltarich}{\ensuremath{\delta_{\lambda}}}
\newcommand{\gammarich}{\ensuremath{\gamma_{\lambda}}}
\newcommand{\sigmarich}{\ensuremath{\sigma_{\lambda}}}
\newcommand{\rhoraterich}{\ensuremath{\rho_{\rate,\rich}}}
\newcommand{\rhoratewl}{\ensuremath{\rho_{\rate,\mathrm{WL}}}}
\newcommand{\rhorichwl}{\ensuremath{\rho_{\rich,\mathrm{WL}}}}
\newcommand{\ratefivez}{\ensuremath{\eta_{50,z}}}
\newcommand{\ratefiveo}{\ensuremath{\eta_{50}}}
\newcommand{\srate}{\ensuremath{s_{\eta}}}
\newcommand{\gammaz}{\ensuremath{\gamma_{z}}}

\newcommand{\ansArate}{\ensuremath{ 0.133^{+0.026}_{-0.020} }}
\newcommand{\ansBrate}{\ensuremath{ 1.86^{+0.20}_{-0.15} }}
\newcommand{\ansdeltarate}{\ensuremath{ -0.58^{+0.43}_{-0.50} }}
\newcommand{\ansgammarate}{\ensuremath{ -0.83^{+0.44}_{-0.50} }}
\newcommand{\anssigmarate}{\ensuremath{ 0.332^{+0.076}_{-0.089} }}

\newcommand{\ansArich}{\ensuremath{ 37.1^{+4.6}_{-5.0} }}
\newcommand{\ansBrich}{\ensuremath{ 1.05^{+0.13}_{-0.12} }}
\newcommand{\ansdeltarich}{\ensuremath{ -0.72^{+0.45}_{-0.37} }}
\newcommand{\ansgammarich}{\ensuremath{ -0.77^{+0.41}_{-0.54} }}
\newcommand{\anssigmarich}{\ensuremath{ 0.291^{+0.133}_{-0.078} }}

\newcommand{\OmegamLCDM}{\ensuremath{ 0.245^{+0.048}_{-0.058} }}
\newcommand{\OmegamLCDMBPL}{\ensuremath{ 0.230^{+0.060}_{-0.043} }}
\newcommand{\OmegamWCDM}{\ensuremath{ 0.234^{+0.048}_{-0.070} }}

\newcommand{\sigmaeightLCDM}{\ensuremath{ 0.833^{+0.075}_{-0.063} }}
\newcommand{\sigmaeightLCDMBPL}{\ensuremath{ 0.847\pm 0.061 }}
\newcommand{\sigmaeightWCDM}{\ensuremath{ 0.846^{+0.092}_{-0.066} }}

\newcommand{\seightLCDM}{\ensuremath{ 0.791^{+0.028}_{-0.031} }}
\newcommand{\seightLCDMBPL}{\ensuremath{ 0.796^{+0.026}_{-0.032} }}
\newcommand{\seightWCDM}{\ensuremath{ 0.784^{+0.034}_{-0.027} }}

\newcommand{\wWCDM}{\ensuremath{ -1.25\pm 0.47 }}

\newcommand{\ansratefiveo}{\ensuremath{ \left( 61.4^{+3.8}_{-3.5} \right) \times 10^{-3} }}
\newcommand{\anssrate}{\ensuremath{ 0.501^{+0.050}_{-0.043} }}
\newcommand{\ansgammaz}{\ensuremath{ 0.06\pm 0.12 }}
\newcommand{\ansratefiveoHghlext}{\ensuremath{ \left( 93.8^{+7.4}_{-6.4} \right) \times 10^{-3} }}
\newcommand{\anssrateHghlext}{\ensuremath{ 0.428^{+0.061}_{-0.054} }}
\newcommand{\ansgammazHghlext}{\ensuremath{ 0.20^{+0.15}_{-0.12} }}

\newcommand{\percent}{\ensuremath{\%}}
\newcommand{\appropto}{\mathrel{\vcenter{
                        \offinterlineskip\halign{\hfil$##$\cr
                        \propto\cr\noalign{\kern2pt}\sim\cr\noalign{\kern-2pt}}}}}
\DeclareMathOperator\erf{erf}
\newcommand{\vect}[1]{\boldsymbol{\mathbf{#1}}}

\title[\eFEDS\ Cosmology]{
Cosmological Constraints from Galaxy Clusters and Groups in the eROSITA Final Equatorial Depth Survey
}

\author[Chiu et al.]{
I-Non~Chiu$^{1,2,3}$\thanks{E-mail: inchiu@phys.ncku.edu.tw},
Matthias~Klein$^{4}$,
Joseph~Mohr$^{4,5}$,
Sebastian~Bocquet$^{4}$
\\
$^{1}$Tsung-Dao Lee Institute, Shanghai Jiao Tong University, Shanghai 200240, China\\
$^{2}$Department of Astronomy, School of Physics and Astronomy, and Shanghai Key Laboratory for Particle Physics and Cosmology, \\
~Shanghai Jiao Tong University, Shanghai 200240, China\\
$^{3}$Department of Physics, National Cheng Kung University, 70101 Tainan, Taiwan\\
$^{4}$University Observatory, Faculty of Physics, Ludwig-Maximilians-Universit\"{a}t, Scheinerstr. 1, 81679 Munich, Germany\\
$^{5}$Max Planck Institute for Extraterrestrial Physics, Giessenbachstrasse 1, 85748 Garching, Germany\\
}

\pubyear{2022}

\hypersetup{urlcolor=[rgb]{0.9,0.4,0.6}, filecolor=magenta, citecolor=cyan}

\begin{document}
\label{firstpage}
\pagerange{\pageref{firstpage}--\pageref{lastpage}}
\maketitle

%
%

\begin{abstract}
We present the first cosmological study of a sample of \eROSITA\ clusters, which were identified in the \eROSITA\ Final Equatorial Depth Survey (\eFEDS).
In a joint selection on X-ray and optical observables, the sample contains $455$ clusters within 
a redshift range of $0.1<\redshift<1.2$, of which $177$ systems are covered by the public data from the Hyper Suprime-Cam (HSC) survey that enables uniform weak-lensing cluster mass constraints.
With minimal assumptions, at each cluster redshift we empirically model 
(1) the scaling relations between the cluster halo mass and the observables, which include
the X-ray count rate, the optical richness, and the weak-lensing mass, and 
(2) the X-ray selection in terms of the completeness function \Comp.
Using the richness distribution of the clusters, we directly measure the X-ray completeness and adopt those measurements as informative priors for the parameters of \Comp.  
In a blinded analysis, we obtain the cosmological constraints
$\OmegaM = \OmegamLCDM$, $\sigmaeight = \sigmaeightLCDM$ and $\seight \equiv \sigmaeight\left(\OmegaM/0.3\right)^{0.3}= \seightLCDM$ in a flat \LCDM\ cosmology.
Extending to a flat \wCDM\ cosmology leads to the constraint on the equation of state parameter of the dark energy of $\w = \wWCDM$.
The \eFEDS\ constraints are in good agreement with the results from the \PLANCK\ mission, the galaxy-galaxy lensing and clustering analysis of the Dark Energy Survey, and the cluster abundance analysis of the SPT-SZ survey at a level of $\lesssim1\sigma$.
With the empirical modelling, this work presents the first fully self-consistent cosmological constraints based on a synergy between wide-field X-ray and weak lensing surveys.
\end{abstract}



%
%

\section{Introduction}
\label{sec:intro}

The abundance, or the number density, of galaxy clusters is a powerful cosmological probe with sensitivity to the nature of dark energy, which is required to explain the accelerating expansion of the universe \citep[see, e.g.,][]{haiman01,allen11,kravtsov12,weinberg13,huterer15}.
Such analyses based on galaxy clusters or groups (hereafter clusters for simplicity) are emerging as a competitive cosmological tool owing to the deployment of wide-field surveys in constructing large samples of clusters in the millimeter wavelengths \citep{staniszewski09,PlanckCollaboration2014,PlanckCollaboration2015b,bleem15,bleem20,hilton21} through the Sunyaev-Zel'dovich effect \citep[SZE;][]{sunyaev72},
in the optical through the identification of overdensities of galaxies \citep{gladders07,yang07,rykoff14,oguri14,oguri18,maturi19}, through the technique of weak gravitational lensing \citep[hereafter weak lensing;][]{wittman06,miyazaki18,oguri21},
and in X-rays \citep{reiprich02,bohringer04,vikhlinin09a,boller16,pacaud16,adami18,finoguenov20} through the emission of intracluster medium (ICM).
To examine our understanding of the universe and fundamental physics, stringent constraints on cosmological parameters have been obtained using these samples \citep{vikhlinin09b,zu14,mantz10a,mantz14,mantz15,PlanckCollaboration2015b,bocquet15,bocquet19,deHaan16,schellenberger17,pacaud18,iderchitham20,garrel21} together with other probes \citep{to21} or across different wavelengths \citep{costanzi21,salvati21}, demonstrating that cluster cosmology is an essential tool in observational cosmology.

Understanding the selection of a cluster sample is of crucial importance in cluster cosmology.
Despite the fact that a large sample of clusters simultaneously spanning a wide range of mass and redshift can be constructed in an optical survey, the modeling of the associated selection function is significantly impacted by projection effects \citep{zu17,costanzi19,sunayama20,zhang22,wu22} and currently challenges the feasibility of employing optically selected samples in cluster cosmology \citep{abbott20desy1clustercosmology}.

Conversely, ICM-selected samples are less affected by projection effects and thus provide cleaner datasets for cosmological studies.
The detection of clusters in the SZE, which at frequencies below the SZE null of $220$~GHz are identified by the shadow they cast (i.e. decrease in flux) on the cosmic microwave background, is primarily impacted by the primary CMB anisotropy and the statistical noise in the mm-wave sky maps. The SZE signature as extracted using angular filtering with arcminute scale angular resolution 
is very sensitive to cluster mass and only weakly dependent on redshift. 
Therefore, samples selected from ground-based SZE surveys with arcminute-scale angular resolution
correspond to nearly mass-limited samples where the limiting mass changes only moderately with redshift.
Because the SZE samples tend to probe the most massive and rarest clusters, the resulting cluster cosmology is currently limited by the relatively small cluster samples.

Meanwhile, X-ray surveys with sufficiently deep imaging can construct a cluster sample with a mass range down to the scale of galaxy groups at low redshift and can identify the more massive clusters out to high redshift.
Despite the ICM-based selection, X-ray selected samples suffer from considerable amounts of contamination at a level of $\gtrsim20$--$50\percent$ due to point sources with energetic emission (e.g., active galactic nuclei or AGNs).
Consequently, the cosmological analyses up until now have been limited to the brightest subsamples with sufficiently high purity and completeness 
\citep[e.g.,][]{vikhlinin09b,mantz10a,mantz14,pacaud18}.
To fully utilize an X-ray selected sample for cosmological studies, a secondary processing using optical data is required to reduce the contamination from the levels delivered by a pure X-ray selection.
Originally, this optical followup included painstaking cluster-by-cluster imaging and spectroscopy \citep{koulouridis21}, but this task has become easier with the accumulation of data from deep and wide-field optical surveys.  Employing optical confirmation significantly suppresses the point-source contamination from AGNs without removing a significant number of clusters \citep{klein18,finoguenov20}.
This methodology has been applied to the \ROSAT\ All-Sky Survey \citep{boller16} with the optical imaging from the Dark Energy Survey \citep{des05,des16}, resulting in the largest sample of X-ray selected clusters out to redshift $\redshift\approx1$  to date with the achieved contamination rate of just a few percent \citep{klein19}.
Such an optical confirmation is effectively equivalent to applying a secondary selection using optical observables (e.g., richness), leading to a joint X-ray and optical selection function that can be modeled using the appropriate mass--observable relations \citep{grandis20}.

A quantum leap has been made in cluster cosmology with the launch of the X-ray telescope \eROSITA\ 
\cite[extended ROentgen Survey with an Imaging Telescope Array;][]{predehl21} 
that will provide the largest ever sample of $\gtrsim100,000$ X-ray selected clusters in the four-year \eROSITA\ All-Sky Survey.
Prior to the all-sky survey, the \eROSITA\ Final Equatorial Depth Survey (\eFEDS) was first carried out as part of the performance verification phase of the \eROSITA\ mission.
There are $\gtrsim540$ X-ray selected clusters in \eFEDS\ with a survey area of $\approx140$~deg$^2$ with systems extending to redshift $\redshift\approx1.3$ \citep{liu21,klein22}. 
The data products and catalogs have been entirely released in the Early Data Release\footnote{\url{https://erosita.mpe.mpg.de/edr}} of the \eROSITA\ mission with a series of studies on the characterization of X-ray properties \cite[e.g.,][]{liu21,ghirardini21,bahar21}.
In \cite{chiu22}, the weak lensing mass calibration of the \eFEDS\ clusters is presented in great detail as part of a study of the baryonic scaling relations of galaxy clusters
\citep[see][for a detailed review of weak lensing studies of clusters]{bartelmann01,hoekstra15,umetsu20b}.
We stress that the weak-lensing mass calibration in \cite{chiu22} was carried out in a forward-modeling framework that is specifically designed to support cosmological analyses.

In this paper, we continue to constrain cosmology by combining the weak-lensing mass calibration presented in \cite{chiu22} with the abundance of the \eFEDS\ clusters.
We stress that the \eFEDS\ clusters used in this work are first selected in X-rays and then confirmed in the optical, resulting in a clean and sizable sample for cosmological study without losing a significant amount of clusters due to a more stringent X-ray selection.
The analysis is carried out in a forward-modeling framework with the empirical modeling of the survey selection and the cluster population \citep{mantz15,mantz16b,bocquet19,grandis19}.

The ``empirical'' modeling here refers to an approach where we allow the data themselves to constrain the underlying properties, e.g., the scaling law between the observables and the cluster halo mass, with only minimal assumptions informed by prior knowledge from theory, simulations and previous observational studies.
The empirical modeling has advantages over other methods, especially those heavily relying on simulations:
First, the empirically calibrated results are not sensitive to the specific configuration or subgrid physics in simulations that are necessary for the modeling.
For instance, an X-ray selection function directly derived from simulations and the resulting cosmological constraints would be biased if the simulated clusters do not accurately and precisely reflect the observed cluster population.
In fact, this is the case for the \eFEDS\ simulation \citep{liuteng21}, where the number of synthetic clusters with low fluxes was over-predicted (we return to discuss this in terms of the X-ray scaling relation and the selection function in Section~\ref{sec:brokenpowerlaw} and Section~\ref{sec:completeness}, respectively).
Second, the modeling is primarily determined by what the data require and therefore fully captures the behavior of the data.
This also allows us to identify and to account for the residual bias or unexpected trends
of the model with respect to the data.
With minimal assumptions and a flexible modeling scheme, the critical and final step in a cosmological analysis is to assess the consistency between the resulting best-fit model and the observed data (e.g., through a goodness-of-fit test), and to incorporate the full systematics into the cosmological constraints (e.g., through a marginalization over so-called nuisance parameters).

In this work, as in \cite{bocquet19}, the minimal assumptions are made in the empirical modeling, as follows.
\begin{itemize}
\item Informed by the knowledge of cosmic structure formation through observations and numerical simulations, we assume that the relation between the observables and the cluster halo mass at each redshift can be well-described by a power-law function with log-normal intrinsic scatter. 
More precisely, the adopted forms of the mass--observable relations encapsulate the freedom to follow the self-similar expectations \citep{kaiser1986,boehringer12} for these relations but at the same time allow for deviations from that behavior \citep[e.g.,][]{mohr97,mohr99,lin04a,chiu16a,chiu16c,chiu18a,bulbul19,chiu22} if preferred by the data.  Moreover, we include possible correlated scatter between the observables. 
\item We assume that the simulation-calibrated bias in the observed count rate \rate\ and weak-lensing mass \Mwl\ can be captured by a mean relation modeled as a power law of the cluster halo mass and redshift with log-normal scatter.
The systematics of this bias have been fully quantified in \cite{chiu22} 
and are accounted for by a marginalization over the associated parameters.
\item We assume that the completeness at a given observed count rate \rate\ at each redshift due to the X-ray selection of the \eFEDS\ clusters can be described by an error function, which is well motivated and informed by the dedicated simulations \citep{clerc18} and what is measured in the \eFEDS\ catalog of point sources \citep{bulbul21}.
The systematics of the X-ray selection function are assessed with 
the completeness that is independently measured in this work (see Section~\ref{sec:completeness_measurements}).
\end{itemize}

This paper is organized, as follows.
The \eFEDS\ cluster sample used in this work and the weak-lensing analysis from \cite{chiu22} are summarized in Section~\ref{sec:sample_and_data}.  
We provide a detailed description of the likelihood for the cosmological analysis in Section~\ref{sec:likelihoods}.
The empirical modeling of the scaling relations and the selection function is provided in Section~\ref{sec:analysis}.
We present the results in Section~\ref{sec:results}, followed by discussion in Section~\ref{sec:discussions}.
The conclusions are drawn in Section~\ref{sec:conclusions}.
Throughout this paper, the uncertainties are quoted as the $68\percent$ confidence level, unless otherwise stated.
In this work, the cluster halo mass, denoted as \mass\ or \Mfiveoo, is defined by a sphere where the interior mass density is $500$ times the cosmic critical density $\rhocrit(\redshift)$ at the cluster redshift \redshift.  
We could just as well adopt another definition of the cluster halo mass.  What is important is that the halo mass definition we adopt is used in the theoretically defined halo mass function, which we use to model the observed abundance of \eFEDS\ clusters.
The notation $\mathcal{N}\left(\mu, \sigma^2\right)$ stands for a Gaussian distribution with a mean $\mu$ and a standard deviation $\sigma$, 
while $\mathcal{U}\left(x, y\right)$ represents a uniform distribution between $x$ and $y$.

%
%

%
\begin{figure*}
\resizebox{0.49\textwidth}{!}{
\includegraphics[scale=1]{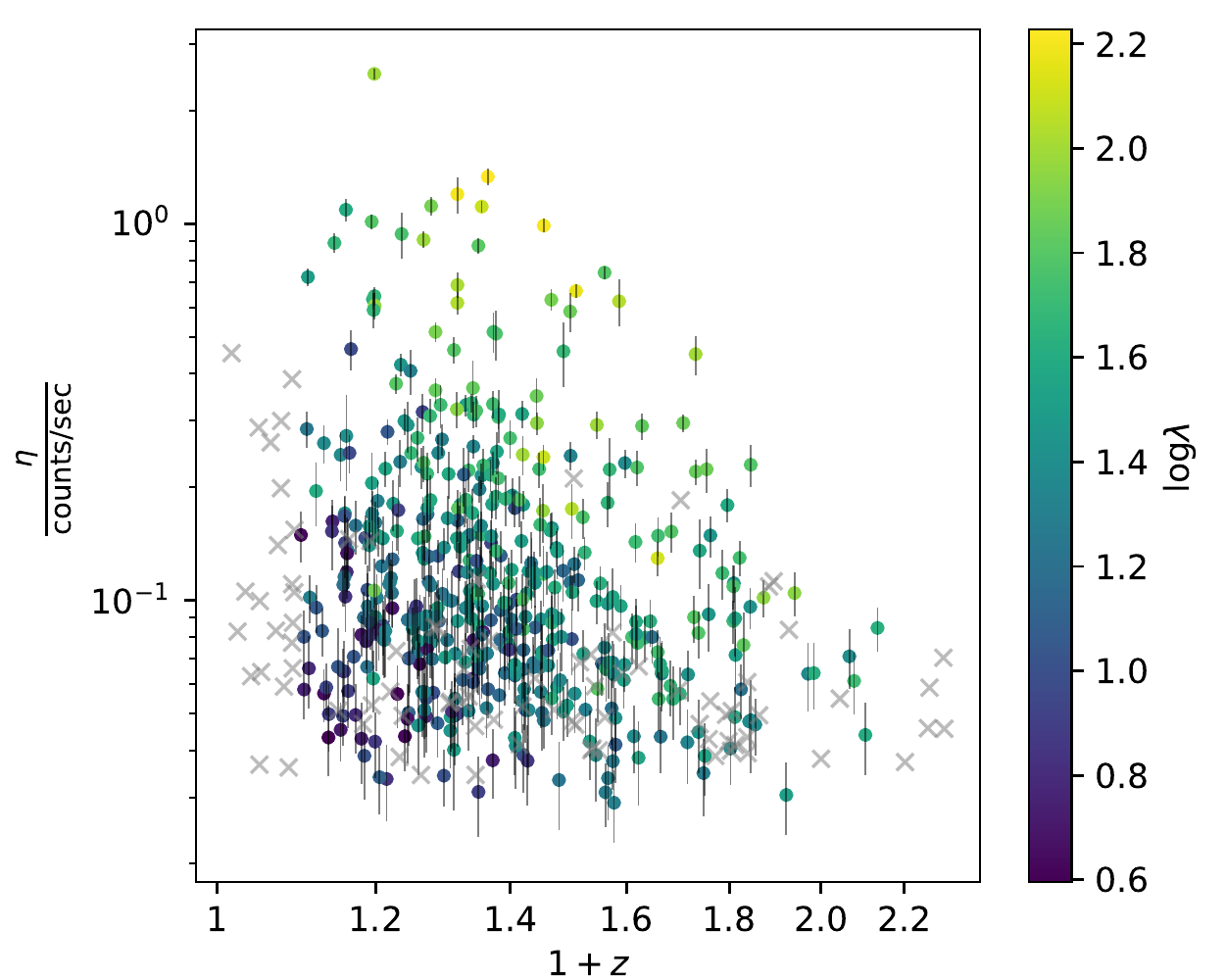}
}
\resizebox{0.49\textwidth}{!}{
\includegraphics[scale=1]{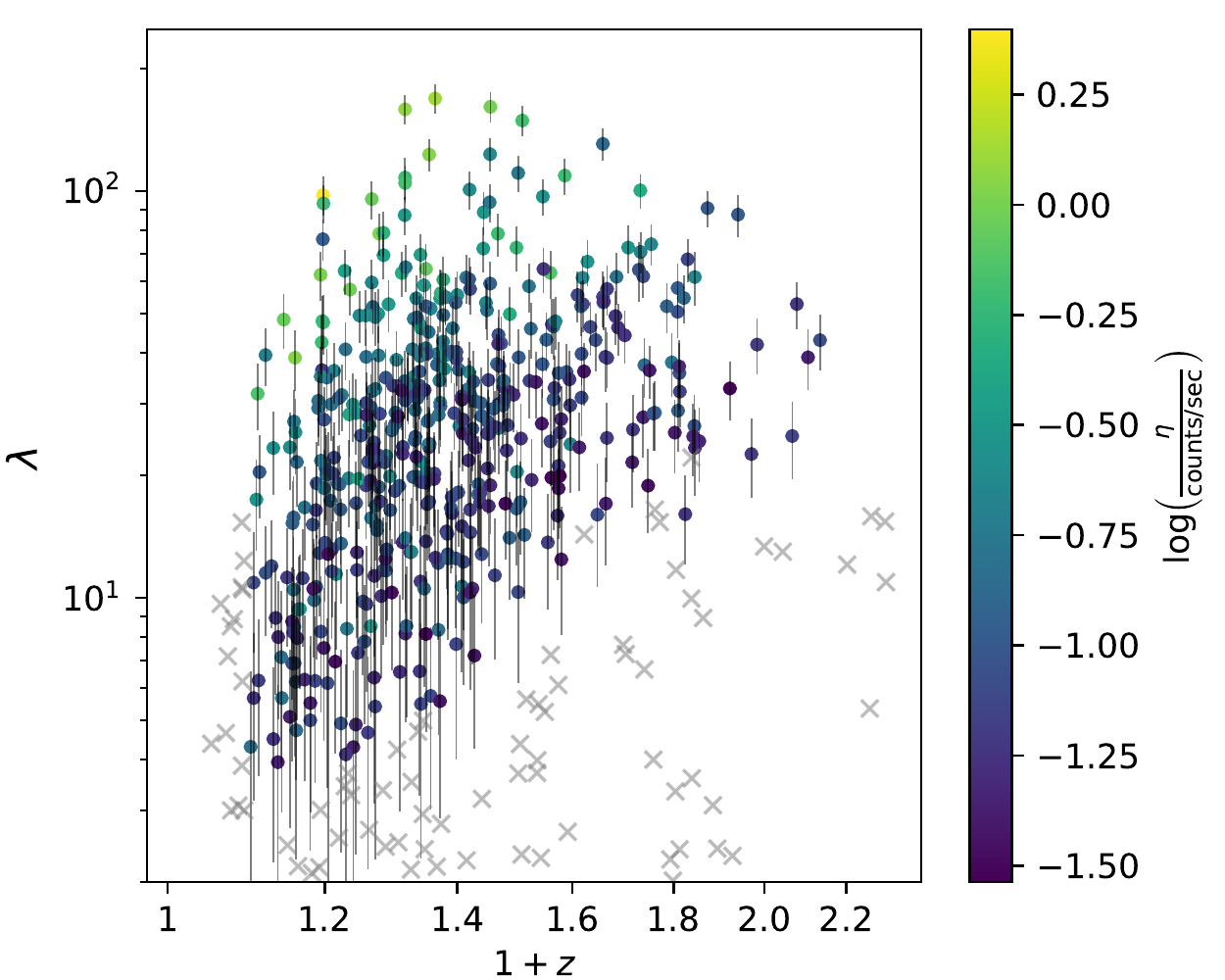}
}
\caption{
The \eFEDS\ cluster sample.
The distribution in the observable space of the redshift \redshift\ and the count rate \rate\ (the optical richness \rich) is shown in the left (right) panel.
The selected clusters used in the analysis are color-coded by their richness and count rate in the left and right panels, respectively, while those are discarded are indicated by the crosses.
}
\label{fig:sample}
\end{figure*}

\section{The Sample and Data}
\label{sec:sample_and_data}

We introduce the cluster sample in Section~\ref{sec:sample} and describe the weak-lensing data and analysis in Section~\ref{sec:wldata}.

\subsection{The \eFEDS\ cluster sample}
\label{sec:sample}

We use the sample of clusters in the \eROSITA\ Final Equatorial Depth Survey (\eFEDS), of which the data and catalogs are 
publicly available within the \eROSITA\ Early Data Release. 
The \eFEDS\ survey is located at Right Ascension (R.A.) between $\approx126^{\circ}$ and $\approx146^{\circ}$ and Declination (Dec.) between $\approx-3^{\circ}$ and $\approx+6^{\circ}$, covering a total solid angle $\Omega_{\mathrm{survey}}\approx140$~deg$^2$.
The imaging depth is expected to match the final depth of the \eROSITA\ All-Sky Survey in the equatorial region, by design, and is uniform with an average exposure time of $\approx2.2$~$k\mathrm{s}$ and $\approx1.2$~$k\mathrm{s}$ before and after the correction for vignetting, respectively.
The data were processed by the pipeline \texttt{eSASSusers\_201009}, which is described in detail in \cite{brunner21}.
In what follows, we provide a brief summary.

The X-ray sources are detected using the X-ray imaging within the energy band of $0.2$--$2.3$~$k\mathrm{eV}$.
Each X-ray source is characterized by, including but not limited to, 
(1) the detection likelihood \Ldet, which describes the likelihood of being a real source,
(2) the extent likelihood \Lext, which presents the likelihood excess of being an extended source rather than a point source, 
(3) the extent \ext, which is the scale of the source extent that maximizes the extent likelihood, and
(4) the observed count rate \rate\ with the observed uncertainty $\delta_{\rate}$.
The parameter \ext\ is limited to a maximum value, $\ext_{\mathrm{up}}=60$~arcsec, which is used if \ext\ exceeds the limit during the maximization of \Lext.
Conversely, \ext\ and \Lext\ are set to zero if the source falls below a threshold in either of these two quantities that are selected to indicate a point source.
This configuration has a profound effect on the feasibility of utilizing these parameters as the X-ray selection observables, which we return to discuss in Section~\ref{sec:discussions}.
After the source detection, the catalog of \eFEDS\ clusters is constructed using the criteria of $\Ldet > 5$, $\Lext > 6$, and $\ext > 0$, resulting in the $542$ clusters that are publicly released in \cite{liu21}.

After the construction of the cluster catalog, each cluster candidate is then further analysed
using the optical imaging from the DESI Legacy Imaging Survey \citep{dey19} and the Hyper Suprime-Cam (HSC) Subaru Strategy Program \citep{aihara18a}.
This optical confirmation analysis is carried out using the Multi-Component Matched Filter \citep[MCMF;][]{klein18} algorithm with the estimated properties fully described and publicly available in \cite{klein22}. 
We refer the reader to \cite{klein22} for a complete description and  only summarize the key results, as follows.
For each cluster candidate centering at the X-ray center, the MCMF algorithm is employed to deliver 
(1) the photometric redshift \zcl\ using the overdensity of red galaxies \citep{gladders00}, 
(2) the optical richness \rich, and 
(3) the optical contamination estimator \fcont, which is related to the probability of detecting a source with the observed \rich\ at the redshift \zcl\ along a random line of sight.
A high value of \fcont\ represents a high probability of observing such a system in the optical simply due to a random superposition of the X-ray source with a physically unassociated optical system \citep[see details in][]{klein19}.
Applying an upper-limit cut in \fcont\ therefore filters out the likely contaminants from the X-ray cluster candidate list and is equivalent to imposing a  redshift-dependent lower threshold on the observed richness.  This cleaning with MCMF reduces the 
contamination level in the initial X-ray selected cluster catalog.
In this work, we impose the threshold $\fcont < 0.3$, reducing the cluster catalog from $542$ to $477$ clusters.
The contamination due to point sources in the initial cluster catalog is estimated to be $\approx20\percent$ \citep{liu21} using the simulations \citep{liuteng21}; therefore, the  residual contamination after the optical confirmation is expected to be at a level of $\approx0.3\times20\percent \approx 6\percent$ \citep{klein22}.
A more stringent cut of $\fcont < 0.2$, as previously used in the weak-lensing mass calibration \citep{chiu22}, would further discard $\approx20$ clusters (corresponding to a $5\percent$ difference in the sample size) and reduce the contamination rate to $\approx4\percent$.
Given the current sample size, it is important to clean the sample using MCMF to reduce the contamination to $4\percent$ or $6\percent$ from an initial $20\percent$, but the difference between $4\percent$ and $6\percent$ is unimportant.

The photometric redshift of each cluster is adopted as the cluster redshift.
The accuracy of the photometric redshift estimate with respect to the spectroscopic redshift in terms of $\Delta\redshift/\left(1 + \redshift \right)$ is at a level of 
$\lesssim0.5\percent$ \citep{klein19}, for which the uncertainty can be safely ignored for the purpose of this work.
In this study, we apply a cut in the cluster redshift by requiring $0.1<\zcl<1.2$, resulting in a final cluster catalog containing $455$ clusters.  This is the cosmological sample we adopt for our analysis.
The redshift upper limit is employed because the optical imaging from the HSC survey does not have the capability to optically confirm cluster candidates beyond such a high redshift; a dedicated treatment to the optical confirmation at $\redshift>1.2$ is required \citep[as done in][]{klein19}.
In the interest of uniformity, we do not use the clusters at redshift $\redshift>1.2$ in this work.
Removing the upper limit of the redshift selection would include two additional clusters at $\redshift\approx1.3$;  we do not expect their exclusion to have a significant impact on our final results.
The choice of the lower-limit selection on the redshift is to discard clusters (mostly low-mass groups) at low redshift where the lensing efficiency is extremely low \citep{chiu22}.

In summary, the low-contamination cosmological sample of \eFEDS\ clusters used in this paper is selected by 
\begin{itemize}
\item the X-ray selection: $\Ldet > 5$, $\Lext > 6$, and $\ext > 0$, 
\item the optical selection: $\fcont < 0.3$, corresponding to a redshift-dependent cut on the richness \rich, and
\item the redshift selection: $0.1<\zcl<1.2$.
\end{itemize}
This results in a sample of $455$ clusters\footnote{
This sample can be directly constructed from the catalog \texttt{eFEDS\_c001\_main\_ctp\_clus\_v2.1.fits} on \url{https://erosita.mpe.mpg.de/edr/}.} with the expected contamination rate of $\approx6\percent$.
The cluster sample is visualized in Figure~\ref{fig:sample}, where the final cosmological sample is shown as the colored points, and the systems discarded from the analysis are the crosses.
As  seen in the right panel,  the redshift-dependent richness selection owing to the optical confirmation efficiently removes the contamination, especially those with low count rates at high redshift.
The sample has a median redshift $\zpiv=0.35$ and spans a mass range of $10^{13}\Msunh \lesssim \mass \lesssim 10^{15}\Msunh$ with a median mass $\mpiv = 1.4\times10^{14}\Msunh$, as previously indicated by the weak-lensing mass calibration \citep{chiu22}.

\subsection{The weak-lensing data and analysis}
\label{sec:wldata}

In this work, we make use of the weak-lensing data from the survey of the Hyper Suprime-Cam (HSC) Subaru Strategy Program \citep{aihara18a}.
By design, the \eFEDS\ survey partially overlaps the HSC survey, enabling us to uniformly measure the weak-lensing observable for each cluster and perform the weak-lensing mass calibration.
The weak-lensing mass calibration, as well as various X-ray observable-to-mass-and-redshift relations, was thoroughly and carefully studied in \cite{chiu22}.   Here we apply the identical method, but with some changes to data and systematics estimates, which
are summarized below.

We use the public weak-lensing shape catalog from the first-year HSC data \citep[S16A;][]{mandelbaum18a}  in the second data release\footnote{\url{https://hsc-release.mtk.nao.ac.jp/doc/index.php/s16a-shape-catalog-pdr2/}}.
This is the main difference between our analysis here and the previous \cite{chiu22} analysis, which used the latest (and proprietary) shape catalog \citep[S19A;][]{li21}.
As a result, only $177$ out of $455$ \eFEDS\ clusters, corresponding to $\approx39\percent$ of the cosmological sample, have weak-lensing measurements.
Roughly $\approx2$ times more clusters are covered by the S19A data, and therefore we expect the statistical power of the weak-lensing mass calibration in this work would decrease by $\approx1-1/\sqrt{2}\approx30\percent$ compared to \cite{chiu22}.

For each cluster, the weak-lensing observable is the individual reduced shear profile \gshear, which is extracted at a range of the projected radius between $\approx0.5\Mpch$ and $\approx3.5\Mpch$ around the X-ray center.
Note that we do not use the cluster core ($<0.5~\Mpch$) in the analysis.
The shear profiles are estimated using background sources that are selected based on the photometric redshift distributions, estimated by the code \texttt{DEmp} \citep{hsieh14}.
The shape measurement of galaxies is rigorously calibrated against intensive image simulations \citep{mandelbaum18b}, delivering the systematic uncertainty of the multiplicative bias at a level of $\lesssim1.7\percent$, sufficiently accurate for cluster lensing.
Various null tests were quantified in \cite{oguri18b} and suggest no significant residual bias that would impact our analysis.
We stress that the shear profile \gshear\ is stored in angular radius, $\gshear\left(\theta\right)$, and is self-consistently compared to the model prediction calculated using the redshift-distance relation given the cosmology in each step of the likelihood exploration.
The lensing covariance matrix, which represents the measurement uncertainty of the observed \gshear, includes not only the shape noise but also the scatter due to the uncorrelated large-scale structure along the line of sight \citep{hoekstra03}.

With the methodology introduced in \cite{grandis21}, the weak-lensing analysis carefully includes the systematic uncertainties associated with
(1) the calibration of the multiplicative bias,
(2) the bias due to the photometric redshift of background sources, as quantified to be $\lesssim1\percent$ and $\lesssim5\percent$ at the cluster redshifts $\redshift\lesssim0.6$ and $0.6\lesssim\redshift\lesssim1.3$, respectively,
(3) the cluster member contamination, which is suggested to be subdominant ($\lesssim6\percent$ at a level of $2\sigma$ at the cluster core),
(4) the miscentering of the X-ray centers, as described by the fraction \fmis\ of miscentered clusters and the characteristic scale \sigmamis\ of the miscentering, 
(5) the inaccurate assumption about the halo profile due to the presence of correlated large-scale structure or triaxiality,
(6) the radial binning used in the fitting, and importantly
(7) the hydrodynamical effects of baryons.
The overall systematic uncertainty of the weak-lensing analysis is quantified by using the hybrid of large dark-matter-only and hydro Magneticum simulations\footnote{\url{http://www.magneticum.org/index.html}} that include the lensing systematics mentioned above.
This results in an empirically calibrated relation between the weak-lensing mass \Mwl\ and the halo mass \mass\ as a function of cluster redshift.
The weak-lensing systematic are then marginalized over in the mass calibration analysis (see Section~\ref{sec:mcalib_likelihood}) through the calibrated weak-lensing mass-to-mass-and-redshift (\Mwl--\mass--\redshift) relation.
 
Note that a similar method of quantifying the weak-lensing systematics was also employed in \cite{schrabback18}, \cite{dietrich19}, and \cite{sommer21}, demonstrating that this is a successful way to empirically calibrate the weak-lensing mass \Mwl\ in the context of a cluster cosmology analysis.

In this work, we further improve the characterization of the uncertainty on the weak-lensing systematics, as the second difference to the analysis in \cite{chiu22}.
Specifically, the uncertainty of the weak-lensing mass bias, $\bwl\equiv\Mwl/\mass$, is reduced to $3\percent$ as opposed to $5.7\percent$ used in \cite{chiu22}.
Unfortunately, this reduction in the systematics floor leads to a negligible difference in our final results, because the constraints are dominated by the cluster abundance instead of the weak-lensing mass calibration in this work.
We provide more details on the updated weak-lensing systematics in Section~\ref{sec:wl_relation}.

In summary, we adopt the same weak-lensing analysis method as in \cite{chiu22}, but there are two differences in the dataset and the systematic estimate.
The first is that we use the public, first-year HSC data, reducing the size of the weak-lensing sample by a factor of $\approx2$.
The second is that we update the uncertainty on the weak-lensing mass systematics from $5.7\percent$ to $3\percent$.  The latter change has a negligible impact on our final results.

%
%

\section{Likelihoods}
\label{sec:likelihoods}

In this section, we describe the likelihood used to constrain cosmology using the \eFEDS\ clusters and HSC weak lensing.
The complete derivation of the likelihood has been given in the appendix of \cite{bocquet15}, which we refer the reader to for more details \citep[see also][]{benson13,liu15a,deHaan16}.
In short, the likelihood compares the number of \eFEDS\ clusters in the multidimensional space of observables with that predicted from our cosmological model, including both the survey selection and the uncertainty in relating the observables to the underlying halo mass.
In this way, constraints on cosmology are obtained.

To carry out cluster cosmology in a survey, we first construct a sample by placing selections on observed quantities, which we call the ``selection observables'' (e.g., X-ray fluxes in an X-ray survey).
In the hope of improving the constraints by adding more information, we typically obtain additional observations of the clusters, resulting in the so-called ``follow-up observables'' (e.g., weak-lensing measurements).
Given a set of selection and follow-up observables, denoted as $\vect{\sobs}$ and $\vect{\fobs}$, respectively, 
the likelihood of observing the number of clusters with $\left\lbrace \vect{\sobs}, \vect{\fobs} \right\rbrace$ at redshift \redshift\ follows a Poisson distribution and can be written as
\begin{multline}
\label{eq:full_like_exact}
\ln L( \vect{p} ) = \sum_{i} 
\ln\left( \frac{\dif N(\vect{\sobs}, \vect{\fobs}, \redshift | \vect{p})}{\dif \vect{\sobs} \dif \vect{\fobs} \dif\redshift }\right) \Bigg|_{i\mathrm{-th~cluster}}
\\
- 
\int\int\int 
\left[ 
\Theta(\vect{\sobs}, \redshift) 
 \frac{\dif N(\vect{\sobs},\vect{\fobs}, \redshift | \vect{p})}{\dif \vect{\sobs} \dif \vect{\fobs} \dif\redshift}  
 \right]
 \dif \vect{\sobs}  \dif \vect{\fobs} \dif\redshift \, ,
\end{multline} 
where $\vect{p}$ is the parameter vector that we want to constrain, 
$i$ runs over the observed clusters, and
$\Theta(\vect{\sobs}, \redshift)$ is the selection function that only depends on $\vect{\sobs}$ and \redshift.
Note that both $\vect{\sobs}$ and $\vect{\fobs}$ can be vectors, in which case we have multiple observables used in the selection and follow-up of the clusters.

By using Bayes' theorem, the differential number in equation~(\ref{eq:full_like_exact}) can be decomposed into two components,
\begin{equation}
\label{eq:like_decomp}
\frac{\dif N(\vect{\sobs}, \vect{\fobs}, \redshift | \vect{p})}{\dif \vect{\sobs} \dif \vect{\fobs} \dif\redshift } = 
P( \vect{\fobs} | \vect{\sobs}, \redshift, \vect{p})
\frac{\dif N(\vect{\sobs}, \redshift | \vect{p})}{\dif \vect{\sobs}  \dif\redshift }
\, ,
\end{equation}
where the first factor $P( \vect{\fobs} | \vect{\sobs}, \redshift, \vect{p})$ describes the probability of observing $\vect{\fobs}$ for a cluster selected by the observable $\vect{\sobs}$ at redshift \redshift, and the second factor $\frac{\dif N(\vect{\sobs}, \redshift | \vect{p})}{\dif \vect{\sobs}  \dif\redshift }$ is the differential number of clusters in the space of $\vect{\sobs}$.

With equation~(\ref{eq:like_decomp}), it is clear that the total number $N_{\mathrm{tot}}$ of clusters expected to be detected in the survey, which is the second term on the right-hand side of equation~(\ref{eq:full_like_exact}), can be rewritten as
\begin{eqnarray}
\label{eq:like_expected_number}
N_{\mathrm{tot}} 
&= 
&\int\int\int 
\Theta(\vect{\sobs}, \redshift) 
 \frac{\dif N(\vect{\sobs},\vect{\fobs}, \redshift | \vect{p})}{\dif \vect{\sobs} \dif \vect{\fobs} \dif\redshift}  
\dif \vect{\sobs}  \dif \vect{\fobs} \dif\redshift 
\, \nonumber \\
&=
&\int \int\int
\Theta(\vect{\sobs}, \redshift) 
\left[
P( \vect{\fobs} | \vect{\sobs}, \redshift, \vect{p}) 
\frac{\dif N(\vect{\sobs}, \redshift | \vect{p})}{\dif \vect{\sobs}  \dif\redshift }
\right]
\dif\vect{\sobs} \dif\vect{\fobs} \dif\redshift
\, \nonumber \\
&=
& \int\int\Theta(\vect{\sobs}, \redshift) 
\left[ 
\int P( \vect{\fobs} | \vect{\sobs}, \redshift, \vect{p})  \dif\vect{\fobs}
\right]
\times
\frac{\dif N(\vect{\sobs}, \redshift | \vect{p})}{\dif \vect{\sobs}  \dif\redshift }
\dif\sobs  \dif\redshift
\, \nonumber \\
&=
&\int\int\Theta(\vect{\sobs}, \redshift) 
\frac{\dif N(\vect{\sobs}, \redshift | \vect{p})}{\dif \vect{\sobs}  \dif\redshift }
\dif\vect{\sobs}   \dif\redshift 
\, .
\end{eqnarray}
The resulting equation~(\ref{eq:like_expected_number}) means that, because no selection is applied to $\vect{\fobs}$, the expected total number of observed clusters in the survey does not depend on the follow-up observable, as expected.

Finally, the full likelihood in equation~(\ref{eq:full_like_exact}) is rewritten as
\begin{eqnarray}
\label{eq:full_like_reduced}
\ln L( \vect{p} ) & = &\left[ 
\sum_{i} \ln\left( 
\frac{\dif N(\vect{\sobs}, \redshift | \vect{p})}{\dif \vect{\sobs}  \dif\redshift }
\right) \Bigg|_{i\mathrm{-th~cluster}}\right.
\nonumber \\
& - & 
\left.\int\int
\Theta(\vect{\sobs}, \redshift) 
 \frac{\dif N(\vect{\sobs},\redshift | \vect{p})}{\dif \vect{\sobs} \dif\redshift}  
 \dif \vect{\sobs} \dif\redshift
\right]
\nonumber \\
&+ &
\sum_{i}  \ln P( \vect{\fobs} | \vect{\sobs}, \redshift, \vect{p}) \Bigg|_{i\mathrm{-th~cluster}}
\, ,
\end{eqnarray} 
where the first two terms in the brackets are the ``number counts'' likelihood for clusters purely selected by $\vect{\sobs}$;
the final term $P( \vect{\fobs} | \vect{\sobs}, \redshift,  \vect{p})$ is the so-called ``mass calibration'' likelihood, which characterizes the probability of obtaining the follow-up observable $\vect{\fobs}$ for the cluster selected by $\vect{\sobs}$ at redshift \redshift.
If the follow-up observations are randomly assigned to the sample selected by $\vect{\sobs}$ (e.g., the coverage of $\vect{\fobs}$ is only subject to the footprint of the survey), the mass calibration likelihood
only need to include the clusters with available $\vect{\fobs}$.

We make the following two remarks.
First, the full likelihood described in equation~(\ref{eq:full_like_reduced}) is exact.
In this way, the mass information can be extracted directly (e.g., from the weak-lensing observable) in the mass calibration likelihood and is incorporated into the number counts likelihood to constrain cosmology.
Second, the mass calibration likelihood does not depend on the survey selection, as long as no extra selection is applied to the follow-up observable $\vect{\fobs}$.
Conversely, if an additional selection is applied to the follow-up observable (e.g., a signal-to-noise cut on the weak-lensing observable), the resulting constraints will be biased \citep[see intensive validations in Appendix~A of][]{umetsu20}.

\subsection{Applications to \eFEDS}
\label{sec:application_to_efeds}

In this study, the sample of \eFEDS\ clusters is first selected through a complex selection on the X-ray observables, namely the detection likelihood \Ldet, the extent likelihood \Lext, and the extent \ext, then followed by a selection on the optical richness \rich\ (see Section~\ref{sec:sample} for more details).

After the construction of the sample, we follow up each cluster in weak lensing and obtain a shear profile \gshear$(\theta)$ if the cluster is covered by the HSC footprint.
That is, we have the follow-up observable, $\fobs = \gshear(\theta)$.

In this work, we choose the observed count rate \rate\ as the X-ray selection observable.
This is because the observed count rate \rate\ is a more physically intuitive quantity than $\left\lbrace \Ldet, \Lext, \ext \right\rbrace$,
it is closely related to the X-ray luminosity of the cluster, and it has previously been demonstrated to be a reliable mass proxy \citep{chiu22}.

A complete modeling of the X-ray selection requires the relations between the X-ray observables, namely $\left\lbrace \Ldet, \Lext, \ext \right\rbrace$, and the underlying true quantities, e.g., the cluster halo mass.
Motivated by \citet[][see their Section~2.3]{grandis20}, 
we empirically model the impact of \Ldet\, \Lext\ and \ext\ on the X-ray count rate \rate\ based selection as introducing incompleteness into a purely \rate\ selected sample.  The resulting completeness function $\Comp\left(\rate, \redshift\right)$ depends on the observed count rate \rate\ and redshift \redshift. This modeling choice is a crucial point that we return to in Section~\ref{sec:completeness}, where we provide an in-depth discussion of the modeling of the completeness $\Comp\left(\rate, \redshift\right)$.

Therefore, our modeling includes the X-ray and optical selection observables $\vect{\sobs} = \left\lbrace \rate, \rich \right\rbrace$.
Incorporating the completeness function $\Comp\left(\rate, \redshift\right)$, the differential number count  in equation~(\ref{eq:full_like_reduced}) then reads
\begin{equation}
\label{eq:completeness_in_diff_n}
\frac{\dif N(\vect{\sobs}, \redshift | \vect{p})}{\dif \vect{\sobs}  \dif\redshift } \rightarrow
\Comp\left( \rate, \redshift \right) \frac{\dif N(\rate, \rich, \redshift | \vect{p})}{\dif \rate \dif \rich \dif\redshift }
\, .
\end{equation}

Substituting equation~(\ref{eq:completeness_in_diff_n}) into equation~(\ref{eq:full_like_reduced}), the full likelihood is written as
\begin{eqnarray}
\label{eq:full_like_final}
\ln L( \vect{p} ) = 
\left[ \sum_{i} \ln\left( 
\Comp\left( \rate, \redshift \right) \frac{\dif N(\rate, \rich, \redshift | \vect{p})}{\dif \rate \dif \rich \dif\redshift }
\right)\Bigg|_{i\mathrm{-th~cluster}}
\right. \nonumber \\
- \left.
\int\limits_{{\redshift}_{\mathrm{min}} }^{ {\redshift}_{\mathrm{max}} }  \dif\redshift 
\int\limits_{\rate = 0}^{\infty}  \dif \rate 
\int\limits_{\rich_{\mathrm{min}(\redshift)}}^{ \infty } \dif \rich ~
\Comp\left( \rate, \redshift \right) 
\frac{\dif N(\rate, \rich, \redshift | \vect{p})}{\dif \rate \dif \rich \dif\redshift }
\right]
\nonumber \\
+
\sum_{j}  \ln P( \gshear(\theta) | \rate, \rich, \redshift, \vect{p}) \Bigg|_{j\mathrm{-th~cluster}}
\, ,
\end{eqnarray} 
where we have employed the selection function $\Theta\left(\vect{\sobs}, \redshift \right)$ as
\begin{multline}
\label{eq:like_selection_function}
\Theta\left(\rate, \rich, \redshift \right) = 
\Theta\left(\rich, \redshift \right) \equiv 
\mathcal{H}(\rich - \rich_{\mathrm{min}}(\redshift)) \times \\
\mathcal{H}(\redshift - {\redshift}_{\mathrm{min}}) \times \left( 1 - \mathcal{H}(\redshift - {\redshift}_{\mathrm{max}}) \right)\, ,
\end{multline}
in which $\mathcal{H}$ is the heaviside step function,
${\redshift}_{\mathrm{min}}$ and ${\redshift}_{\mathrm{max}}$ are fixed to $0.1$ and $1.2$, as the minimum and maximum redshifts used in this work, and
$\rich_{\mathrm{min}}(\redshift)$ is the redshift-dependent richness threshold that enforces the MCMF filtering criterion of $\fcont < 0.3$ \citep{klein22}.
Although the observed count rate \rate\ is used as the X-ray selection observable, we stress that the selection function $\Theta$ does not depend on \rate.
This is because the X-ray selection in \rate\ has been incorporated into the completeness $\Comp(\rate,\redshift)$.  Therefore, in equation~(\ref{eq:full_like_final}) we
integrate over the full range of \rate\ when accounting for the X-ray selection.

Finally, we note that because
the weak-lensing data are available only for a subset of our clusters that lie within the HSC survey footprint,
the index $j$ in equation~(\ref{eq:full_like_final}) runs over only those clusters with available weak-lensing profiles \gshear$(\theta)$.

\subsection{Number counts likelihood}
\label{sec:nbc}

To evaluate the number counts likelihood, we calculate the differential number of clusters in the observable space as 
\begin{equation}
\label{eq:sobs_mass_function}
\frac{\dif N(\rate, \rich, \redshift | \vect{p})}{\dif \rate \dif \rich \dif\redshift } = 
\int\dif\mass~
P\left(\rate, \rich | \mass, \redshift, \vect{p} \right)
\frac{\dif N \left( \mass, \redshift | \vect{p} \right) }{\dif \mass\dif\redshift}
\, ,
\end{equation}
in which $\frac{\dif N \left( \mass, \redshift | \vect{p} \right) }{\dif \mass\dif\redshift}$ is related to the differential comoving volume $V_{\mathrm{c}}\left(\redshift | \vect{p}\right)$ at redshift \redshift\ per solid angle and the survey solid angle $\Omega_{\mathrm{survey}}$ as
\begin{equation}
\label{eq:survey_volume}
\frac{\dif N \left( \mass, \redshift | \vect{p} \right) }{\dif \mass\dif\redshift} = 
\frac{\dif n \left( \mass, \redshift | \vect{p} \right) }{\dif \mass\dif\redshift}
\times V_{\mathrm{c}}\left(\redshift | \vect{p}\right) 
\times \Omega_{\mathrm{survey}}
\, ,
\end{equation}
where $\frac{\dif n \left( \mass, \redshift | \vect{p} \right) }{\dif \mass\dif\redshift}$ is the halo mass function following the dark-matter-only formula in \cite{bocquet16}, 
and $P\left(\rate, \rich | \mass, \redshift, \vect{p} \right)$ is the joint distribution of \rate\ and \rich\ given the cluster halo mass \mass\ and redshift \redshift.
We note that the halo mass function is expressed as 
$
\frac{\dif n }{\dif\mass\dif\redshift} = 
f\left(\sigma,\redshift\right) \frac{\bar{\rho}_{\mathrm{m}}}{\mass} \frac{\dif \ln \sigma^{-1}}{ \dif \mass} \, ,
$
where $\bar{\rho}_{\mathrm{m}}$ is the mean comoving matter density, $\sigma$ is the root mean square of the density field smoothed by a kernel with a comoving scale of $R_{\mathrm{sm}} = \left(\frac{3\mass}{ 4\pi \bar{\rho}_{\mathrm{m}} }\right)^{1/3}$, and the function $f\left(\sigma, \redshift\right)$ is calibrated against the simulation \citep[see equation~(4) in][]{bocquet16}.
The cosmological dependence of the halo mass function is largely encoded in the terms $\sigma$ and $\bar{\rho}_{\mathrm{m}}$, while the shape of $\frac{\dif n }{\dif\mass\dif\redshift}$ is nearly universal across a wide range of redshifts and cosmologies.

The joint distribution $P\left(\rate, \rich | \mass, \redshift, \vect{p} \right)$ includes two components: the uncertainty in observable due to the measurement noise, and the intrinsic scatter of the observables at a fixed cluster mass and redshift.
Specifically, this can be written as
\begin{equation}
\label{eq:joint_distribution_sobs}
P\left(\rate, \rich | \mass, \redshift, \vect{p} \right) = 
\int\dif\hat{\rate}
\int\dif\hat{\rich}~
P\left(\rate | \hat{\rate}, \redshift \right)
P\left(\rich | \hat{\rich} \right)
P\left(\hat{\rate}, \hat{\rich} | \mass, \redshift, \vect{p} \right) 
\, ,
\end{equation}
where $\hat{\rate}$ and $\hat{\rich}$ are the intrinsic observables of the count rate and richness, respectively,
$P\left(\hat{\rate}, \hat{\rich} | \mass, \redshift, \vect{p} \right)$ characterizes the intrinsic scatter of the observables, 
and $P\left(\rate | \hat{\rate}, \redshift \right)$ ($P\left(\rich | \hat{\rich} \right)$) describes the dispersion of observed count rate (richness) with respect to the intrinsic observable $\hat{\rate}$ ($\hat{\rich}$) due to the measurement uncertainty.
Note that we additionally include a redshift dependence in characterizing the measurement uncertainty of the count rate, $P\left(\rate | \hat{\rate}, \redshift \right)$.
This redshift dependence is suggested and empirically calibrated by the data, as we describe in more detail in Appendix~\ref{app:meas_observed_rate}.

We assume a bivariate log-normal distribution for $P\left(\hat{\rate}, \hat{\rich} | \mass, \redshift, \vect{p} \right)$ around the mean value of 
$\left( \left\langle\ln\hat{\rate} | \mass, \redshift \right\rangle, \left\langle\ln\hat{\rich} | \mass, \redshift \right\rangle \right)$ 
at a given cluster mass and redshift, characterized by the intrinsic scatter covariance matrix
\begin{equation}
\label{eq:covar_intrinsic}
\Sigma_{\rate,\rich} = \left( 
\begin{array}{cc}
{\sigmarate}^2                                             & \sigmarate~\sigmarich~\rhoraterich \\
\sigmarate~\sigmarich~\rhoraterich            & {\sigmarich}^2 \\
\end{array}
\right)
\, ,
\end{equation}
where \rhoraterich\ is the correlation coefficient between the count rate and richness.
Note that the variables in equation~(\ref{eq:covar_intrinsic}) denote the scatter of and the correlation between the intrinsic quantities ($\hat{\rate}$ and $\hat{\rich}$), despite that the notations of the observed quantities (\rate\ and \rich) are used in the subscripts for simplicity.
The mean values of the count rate and richness, $\left\langle\ln\hat{\rate} | \mass, \redshift \right\rangle$ and $\left\langle\ln\hat{\rich} | \mass, \redshift \right\rangle$, at a given the mass and redshift are determined by the scaling relations given in Sections~\ref{sec:rate_relation} and \ref{sec:richness_relation}, respectively.

We model the distribution of the observed richness due to the measurement uncertainty as a Poisson distribution in the Gaussian limit.
That is, 
\begin{equation}
\label{eq:measurement_uncertainty_richness}
P\left(\rich | \hat{\rich} \right) = 
\frac{1}{\sqrt{2\pi} \delta_{\rich} }
\exp\left(
-
\frac{ 
\left( \rich - \hat{\rich} \right)^2
}{
{2\delta_{\rich}}^2
} \right) \, ,
\end{equation}
where $\delta_{\rich} = 1.1\times\sqrt{\hat{\rich}}$, in which we increase the measurement uncertainty by a factor of $1.1$.
We obtain this factor by fitting the parameter $a$ in the functional form of $\delta_{\rich} = a\times \sqrt{\rich}$ between the observed richness \rich\ and its observed uncertainty $\delta_{\rich}$ for the \eFEDS\ clusters.

Meanwhile, the distribution of the observed count rate at a given intrinsic rate is modeled as a Gaussian,
\begin{equation}
\label{eq:measurement_uncertainty_rate}
P\left(\rate | \hat{\rate}, \redshift \right) = 
\frac{1}{\sqrt{2\pi} \delta_{\rate}\left( \rate,\redshift \right) }
\exp\left(
-
\frac{ 
\left( \rate - \hat{\rate} \right)^2
}{
2
{\delta_{\rate} \left( \rate,\redshift \right) }^2
} \right) \, ,
\end{equation}
where we empirically characterize the measurement uncertainty $\delta_{\rate}\left( \rate,\redshift \right)$ as a function of the ``observed'' \rate\ and \redshift\ based on the fitting results of the \eFEDS\ clusters \citep[see also][]{klein19,grandis20}.
We refer the reader to Appendix~\ref{app:meas_observed_rate} for more discussions of $\delta_{\rate}\left( \rate,\redshift \right)$.

\subsection{Mass calibration likelihood}
\label{sec:mcalib_likelihood}

We evaluate the mass calibration likelihood in the same manner as in \citet[][see their Section~5.2]{chiu22}, where we refer the reader for a complete description.
In what follows, we provide a brief summary.

For each cluster that is covered by the HSC footprint, we have a shear profile \gshear$(\theta)$, as the follow-up observable, and the  two selection observables, the observed count rate \rate\ and richness \rich.
The goal is to calculate the probability of observing \gshear$(\theta)$ for the cluster at redshift \redshift\ given the selection observables, i.e., the last term in equation~(\ref{eq:full_like_final}).
In this work, as well as in \cite{chiu22}, we approximate this probability as
\begin{eqnarray}
\label{eq:approximated_mass_calib}
P( \gshear(\theta) | \rate, \rich, \redshift, \vect{p}) 
&\equiv 
&\frac{ P( \gshear(\theta), \rate, \rich, \redshift, \vect{p}) }{  P( \rate, \rich, \redshift, \vect{p})  }   \, \nonumber \\
&\approx
&\frac{ \Comp_{\rich}\left(\rate,\redshift\right) P( \gshear(\theta), \rate, \redshift, \vect{p}) }{  \Comp_{\rich}\left(\rate,\redshift\right)   P( \rate, \redshift, \vect{p})  }   \, \nonumber \\
&=
&\frac{ P( \gshear(\theta), \rate, \redshift, \vect{p}) }{  P( \rate, \redshift, \vect{p})  }   \, \nonumber \\
&=
&P( \gshear(\theta) | \rate, \redshift, \vect{p})   \,  ,
\end{eqnarray}
where $\Comp_{\rich}\left(\rate,\redshift\right)$ is the completeness of the cluster at a given \rate\ and redshift \redshift\ in the presence of the richness selection, and is cancelled out in both numerator and denominator.
Physically, it means that the richness selection does not introduce any significant additional selection on the weak-lensing observables, and there the Malmquist bias due to the richness cut is negligible.
That is, $\Comp_{\rich}$ does not have a dependence on \gshear.
It is a reasonable assumption, because the richness selection in this work is conservatively chosen to only remove the contamination (i.e., the false-positive detection) in the purely X-ray selected sample based on the observed richness (see Section~\ref{sec:sample}).
Thus, the overall selection of a contamination-free sample can be described by the X-ray selection, to first order.

It is straightforward to see $\Comp_{\rich}\left(\rate,\redshift\right) = \int_{\mathcal{I}_{\rich}(\redshift)}\dif\rich P(\rich | \rate, \redshift)$, where the interval $\mathcal{I}_{\rich}(\redshift)$ defines the selection of the richness at a given redshift \redshift.
To have the completeness factors cancelled out in equation~(\ref{eq:approximated_mass_calib}), we have implicitly made an assumption of 
$P(\rich | \rate, \gshear(\theta), \redshift) \approx P(\rich | \rate, \redshift)$, such that 
$\Comp_{\rich}\left(\rate,\gshear(\theta),\redshift\right) = \int_{\mathcal{I}_{\rich}(\redshift)}\dif\rich P(\rich | \rate, \gshear(\theta), \redshift) \approx \int_{\mathcal{I}_{\rich}(\redshift)}\dif\rich P(\rich | \rate, \redshift) = \Comp_{\rich}\left(\rate,\redshift\right)$.
Equivalently, this assumes that the intrinsic correlation between the weak-lensing mass \Mwl\ and richness \rich\ is subdominant, i.e., $\rhorichwl \approx 0$.

It is worth mentioning that great efforts have been made in studying the projection effect of optically selected clusters, mostly based on simulations \citep[e.g.][]{song12a,sunayama20,zhang22,wu22} but also with observations \citep{sunayama22,zu22}.
The projection is expected to have a major impact on optically selected clusters, because it simultaneously affects and entangles the weak-lensing 
signals and the richness selection at a fixed cluster mass.
However, this is not the case in this study, where the main selection of \eFEDS\ clusters relies on the X-ray observables instead of the optical richness that acts as an auxiliary observable, which is effectively used only to clean the \eFEDS\ sample.
The assessment of the projection effect based on observed clusters is subject to various systematics \citep[such as the cluster detection algorithm, e.g.,][]{murata20} 
and thus remains quantitatively inconclusive.
Moreover, a recent cosmological analyses \citep[e.g.,][]{costanzi21} found that the posterior of the 
correlated scatter \rhorichwl\ between the weak-lensing mass and optical richness is statistically consistent with zero 
and has no significant impact on the final results.
Therefore, the approximation of $\rhorichwl = 0$ is expected to be a subdominant effect in this work.
We defer the inclusion of the possible correlated scatter \rhorichwl\ to a future work.

Equation~(\ref{eq:approximated_mass_calib}) is then evaluated as
\begin{equation}
\label{eq:mass_calib_evaluate}
P( \gshear(\theta) | \rate, \redshift, \vect{p}) = 
\frac{ 
\int \dif\mass P(\gshear(\theta), \rate | \mass, \redshift, \vect{p}) \frac{\dif N \left( \mass, \redshift | \vect{p} \right) }{\dif \mass\dif\redshift}
}{
\int \dif\mass P(\rate | \mass, \redshift, \vect{p}) \frac{\dif N \left( \mass, \redshift | \vect{p} \right) }{\dif \mass\dif\redshift}
}
\, ,
\end{equation}
where $P(\rate | \mass, \redshift, \vect{p})$ ($P(\gshear(\theta), \rate | \mass, \redshift, \vect{p})$) is the probability of observing \rate\ (\gshear$(\theta)$ and \rate) given the cluster mass and redshift.
Note that the inclusion of the halo mass function in equation~(\ref{eq:mass_calib_evaluate}) is needed to account for the Eddington bias raised from the 
steeply falling distribution of the underlying halo mass.

We calculate the probability $P(\rate | \mass, \redshift, \vect{p})$ as
\begin{equation}
\label{eq:decompose_Prate}
P(\rate | \mass, \redshift, \vect{p}) = \int  \dif\hat{\rate}\, P(\rate | \hat{\rate}, \redshift) P(\hat{\rate} | \mass, \redshift, \vect{p})
 ,
\end{equation}
in which the first factor $P(\rate | \hat{\rate}, \redshift)$ describes the distribution of observed \rate\ at a given intrinsic count rate $\hat{\rate}$ due to the measurement uncertainty, following equation~(\ref{eq:measurement_uncertainty_rate}),
and the second factor $P(\hat{\rate} | \mass, \redshift, \vect{p})$ characterizes the distribution of the intrinsic rate $\hat{\rate}$ at the given mass and redshift.  The intrinsic rate $\hat{\rate}$ at a given mass and redshift is assumed to exhibit an intrinsic scatter, which we model to be log-normal
\begin{equation}
\label{eq:primitive_rate_distribution}
P(\hat{\rate} | \mass, \redshift, \vect{p}) = \frac{1}{\sqrt{2\pi}\sigmarate\hat{\rate}}
\exp\left(
\frac{ \left( \ln\hat{\rate} - \left\langle\ln\hat{\rate}|\mass,\redshift\right\rangle\right)^2 }{
2{\sigmarate}^2
}
\right)
\, ,
\end{equation}
where $\left\langle\ln\hat{\rate}|\mass,\redshift\right\rangle$ is the mean count rate at the given mass and redshift predicted with the parameters $\vect
p$.

The measurement uncertainty of the count rate is accounted for by the factor $P(\rate | \hat{\rate}, \redshift)$.
In the number counts likelihood, the factor $P(\rate | \hat{\rate}, \redshift)$ is calculated in equation~(\ref{eq:measurement_uncertainty_rate}) using the mean count rate uncertainty
given the observed \rate\ at the cluster redshift \redshift, predicted by the empirically modelled relation $\delta_{\rate}(\rate, \redshift)$ (see details in Section~\ref{app:meas_observed_rate}).
While evaluating $P(\rate | \hat{\rate}, \redshift)$ in the mass calibration likelihood in equation~(\ref{eq:decompose_Prate}), we use the exact observed uncertainty $\delta_{\rate}$ of the cluster in the catalog.

Similarly to $P\left(\rate,\rich | \mass, \redshift, \vect{p} \right) $ in equation~(\ref{eq:joint_distribution_sobs}), the probability $P(\gshear(\theta) , \rate | \mass, \redshift, \vect{p})$ is written as
\begin{multline}
\label{eq:decompose_Pratewl}
P(\gshear(\theta) , \rate | \mass, \redshift, \vect{p}) = 
\int \int 
P\left(\rate | \hat{\rate}, \redshift\right) P\left(\gshear(\theta) | \Mwl, \redshift, \vect{p}\right) 
\\
 P\left( \Mwl, \hat{\rate} | \mass, \redshift, \vect{p} \right)
~\dif\hat{\rate}~\dif\Mwl~
\, ,
\end{multline}
where $P(\Mwl, \hat{\rate} | \mass, \redshift, \vect{p})$ is the joint distribution of the weak-lensing mass \Mwl\ and the count rate $\hat{\rate}$ at a given mass \mass\ and redshift \redshift, and 
$P(\gshear(\theta) | \Mwl, \redshift, \vect{p})$ accounts for the measurement uncertainty of the observed shear profile \gshear$(\theta)$ given the weak-lensing mass \Mwl\ and the redshift \redshift.
The halo concentration is fixed to that predicted by the concentration-to-mass relation from \cite{diemer15} at the given \Mwl\ and cluster redshift in each step of the likelihood exploration.
Moreover, the miscentering is taken into account in $P(\Mwl, \hat{\rate} | \mass, \redshift, \vect{p})$ when predicting the shear profile at a given weak-lensing mass \Mwl\ and redshift \redshift.

We assume that $P(\Mwl, \hat{\rate} | \mass, \redshift, \vect{p})$ follows a bivariate log-normal distribution around the mean value  
($\left\langle\ln\Mwl | \mass, \redshift\right\rangle$, 
$\left\langle\ln\hat{\rate} | \mass, \redshift\right\rangle$) at a given mass and redshift 
with the intrinsic scatter covariance matrix,
\[
\Sigma_{\rate,\wl} = \left( 
\begin{array}{cc}
{\sigmarate}^2                                           & \sigmarate~\sigmawl~\rhoratewl \\
\sigmarate~\sigmawl~\rhoratewl              & {\sigmawl}^2 \\
\end{array}
\right)
\, ,
\]
where \rhoratewl\ is the intrinsic correlation coefficient between the weak-lensing mass and count rate.
We note that the mean weak-lensing mass $\left\langle\ln\Mwl | \mass, \redshift\right\rangle$ and scatter \sigmawl\ are presented in Section~\ref{sec:wl_relation} and are carefully calibrated against both dark-matter-only and hydro simulations \citep[see][for detailed methodology]{grandis21}, accounting for various weak-lensing systematics as quantified in \cite{chiu22}.
Readers are referred to \cite{chiu22} for a complete description of the weak-lensing analysis.
The weak-lensing systematics are marginalized over in the statistical inference (see Section~\ref{sec:statistics}).

We stress here again that the mass calibration method is the same as that in \cite{chiu22}, but the analysis has two differences of note.
First, we use the public weak-lensing data (S16A) from the second data release instead of the proprietary data (S19A) as used in \cite{chiu22}.
This results in a $\approx50\percent$ smaller sample size, which corresponds to a reduction of the statistical power in the weak-lensing mass calibration at a level of $\approx30\percent$.  Second, we improve the modeling the weak lensing systematic floor, reducing it from $5.7\percent$ to $3\percent$.
This reduction in the systematic floor has only a negligible impact on our results, because the so-called ''self-calibration'' mass information coming from the cluster counts likelihood \citep[see discussion in][]{majumdar04} is comparable to or greater than the information provided by the HSC weak lensing.

%
%

\section{Analysis}
\label{sec:analysis}

In this section, we present the analysis in detail.
The modeling of the cluster population in terms of the scaling relations is described in Section~\ref{sec:scaling_relation}.
The modeling of the X-ray completeness \Comp, which accounts for the X-ray selection of the \eFEDS\ clusters, is described in Section~\ref{sec:completeness}.
The statistical inference of the analysis and our blinding strategy are contained in Sections~\ref{sec:statistics} and \ref{sec:blinding}, respectively.

\subsection{Modeling of scaling relations}
\label{sec:scaling_relation}

Three scaling relations, which relate the observable and the underlying halo mass at the cluster redshift, are used in this work:
(1) the count rate-to-mass-and-redshift (\rate--\mass--\redshift) relation, 
(2) the richness-to-mass-and-redshift (\rich--\mass--\redshift) relation, and 
(3) the weak-lensing mass-to-mass-and-redshift (\Mwl--\mass-\redshift) relation.
Each of them\footnote{
The observable-to-mass-and-redshift relation here describes the scaling between the mean ``intrinsic'' observable $\left\langle\ln\hat{\mathcal{O}}\right\rangle$ (instead of the mean observed quantity $\left\langle\ln\mathcal{O}\right\rangle$) and the mass \mass\ at each redshift \redshift.
Therefore, it should be denoted as the $\hat{\mathcal{O}}$--\mass--\redshift\ relation, rather than the $\mathcal{O}$--\mass--\redshift\ relation.
For simplicity, however, we use the notations of the \rate--\mass--\redshift\ relation, the \rich--\mass--\redshift\ relation and the \Mwl--\mass--\redshift\ relation to describe the scaling of the intrinsic observables in this work.
}$^,$\footnote{
For the \rate--\mass--redshift\ relation, we not only use the same parameterization as in \cite{chiu22} but also additionally generalize the functional form to probe the mass scaling of low-mass clusters.
This is a novel perspective in this work, since this is the first attempt to constrain the feature of a broken power-law mass scaling based upon both the weak-lensing mass calibration and the cluster abundance.
}
follows exactly the same parameterization used in \cite{chiu22}. 
 In all cases the underlying halo mass is selected to be the mass used in describing the halo mass function adopted for this analysis \citep{bocquet16}, which is \Mfiveoo.

\subsubsection{The \rate--\mass--\redshift\ relation}
\label{sec:rate_relation}

As described in the following two subsections, we adopt two different forms for the mass trend of the rate-to-mass-and-redshift relation.  One is a single 
power law
in mass, which is discussed immediately below, and the other is a broken 
power law
in mass, which is discussed thereafter.

\subsubsection*{The single power-law mass scaling}
\label{sec:singlepowerlaw}

The count rate-to-mass-and-redshift relation is parameterized as
\begin{multline}
\label{eq:countrate_to_mass}
\left\langle\ln\left(\frac{\rate}{\mathrm{counts}/\mathrm{sec}} \Bigg|\mass,\redshift \right)\right\rangle 
= \ln \Arate + \\
\left[ \Brate + \deltarate\ln\left(\frac{1 + \redshift}{1 + \zpiv}\right) \right] \times
\ln\left(\frac{\mass}{\mpiv}\right) +
2 \times \ln\left(\frac{\Ez}{\Ezpiv}\right) + \\
\gammarate \times 
\ln \left(\frac{1 + \redshift}{1 + \zpiv}\right)
-2\times\ln\left(\frac{D_{\mathrm{L}}\left(\redshift\right)}{D_{\mathrm{L}}\left(\zpiv\right)}\right)
+ \ln\left( \brate\left(\mass, \redshift\right) \right) 
\, ,
\end{multline}
with log-normal intrinsic scatter at fixed mass and redshift \redshift\
\begin{equation}
\label{eq:intrinsic_scatter_rate}
\sigmarate \equiv \left(\mathrm{Var}\left[\ln \rate|\mass, \redshift\right]\right)^{\frac{1}{2}} \, ,
\end{equation}
where 
(1) \Arate\ is the characteristic count rate at the pivot mass $\mpiv = 1.4\times10^{14}\Msunh$ and the pivot redshift $\zpiv = 0.35$, 
(2) \Brate\ and \gammarate\ are the power-law indices of the mass and redshift scaling, respectively,
(3) \deltarate\ is the index of the cross-scaling between the mass and redshift,
(4) \Ez\ describes the evolution of the Hubble parameter,
(5) the factor $2$ in front of the \Ez\ ratio captures the expected scaling of the count rate under the assumption of self-similarity \citep{kaiser1986}, which is assumed to be the same as for the soft-band luminosity, 
(6) \gammarate\ characterizes the departure from the self-similar redshift trend,
(7) the second factor $-2$ with the ratio of luminosity distances $D_{\mathrm{L}}$ accounts for the distance dependence, 
and 
(8) the factor \brate\ characterizes the bias of the observed count rate \rate\ with respect to the ``true'' count rate due to the bias in the process of the \eROSITA\ cluster detection \citep[see Section~4.1 in][for more details]{chiu22}.
We note that because \gammarate\ characterizes the departure from the self-similar redshift trend of \rate, a $\gammarate = 0$ means that the count rate perfectly follows the self-similar prediction.

The factor \brate\ is expressed as
\begin{multline}
\label{eq:bias_rate}
\ln\brate \left(\mass,\redshift\right) = 
\Af +
\left[ \Bf + \deltaf\ln\left(\frac{\redshift}{\zpiv}\right) \right] \times \ln\left(\frac{\mass}{\mpiv}\right)  + 
\gammaf\times\ln\left(\frac{\redshift}{\zpiv}\right)\, .
\end{multline}
Equation~(\ref{eq:bias_rate}) is calibrated using the simulations \citep{comparat19,comparat20}, resulting in the constraints fully quantified in \cite{chiu22} as 
\begin{align}
\Af            &=    0.18\pm0.02        \, \nonumber  \\
\Bf             &=  -0.16\pm0.03      \, \nonumber  \\
\deltaf       &=  -0.015\pm0.05    \, \nonumber  \\
\gammaf   &=    0.42\pm0.03         \, .
\end{align}
These constraints are marginalized over in the statistical inference to empirically account for the bias of the observed count rate.

\subsubsection*{The broken power-law mass scaling}
\label{sec:brokenpowerlaw}

In what follows, we introduce one novel aspect of this work regarding the \rate--\mass--\redshift\ relation.
It is known that there exists a deficit of the cluster population with low extent likelihoods ($\Lext\lesssim10$) compared to that predicted by the \eFEDS\ simulations \citep{liuteng21}.
Specifically, the distribution of \Lext\ in the \eFEDS\ catalog shows a peak at $\Lext\approx10$ and decays at $\Lext\lesssim10$, which is distinctly different  from the simulations that show a power-law distribution of \Lext\ without a turnover.
This discrepancy cannot be explained by the systematics due to the \eROSITA\ detection pipeline, because the same end-to-end processing has been employed on both the \eFEDS\ observations and the simulations.
This suggests that either the synthetic clusters that were injected into the simulations are different from those of observed in \eFEDS---perhaps due to the recipes used to generate the X-ray emissions of synthetic clusters being inaccurate.
In such a case, the X-ray selection (or the completeness as defined in Section~\ref{sec:completeness}) quantified from the simulations would not be expected to accurately reflect the true selection of the observed sample.  Such an error in the selection model would generally result in biased posteriors on cosmological parameters.

On possible explanation for the deficit seen at the low-\Lext\ end could come from the different impact of, e.g., feedback processes in the low-mass systems relative to the high-mass clusters.
It is important to note that there exists a scaling between \Lext\ and \Ldet\ \citep[see Figure~16 in][]{klein22}, in which \Ldet\ again strongly scales with the observed count rate \rate;
this suggests that the missing population of low-\Lext\ clusters could be dominated by those with low \rate, which would mostly be low-mass clusters with $\mass\lesssim10^{14}\Msunh$ at $\redshift\lesssim0.4$ given the nature of an X-ray selected sample. 
Such a deficit of clusters at low-\Lext\ could occur if the mass scaling of the \rate--\mass--\redshift\ relation at the low-mass end is steeper than the mass scaling exhibited by more massive clusters.
This is also physically intuitive: halos at the group mass scale have shallower gravitational potential wells, and therefore the ICM is more easily expelled from the potential wells due to energetic feedback within the virial regions or even an early preheating due to AGNs and star formation activities prior to the cluster formation \citep[see, e.g.,][as this relates to the overall mass trend]{mohr97}.
This would result in a smaller observed flux per halo mass for groups than massive clusters.
In the \eFEDS\ simulations \citep{liuteng21}, the low-mass clusters were injected into the X-ray imaging as though they are scaled-down versions of massive clusters \citep[see  Section~3.1.1 in][]{comparat20}, hence the population of synthetic galaxy groups might be neither complete nor accurately capturing the changing impact of ICM physics at the lowest mass scales.
Motivated by this, we introduce a broken power-law feature at the group scale as
\begin{multline}
\label{eq:broken_power_law}
\left\langle\ln\left(\frac{\rate}{\mathrm{counts}/\mathrm{sec}} \Bigg|\mass,\redshift \right)\right\rangle\Bigg|_{\mathrm{broken}} = 
\left\langle\ln\left(\frac{\rate}{\mathrm{counts}/\mathrm{sec}} \Bigg|\mass,\redshift \right)\right\rangle + \\
\left(\left(\Brate - \Bgrp\right) \times \Delta \right) \times \ln\left(1 + \left(\frac{\mass}{\mgrp}\right)^{\frac{-1}{\Delta}}\right)
\, ,
\end{multline}
where the bracket on the right side follows the single power-law \rate--\mass--\redshift\ relation as defined in equation~(\ref{eq:countrate_to_mass}), 
\Bgrp\ is the index of the ``broken'' power law at the mass range of $\mass\lesssim\mgrp\equiv10^{14}\Msunh$,
and the term $\Delta$ is the smoothing factor which is fixed as $\Delta \equiv \ln\left(\sqrt{\mpiv/\mgrp}\right)$.
In such a parameterization, the power-law index of the mass scaling remains intact as \Brate\ at $\mass\gtrsim\mgrp$ and becomes \Bgrp\ if $\mass\lesssim\mgrp$.
The case of $\Bgrp > \Brate$ implies a scenario where the X-ray luminosity of groups is less than what would have been expected given the behavior of the more massive clusters. 
Such a trend of a steepening mass scaling could explain the over-predicted number of the simulated clusters at the low-\Lext\ end.

It is worth mentioning that few investigations have been made to study the broken power-law feature in terms of the X-ray luminosity-to-mass relation \citep{lovisari15,barnes17,schellenberger17,pop22}.
With the unique and relatively large \eFEDS\ sample, which probes over two orders of magnitude in cluster mass, the joint analysis of the weak-lensing mass calibration and the cluster abundance provides a good opportunity to observationally examine the assumption of a single power-law relation in mass.
This novel characterization of the \rate--\mass--\redshift\ relation is incorporated into our blinded analysis, where we ask how sensitive the blinded posteriors are to the choice of single versus broken power law (see Section~\ref{sec:blinding} for more details).

\subsubsection{The \rich--\mass--\redshift\ relation}
\label{sec:richness_relation}

The richness-to-mass-and-redshift relation is characterized as
\begin{multline}
\label{eq:richness_to_mass}
\left\langle\ln\rich|\mass,\redshift\right\rangle 
= \ln \Arich + \\
\left[ \Brich + \deltarich\ln\left(\frac{1 + \redshift}{1 + \zpiv}\right) \right] \times
\ln\left(\frac{\mass}{\mpiv}\right) +
\gammarich \times 
\ln \left(\frac{1 + \redshift}{1 + \zpiv}\right)
\, ,
\end{multline}
with log-normal intrinsic scatter at fixed mass and redshift
\begin{equation}
\label{eq:intrinsic_scatter_richness}
\sigma_{\rich} \equiv \left(\mathrm{Var}\left[\ln\rich|\mass,\redshift\right] \right)^{\frac{1}{2}} \, ,
\end{equation}
where 
\Arich\ is the normalization, 
\Brich\ and \gammarich\ describe the power-law indices of the mass and redshift scaling, respectively,
and \deltarich\ characterizes the potential cross-scaling between the mass and redshift.
The scattering in the observed richness due to the Poisson noise is accounted for by the measurement uncertainty $\delta_{\rich}$ in equation~(\ref{eq:measurement_uncertainty_richness}).
Note that the parameter \gammarich\ fully describes the redshift trend of the richness-to-mass-and-redshift relation in equation~(\ref{eq:richness_to_mass}), which reflects our baseline assumption that the self-similar redshift behavior would be a lack of a redshift trend in the richness.

\subsubsection{The \Mwl--\mass--\redshift\ relation}
\label{sec:wl_relation}

We parameterize the weak-lensing mass-to-mass-and-redshift relation in terms of the weak-lensing mass bias $\bwl\equiv \Mwl/\mass$, as follows.
\begin{multline}
\label{eq:bwl}
\left\langle\ln\left(\bwl | \mass, \redshift\right)\right\rangle =
\ln \Awl + \\
\Bwl\times \ln\left(\frac{\mass}{2\times10^{14}\Msunh}\right) +
\gammawl \times \ln\left(\frac{1 + \redshift}{1 + 0.6}\right)
\, ,
\end{multline}
with log-normal intrinsic scatter \sigmawl\ at fixed mass and redshift,
\begin{equation}
\label{eq:bwl_sigma}
\sigmawl \equiv \left(\mathrm{Var}\left[\ln\left(\bwl|\mass,\redshift\right)\right] \right)^{\frac{1}{2}}
\, .
\end{equation}
Following the methodology in \cite{grandis21}, the parameters, $\left\lbrace\Awl, \Bwl, \gammawl, \deltawl\right\rbrace$, are calibrated against large simulations that include the weak-lensing systematics observed in the HSC survey \citep[see Section~\ref{sec:wldata} and also][]{chiu22}.

In \cite{grandis21}, the simulation calibration was done at four snapshots of redshifts, $\redshift = 0.25, 0.47, 0.78, 1.18$, resulting in four sets of the normalization \Awl, the mass scaling \Bwl, and the scatter \sigmawl.
Moreover, the mass trend was constrained to be the same at the four snapshots.
Then, in \cite{chiu22} the overall scaling relation parameters $\left\lbrace\Awl, \Bwl, \gammawl, \sigmawl\right\rbrace$ were obtained by fitting the functional form of equation~(\ref{eq:bwl}) to the constraints at the four snapshots including the correlation among them.
This resulted in the uncertainties of \Awl, \gammawl, and \sigmawl\ at a level of $0.057$, $0.080$, and $0.032$, respectively.

In this work, we directly fit the functional form to the data points randomly sampled from the chains of the weak-lensing mass bias and scatter.
Specifically, we fit the functional form (with \Bwl\ fixed to the constraint from \citealt{grandis21}) to 
a set of four data points of the weak-lensing bias measurements, one taken from each of the four redshift snapshots, and derive a set of the best-fit parameters of $\left(\Awl, \gammawl\right)$.
This fitting is repeated for $5000$ sets of the data points randomly sampled from the collection of weak-lensing mass bias results measured from simulations, resulting in a collection of best-fit parameters \Awl\ and \gammawl.
Finally, the best-fit parameters and the uncertainty of \Awl\ and \gammawl\ are obtained from the collection of best-fit parameters.
We perform the same fitting procedure to obtain the scatter \sigmawl.
Note that the redshift dependence of the scatter \sigmawl\ is constrained to be consistent with zero and hence is ignored in this work.
This fitting strategy leads to the improved uncertainties at a level of $0.03$, $0.062$ and $0.037$ for the parameters of \Awl, \gammawl, and \sigmawl, respectively, across the whole redshift range.
In this work, we stress that the best-fit values of $\left\lbrace\Awl, \Bwl, \gammawl, \sigmawl\right\rbrace$ are unchanged with respect to \citet[][see their equation~(45)]{chiu22} and that only the uncertainties are updated.

The resulting constraints on the parameters of the \Mwl--\mass--\redshift\ relation are
\begin{align}
\label{eq:wlbias_constraints}
\Awl               &= 0.903   \pm 0.030    \, , \nonumber \\
\Bwl                &= -0.057 \pm 0.022   \, , \nonumber \\
\gammawl      &= -0.474 \pm 0.062   \, , \nonumber \\
\sigmawl        &= 0.238   \pm 0.037  \, ,
\end{align}
which are marginalized in the likelihood analysis to account for the weak-lensing systematics.
It is worth mentioning that the updated systematic uncertainty in the absolute mass scale without the core ($>0.5\Mpch$) is at a level of $3\percent$, which is smaller than that including the cluster core ($<0.5\Mpch$), which would be $6\percent$.

\subsection{Modeling of the X-ray selection}
\label{sec:completeness}

\subsubsection{Theoretical framework}
\label{sec:theory_xray_selection}

We denote $\vect{\Xlabel}$ as the X-ray observables that are actually used in selecting the \eFEDS\ clusters, $\vect{\Xlabel} \equiv \left\lbrace \Ldet, \Lext, \ext \right\rbrace$.
Then, the total number of clusters that can be detected in the survey is
\begin{equation}
\label{eq:total_number_of_all_xobs}
N_{\mathrm{tot}} = 
\int\dif\redshift
\int\limits_{\mathcal{I}_{\vect{\Xlabel}}}\dif\vect{\Xlabel}
\frac{ \dif N }{ \dif \vect{\Xlabel}\dif\redshift }
\, ,
\end{equation}
where the interval $\mathcal{I}_{\vect{\Xlabel}}$ defines the observable space of the X-ray selection, i.e., $\mathcal{I}_{\vect{\Xlabel}} = \left\lbrace\left(\Ldet, \Lext, \ext\right) | \Ldet > 5, \Lext > 6, \ext > 0\right\rbrace$, and the differential number of clusters reads
\begin{equation}
\label{eq:xobs_function}
\frac{
\dif N
}{
\dif \vect{\Xlabel} \dif\redshift
}
= 
\int \dif\mass 
\left[
P\left(\vect{\Xlabel} | \mass, \redshift \right)
\frac{\dif N \left(\mass,\redshift\right)}{\dif \mass\dif\redshift}
\right]
\, .
\end{equation}
As seen in equation~(\ref{eq:xobs_function}), one needs to calculate $P\left(\vect{\Xlabel} | \mass, \redshift \right)$ that requires 
the modeling of the $\vect{\Xlabel}$--\mass--\redshift\ relation, which describes the relation between the X-ray observables and the underlying halo mass at each redshift.
A challenge is that none of these selection observables, which have been applied to define the \eFEDS\ sample, have a well established connection to the cluster halo mass and redshift \citep[see, e.g.,][]{pacaud06}.  Of course one can attempt to use image simulations to characterize these relations, but then the accuracy of the resulting selection function will only be good in the limit that the simulated and real cluster properties are approximately identical.  While this is a worthy goal over the long term, here we proceed using an empirical calibration of the X-ray selection that uses the observed sample characteristics to inform our selection model.  We describe this below.

Following previous X-ray, SZE and optical cluster forecasts and analyses \citep[e.g.,][]{haiman01,vikhlinin09b,vanderlinde10,rykoff14,bocquet19,klein19,grandis20,chiu22}, we adopt a cluster observable for our \eFEDS\ analysis---the X-ray count rate \rate---that has a well defined relation with cluster mass and redshift.  This observable is the \eROSITA\ photon based flux of a cluster and is therefore reflecting the cluster luminosity distance, X-ray spectral energy distribution (SED) and X-ray luminosity.  The X-ray SED variation is driven by ICM temperature, which along with X-ray luminosity closely tracks mass and redshift.  In particular, the X-ray luminosity has been extensively studied and is known to provide a  cluster mass proxy that is comparable in scatter properties to SZE signal-to-noise ratios and optical richness.   Finally, the \eFEDS\ rate-to-mass-to-redshift relation has been previously studied \citep{chiu22} and shown to be a high-quality mass proxy.

Given this choice, we can use Bayes' theorem, we then have
\begin{eqnarray}
\label{eq:selection_integral}
\hspace{-0.8cm}
\int\limits_{\mathcal{I}_{\vect{\Xlabel}}}\dif\vect{\Xlabel}
P(\vect{\Xlabel} | \mass, \redshift)
&= 
&\int\limits_{\mathcal{I}_{\vect{\Xlabel}}}\dif\vect{\Xlabel}
\left[
\int\limits\dif\rate
P(\vect{\Xlabel} | \rate,  \mass, \redshift) P(\rate |  \mass, \redshift ) 
\right] \, \nonumber \\
&=
&\int\limits\dif\rate
\left[
\int\limits_{\mathcal{I}_{\vect{\Xlabel}}}\dif\vect{\Xlabel}
P(\vect{\Xlabel} | \rate,   \mass, \redshift) 
\right]  
P(\rate | \mass, \redshift )  \, . 
\end{eqnarray}
Thus, the modeling of the X-ray selection is then equivalent to understanding the impact of the $\vect{\Xlabel}$-selection on a count rate selected cluster sample.
In this work, we make an assumption that the distribution of $\vect{\Xlabel}$ can be empirically described the observed count rate \rate\ at the cluster redshift \redshift\ without residual dependence on the cluster mass \mass, i.e.,
\begin{equation}
\label{eq:empirical_calib_xselection}
P(\vect{\Xlabel} | \rate,   \mass, \redshift) \approx P(\vect{\Xlabel} | \rate, \redshift)
\, .
\end{equation}
Physically, equation~(\ref{eq:empirical_calib_xselection}) is a statement that the cluster halo mass can be represented by the count rate \rate\ at a particular redshift.  This in effect is just a restatement of equation~(\ref{eq:countrate_to_mass}). 
Indeed, the two most important factors affecting whether a cluster is selected in X-rays are
(1) the total flux, which determines \Ldet\ and \rate\ and has a direct relation with the X-ray luminosity and luminosity distance  $D_{\mathrm{L}}$ and hence the mass and redshift, and 
(2) the angular scale of the cluster, or the extendedness, which directly determines \Lext\ and \ext\ and has a dependence on the cluster mass as well as the angular diameter distance $D_{\mathrm{A}}$ and hence the redshift \redshift.

We therefore model the completeness of \rate\ in the presence of the $\vect{\Xlabel}$-selection, 
\begin{equation}
\label{eq:empirical_calib_xselection_on_rate}
\Comp\left(\rate, \redshift\right)  \equiv
\int\limits_{\mathcal{I}_{\vect{\Xlabel}}}\dif\vect{\Xlabel}
P(\vect{\Xlabel} | \rate, \redshift) 
\, .
\end{equation}
Substituting equations~(\ref{eq:xobs_function})--(\ref{eq:empirical_calib_xselection_on_rate}) into equation~(\ref{eq:total_number_of_all_xobs}), the total number of the clusters is evaluated as
\begin{eqnarray}
\label{eq:total_number_of_rate}
N_{\mathrm{tot}}
&= 
&
\int\limits\dif\redshift
\int\limits\dif\mass
\int\limits\dif\rate
\Comp\left(\rate, \redshift\right)
P\left(\rate | \mass, \redshift\right)
\frac{ \dif N\left(\mass,\redshift\right) }{ \dif\mass\dif\redshift } \, \nonumber \\
&=
&
\int\limits\dif\redshift
\int\limits\dif\rate\,
\Comp\left(\rate, \redshift\right)
\left[
\int\limits\dif\mass
P\left(\rate | \mass, \redshift\right)
\frac{ \dif N\left(\mass,\redshift\right) }{ \dif\mass\dif\redshift } 
\right] \, \nonumber \\
&=
&
\int\limits\dif\redshift
\int\limits\dif\rate\,
\Comp\left(\rate, \redshift\right)
\frac{ \dif N }{ \dif\rate\dif\redshift }
\, ,
\end{eqnarray}
which results in the modeling in equation~(\ref{eq:completeness_in_diff_n}).
It is seen in equation~(\ref{eq:total_number_of_rate}) that the modeling of the X-ray selection has become a task to characterize the completeness $\Comp\left(\rate, \redshift\right)$.

Motivated by \cite{clerc18}, where they modeled the completeness as a function of the flux, the angular scale of core radii, and the exposure time,
we model the completeness function $\Comp\left(\rate, \redshift\right)$ as
\begin{equation}
\label{eq:completeness_function}
\Comp\left(\rate, \redshift\right) = 
\frac{1}{2} \left( 1 + 
\erf\left( 
\frac{\ln\rate - \ln\ratefivez}{\srate}
\right) \right)
\, ,
\end{equation}
where $\erf$ is the error function with the scaling factor \srate, and \ratefivez\ is the count rate which has $50\percent$ completeness at the redshift \redshift, characterized as
\begin{equation}
\label{eq:rate50_scaling}
\ratefivez = \ratefiveo \times \left(\frac{D_{\mathrm{A}}(\redshift)}{D_{\mathrm{A}}(\zpiv) }\right)^{\gammaz} 
\, ,
\end{equation}
in which \ratefiveo\ is the count rate with $50\percent$ completeness at the pivot redshift $\zpiv = 0.35$, 
and \gammaz\ is the scaling index of the angular diameter distance dependence.
We ignore the dependence on the exposure time in this work, given that the imaging depth is uniform in the \eFEDS\ survey \citep[see Section~\ref{sec:sample} and also][]{liu21}.
The inclusion of an exposure-time dependence will likely be helpful in the modeling of the \eROSITA\ All-Sky Survey sample at some future date.

Note that \cite{bulbul21} recently derived the completeness of the \eFEDS\ clusters in terms of the count rate
by analyzing the population of mis-identified clusters in the point-source catalog \citep{salvato21} and the simulations \citep{liuteng21}, showing a similar functional form as in equation~(\ref{eq:completeness_function}). 
However, we stress that the completeness presented in \cite{bulbul21} cannot be directly utilized in a cosmological analysis.
This is because the observed count rate of an \eFEDS\ cluster mis-identified as a point source is estimated in the point-source mode (specifically with $\ext = 0$) of the \eROSITA\ pipeline, which is significantly different from that used for an extended source, which is based upon a $\beta$-modeling fitting.
That is, the observed count rate of a mis-identified cluster is estimated within an aperture largely following the point spread function of \eROSITA\ instead of the cluster X-ray surface brightness profile.
As a result, the observed count rate of the population of the mis-identified clusters is expected to have a systematic offset with respect to that would have been measured for an extended source. 
Hence, the completeness derived based on a mix of these two kinds of count rates could merely serve as a qualitative gauge of the X-ray selection but not as a sufficiently accurate modeling to support a cosmological analysis.
In the next section, we present the direct measurement of the X-ray completeness. 

\begin{figure}
\resizebox{0.5\textwidth}{!}{
\includegraphics[scale=1]{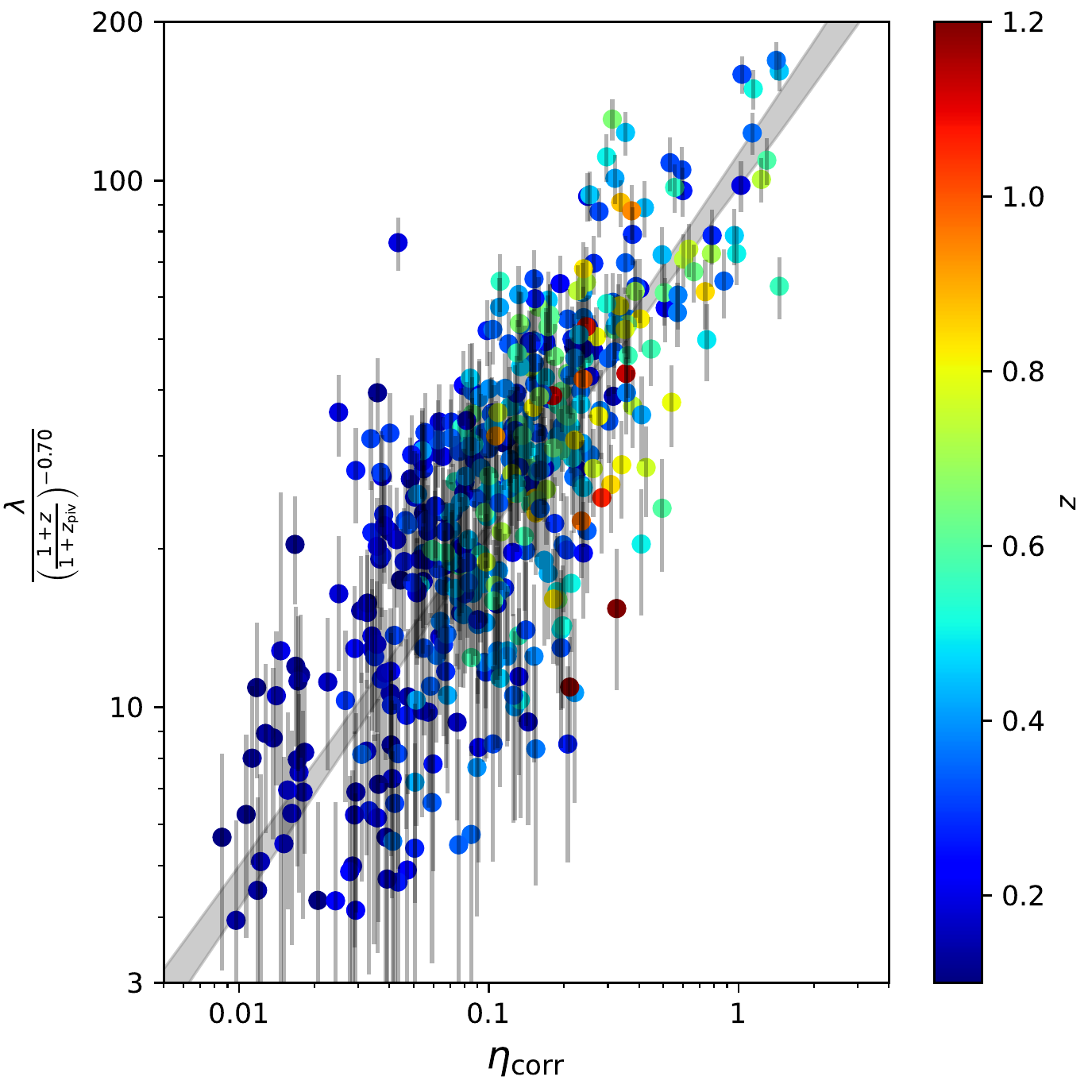}
}
\caption{
The richness-to-rate-and-redshift (\rich--\rate--\redshift) relation.
The cluster sample is color-coded by their redshift, while the $68\percent$ confidence level of the best-fit model, i.e., equation~(\ref{eq:richness_to_rate}),  is indicated by the grey region.
Note that the observed count rate \rate\ is normalized to ${\rate}_{\mathrm{corr}}$ at the pivotal redshift $\zpiv\equiv0.35$ following equation~(\ref{eq:corrected_count_rate}).
}
\label{fig:rich_rate_relation}
\end{figure}

\subsubsection{Measurements of the completeness}
\label{sec:completeness_measurements}

\paragraph{Measurements in the \eFEDS\ data}

Our goal is to quantify the completeness of the \eFEDS\ sample in terms of the observed count rate \rate.
We do this by leveraging the fact that the \eFEDS\ clusters missing in the extended-source catalog are detected in the point-source catalog. 
In other words, the joint sample of the clusters in the extended-source and point-source catalogs is complete down to a threshold in the count rate. 
Therefore, we can derive the completeness function as the ratio of the number of the \eFEDS\ clusters in the extended-source catalog to that of the total number of clusters in the joint catalog.

In Section~\ref{sec:theory_xray_selection}, we have stressed that the estimated count rate of the clusters mis-identified in the point-source catalog cannot be directly compared with those in the extended-source catalog, due to the fact that different schemes are used to estimate the count rate for these two populations.
This prevents us from directly using the count rate estimated in the point-source catalog.
However, the optical richness \rich\ of the clusters in both the extended-source and point-source catalogs is consistently estimated with the MCMF algorithm \citep{klein22}.
This uniform estimation of the richness enables us to quantify the X-ray completeness by jointly studying the richness distribution of the cluster populations in both the extended-source and point-source catalogs.

The procedure to measure the completeness is described, as follows.
\begin{enumerate}
\item We first characterize the relation between the observed richness \rich\ and count rate \rate\ at the cluster redshift \redshift\ based on the sample of \eFEDS\ clusters that are correctly identified in the extended-source catalog.
This results in the richness-to-rate-and-redshift (\rich--\rate--\redshift) relation.
\item We derive the richness distributions of MCMF counterparts around all sources in the point-source catalog in seven redshift bins.
We then repeat the same analysis along random lines of sight to obtain the random-LOS richness distribution, representing the sample contamination, which is also employed for the optical confirmation of the \eFEDS\ candidates  \citep[see Section~\ref{sec:sample} and also][]{klein22}.
\item With a proper normalization, we subtract the  random-LOS richness distribution 
from that of the MCMF counterparts around the \eFEDS\ point sources.
The residual richness distribution, denoted as $N_{\mathrm{res}}(\rich, \redshift)$, represents the population of \eFEDS\ clusters that are mis-identified in the point-source catalog.
\item We convert the residual richness distribution $N_{\mathrm{res}}(\rich, \redshift)$ to the corresponding distribution of the count rate, denoted as $N_{\mathrm{ps}}^\mathrm{\prime}(\rate, \redshift)$, 
using the \rich--\rate--\redshift\ relation described in step (i) above.
In this step, we make the assumption that the clusters in both the extended-source and point-source catalogs represent the same population in terms of their optical properties.
This assumption is validated by the \eFEDS\ data, because the richness distributions of these two populations are consistent with each other \citep{klein22,bulbul21}.
\item The resulting distribution $N_{\mathrm{ps}}^\mathrm{\prime}(\rate, \redshift)$ cannot be directly compared to the count rate distribution estimated from the clusters in the extended-source catalog, because the distribution $N_{\mathrm{ps}}^\mathrm{\prime}(\rate, \redshift)$ is expressed in the count rate that is obtained by directly inverting the observed richness through the \rich--\rate--\redshift\ relation.
That is, the distribution $N_{\mathrm{ps}}^\mathrm{\prime}(\rate, \redshift)$ is impacted by the scatter in the richness at a fixed count rate.
We remove this scatter by deconvolving $N_{\mathrm{ps}}^\mathrm{\prime}(\rate, \redshift)$ by a kernel that is calibrated against the cluster sample in the extended-source catalog (see more details below).
This results in the ``richness-inferred'' count rate distribution $N_{\mathrm{ps}}(\rate, \redshift)$ of the clusters that are mis-identified in the point-source catalog.
We stress that the 
distribution 
$N_{\mathrm{ps}}(\rate, \redshift)$
is expressed in terms of the observed count rate \rate\ that could have been estimated in the extended-source mode of the \eROSITA\ pipeline, because the \rich--\rate--\redshift\ relation is derived based on the sample in the extended-source catalog.

\item We account for the over sampling of the \eFEDS\ field by X-ray sources. 
Due to the high density of X-ray sources in \eFEDS\ and the typical size of clusters (which have a radius of $\Rfiveoo\approx3$ arcmin), it is possible that the same cluster or group contributes multiple times to the residual richness distribution and hence in $N_{\mathrm{ps}}(\rate, \redshift)$. 
In \cite{salvato21}, they accounted for this effect by only selecting those X-ray sources as clusters that have the closest positional match to the optical center of the MCMF counterparts. 
This reduces the sample size by $\approx25\percent$.
We therefore re-scale $N_{\mathrm{ps}}(\rate, \redshift)$ by $0.75$, i.e., $N_{\mathrm{ps}}(\rate, \redshift) \rightarrow 0.75 \times N_{\mathrm{ps}}(\rate, \redshift)$.
Without including the factor, the best-fit parameters of the resulting completeness change by less than $1\sigma$.
Because the posteriors are adopted as the priors on the parameters of the completeness that will be self-calibrated in the analysis, this oversampling correction does not have significant impact on our final results.
Moreover, we stress that the resulting completeness, which characterizes the X-ray selection, only affects the modeling of the cluster abundance, and does not enter into the weak-lensing mass calibration.

\item By combining with the count rate distribution $N_{\mathrm{clu}}(\rate, \redshift)$ of the \eFEDS\ clusters that are correctly identified in the extended-source catalog, we derive the completeness as 
\begin{equation}
\label{eq:completeness_measurement_final}
\Comp(\rate,\redshift) = \frac{ N_{\mathrm{clu}}(\rate, \redshift) }{ N_{\mathrm{clu}}(\rate, \redshift) + N_{\mathrm{ps}}(\rate, \redshift)} \, .
\end{equation}
\end{enumerate}

\paragraph*{}
In what follows, we provide details of the modeling of the richness-to-rate-and-redshift relation and the deconvolution procedure used in the step (v).

\paragraph*{}
In concordance with the richness-to-mass-and-redshift (\rich--\mass--\redshift) and the count rate-to-mass-and-redshift (\rate--\mass-\redshift) relations, the \rich--\rate--\redshift\ relation is obtained by modeling $P(\rich|\rate,\redshift)$ as a log-normal distribution with intrinsic scatter $\sigma_{\rich|\rate,\redshift}$ around the mean richness characterized as a power-law function of \rate\ and $(1 + \redshift)$.
Specifically, we maximize the likelihood
\begin{equation}
\label{eq:P_richness_rate}
\mathrm{L}(\vect{p}_{\rich|\rate}) = 
\prod\limits_{i} \frac{ 
P(\rich_{i} | \rate_{i}, \redshift_{i}, \vect{p}_{\rich|\rate})
}{ 
P(\rich > \rich_{\mathrm{min}}(\redshift) | \rate_{i}, \redshift_{i}, \vect{p}_{\rich|\rate})
}
\, ,
\end{equation}
where $i$ runs over the \eFEDS\ clusters,
$\vect{p}_{\rich|\rate}$ records the parameters of the \rich--\rate--\redshift\ relation, and
$\rich_{\mathrm{min}}(\redshift)$ is the richness threshold due to the cut of $\fcont < 0.3$ (see Section~\ref{sec:sample}).
The inclusion of the denominator in equation~(\ref{eq:P_richness_rate}) is needed to account for the Malmquist bias caused by the redshift-dependent richness cut.
The resulting \rich--\rate--\redshift\ relation is then obtained as
\begin{multline}
\label{eq:richness_to_rate}
\left\langle \ln \left( \frac{\rich}{  \mathrm{counts}/\mathrm{sec} } \Bigg| \rate_{\mathrm{corr}}, \redshift \right) \right\rangle = 
\ln \left(21.81\pm0.66\right) + \\
 \left(0.68 \pm 0.03\right) \times \ln\left(\frac{\rate_{\mathrm{corr}}}{0.1 ~ \mathrm{counts}/\mathrm{sec} }\right) + \\
 \left(-0.70 \pm 0.27\right) \times \ln\left(\frac{1 + \redshift}{1 + \zpiv}\right) 
 \, ,
\end{multline}
with intrinsic scatter of $\sigma_{\rich|\rate,\redshift} = 0.45\pm0.02$ and the normalized count rate $\rate_{\mathrm{corr}}$ at $\zpiv = 0.35$ defined as
\begin{multline}
\label{eq:corrected_count_rate}
\ln\left( \frac{ \rate_{\mathrm{corr}} }{ \mathrm{counts}/\mathrm{sec} }  \right)
\equiv \ln \left( \frac{ \rate }{ \mathrm{counts}/\mathrm{sec} } \right)
- 2 \ln\left(\frac{\Ez}{\Ezpiv}\right) \\
+ 2 \ln\left(\frac{D_{\mathrm{L}}(\redshift)}{D_{\mathrm{L}}(\zpiv)}\right) 
- \gammaf \ln\left(\frac{\redshift}{\zpiv}\right)
\, ,
\end{multline}
where we fix $\gammaf = 0.42$ (see Section~\ref{sec:rate_relation}).
This result is visualized in Figure~\ref{fig:rich_rate_relation}, where we can see that the resulting \rich--\rate--\redshift\ relation provides a good description of the data.
It is worth mentioning that the parameterization of equation~(\ref{eq:richness_to_rate}) 
assumes the \rate--\mass--\redshift\ relation 
is described by a single power-law scaling.
Indeed, as seen in Figure~\ref{fig:rich_rate_relation}, no significant broken power-law feature is revealed in the scaling between \rich\ and \rate.

\begin{figure}
\resizebox{0.45\textwidth}{!}{
\includegraphics[scale=1]{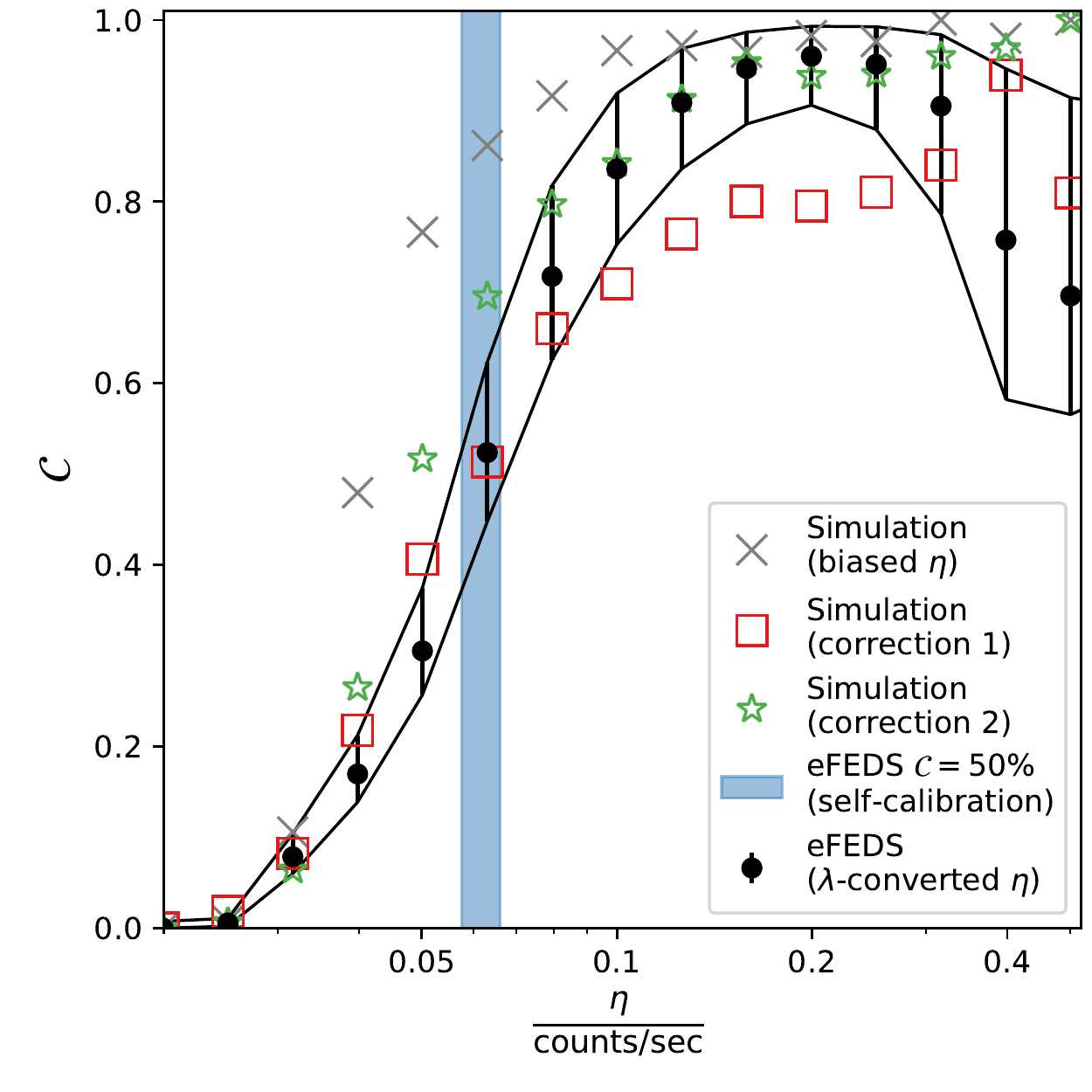}
}
\caption{
The measurements of the X-ray completeness \Comp.
The completeness directly estimated by the count rate of the \eFEDS\ clusters from both the extended and point-source catalogs
without a count-rate correction for the mis-identified population in the point-source catalog is marked by the grey crosses.
The measurements 
with a count-rate correction for 
the clusters identified in the point-source catalogs are shown by the squares and stars, using the first and second correction methods (see more details in the text), respectively. 
The blue vertical band represents the count rate at $\rate=\ansratefiveo\mathrm{counts}/\mathrm{sec}$ where the $50\percent$ completeness occurs, constrained by the joint modeling of the weak-lensing mass calibration and the cluster abundance.
The richness-inferred measurements of the completeness are shown by the black circles with the $68\percent$ confidence levels marked by the open intervals.
The information of the richness-inferred \Comp\ is extracted as the priors applied to the parameters of the completeness function in this work.
}
\label{fig:completeness_data}
\end{figure}

With the \rich--\rate--\redshift\ relation, we perform the deconvolution described in step (v).
Specifically, the distribution $N_{\mathrm{ps}}^\mathrm{\prime}(\rate, \redshift)$ is deconvolved using a 
log-normal kernel 
that remains to be derived.
We derive this log-normal kernel by
using the sample of clusters detected in the extended-source catalog, under the assumption that the populations of the clusters detected in the extended-source and point-source catalogs follow a consistent \rich--\rate--\redshift\ relation with the count rate \rate\ estimated in the extended mode of the \eROSITA\ pipeline.
To do so, we follow the same procedure described in the step (iv) to derive the count rate distribution, denoted as $N_{\mathrm{clu}}^{\prime}(\rate, \redshift)$, of the extended sample using the count rate directly converted through the \rich--\rate--\redshift\ relation.
With a fitting procedure, we derive 
the kernel by requiring that the convolution of the kernel and the distribution $N_{\mathrm{clu}}^{\prime}(\rate, \redshift)$ results in the count rate distribution $N_{\mathrm{clu}}(\rate, \redshift)$, which is derived using the count rate estimated in the ``extended-source" mode. 
After obtaining 
the kernel, we perform the deconvolution of the distribution $N_{\mathrm{ps}}^{\prime}(\rate, \redshift)$ to obtain the distribution $N_{\mathrm{ps}}(\rate, \redshift)$ for the cluster sample detected in the point-source catalog.
Note that the scatter parameter of the resulting log-normal kernel is slightly larger than the value of $\sigma_{\rich|\rate,\redshift} / 0.68 $, where $\sigma_{\rich|\rate,\redshift}$ is the measured intrinsic scatter of the richness at a fixed count rate and redshift, and the factor $0.68$ is the best-fit count rate scaling of the richness in equation~(\ref{eq:richness_to_rate}).
This is expected, because the dispersion in \rich\ at a given \rate\ contains both contributions from the measurement uncertainty and the intrinsic scatter.
By calibrating using the cluster sample in the extended-source catalog, this deconvolution process ensures that (1) we take into account not only the measurement uncertainty but also the intrinsic scatter of \rich\ at a given count rate, and that (2) the resulting $N_{\mathrm{ps}}(\rate, \redshift)$ of the clusters mis-identified in the point-source catalog is unbiased with respect to that which could have been estimated using the count rate estimate from the extended-source mode of the \eROSITA\ pipeline.

\paragraph*{}
We find no compelling evidence for a redshift dependence in the completeness measurements \Comp\ inferred from the richness distribution.
We therefore derive the richness-inferred completeness without any redshift binning, as the final estimate.
That is, the completeness is effectively derived at the pivotal redshift of the sample, i.e., $\Comp\left(\rate,\zpiv\right)$, where $\zpiv = 0.35$.
In Figure~\ref{fig:completeness_data}, we show the resulting  $\Comp\left(\rate,\zpiv\right)$ as the black points with the $68\percent$ confidence level indicated by the open interval.
The blue vertical band indicates the count rate where the completeness is $50\percent$, constrained by the joint modeling of the weak-lensing mass calibration and the cluster abundance (see Section~\ref{sec:results_completeness}).
As seen, the results show that the richness-inferred completeness is $\approx50\percent$ at $\rate\approx0.06$~$\mathrm{counts}/\mathrm{sec}$, in excellent agreement with that from the joint modeling, and decreases to zero at $\rate\lesssim0.02$~$\mathrm{counts}/\mathrm{sec}$.
We note that there exists a mildly decreasing completeness at  the high-\rate\ end ($\rate\gtrsim0.3$~$\mathrm{counts}/\mathrm{sec}$).
In this work, we are not able to further explore the cause of this trend at such a high count rate, because more precise measurements of \Comp\ are limited by the large Poisson noise.
A significantly larger sample from the first-year \eROSITA\ survey will shed further light on this.

After deriving the completeness, we fit the functional form of equation~(\ref{eq:completeness_function}) to the measured completeness (equation~(\ref{eq:completeness_measurement_final})) and obtain
\begin{equation}
\label{eq:completeness_prior}
\begin{array}{ccc}
\frac{ \ratefiveo }{ \mathrm{counts}/\mathrm{sec} } &= &0.0624 \pm 0.0057  \, ,  \\
\srate       &= &0.6514 \pm 0.1687 \, .
\end{array}
\end{equation}
Note that we fix $\gammaz$ to be zero in deriving equation~(\ref{eq:completeness_prior}), given that we do not see a significant difference 
in the measured completeness in different redshift bins.
That is, the best-fit results in equation~(\ref{eq:completeness_prior}) represent the overall completeness of the \eFEDS\ clusters at 
all redshifts.
The parameter measurements in equation~(\ref{eq:completeness_prior}) are used as priors on the parameters of $\left(\ratefiveo,\srate\right)$ in the number counts likelihood (see Section~\ref{sec:likelihoods}) to empirically account for the X-ray selection.
Although equation~(\ref{eq:completeness_prior}) is obtained with $\gammaz = 0$, we stress that the parameter \gammaz\ is left free and self-calibrated in the number counts likelihood.
In Section~\ref{sec:results}, indeed, we observe that the self-calibrated constraint on \gammaz\ is consistent with zero.

We make one remark regarding the redshift scaling of the completeness function.
The redshift-independent completeness function obtained in this work suggests that the quantity \ext\ contains very little information about the X-ray core size of the cluster.
This is not surprising, given that (1) \eROSITA\ is not capable of resolving the ICM structure of the majority of clusters due to the relatively large PSF (FWHM$\approx28\arcsec$), and (2) \ext\ is designed to act as a nuisance parameter to maximize the extent likelihood \Lext\ under the detection configuration, rather than actually characterizing the angular size of the ICM core.
This conclusion is supported by the fact that no clear scaling between \ext\ and quantities other than \Lext\ is seen in the \eFEDS\ sample.

\paragraph{Measurements in the \eFEDS\ simulations}

Finally, we turn to derive the completeness from the \eFEDS\ simulations \citep{liuteng21}.
It is important to stress that we independently derive the completeness from the \eFEDS\ simulations only for the purpose of a comparison with that inferred using the richness distribution; we do not use the simulation-inferred completeness in the analysis.

By directly using the count rate estimates \rate\
provided in 
the output catalogs\footnote{\url{https://erosita.mpe.mpg.de/edr/eROSITAObservations/Catalogues/liuT/eFEDS_catalog_V3.3.html}} of the \eFEDS\ simulations, we derive 
$N_{\mathrm{clu}}(\rate,\redshift)$ and $N_{\mathrm{ps}}(\rate,\redshift)$ for the clusters 
in the extended-source and point-source catalogs, respectively.
Then, the resulting completeness is derived using equation~(\ref{eq:completeness_measurement_final}).
However, the completeness estimated in this way \citep[also as in][]{bulbul21} is biased, as mentioned before.
This is because we are mixing the count rates in both the ``extended-source'' and ``point-source'' modes 
to derive the completeness \Comp.
The resulting completeness \Comp\ is marked as the grey crosses in Figure~\ref{fig:completeness_data}, showing a significant overestimation of the X-ray completeness at a given count rate as opposed to our richness-inferred measurements (black points).
This is expected, since the count rates of the mis-identified clusters provided in the output catalog are estimated in a point-source aperture and hence systematically underestimated than what would have been measured as extended sources.

With a further attempt, we account for the bias in the ``point-source-mode'' count rate of the clusters 
mis-identified in the point-source catalog by applying a correction factor.
We use two methods to empirically derive the correction.
In the first method, referred to as ``correction 1'', we empirically derive the relation between the input X-ray flux $f_{\mathrm{X}}$ and the measured ``extended-source-mode'' count rate \rate\ based on the sample of the clusters in the extended-source catalog---this gives the $f_{\mathrm{X}}$--\rate\ relation.
This relation is straightforward to obtain, because both the quantities  $f_{\mathrm{X}}$ and \rate\ are provided in the output catalogs of the \eFEDS\ simulations.
Then, for a simulated cluster mis-identified as a point source, we directly convert the input X-ray flux $f_{\mathrm{X}}$ of the cluster to the count rate using the $f_{\mathrm{X}}$--\rate\ relation (ignoring the scatter).
This resulting $f_{\mathrm{X}}$-inferred count rate then represents the count rate that would have been measured in the extended mode of the \eROSITA\ pipeline, under the assumption that the relation between $f_{\mathrm{X}}$ and \rate\ is the same for the clusters in the extended-source and point-source catalogs.
With this $f_{\mathrm{X}}$-inferred count rate for the mis-identified clusters in the point-source catalog, the resulting completeness is derived and shown as the red squares in Figure~\ref{fig:completeness_data}.

In the second method, referred to  as``correction 2'', we empirically fit an angular size-dependent relation for the correction $C_{\rate}(\rate, \theta_{500})$, where the quantity $\theta_{500}$ is the angular size of the clusters and directly provided in the output catalogs of the \eFEDS\ simulations.
The dependence on $\theta_{500}$ is included to account for the bias due to the inappropriate choice of the aperture size used by the \eROSITA\ pipeline  for the clusters in the point-source catalog.
The relation $C_{\rate}\left(\rate, \theta_{500}\right)$ is parameterized as $C_{\rate}\left(\rate, \theta_{500}\right) \propto a {\theta_{500}}^b$, where $a$ and $b$ are to be fitted.
With a $\chi^2$ fitting, the best-fit relation $C_{\rate}\left(\rate, \theta_{500}\right)$ is obtained by requiring that the distribution of the mis-identified clusters in the space of $\left(f_{\mathrm{X}}, \rate\right)$ follows the $f_{\mathrm{X}}$--\rate\ relation of the clusters in the extended-source catalog.
In this way, we obtain the corrected count rate of the mis-identified clusters by multiplying the correction factor $C_{\rate}\left(\rate, \theta_{500}\right)$ to the ``point-source-mode'' count rate provided in the catalog.
The resulting completeness function is shown by the green stars in Figure~\ref{fig:completeness_data}.

If the simulations are a perfect representation of the \eFEDS\ observations, we would expect that the X-ray completeness with the correction methods applied could reproduce the richness-based measurements.
As seen, both correction methods (red squares and green stars) give the completeness functions that are not in good agreement with our measurements (black points).
The methods of ``correction 1'' and `correction 2'', respectively, underestimate and overestimate the completeness at $\rate\gtrsim0.1~\mathrm{counts}/\mathrm{sec}$ and $\rate\lesssim0.07~\mathrm{counts}/\mathrm{sec}$.
This suggests that the simulations do not perfectly reproduce the \eFEDS\ observations.
Meanwhile, both correction methods increase the $50\percent$ completeness \ratefiveo\ with respect to that without any corrections, showing that
the ``raw'' simulation-based completeness estimate (grey crosses) is significantly overestimated at a given count rate.
Despite the efforts in making the simulations as identical to the \eFEDS\ observations as possible \citep{liuteng21}, the ``raw'' simulation-based X-ray completeness \citep[as also in][]{bulbul21} is significantly different to the richness-based measurements.
Even with the two correction methods, the corrected completeness estimates from the simulations still do not perfectly agree with the measurements.
Our results pose a concern for a study of \eFEDS\ clusters that heavily relies on the completeness function derived from the simulations.

Again, we stress that we present these two correction methods merely for the purpose of the comparison; we do not use them in any analyses presented in this work.

\subsection{Modeling of cosmology}
\label{sec:modeling_cosmology}

In this work, we focus on constraining two models of cosmology, namely the flat \LCDM\ and \wCDM\ models.
Each model is characterized by the mean matter density \OmegaM, the mean baryon density \Omegab, 
the degree of fluctuations in the density field at the present epoch \sigmaeight, the spectral index of the primordial power spectrum \ns, the current Hubble parameter 
\Hnow\ in an unit of $\mathrm{km}/\mathrm{sec}/\Mpc$, and the dark energy equation of state parameter \w.
The curvature is fixed to zero ($\Omegak = 0$) for both models, and \w\ is fixed to $-1$ for the \LCDM\ model.
The temperature of the Cosmic Microwave Background today is fixed to $2.7255$~Kelvin \citep{fixsen09}, 
while the effective number of neutrino species is set to be $3.046$ \citep{PlanckCollaboration20}.  
We make no attempt to constrain the sum of neutrino masses, 
which are not included in evaluating the predictions from the cosmological models.
We do not expect significant impact raising from neglecting massive neutrinos given the current sample size; this statement is also supported by the analysis in the SPT-SZ survey, where the inclusions of massive neutrinos did not alter the modeling of the cluster abundance in any significant ways \citep{bocquet19}.

The overall parameter vector $\vect{p}$ is defined as 
\[
\vect{p} \equiv  \vect{p}_{\mathrm{c}}  \cup \vect{p}_{\rate} \cup \vect{p}_{\rich} \cup \vect{p}_{\wl} \cup \vect{p}_{\mathrm{mis}} \cup \vect{p}_{\rho} \cup \vect{p}_{\Comp}  \, ,
\] 
where
\begin{itemize}
\item $\vect{p}_{\mathrm{c}} \equiv \left\lbrace\OmegaM, \Omegab, \sigmaeight, \ns, \Hnow, \w \right\rbrace$ for the cosmological parameters,
\item $\vect{p}_{\rate} \equiv \left\lbrace\Arate, \Brate, \deltarate, \gammarate, \sigmarate, \Af, \Bf, \deltaf, \gammaf, \Bgrp\right\rbrace$ for the \rate--\mass--\redshift\ relation in equation~(\ref{eq:countrate_to_mass}), 
\item $\vect{p}_{\rich} \equiv \left\lbrace\Arich, \Brich, \deltarich, \gammarich, \sigmarich\right\rbrace$ for the \rich--\mass--\redshift\ relation in equation~(\ref{eq:richness_to_mass}),
\item $\vect{p}_{\wl} \equiv  \left\lbrace\Awl, \Bwl, \gammawl, \sigmawl\right\rbrace$ for the \Mwl--\mass--\redshift\ relation in equation~(\ref{eq:bwl}),
\item $\vect{p}_{\mathrm{mis}} \equiv \left\lbrace\fmis, \sigmamis\right\rbrace$ for the distribution of cluster miscentering \citep[see Section~3.2 in][]{chiu22},
\item $\vect{p}_{\rho} \equiv \left\lbrace \rhoraterich, \rhoratewl\right\rbrace$ for the intrinsic correlated scatter between the richness and count rate (\rhoraterich) and between the weak-lensing mass and count rate (\rhoratewl).
\item $\vect{p}_{\Comp} \equiv \left\lbrace\ratefiveo, \srate, \gammaz\right\rbrace$ for the X-ray completeness in equation~(\ref{eq:completeness_function}),
\end{itemize}

Before the joint modeling, we first perform the modeling of the cluster abundance and the weak-lensing mass calibration separately to ensure the consistency between them (especially in the \rate--\mass--\redshift\ relation).
Finally, after assuring the consistency, the joint constraint on cosmology is obtained in the combined likelihood of the cluster abundance and the weak-lensing mass calibration.
In the next section we describe the modeling and blinding strategies.

\subsubsection*{Statistical inference}
\label{sec:statistics}

Using Bayes' theorem, the posterior $\mathcal{P}\left(\vect{p}| \mathcal{D}\right)$ of the parameter vector $\vect{p}$ given the data vector $\mathcal{D}$ can be written in the form
\begin{equation}
\label{eq:bayesian}
P\left(\vect{p}| \mathcal{D}\right) \propto \mathcal{L}(\mathcal{D}| \vect{p}) \times \mathcal{P(\vect{p})} \, ,
\end{equation}
where $\mathcal{L}(\mathcal{D}| \vect{p})$ is the likelihood of observing $\mathcal{D}$ given $\vect{p}$,
and $\mathcal{P(\vect{p})}$ is the prior on $\vect{p}$.
We explore the parameter space using the Affine Invariant Markov Chain Monte Carlo sampler, \texttt{emcee} \citep{foreman13,foreman19}.

\begin{table}
\centering
\caption{
A summary of the priors used in the modeling.
The first (second) column represents the name (prior) of the parameters.
}
\label{tab:priors}
\resizebox{!}{0.48\textheight}{
\begin{tabular}{cc}
\hline\hline
Parameter &Prior \\
\hline
\multicolumn{2}{c}{Cosmology} \\
\hline
\OmegaM      &$\mathcal{U}(0.1,0.5)$          \\
\Omegab      &$\mathcal{U}(0.042,0.049)$   \\
\Omegak      &Fixed to $0$          \\
\sigmaeight  &$\mathcal{U}(0.40,1.20)$       \\
\ns          &$\mathcal{U}(0.92,1.0)$         \\
$\frac{\Hnow}{\mathrm{km}/\mathrm{sec}/\Mpc}$          &$\mathcal{U}(50,90)$             \\
\w           &Fixed to $-1$ or $\mathcal{U}(-2.5,-1/3)$        \\
\hline
\multicolumn{2}{c}{Completeness} \\
\multicolumn{2}{c}{Equation~(\ref{eq:completeness_function})
} \\
\hline
$\frac{\ratefiveo}{\mathrm{counts}/\mathrm{sec}}$            &$\mathcal{U}(0,0.15)$ or $\mathcal{N}(0.062,0.0057^2)$ \\
\srate\      &$\mathcal{U}(0.01,1)$  or $\mathcal{N}(0.651,0.168^2)$\\
\gammaz\     &$\mathcal{U}(-3,3)$     \\
\hline
\multicolumn{2}{c}{The \rate--\mass--\redshift\ relation} \\
\multicolumn{2}{c}{Equation~(\ref{eq:countrate_to_mass}) } \\
\hline
\Arate       &$\mathcal{U}(0,0.5)$              \\
\Brate       &$\mathcal{U}(0,5)$                 \\
\deltarate   &$\mathcal{U}(-3,3)$               \\
\gammarate   &$\mathcal{U}(-3,3)$               \\
\sigmarate   &$\mathcal{N}(0.3, 0.08^2) \times \mathcal{U}(0.05, 0.8)$ \\
\Af          &$\mathcal{N}(0.18,0.02^2)$    \\
\Bf          &$\mathcal{N}(-0.16,0.03^2)$   \\
\deltaf      &$\mathcal{N}(-0.015,0.05^2)$ \\
\gammaf      &$\mathcal{N}(0.42,0.03^2)$     \\
\Bgrp        &$\mathcal{U}(0,5)$ or fixed to \Brate  \\
\hline
\multicolumn{2}{c}{The \rich--\mass--\redshift\ relation} \\
\multicolumn{2}{c}{Equation~(\ref{eq:richness_to_mass}) } \\
\hline
\Arich        &$\mathcal{U}(1,100)$       \\
\Brich\       &$\mathcal{U}(0,5)$          \\
\deltarich\   &$\mathcal{U}(-3,3)$        \\
\gammarich\   &$\mathcal{U}(-3,3)$        \\
\sigmarich\   &$\mathcal{U}(0.05, 0.8)$ \\
\hline
\multicolumn{2}{c}{The \Mwl--\mass--\redshift\ relation} \\
\multicolumn{2}{c}{Equation~(\ref{eq:bwl})} \\
\hline
\Awl          &$\mathcal{N}(0.903,0.030^2)$    \\
\Bwl          &$\mathcal{N}(-0.057,0.022^2)$  \\
\gammawl      &$\mathcal{N}(-0.474,0.062^2)$   \\
\sigmawl      &$\mathcal{N}(0.238,0.037^2)$   \\
\hline
\multicolumn{2}{c}{Miscentering} \\
\multicolumn{2}{c}{Section~\ref{sec:wldata}} \\
\hline
\fmis         &$\mathcal{N}(0.54,0.02^2)$  \\
\sigmamis     &$\mathcal{N}(0.61,0.03^2)$  \\
\hline
\multicolumn{2}{c}{Correlated scatter} \\
\hline
\rhoraterich  & $\mathcal{U}(-0.9,0.9)$ \\
\rhoratewl    & $\mathcal{U}(-0.9,0.9)$ \\
\hline
\end{tabular}
}
\end{table}

We adopt the following priors on the parameters and tabulate them in Table~\ref{tab:priors}.
In what follows, a brief summary is provided.
The uniform priors of $\mathcal{U}(0.1,0.5)$, $\mathcal{U}(0.042,0.049)$, $\mathcal{U}(0.40,1.20)$, $\mathcal{U}(0.92,1.0)$, and $\mathcal{U}(50,90)\times\mathrm{km}/\mathrm{sec}/\Mpc$ are applied to \OmegaM, \Omegab, \sigmaeight, \ns, and \Hnow, respectively.
For the \wCDM\ model, a uniform prior of $\mathcal{U}(-2.5,-1/3)$ is applied to \w.
We are not able to put meaningful constraints on \Omegab, \ns, and \Hnow, such that we are effectively marginalizing these parameters over the range of the priors in this work.

For the parameters of the scaling relations, we adopt the same priors used in \cite{chiu22}.
For the \rate--\mass--\redshift\ relation, a simulation-informed Gaussian prior $\mathcal{N}(0.3, 0.08^2)$ allowing to be in a range between $0.05$ and $0.8$ are used for the intrinsic scatter \sigmarate.
The Gaussian priors, which are calibrated against the \eFEDS\ simulations, are applied to the parameters of $\left\lbrace\Af, \Bf, \deltaf, \gammaf\right\rbrace$ to account for the bias in the observed count rate \citep[see Section~4.1 in][and also Section~\ref{sec:rate_relation}]{chiu22}.
For the weak-lensing analysis, the simulation-calibrated constraints on $\vect{p}_{\wl}$ are used as the priors to marginalize over the weak-lensing systematics \citep[see Section~4.2 in][and also Section~\ref{sec:wl_relation}]{chiu22}.
The miscentering parameters $\vect{p}_{\mathrm{mis}}$ are also marginalized over the priors to account for the cluster miscentering.
Flat priors are applied to other parameters of the mass and redshift scaling, the intrinsic scatter of richness \sigmarich, and 
the intrinsic correlations \rhoraterich\ and \rhoratewl.

We compare two different approaches to model the completeness.
First, the informative priors $\mathcal{N}(0.062,0.0057^2)\times\mathrm{counts}/\mathrm{sec}$ and $\mathcal{N}(0.651,0.168^2)$, which are the constraints independently measured in Section~\ref{sec:completeness_measurements}, are used for \ratefiveo\ and \srate, respectively.
Second, the uniform priors $\mathcal{U}(0,0.15)\times\mathrm{counts}/\mathrm{sec}$  and $\mathcal{U}(0.01,1)$ are applied to \ratefiveo\ and \srate, respectively.
That is, these parameters are self-calibrated in the modeling of the cluster abundance.
This is part of the blinding analysis in this work, which we return to 
in the next section.

For modeling the weak-lensing mass calibration alone, we adopt cosmological priors that constrain the distance-redshift relation.  With this distance-redshift relation information, we can probe how the weak lensing observables constrain the \rate--\mass--redshift\ relation.  These priors are identical to those adopted in \cite{chiu22}\footnote{
The priors of
$\mathcal{N}\left(0.30, 0.016^2\right)$,
$\mathcal{N}\left(0.045, 0.0002^2\right)$,
$\mathcal{N}\left(0.80, 0.014^2\right)$,
$\mathcal{N}\left(0.965, 0.004^2\right)$, and
$\mathcal{N}\left(70, 5.6^2\right)\times\mathrm{km}/\mathrm{s}/\Mpch$
are applied to
\OmegaM, \Omegab, \sigmaeight, \ns, and \Hnow, respectively, with
$\w$ fixed to $-1$.
}.
With these priors, the cosmology is effectively fixed to the \LCDM\ model with 
$\left(\OmegaM, \Omegab, \sigmaeight, \ns, \Hnow\right) = \left(0.30, 0.045, 0.80, 0.965, 70~\mathrm{km}/\mathrm{s}/\Mpch\right)$.
We stress that these Gaussian priors on the cosmological parameters are removed when including the modeling of the cluster abundance.

For the modeling of the cluster abundance alone, we apply the Gaussian priors of 
$\mathcal{N}\left(0.13,    0.03^2\right)$,
$\mathcal{N}\left(1.65,    0.20^2\right)$,
$\mathcal{N}\left(0.00,    1.50^2\right)$, and
$\mathcal{N}\left(0.00,    1.50^2\right)$
to the parameters of 
\Arate, \Brate, \deltarate, and \gammarate, respectively.
These priors are conservatively chosen to mimic the constraining power from the weak-lensing mass calibration to break the degeneracy between $\vect{p}_{\rate}$ and $\vect{p}_{\mathrm{c}}$.

Under this framework, it is important to stress that the modeling of the abundance alone includes the external information of the \rate--\mass--\redshift\ relation from the adopted priors on $\vect{p}_{\rate}$, especially the normalization \Arate.
The priors on $\left(\Arate, \Brate, \deltarate, \gammarate\right)$ are informed by the weak-lensing analysis.
Therefore, a comparison between the modeling of the weak-lensing mass calibration and the cluster abundance is not expected to provide an independent consistency test of the \rate--\mass--\redshift\ relation.
However, by adopting the informative priors on $\vect{p}_{\rate}$,
this strategy allows us to examine whether the empirical modeling is adequately accurate to simultaneously model both the scaling relations and cosmological parameters, and to assess the systematics on the final constraints on the cosmological parameters $\vect{p}_{\mathrm{c}}$.

To provide an independent consistency test on the \rate--\mass--\redshift\ relation, in a post-unblinding analysis we perform the modeling of the cluster abundance alone while (1) removing the Gaussian priors on \Arate, \Brate, \deltarate, and \gammarate\ and (2) applying the informative priors to the cosmological parameters $\vect{p}_{\mathrm{c}}$ as identical as in the modeling of the weak-lensing mass calibration alone.
This modeling gives an independent and fair comparison of the \rate--\mass--\redshift\ relation with the modeling of the weak-lensing mass calibration, since the scatter \sigmarate\ and the cosmological parameters are effectively fixed to the same values in the both modeling.
We discuss this comparison in Section~\ref{sec:results_abundance}.

\subsection{Blinding}
\label{sec:blinding}

In this work, we constrain cosmology in a blind analysis to not only avoid confirmation bias but also to assess the adequacy of the modeling strategy we have chosen.
Because the weak-lensing mass calibration had been carried out blindly in \cite{chiu22} and the method identical to that in the present analysis, 
the same blinding scheme on the mass calibration shall not be used in this work.
Taking this opportunity, we employ a blinding strategy to validate the two most important and novel aspects of our analysis, which are (1) the empirical calibration of the completeness due to the X-ray selection and (2) the broken-power law feature in the \rate--\mass--\redshift\ relation.
We describe the blinding strategy, as follows.

The blinding takes place at the parameter level.
Specifically, the actual mean values and the uncertainties of the cosmological parameters $\vect{p}_{\mathrm{c}}$, except for the modeling of the weak-lensing mass calibration alone, are entirely hidden from the analyst by a random scaling.  Therefore, there is no information about the mean values 
and uncertainties 
of any parameters during the blinded analysis.

Our requirement for unblinding is that the following criteria are met:
\begin{itemize}
\item In a validation test, the same analysis code can recover (i.e., within the $68\percent$ confidence levels) the input values that are used to generate the mock sample of clusters in a sky footprint of $1400$~deg$^2$ (corresponding to a size at least ten times larger than \eFEDS). This validates the accuracy of the code (limits systematics to $\approx1/3$ the scale of the statistical uncertainties reflected in the final posteriors).
\item The best-fit model provides a good description of the data in the observable space of the count rate \rate, richness \rich, and the redshift \redshift; that is, the goodness-of-fit test shows no sign of significant tension between the data and the best-fit model.
See more details of the goodness-of-fit tests in Section~\ref{sec:results_gof}.
\item The constraints on cosmology in the modeling of the cluster abundance alone without the Gaussian priors (inferred from equation~(\ref{eq:completeness_prior})) on $\vect{p}_{\Comp}$ are consistent at a level of $1\sigma$ when derived from the full sample and the other sub-samples.
These subsamples include (1) the low-\redshift\ sample with $0.1<\redshift < 0.35$, (2) the high-\redshift\ sample with $0.35<\redshift<1.2$, and (3) the high-\Lext\ sample with $\Lext > 15$.
Except the comparison between the low-\redshift\ and high-\redshift\ sub-samples, we note that other comparisons contain non-trivial correlations in the sub-samples that we do not account for.
However, the consistency among them indicates that the empirical modeling of the completeness (see Section~\ref{sec:theory_xray_selection}) is sufficiently accurate for different samples of \eFEDS\ clusters and does not have a significant impact on the resulting posteriors of cosmological parameters.
\item With the weak-lensing informed priors on the parameters $\vect{p}_{\rate}$ of the \rate--\mass--\redshift\ relation, the resulting constraints in the modeling of the cluster abundance alone are consistent with and without the Gaussian priors 
applied to the completeness parameters $\vect{p}_{\Comp}$.
This implies that our independent measurement of the completeness is consistent with those obtained from the self-calibration of the completeness model using the cluster abundance.
\item The resulting constraints on both the \rate--\mass--\redshift\ relation and cosmological parameters are consistent with and without the broken power-law feature of the observed count rate (see Section~\ref{sec:brokenpowerlaw}).
This means that the single power-law relation of previously assumed relation between \rate\ and \mass\ provides a sufficiently accurate description for both galaxy clusters and groups, and does not significantly bias the cosmological constraints.
\end{itemize}

We confirm that these criteria are met before the unblinding.
No post-unblinding changes are made in this work, except for an additional test.
In a post-unblinding stage, we additionally performed a test to examine the impact from the weak-lensing informed priors applied to the parameters $\left(\Arate, \Brate, \deltarate, \gammarate\right)$ of the \rate--\mass--\redshift\ relation in the modeling of the cluster abundance.
See more details in the last paragraph of Section~\ref{sec:statistics}.

\begin{figure}
\centering
\resizebox{0.52\textwidth}{!}{
\includegraphics[scale=1]{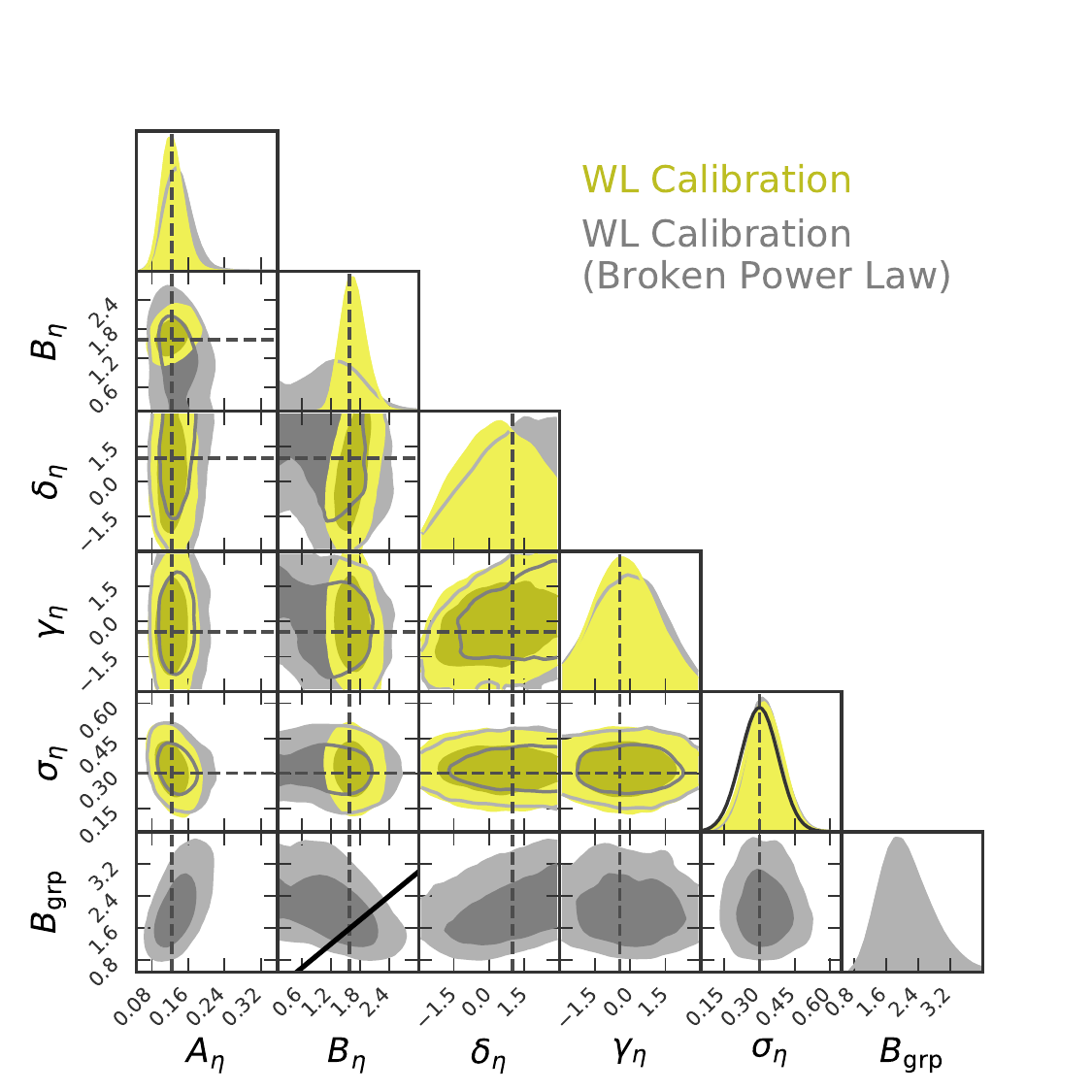}
}
\vspace{-0.5cm}
\caption{
The fully marginalized and joint posteriors 
of the parameters in the modeling of the weak-lensing mass calibration alone.
The constraints on the \rate--\mass--\redshift\ relation characterized by the single and broken power-law functions are in yellow and grey, respectively.
The solid line in the parameter space spanned by \Brate\ and \Bgrp\ indicates the mass scaling of the count rate with a single power-law form ($\Brate = \Bgrp$).
The dash lines indicate the parameters of the best-fit \rate--\mass--\redshift\ relation obtained in the previous work \citep{chiu22} using the third-year weak-lensing data from the HSC survey.
The Gaussian prior applied to the intrinsic scatter \sigmarate\ is shown with the solid black line in the on-diagonal plot.
The contours indicate the $68\percent$ and $95\percent$ confidence levels.
}
\label{fig:mcalib_gtc_smf}
\end{figure}
\begin{figure*}
\resizebox{0.9\textwidth}{!}{
\includegraphics[scale=1]{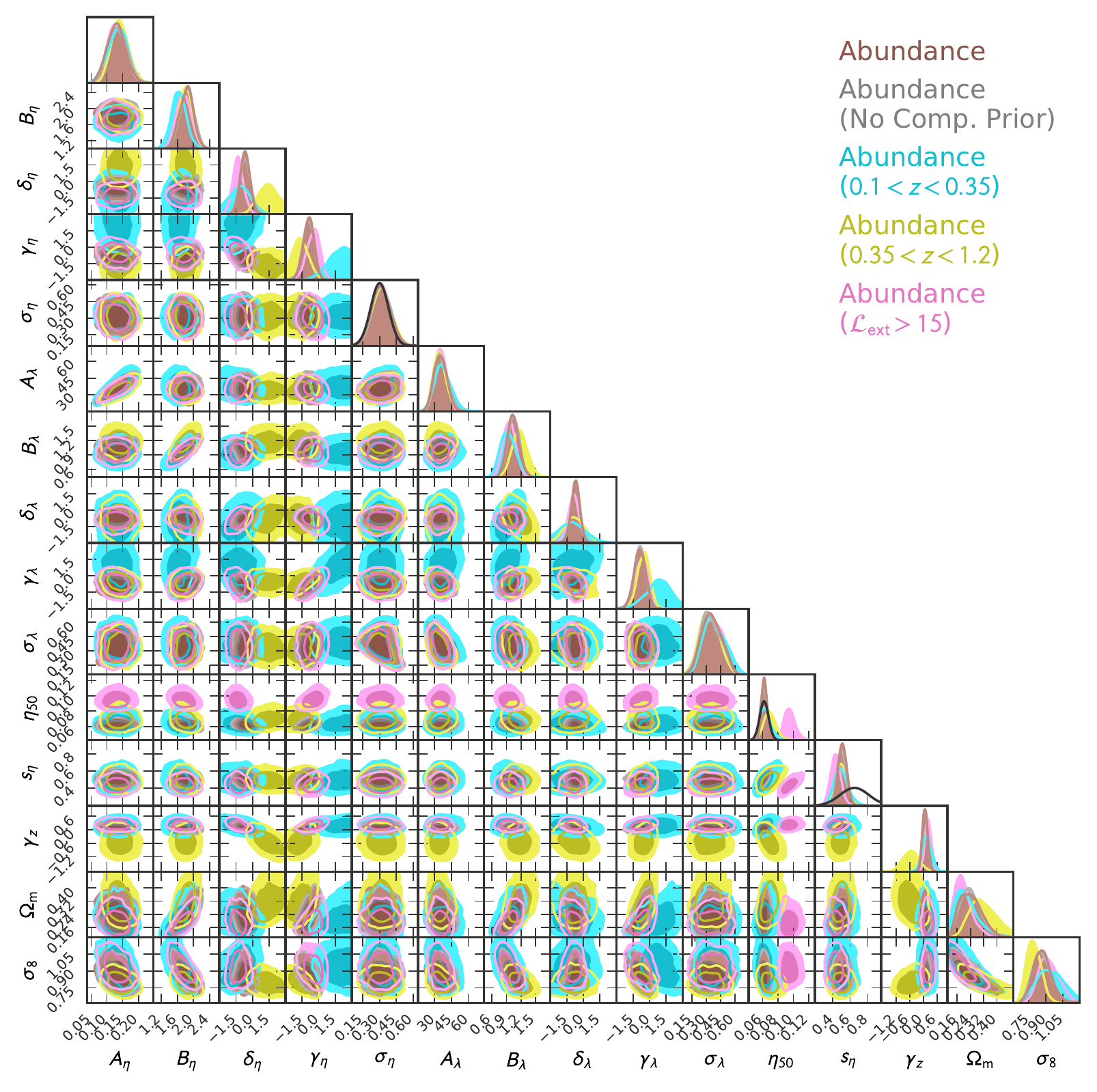}
}
\vspace{-0.5cm}
\caption{
The fully marginalized and joint posteriors of 
the parameters from the modeling of the cluster abundance with the priors on the parameters of the \rate--\mass--\redshift\ relation.
The results based on the full \eFEDS\ sample with Gaussian priors applied to the parameters 
$\left(\ratefiveo, \srate\right)$ of the X-ray completeness $\Comp(\rate,\redshift)$ are shown in brown.
The Gaussian priors are independently measured from the richness distribution (see Section~\ref{sec:completeness_measurements}) and 
are indicated by the solid lines in the on-diagonal plots of \ratefiveo\ and \srate.
The results based on the full sample with a self-calibration of the completeness parameters $\left(\ratefiveo, \srate\right)$, i.e., without applying the Gaussian priors, are in grey.
The results of the subsamples at low redshift ($0.1<\redshift<0.35$), high redshift  ($0.35<\redshift<1.2$), and the high-\Lext\ end ($\Lext > 15$) are shown in cyan, yellow, and pink, respectively.
For the modeling of the low-\redshift, high-\redshift, and high-\Lext\ subsamples, the Gaussian priors are not applied to $\left(\ratefiveo, \srate\right)$.
The contours indicate the $68\percent$ and $95\percent$ confidence levels.
}
\label{fig:nbc_gtc_smf}
\end{figure*}
%

%
%

\section{Results}
\label{sec:results}

In this section, we first present the results obtained from independently modeling the mass calibration (Section~\ref{sec:results_mcalib}) and the cluster abundance (Section~\ref{sec:results_abundance}), and then the results of the joint analysis (Section~\ref{sec:results_joint}). 
Finally, the cosmological constraints are presented in Section~\ref{sec:results_cosmos}.  In Sections~\ref{sec:results_mcalib} to~\ref{sec:results_joint}, we focus on the modeling assuming a flat \LCDM\ cosmology and then extend the results to the flat \wCDM\ cosmology in Section~\ref{sec:results_cosmos}.

\subsection{Weak-lensing mass calibration}
\label{sec:results_mcalib}

We stress again that the modeling of the weak-lensing mass calibration is nearly identical to that in \cite{chiu22}, therefore we do not expect any significant difference in this work except for a weaker constraining power.
Indeed, we recover the results of \cite{chiu22} by using the public weak-lensing data (S16A), as seen in Figure~\ref{fig:mcalib_gtc_smf} showing the consistency between this work (yellow contours) and the best-fit parameters of the \rate--\mass--\redshift\ relation obtained in \citet[][dashed lines]{chiu22}.
We refer the reader to Appendix~\ref{app:detailed_results} for the complete results from the weak-lensing mass calibration.

We then explore the weak-lensing mass calibration with the broken power-law scaling of the \rate--\mass--\redshift\ relation (see equation~(\ref{eq:broken_power_law})).
We find that the inclusion of the broken power-law scaling 
does not significantly alter the results of the weak-lensing mass calibration.
This is visualized in Figure~\ref{fig:mcalib_gtc_smf}, showing good consistency between the results obtained with the single (yellow contours) and broken (grey contours)  power-law scaling.
One noticeable difference is that the mass trend at the group scale of $\mass\lesssim\mgrp\equiv10^{14}\Msunh$ shows a tendency to be steeper than that at the cluster scale of $\mass\gtrsim\mgrp$, i.e., $\Bgrp \gtrsim \Brate$; however, this difference is not statistically significant.
This finding is also highlighted in the parameter space spanned by \Brate\ and \Bgrp\ in Figure~\ref{fig:mcalib_gtc_smf}, where the mass calibration with the broken power-law \rate--\mass--\redshift\ relation (grey contours) is consistent with the single power law ($\Brate = \Bgrp$; black solid lines). 
In short, the weak-lensing mass calibration alone does not offer statistically significant evidence of a broken power-law behavior of the mass scaling of the count rate \rate.

\begin{figure}
\centering
\resizebox{0.5\textwidth}{!}{
\includegraphics[scale=1]{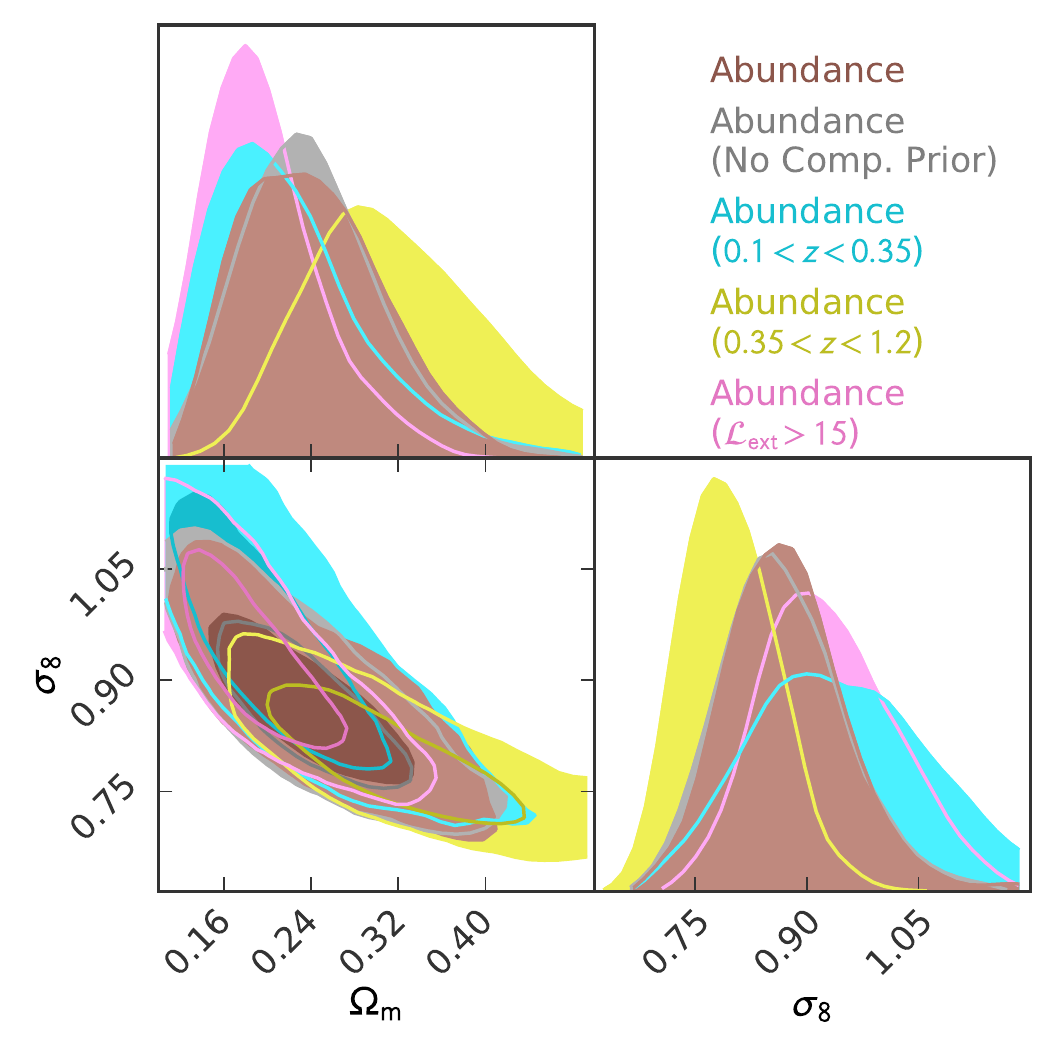}
}
\vspace{-0.5cm}
\caption{
The comparisons of the constraints on \OmegaM\ and \sigmaeight\ from the different samples and with and without Gaussian priors applied to the parameters of the X-ray completeness \Comp(\rate,\redshift).
The color coding is the same as in Figure~\ref{fig:nbc_gtc_smf}.
}
\label{fig:omega_sigma8_nbcalone}
\end{figure}
\begin{figure}
\centering
\resizebox{0.5\textwidth}{!}{
\includegraphics[scale=1]{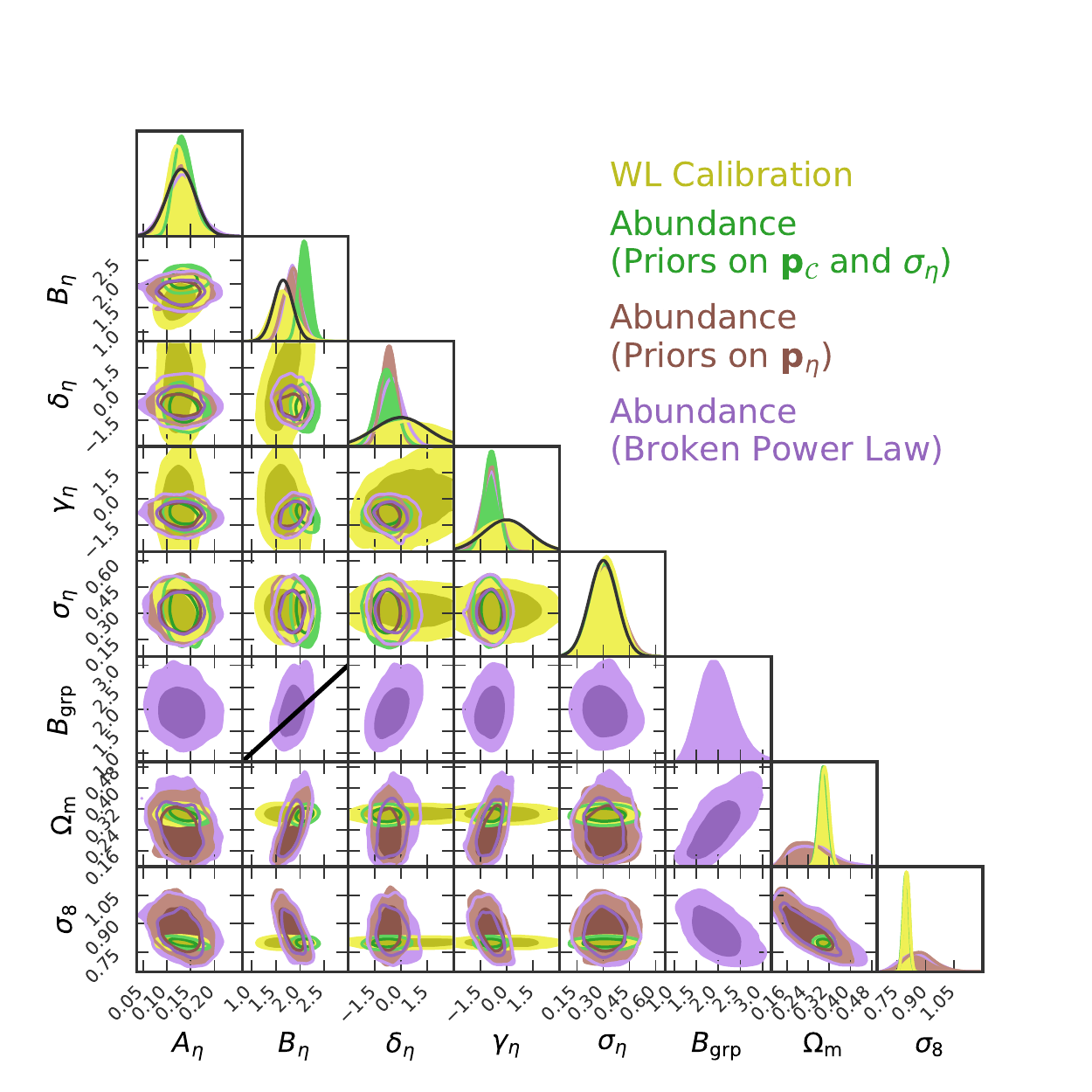}
}
\vspace{-0.5cm}
\caption{
The comparison of the constraints on the \rate--\mass--\redshift\ relations and the cosmological parameters $\left(\OmegaM, \sigmaeight\right)$ for the modeling of the weak-lensing mass calibration and the cluster abundance.
The fully marginalized and joint posteriors of 
the parameters are shown in the on-diagonal and off-diagonal plots, respectively.
The results from the modeling of the weak-lensing mass calibration with the informative priors on the cosmological parameters $\mathbf{p}_{\mathrm{c}}$ are in yellow.
The modeling of the cluster abundance with the informative priors on $\mathbf{p}_{\mathrm{c}}$
and with the Gaussian prior on the intrinsic scatter \sigmarate\ of the count rate is presented in green.
When removing the informative priors on 
$\mathbf{p}_{\mathrm{c}}$ and applying the weak-lensing informed priors on the parameters $(\Arate, \Brate, \deltarate, \gammarate)$ of the \rate--\mass--\redshift\ relation, the results from the modeling of the cluster abundance are shown in brown.
For the modeling of the cluster abundance (green and brown contours), the mass scaling of the \rate--\mass--\redshift\ relation is assumed to be a single power law.
With the weak-lensing informed priors on $(\Arate, \Brate, \deltarate, \gammarate)$ and the assumption of a broken power-law mass scaling in the count rate, the modeling of the cluster abundance is presented in purple.
The black solid line shown in the on-diagonal plot represents the Gaussian prior applied to the parameters $\mathbf{p}_{\eta}$ of the \rate--\mass--\redshift\ relation.
The black line shown in the parameter space of \Brate\ and \Bgrp\ indicates the mass scaling of the count rate with a single power law ($\Brate = \Bgrp$).
The contours indicate the $68\percent$ and $95\percent$ confidence levels.
}
\label{fig:comparisons_mcalib_nbc}
\end{figure}

\subsection{Cluster abundance}
\label{sec:results_abundance}

While blinding the constraints on the cosmological parameters (see Section~\ref{sec:blinding}), we carry out the modeling of the cluster abundance alone with special interest in examining our empirical modeling of the X-ray completeness (see Section~\ref{sec:completeness}) and the broken power-law feature of the \rate--\mass--\redshift\ relation (see Section~\ref{sec:brokenpowerlaw}).
Note that we additionally apply the conservative Gaussian priors to the parameters ($\Arate, \Brate, \deltarate, \gammarate$) of the \rate--\mass--\redshift\ relation to mimic the constraining power from the weak-lensing mass calibration (see Section~\ref{sec:statistics}).

We first assess the empirical modeling of the X-ray completeness by testing the modeling of the cluster abundance against (1) the different subsamples and (2) the Gaussian priors applied to the parameters of the completeness function.
The purpose of this test is to examine whether our empirical modeling provides a flexible scheme to capture the data behavior of different subsamples and to further recover the underlying cosmology.

In Figure~\ref{fig:nbc_gtc_smf}, we show the constraints of highlighted parameters obtained with
\begin{enumerate}
\item the full sample with the Gaussian priors on the X-ray completeness $\Comp\left(\rate,\redshift\right)$ (parameters \ratefiveo\ and \srate), where the Gaussian priors are the constraints on the X-ray completeness independently measured using the richness distribution (see Section~\ref{sec:completeness_measurements}),
\item the full sample with a self-calibration of the X-ray completeness $\Comp\left(\rate,\redshift\right)$
i.e., without the Gaussian priors,
\item the subsample of low-\redshift\ clusters at $0.1<\redshift<0.35$ with a self-calibration of the X-ray completeness $\Comp\left(\rate,\redshift\right)$,
\item the subsample of high-\redshift\ clusters at $0.35<\redshift<1.2$ with a self-calibration of the X-ray completeness $\Comp\left(\rate,\redshift\right)$, and
\item the subsample of high-\Lext\ clusters with $\Lext > 15$ and a self-calibration of the X-ray completeness $\Comp\left(\rate,\redshift\right)$.
\end{enumerate}
The constraints on the full parameter list are contained in Appendix~\ref{app:detailed_results}, where reader can find more details.
As seen in Figure~\ref{fig:nbc_gtc_smf}, we find that the self-calibrated constraints  on 
the X-ray completeness $\Comp\left(\rate,\redshift\right)$ parameters $\left(\ratefiveo, \srate\right)$, except for the subsample with $\Lext>15$, are consistent with the adopted Gaussian priors, which are indicated by the black solid lines on the on-diagonal plots.
This suggests that the self-calibrated completeness of the count rate is in good agreement with the measurements that are independently inferred using the richness distribution of the extended and full sample of clusters.
Due to the more stringent X-ray extent selection (fiducial cut is $\Lext>6$), we note that a significantly higher value of \ratefiveo\ is obtained for the subsample with $\Lext>15$ (pink contours), as expected.
In addition, no sign of a redshift dependence in the completeness function is suggested, given that \gammaz\ is constrained to be statistically consistent with zero at a level of $\lesssim1\sigma$ in all tests.
We discuss the resulting X-ray completeness function $\Comp\left(\rate,\redshift\right)$ in Section~\ref{sec:results_completeness}.

Meanwhile, as highlighted in Figure~\ref{fig:omega_sigma8_nbcalone}, we find that the constraints on the cosmological parameters of \OmegaM\ and \sigmaeight\ obtained from the modeling of the cluster abundance alone are in good agreement among the different samples.
Moreover, the constraints on \OmegaM\ and \sigmaeight\  are insensitive to the Gaussian priors applied to the X-ray completeness $\Comp\left(\rate,\redshift\right)$ parameters $\left(\ratefiveo, \srate\right)$.
This suggests that our empirical modeling provides (1) a flexible scheme to describe the completeness and (2) an accurate characterization of the X-ray completeness.
This validates our empirical modeling of the X-ray completeness of the \eFEDS\ extent selected sample.

Owing to the lack of sufficient coverage in both the mass and redshift ranges, we make one remark that the redshift-dependent parameters of the \rate--\mass--\redshift\ relation, specifically the scaling of the redshift (\gammarate) and the mass-redshift cross term (\deltarate),
are poorly constrained in the low-\redshift\ (cyan contours) and high-\redshift\ (yellow contours) samples, as seen in Figure~\ref{fig:nbc_gtc_smf}.
However, the degeneracy between the parameters of \deltarate\ and \gammarate\ does not significantly impact the constraints on \OmegaM\ and \sigmaeight, showing good agreement with those based on the full sample.

Next, we examine the consistency between the weak-lensing mass calibration and the cluster abundance in terms of the \rate--\mass--\redshift\ relation in Figure~\ref{fig:comparisons_mcalib_nbc}.
Note that the Gaussian priors on the X-ray completeness \Comp(\rate,\redshift) inferred using the richness distribution (see Section~\ref{sec:completeness_measurements}) are applied to the parameters $\left(\ratefiveo, \srate\right)$ in obtaining the results shown in Figure~\ref{fig:comparisons_mcalib_nbc}.
The modeling of the cluster abundance shown by the green contours is performed (1) with the informative priors applied to the intrinsic scatter \sigmarate\ of the count rate and the cosmological parameters $\vect{p}_{\mathrm{c}}$ and (2) without the weak-lensing informed priors on $\left(\Arate, \Brate, \deltarate, \gammarate\right)$.
This effectively fixes the scatter \sigmarate\ and the cosmology as identical as in the modeling of the weak-lensing mass calibration (yellow contours), which gives an independent comparison of the \rate--\mass--\redshift\ relations (see the last paragraph in Section~\ref{sec:statistics}). 
As seen, there exists good agreement in the resulting \rate--\mass--\redshift\ relations obtained between the modeling of the weak-lensing mass calibration (yellow contours) and the cluster abundance (green contours).
The excellent agreement in the posterior of the normalization \Arate\ suggests that there is no significant tension in the absolute mass scale of \eFEDS\ clusters inferred between the weak lensing and the cluster abundance.
Meanwhile, the cluster abundance prefers a higher \Brate\ than that from the weak-lensing mass calibration at a level of $\lesssim1\sigma$.
Overall, there is no strong tension 
between the weak-lensing mass calibration and cluster abundance of the \eFEDS\ clusters.
This enables us to further combine them to obtain a joint constraint on cosmological parameters, which is what we discuss in the next section.
Note that the modeling of the cluster abundance (green contours) is carried out for this consistency test in a post-unblinding stage.

Without the informative priors on both the cosmological parameters and the parameters $(\Arate, \Brate, \deltarate, \gammarate)$ of the \rate--\mass--\redshift\ relation, we find that the result yields a highly degenerate constraint among \OmegaM, \sigmaeight, and \Arate.
This suggests 
that (1) the cluster abundance alone does not have constraining power on \Arate, unless informative priors are applied to the cosmological parameters, and 
that (2) the inclusion of the weak-lensing mass calibration is crucially important for obtaining cosmological constraints, since it anchors the absolute mass scale.
Removing the strong priors on $\vect{p}_{\mathcal{C}}$ and adding the weak-lensing informed priors on the parameters $(\Arate, \Brate, \deltarate, \gammarate)$ of the \rate--\mass--\redshift\ relation leads to the modeling of the cluster abundance presented by the brown contours in Figure~\ref{fig:comparisons_mcalib_nbc}.

In what follows, we move to assess the impact raised from the broken power-law behavior of the \rate--\mass--\redshift\ relation on the cosmological constraints.
This result is visualized in Figure~\ref{fig:comparisons_mcalib_nbc}, showing the comparison of the constraints from the cluster abundance between the single (brown contours) and broken (purple contours) power law in the \rate--\mass--\redshift\ relation.
We find excellent agreement between the cosmological parameters of $\left(\OmegaM, \sigmaeight\right)$ with and without the broken power-law feature.
Moreover, the constraints in the parameter space spanned by \Brate\ and \Bgrp\ shown in Figure~\ref{fig:comparisons_mcalib_nbc} are fully consistent with the scenario of a single power law $\Brate = \Bgrp$ (the solid line).
This suggests that the modeling of the cluster abundance alone does not prefer a broken power-law feature for the \rate--\mass--\redshift\ relation at the low-mass end.
This validates the robustness of the cosmological constraints to the details of the functional form assumed for the mass scaling of the \rate--\mass--\redshift\ relation.

\begin{figure*}
\centering
\resizebox{0.9\textwidth}{!}{
\includegraphics[scale=1]{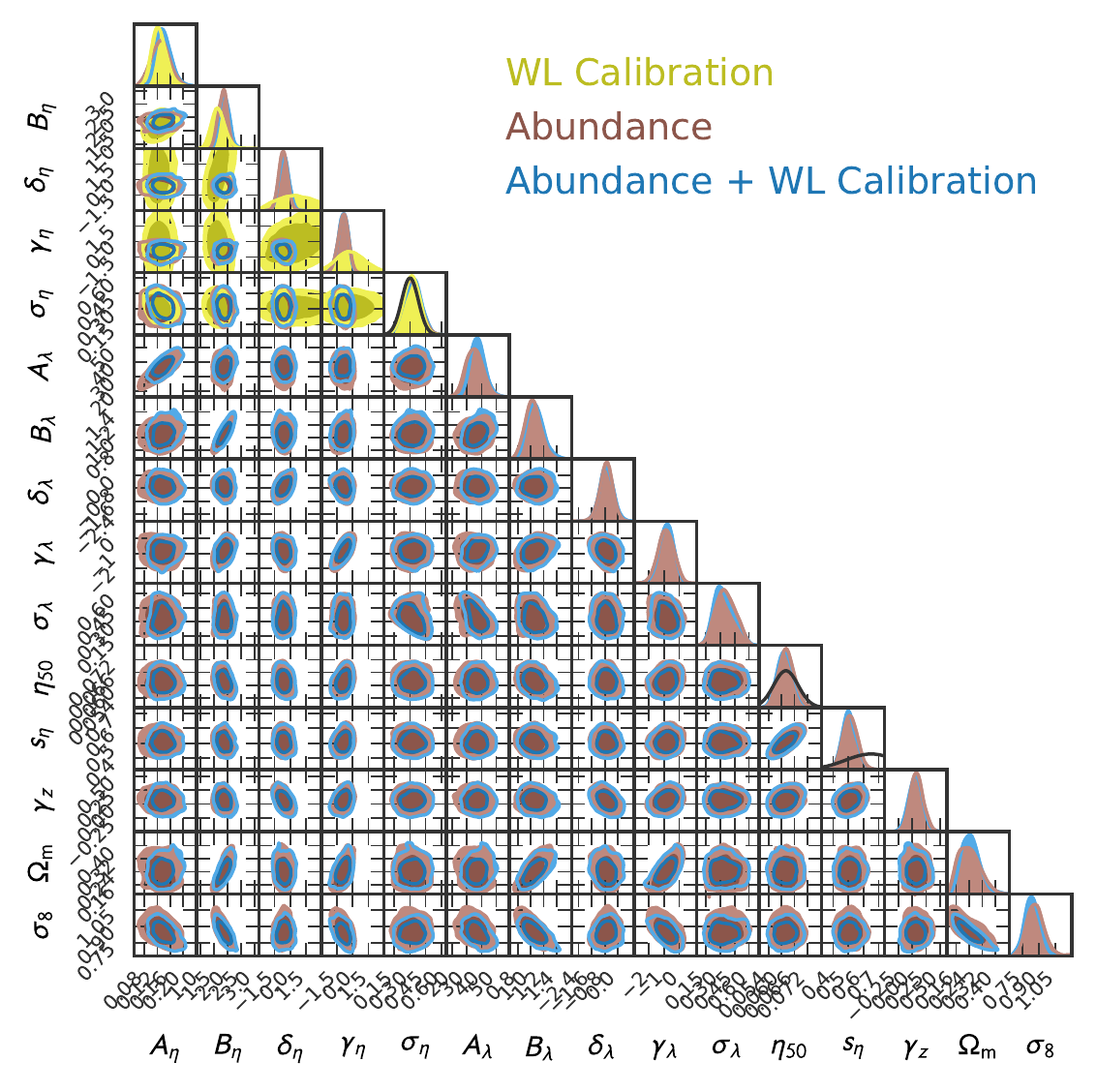}
}
\vspace{-0.5cm}
\caption{
The comparison of the constraints obtained from the modeling of the mass calibration (yellow contours), the cluster abundance (brown contours), and the joint modeling of them (blue contours).
These results are consistently obtained with the single power-law \rate--\mass--\redshift\ relation and the Gaussian priors applied to the parameters of the X-ray completeness $\Comp\left(\rate,\redshift\right)$ when including the modeling of the cluster abundance.
For the modeling of the cluster abundance (brown contours), the informative priors are applied to the parameters of the \rate--\mass--\redshift\ relation (see Section~\ref{sec:statistics}).
The posteriors of and the covariance between the parameters are shown in the on-diagonal and off-diagonal plots, respectively.
The Gaussian priors applied to the parameters are shown as the black solid lines in the off-diagonal plots.
The contours indicate the $68\percent$ and $95\percent$ confidence levels.
}
\label{fig:mcalib_nbc_smf}
\end{figure*}
\begin{figure*}
\centering
\resizebox{1\textwidth}{!}{
\includegraphics[scale=1]{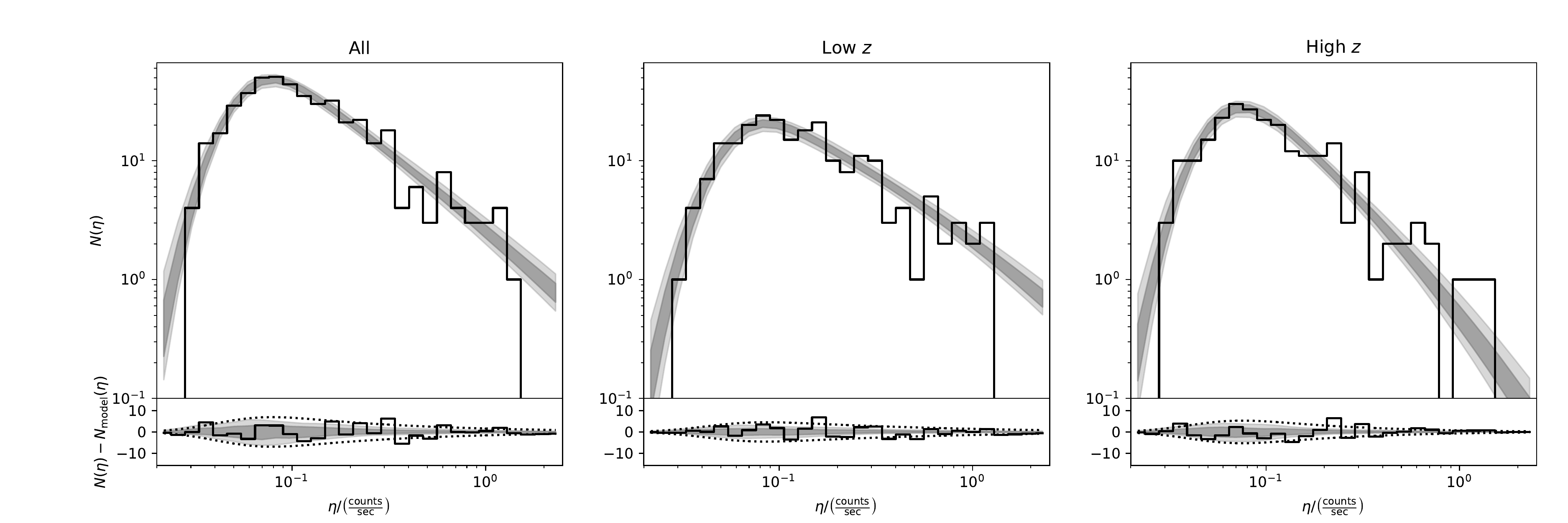}
}
\resizebox{1\textwidth}{!}{
\includegraphics[scale=1]{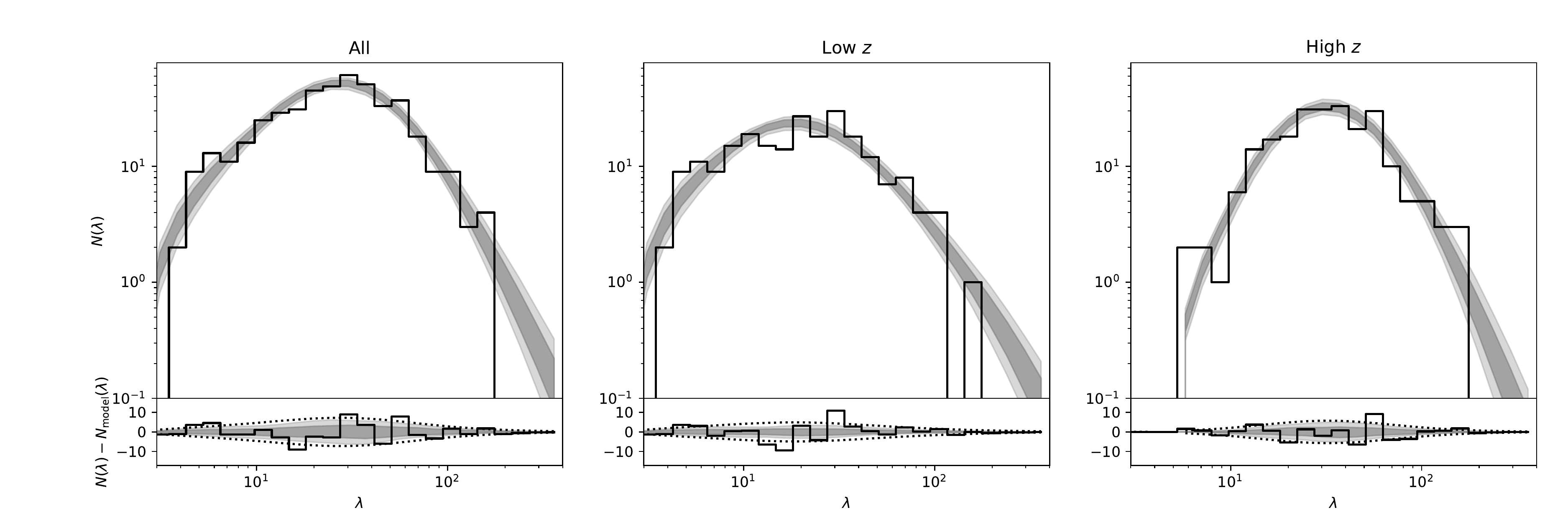}
}
\resizebox{1\textwidth}{!}{
\includegraphics[scale=1]{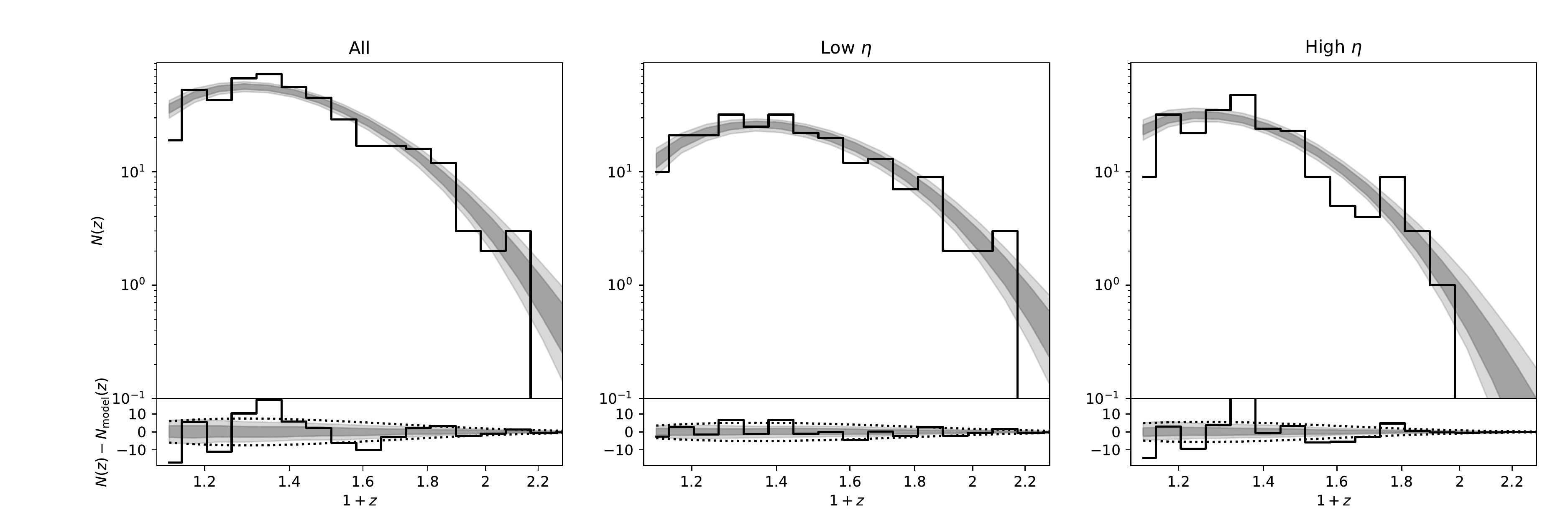}
}
\vspace{-0.15cm}
\caption{
The distributions of the observed count rate \rate\ (the first row), the observed richness \rich\ (the second row), and the observed redshift (the third row).
In the first and second rows, we additionally show the full (left),  the low-\redshift\ (middle) and the high-\redshift\ (right) samples.
The low-\redshift\ (high-\redshift) clusters are selected with an additional criterion of $0.1<\redshift<0.35$ ($0.35<\redshift<1.2$).
In the last row, the redshift distributions of the full (the left panel),  the low-count rate (low-\rate; the middle panel) and the high-count rate (high-\rate; the right panel) samples are shown.
The low-\rate\ (high-\rate) sample is further selected with $\rate<0.1~\mathrm{counts}/\mathrm{sec}$ ($\rate>0.1~\mathrm{counts}/\mathrm{sec}$).
In the upper panel of each subplot, the distributions of the observed \eFEDS\ clusters are shown by the black histograms, while the best-fit models with the $68\percent$ and $95\percent$ confidence levels are indicated by the dark and light grey regions, respectively.
The modeling residuals, which are defined as the observed distributions $N$ subtracting the best-fit models $N_{\mathrm{model}}$, are shown in the lower panel of each subplot, wherein the dotted lines indicate the $1\sigma$ Poisson noises expected by the best-fit models, i.e., $\sqrt{N_{\mathrm{model}}}$.
The best-fit models and the confidence levels are calculated using the chains from the joint modeling of the weak-lensing mass calibration and the cluster abundance assuming the single power-law \rate--\mass--\redshift\ relation and with the Gaussian priors applied to the parameters of $\left(\ratefiveo, \srate\right)$.
}
\label{fig:histograms_observed_modeled}
\end{figure*}
\begin{table*}
\centering
\caption{
The constraints on the \rate--\mass--\redshift\ and the \rich--\mass--\redshift\ relations.
The first column contains the parameter name.
The columns (2) and (3) present the constraints obtained in the modeling of the weak-lensing mass calibration alone with the single and broken power-law mass scaling of the \rate--\mass--\redshift\ relation, respectively.
From the modeling of the cluster abundance alone with the weak-lensing informed priors on the parameters of the single power-law \rate--\mass--\redshift\ relation, the columns (4) and (5) contain the constraints obtained without and with the Gaussian priors applied to the parameters $\left(\ratefiveo, \srate\right)$ of the X-ray completeness, respectively.
The column (6) shows the results from the modeling of the cluster abundance alone with the Gaussian priors applied to $\left(\ratefiveo, \srate\right)$ and with the broken power-law feature in the \rate--\mass\ relation.
Under the \LCDM\ model, the columns (7) and (8) present the results of the joint modeling with the single and broken power-law mass scaling of the \rate--\mass--\redshift\ relations, respectively.
The results of the same joint modeling in the \wCDM\ model are contained in the columns (9) and (10).
Note that columns (7) to (10) contain results that are obtained with the Gaussian priors applied to the parameters of the X-ray completeness $\Comp(\rate,\redshift)$.
}
\label{tab:sr}
\resizebox{\textwidth}{!}{
\begin{tabular}{cccccccccc}
\hline
\multirow{3}{*}{Parameters} 
&\multicolumn{2}{c}{WL Mass Calibration} 
&\multicolumn{3}{c}{Cluster Abundance}
&\multicolumn{4}{c}{Cluster Abundance + WL Calibration} \\
\cmidrule(lr){2-3}\cmidrule(lr){4-6}\cmidrule(lr){7-10}
&\multirow{2}{*}{$\Brate = \Bgrp$} & \multirow{2}{*}{$\Brate \neq \Bgrp$}
&Without Comp. Prior &With Comp. Prior & With Comp. Prior
&\multicolumn{2}{c}{\LCDM} &\multicolumn{2}{c}{\wCDM} \\
&&& and $\Brate = \Bgrp$ & and $\Brate = \Bgrp$ & and $\Brate \neq \Bgrp$ &$\Brate = \Bgrp$ & $\Brate \neq \Bgrp$& $\Brate = \Bgrp$ & $\Brate \neq \Bgrp$ \\
(1) &(2) &(3) &(4) &(5) &(6) &(7) &(8) &(9) &(10) \\
\hline
		$A_{\eta}$ & $0.121^{+0.025}_{-0.020}$ & $0.133^{+0.028}_{-0.029}$ & $0.129^{+0.033}_{-0.028}$ & $0.131^{+0.027}_{-0.031}$ & $0.133^{+0.032}_{-0.035}$ & $0.133^{+0.026}_{-0.020}$ & $0.128^{+0.027}_{-0.021}$ & $0.116^{+0.037}_{-0.024}$ & $0.129^{+0.025}_{-0.028}$ \\ [3pt]
		$B_{\eta}$ & $1.61^{+0.26}_{-0.22}$ & $1.29^{+0.60}_{-0.74}$ & $1.87^{+0.14}_{-0.21}$ & $1.86^{+0.14}_{-0.21}$ & $1.84\pm 0.17$ & $1.86^{+0.20}_{-0.15}$ & $1.93^{+0.19}_{-0.20}$ & $1.86^{+0.20}_{-0.16}$ & $1.92^{+0.21}_{-0.17}$ \\ [3pt]
		$\delta_{\eta}$ & $0.4^{+1.7}_{-1.6}$ & $1.7^{+1.1}_{-2.1}$ & $-0.72^{+0.52}_{-0.44}$ & $-0.68^{+0.44}_{-0.42}$ & $-0.53^{+0.64}_{-0.63}$ & $-0.58^{+0.43}_{-0.50}$ & $-1.01^{+0.80}_{-0.62}$ & $-0.65^{+0.56}_{-0.54}$ & $-0.87^{+0.73}_{-0.75}$ \\ [3pt]
		$\gamma_{\eta}$ & $-0.4^{+1.5}_{-1.3}$ & $0.0^{+1.4}_{-1.5}$ & $-0.88^{+0.49}_{-0.53}$ & $-0.83^{+0.43}_{-0.53}$ & $-0.84^{+0.45}_{-0.60}$ & $-0.83^{+0.44}_{-0.50}$ & $-0.87^{+0.41}_{-0.45}$ & $-0.38^{+0.72}_{-0.82}$ & $-0.41^{+0.52}_{-1.06}$ \\ [3pt]
		$\sigma_{\eta}$ & $0.314^{+0.082}_{-0.071}$ & $0.312^{+0.080}_{-0.069}$ & $0.307^{+0.090}_{-0.074}$ & $0.308^{+0.086}_{-0.083}$ & $0.299^{+0.087}_{-0.074}$ & $0.332^{+0.076}_{-0.089}$ & $0.350^{+0.087}_{-0.086}$ & $0.308^{+0.088}_{-0.066}$ & $0.341^{+0.070}_{-0.095}$ \\ [3pt]
		$B_{\rm{grp}}$ & -- & $1.89^{+0.69}_{-0.52}$ & -- & -- & $1.86^{+0.44}_{-0.32}$ & -- & $1.75^{+0.27}_{-0.24}$ & -- & $1.76^{+0.32}_{-0.24}$ \\ [3pt]
		$A_{\lambda}$ & -- & -- & $34.5^{+6.8}_{-5.1}$ & $34.8\pm 5.8$ & $34.7^{+6.8}_{-6.7}$ & $37.1^{+4.6}_{-5.0}$ & $37.5^{+4.6}_{-4.8}$ & $35.1^{+5.1}_{-6.8}$ & $36.8^{+5.0}_{-5.7}$ \\ [3pt]
		$B_{\lambda}$ & -- & -- & $1.03^{+0.14}_{-0.10}$ & $1.03\pm 0.12$ & $1.06\pm 0.13$ & $1.05^{+0.13}_{-0.12}$ & $1.07^{+0.10}_{-0.12}$ & $1.06^{+0.11}_{-0.14}$ & $1.07^{+0.12}_{-0.11}$ \\ [3pt]
		$\delta_{\lambda}$ & -- & -- & $-0.69^{+0.36}_{-0.41}$ & $-0.68^{+0.37}_{-0.39}$ & $-0.62^{+0.36}_{-0.47}$ & $-0.72^{+0.45}_{-0.37}$ & $-0.70^{+0.41}_{-0.45}$ & $-0.68^{+0.53}_{-0.42}$ & $-0.60^{+0.45}_{-0.49}$ \\ [3pt]
		$\gamma_{\lambda}$ & -- & -- & $-0.85^{+0.52}_{-0.51}$ & $-0.90^{+0.56}_{-0.48}$ & $-0.83^{+0.56}_{-0.62}$ & $-0.77^{+0.41}_{-0.54}$ & $-0.79^{+0.47}_{-0.53}$ & $-0.79^{+0.46}_{-0.47}$ & $-0.75^{+0.44}_{-0.50}$ \\ [3pt]
		$\sigma_{\lambda}$ & -- & -- & $0.307^{+0.135}_{-0.081}$ & $0.334^{+0.122}_{-0.093}$ & $0.334^{+0.140}_{-0.080}$ & $0.291^{+0.133}_{-0.078}$ & $0.279^{+0.148}_{-0.077}$ & $0.320^{+0.129}_{-0.089}$ & $0.290^{+0.135}_{-0.079}$
 \\
\hline
\end{tabular}
}
\end{table*}

\subsection{Joint weak-lensing and the cluster abundance analysis}
\label{sec:results_joint}

In Sections~\ref{sec:results_mcalib} and~\ref{sec:results_abundance}, we find that 
(1) the constraints on the \rate--\mass--\redshift\ relation from both the modeling of the weak-lensing mass calibration and the cluster abundance are in excellent agreement, and that
(2) neither the modeling of the weak-lensing mass calibration nor the cluster abundance prefer a broken power-law feature of the \rate--\mass--\redshift\ relation at the low-mass end of $\mass\approx\mgrp$.
Therefore, we can combine the two datasets to obtain a joint constraint on the scaling relations and cosmology.
Because the data do not provide compelling evidence of a broken power-law scaling of the count rate, we use the \rate--\mass--\redshift\ relation with a single power-law form as the fiducial analysis in this work.
Moreover, given that the X-ray completeness $\Comp\left(\rate,\redshift\right)$ self-calibrated by the cluster abundance is fully consistent with that independently measured using the richness distribution (see Figure~\ref{fig:nbc_gtc_smf} and Section~\ref{sec:results_abundance}), the constraints in equation~(\ref{eq:completeness_prior}) are consistently applied as the Gaussian priors to the X-ray completeness parameters of $\left(\ratefiveo, \srate\right)$ in the joint modeling.  Note that the results presented in this section are obtained assuming the flat \LCDM\ model, and the cosmological constraints for the extension to the \wCDM\ model are given in Section~\ref{sec:results_cosmos}.

We first show the joint constraints of highlighted parameters in Figure~\ref{fig:mcalib_nbc_smf} and refer the reader to Appendix~\ref{app:detailed_results} for the complete results.
As can be seen (and as expected), the joint constraints (blue contours) are in good agreement with those previously obtained from the independent modeling of the weak-lensing mass calibration (yellow contours) and the cluster abundance (brown contours).
Moreover, we find that the joint constraints on the parameters of the \rate--\mass--\redshift\ are dominated by the cluster abundance, as the constraints on the scaling parameters $\left(\Brate, \deltarate, \gammarate\right)$ tighten significantly when including the number counts likelihood (blue contours) in comparison to that from the modeling of the weak-lensing mass calibration alone (yellow contours).
The uncertainties on \Brate, \deltarate, and \gammarate\ inferred from the weak-lensing calibration alone are 
reduced by a factor of $1.6$, $2.5$ and $3.25$, respectively, when including the cluster abundance.

Next, we show the distributions of the observables (namely, the count rate \rate, richness \rich, and redshift \redshift) predicted by the best-fit model in comparison with the observations in Figure~\ref{fig:histograms_observed_modeled}.
In the top (middle) row of Figure~\ref{fig:histograms_observed_modeled}, we additionally show the distribution of the count rate (richness) at low and high redshift in the middle and right columns, respectively.
The redshift distribution of the full (left), low-\rate\ ($\rate<0.1~\mathrm{counts}/\mathrm{s}$; middle) and high-\rate\ ($\rate>0.1~\mathrm{counts}/\mathrm{s}$; right) samples are shown in the bottom row of Figure~\ref{fig:histograms_observed_modeled}.  Each panel contains also a residual subpanel.
All these comparisons show no sign of bias in the best-fit model with respect to the observations, suggesting that our empirical modeling analysis including weak lensing mass calibration has accurately captured the characteristics of the eFEDS abundance data.
The reader is referred to Section~\ref{sec:results_gof} for a more quantitative assessment of the goodness of fit.

\begin{figure*}
\centering
\resizebox{\textwidth}{!}{
\includegraphics[scale=1]{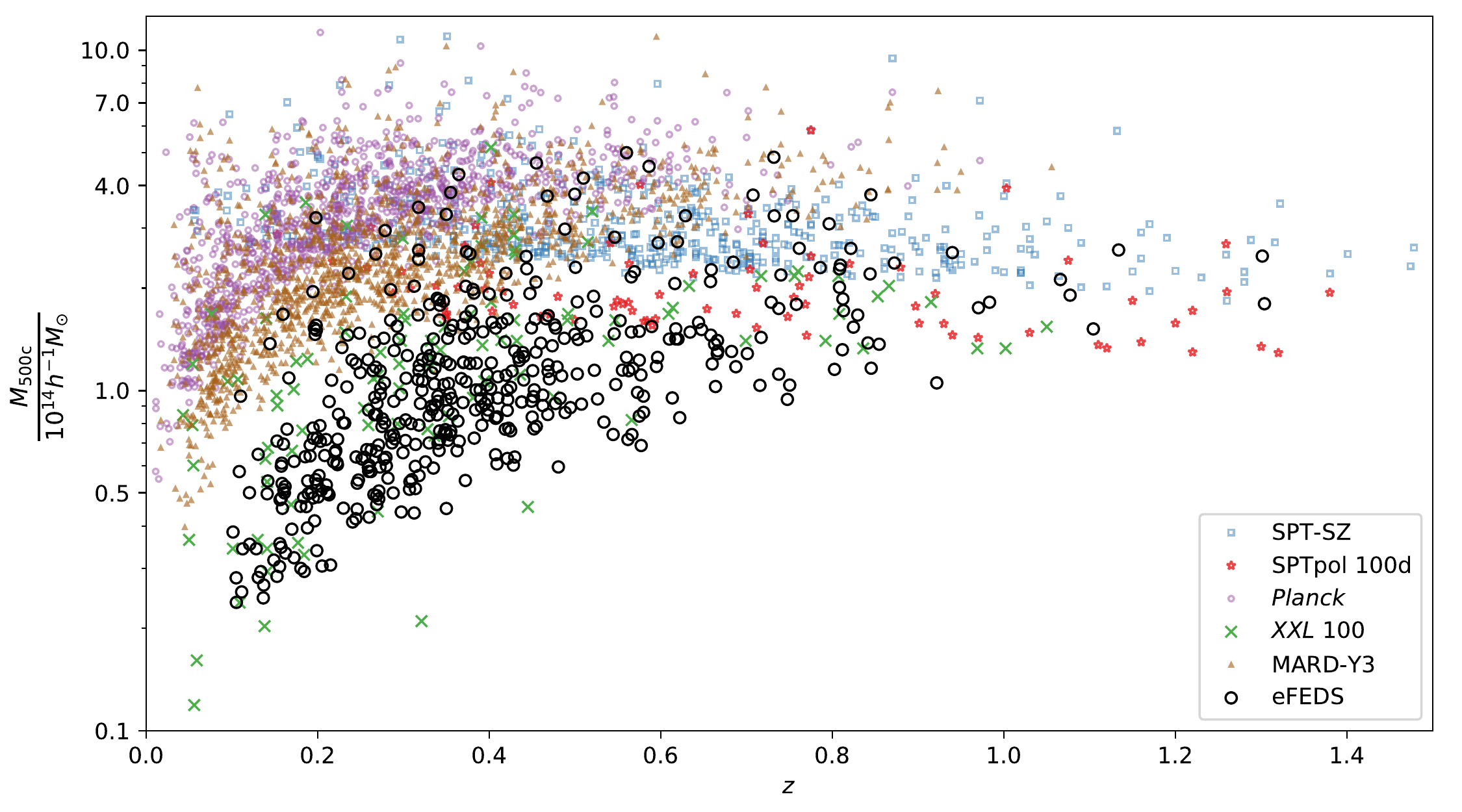}
}
\vspace{-0.5cm}
\caption{
The mass and redshift of the \eFEDS\ clusters (black circles), and those in the SPT-SZ survey \citep[blue squares;][]{bleem15}, the SPTpol 100~degree$^2$ survey \citep[red stars;][]{huang20}, the \PLANCK\ mission \citep[purple circles;][]{PlanckCollaboration2015a}, the brightest sample in the \XXL\ survey \citep[green crosses;][]{pacaud16}, and the X-ray MARD-Y3 sample \citep[brown triangles;][]{klein19}.
When plotting the \eFEDS\ sample, we additionally include the two clusters at $\redshift\approx1.3$ that satisfy both the X-ray and optical selections. 
}
\label{fig:mz}
\end{figure*}

\subsubsection{Individual cluster mass estimates}
\label{sec:individualmasses}

Using Bayes' theorem and following the procedure in \cite{chiu22}, we compute the mass posterior $P(\mass | \rate, \redshift, \vect{p}_{\rate})$ of individual clusters inferred using the best-fit \rate--\mass--\redshift\ relation based on the observed count rate \rate\  at the cluster redshift \redshift\ as
\begin{equation}
\label{eq:mass_post}
P(\mass | \rate, \redshift, \vect{p}_{\rate}) \propto P(\rate | \mass, \redshift, \vect{p}_{\rate})~P(\mass | \redshift, \vect{p}_{\rate}) \, ,
\end{equation}
where $\vect{p}_{\rate}$ is the best-fit parameter vector of the \rate--\mass--\redshift\ relation.
Note that we do not marginalize over the chain of $\vect{p}_{\rate}$ and hence do not include the systematic uncertainty in the mass posterior.
That is, $P(\mass | \rate, \redshift, \vect{p}_{\rate})$ only includes the dispersion introduced by the measurement uncertainty and intrinsic scatter about the count rate to mass relation.
The mean masses of individual clusters, estimated as $\Mfiveoo \equiv \left\langle \mass \right\rangle = \int \dif \mass P(\mass | \rate, \redshift, \vect{p}_{\rate})$,  are plotted versus redshift in Figure~\ref{fig:mz}, where we also plot the samples from the South Pole Telescope survey \citep{bleem15,huang20}, the \PLANCK\ catalog \citep{PlanckCollaboration2015a}, the \XXL\ survey \citep{pacaud16}, and the \ROSAT-based MARD-Y3 sample  \citep{klein19}.
As seen, the \eFEDS\ sample used in this work spans a mass range between $\approx10^{13.3}\Msunh\approx2\times10^{13}\Msunh$ and $\approx10^{14.7}\Msunh\approx5\times10^{14}\Msunh$.
The sample contains a significant fraction ($\approx50\percent$) of galaxy groups with mass of $\mass\lesssim 10^{14}\Msunh$ at redshifts $\redshift\lesssim0.5$, which is an important part of the cluster phase space for the \eROSITA\ survey.
Surpassing other cluster surveys, as seen in Figure~\ref{fig:mz}, the \eFEDS\ survey provides a unique sample that delivers clusters with $\mass\gtrsim10^{14}\Msunh$ out to a redshift beyond unity but also that probes a mass range of over an order of magnitude in a broad range of redshift.

With the mass posterior $P(\mass | \rate, \redshift, \vect{p}_{\rate})$ for each cluster we randomly 
sample a mass estimate from the posterior as the ``ensemble'' mass.
The ensemble mass here is referred to as the individual mass estimate that acts as an 
ensemble statistic inferred from the overall population modeling \citep[see also][]{bocquet19, chiu20b}.
The uncertainty of the cluster mass, which is derived as the dispersion of the mass posterior, 
is at a level of $\approx20\percent$ for a cluster with $\Mfiveoo\approx10^{14}\Msunh$.
We provide these mass measurements in Table~\ref{tab:mass} in Appendix~\ref{app:detailed_results}.

\subsubsection{Constraints on cluster scaling relations}
\label{sec:scalingrelations}

\begin{figure*}
\centering
\resizebox{0.49\textwidth}{!}{
\includegraphics[scale=1]{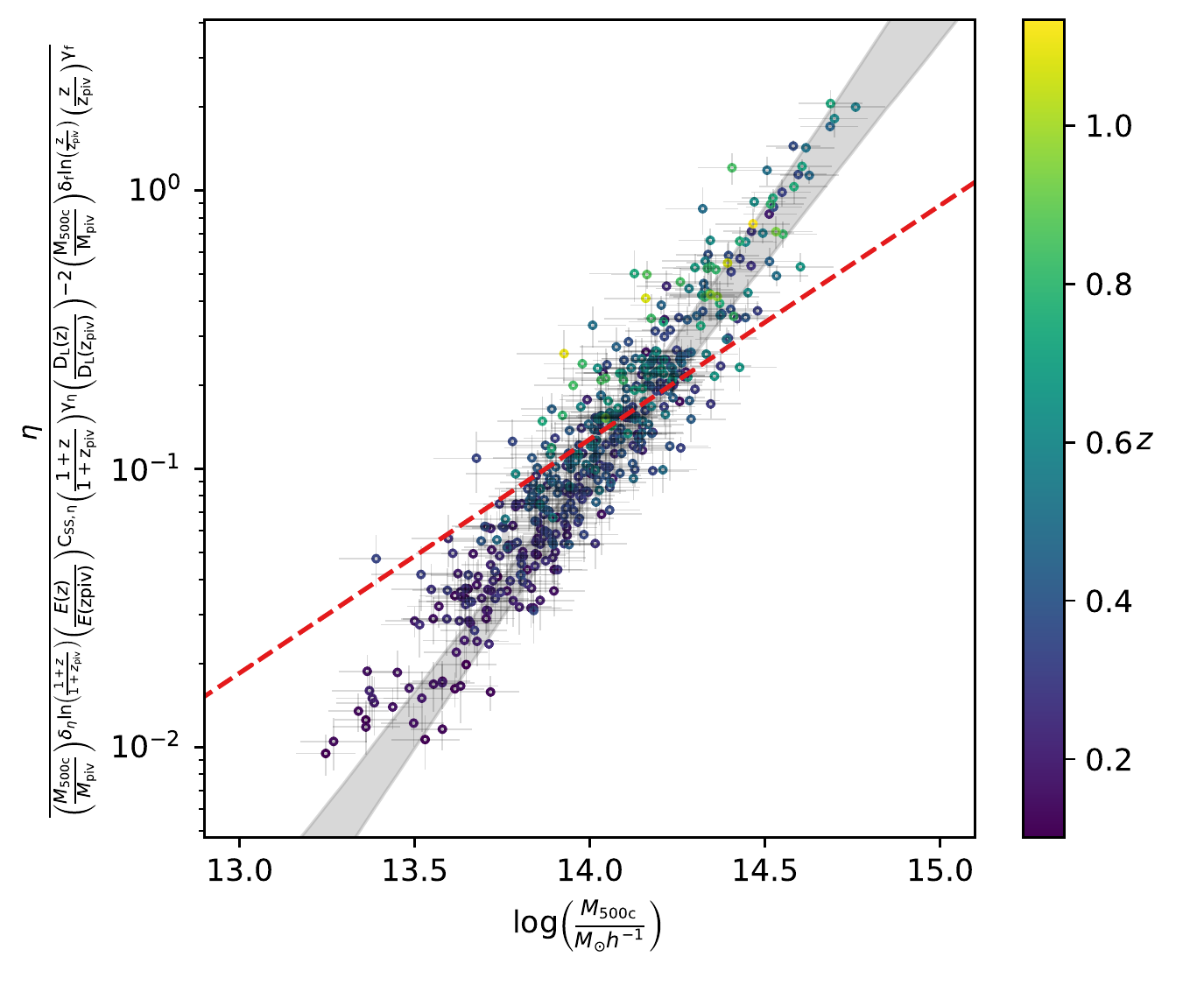}
}
\resizebox{0.49\textwidth}{!}{
\includegraphics[scale=1]{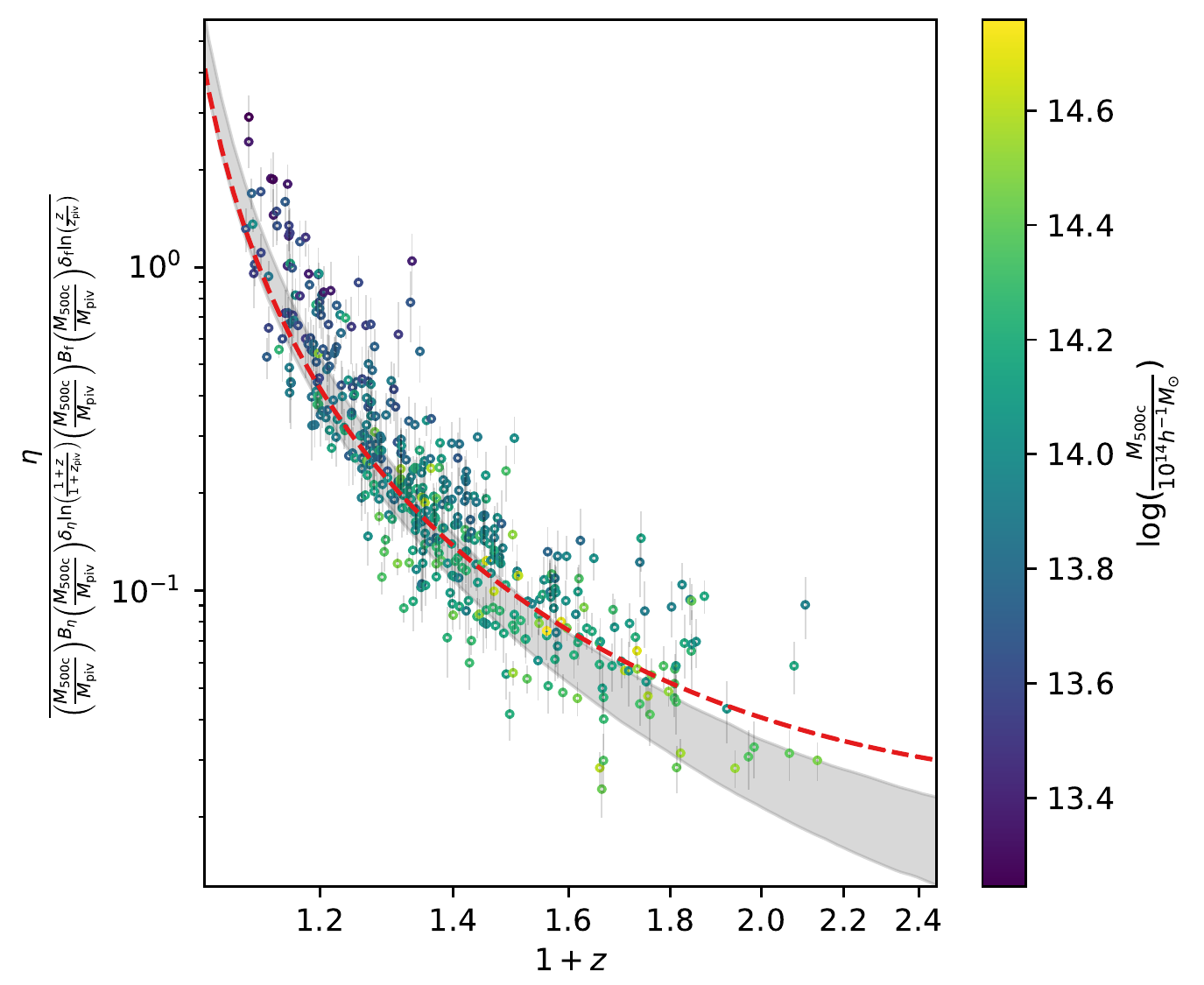}
}
\vspace{-0.2cm}
\caption{
The observed count rate of the \eFEDS\ clusters as a function of the cluster halo mass \mass\ (left) and redshift (right).
The \eFEDS\ clusters are shown by the circles color-coded by their redshift (halo mass) in the left (right) panel, wherein the $y$-axis shows the count rate \rate\ normalized at the pivotal redshift $\zpiv = 0.35$ (the pivotal mass $\mpiv = 1.4\times10^{14}\Msunh$) and the $x$-axis indicates the cluster halo mass (the cluster redshift).
In both the left and right panels, the $68\percent$ confidence levels of the best-fit model are represented by the grey regions.
The confidence levels of the best-fit models are calculated using the chains from the joint modeling of the weak-lensing mass calibration and the cluster abundance assuming the single power-law \rate--\mass--\redshift\ relation and with the Gaussian priors applied to the X-ray completeness \Comp(\rate,\redshift) parameters. 
The red dashed lines represent the self-similar scaling for the count rate \rate\ assuming the self-similar mass and redshift trends in the soft-band luminosity $L_{\mathrm{X}}$, namely $\rate\propto\mass^{ B_{L_{\mathrm{X}}} + \Bf}$  with $B_{L_{\mathrm{X}}} = 1$ and $\rate\propto \Ez^2 D_{\mathrm{L}}(\redshift)^{-2}\redshift^{\gammaf}$ in the left rand right panels, respectively.
For plotting the best-fit model, we make an assumption of $\left\langle\ln\hat{\rate}|\mass,\redshift\right\rangle \approx \left\langle\ln\rate|\mass,\redshift\right\rangle$.
}
\label{fig:rate_m_z}
\end{figure*}

Based on the joint modeling, the resulting best-fit relation of the \rate--\mass--\redshift\ relation reads
\begin{multline}
\label{eq:results_rate_m_z}
\left\langle\ln\left(\frac{\rate}{\mathrm{counts}/\mathrm{sec}} \Bigg|\mass,\redshift \right)\right\rangle 
= \ln \left(\ansArate\right) + \\
\left[ \left( \ansBrate \right) + \left( \ansdeltarate \right) \ln\left(\frac{1 + \redshift}{1 + \zpiv}\right) \right] \times
\ln\left(\frac{\mass}{\mpiv}\right) + \\
2 \times \ln\left(\frac{\Ez}{\Ezpiv}\right) + 
\left(\ansgammarate \right) \times 
\ln \left(\frac{1 + \redshift}{1 + \zpiv}\right)
\\
-2\times\ln\left(\frac{D_{\mathrm{L}}\left(\redshift\right)}{D_{\mathrm{L}}\left(\zpiv\right)}\right)
+ \ln\left( \brate\left(\mass, \redshift\right) \right) 
\,  ,
\end{multline}
with the log-normal intrinsic scatter $\sigmarate = \anssigmarate$.
We find that the mass slope ($\Brate = \ansBrate$) is steeper than that obtained in \citet[][$\Brate = 1.58^{+0.17}_{-0.14}$]{chiu22} at a level of $\approx1\sigma$, in which they constrained the \rate--\mass--\redshift\ relation in the weak-lensing mass calibration alone using the latest HSC weak-lensing data \citep{li21}.
Meanwhile, the resulting redshift scaling ($\gammarate = \ansgammarate$) reveals a negative deviation from the self-similar prediction ($\gammarate = 0$) at a level of $\lesssim2\sigma$, which is different from what was found in \citet[][$\gammarate = -0.44^{+0.81}_{-0.85}$]{chiu22}.
We note that the differences of the constraints on the \rate--\mass--\redshift\ relation between this work and \cite{chiu22} are driven by the modeling of the cluster abundance and are not statistically significant due to the large errorbars.
In Figure~\ref{fig:rate_m_z}, we plot the count rate \rate\ of the \eFEDS\ clusters as a function of the ensemble mass (the left panel) and the cluster redshift (the right panel) with the $68\percent$ confidence level of the best-fit model indicated by the grey area.
Note that the grey area, which represents the uncertainty of the best-fit mean relation in equation~(\ref{eq:results_rate_m_z}), is calculated using the chains of the scaling relation parameters marginalizing over the systematics.
When producing the mass (redshift) trend in the left (right) panel, we normalize the observed data to the pivotal redshift $\zpiv=0.35$ (pivotal mass $\mpiv=1.4\times10^{14}\Msunh$) after dividing them by the known redshift (mass) scaling predicted by the best-fit model.
As seen, no sign of bias in the best-fit model is revealed with respect to the data.
Moreover, the mass scaling of the count rate is well described by a single power-law function without any clear feature that might hint at a broken power law at the low-mass end.

We make one remark regarding the mass trend of X-ray observable-to-mass-and-redshift relations, as follow.
In \cite{chiu22}, we constrained various X-ray observable scaling relations, including the X-ray temperature $T_{\mathrm{X}}$, luminosity $L_{\mathrm{X}}$, ICM mass $M_{\mathrm{gas}}$, and $Y_{\mathrm{X}}\equiv T_{\mathrm{X}}\times M_{\mathrm{gas}}$.
Overall, their mass trends were all found to be statistically consistent with or mildly steeper than the self-similar prediction at a level of $\lesssim1.7\sigma$.
On the other hand, the analysis with the same methodology was adopted in deriving the X-ray observation-to-mass-and-redshift relations for a sample of SZE-selected clusters at a similar redshift range of $0.2<\redshift<1.5$ in the SPT-SZ survey \citep[][see also \citealt{chiu18a}]{bulbul19}, where it was found that the mass trends are generally steeper than the self-similar prediction at a level of $\gtrsim3\sigma$.
This mild inconsistency is intriguing, given the fact that the two samples (\eFEDS\ and SPT) are both selected by their ICM properties but span different mass ranges.
However, we stress that the cluster mass \mass\ in \cite{bulbul19} was calibrated in the joint modeling of the weak-lensing mass calibration and the cluster abundance \citep{bocquet19}, as opposed to only the weak-lensing mass calibration as in \cite{chiu22}.
When including the cluster abundance of the \eFEDS\ clusters, as done in this work, we find that the mass trend of the count rate shows a steeper slope than that from the weak-lensing mass calibration alone (at the $\approx1\sigma$ level).
Considering the degeneracy of the mass trends between the count rate and the other X-ray observables \citep{chiu22}, the steepening mass trend of the count rate when including the cluster abundance would pull the mass slope of the X-ray observables toward a higher value. 
This would result in an even steeper slope than the self-similar prediction, bringing  the \eFEDS\ results into better agreement with the SPT sample results \citep{bulbul19}.
A study using a larger sample of clusters in the first-year \eROSITA\ survey with a wide weak-lensing coverage will shed light on this in the future.

The best-fit \rich--\mass--\redshift\ relation at a fixed mass and redshift inferred from the joint modeling yields
\begin{multline}
\label{eq:results_rich_m_z}
\left\langle\ln\rich|\mass,\redshift\right\rangle 
= \ln \left( \ansArich \right) + \\
\left[ \left( \ansBrich \right) + \left( \ansdeltarich \right) \ln\left(\frac{1 + \redshift}{1 + \zpiv}\right) \right] \times
\ln\left(\frac{\mass}{\mpiv}\right) \\
+ \left( \ansgammarich \right)
 \times \ln \left(\frac{1 + \redshift}{1 + \zpiv}\right)
 \, ,
\end{multline}
with the log-normal intrinsic scatter of \anssigmarich.
We find that the mass trend ($\Brich = \ansBrich$) is in good agreement with the self-similar prediction ($\Brich = 0$) with a mild cross scaling with redshift ($\deltarich = \ansdeltarich$) at a level of $\approx1.6\sigma$.
Interestingly, the redshift scaling of the richness reveals a decreasing trend ($\gammarich = \ansgammarich$) at a level of $\lesssim2\sigma$.
The resulting redshift trend is statistically consistent with \citet[][$\gammarich = -0.46^{+0.54}_{-0.51}$]{chiu22} obtained with the weak lensing mass calibration alone at a level of $\approx1\sigma$.
It is worth mentioning that, as seen in Figure~\ref{fig:mcalib_nbc_smf}, the parameters of the \rich--\mass--\redshift\ relation have strong degeneracies with their corresponding parameters in the \rate--\mass--\redshift\ relation, because both sets of parameters are subject to the same weak lensing and cluster abundance constraints.
Note that we do expect a steeper mass trend ($\Brich = \ansBrich$) than the value ($\Brich = 0.881^{+0.077}_{-0.088}$) obtained in \cite{chiu22}, given that in this work we properly account for the Malmquist bias caused by the richness selection in deriving the \rich--\mass--\redshift\ relation \citep[see Appendix~A in][]{chiu22}.
In Figure~\ref{fig:rich_m_z}, we plot the richness \rich\ in terms of the mass and redshift scaling in the left and right panels, respectively.
As seen, the best-fit model (grey area) well describes the data behavior and reveals a weak deviation from the self-similar prediction (red dashed lines) in the redshift dimension.

\begin{figure*}
\centering
\resizebox{0.49\textwidth}{!}{
\includegraphics[scale=1]{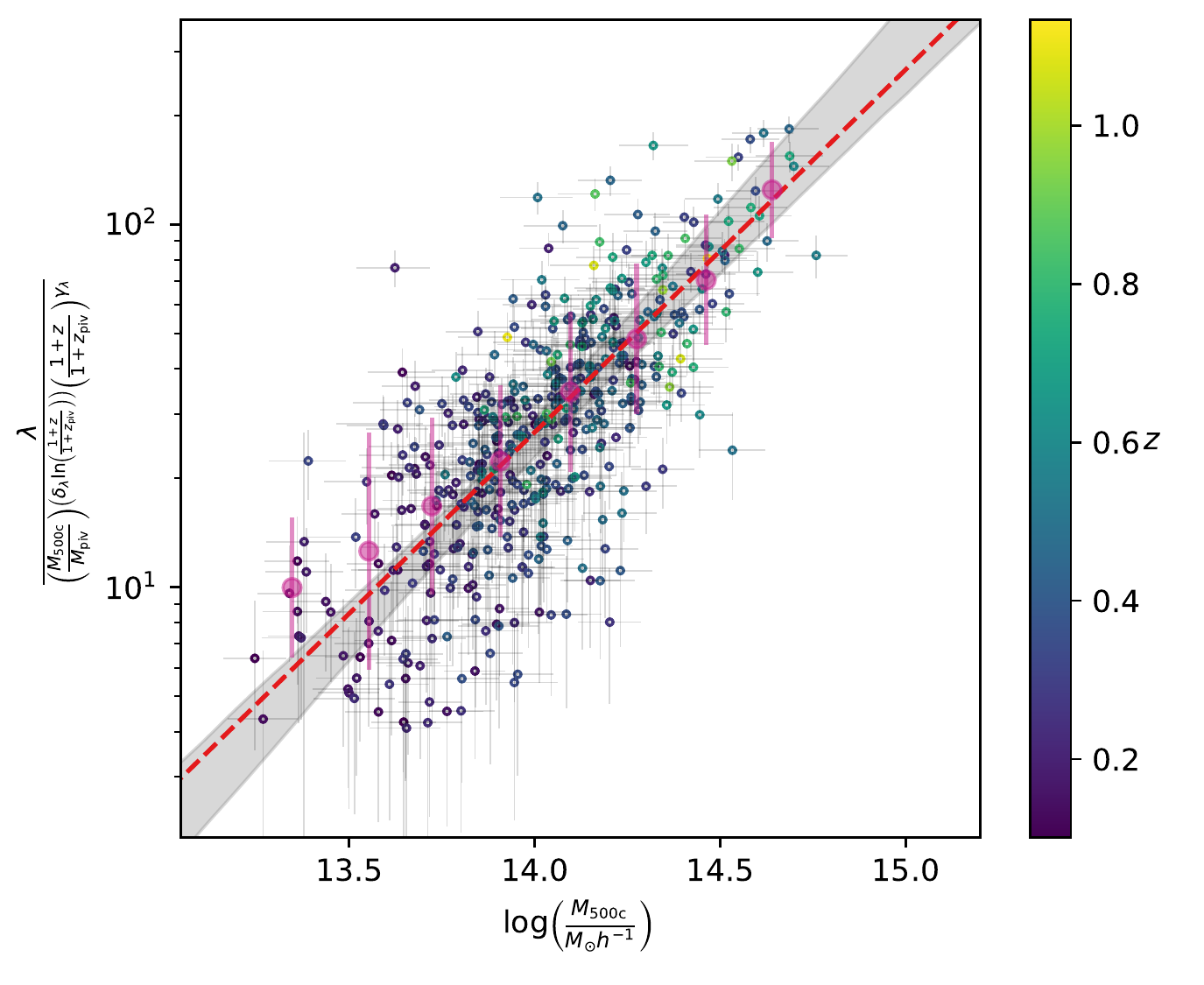}
}
\resizebox{0.49\textwidth}{!}{
\includegraphics[scale=1]{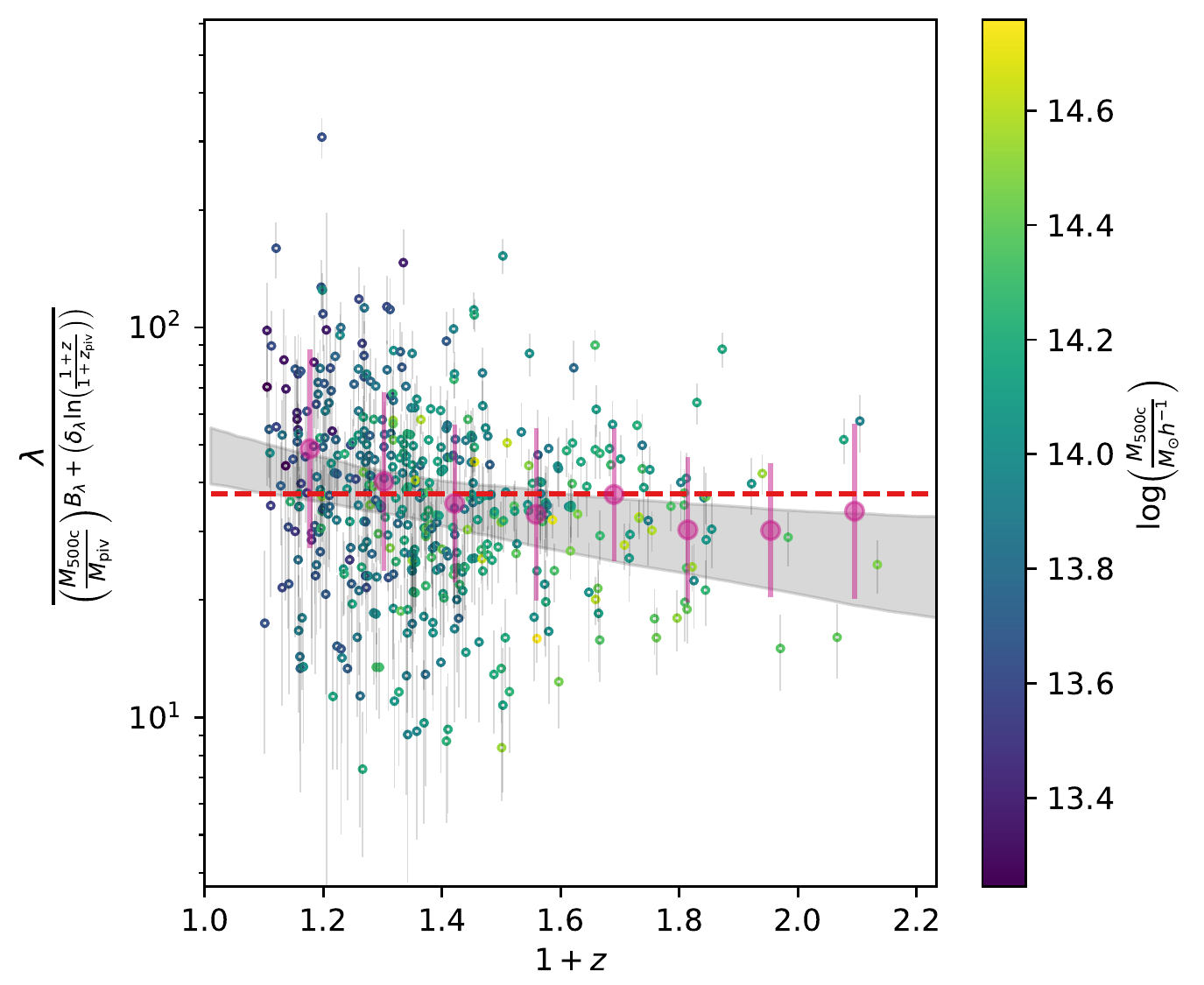}
}
\vspace{-0.2cm}
\caption{
The optical richness of the \eFEDS\ clusters as a function of the cluster halo mass \mass\ (left) and redshift (right).
The same plotting scheme  as in Figure~\ref{fig:rate_m_z} is used.
The red dashed lines represent the self-similar scaling for the richness, namely $\rich\propto\mass^{1}\left(1 + z\right)^{0}$ in the left rand right panels, respectively.
The pink circles are the mean richnesses in eight logarithmic bins of mass and redshift in the left and right panels, respectively.
The errorbars of the pink circles are derived as the standard deviation of the richness distribution within the bin with a weighting factor of $1/\left({\delta_{\rich}}^2 + {\sigmarich}^2\right)$, which includes both the measurement uncertainty and the intrinsic scatter of the richness.
}
\label{fig:rich_m_z}
\end{figure*}

The constraints on the \rate--\mass--\redshift\ and \rich--\mass--\redshift\ relations from the different modeling we have done are tabulated in Table~\ref{tab:sr}.

\begin{figure}
\centering
\resizebox{0.45\textwidth}{!}{
\includegraphics[scale=1]{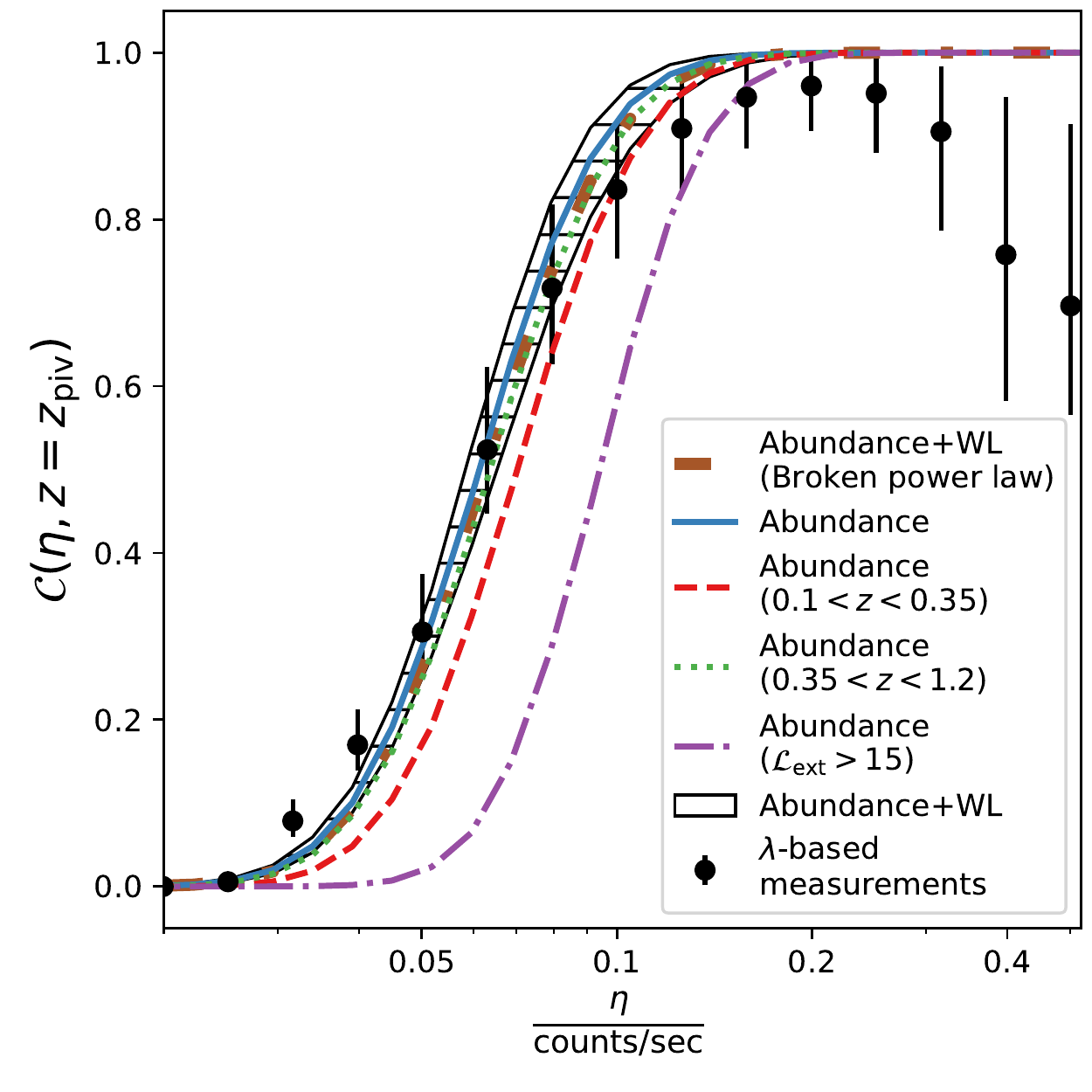}
}
\vspace{-0.5cm}
\caption{
The X-ray completeness function \Comp\ of the count rate at the pivotal redshift $\zpiv = 0.35$.
Without the Gaussian priors applied to $\left(\ratefiveo, \srate\right)$ and with the informative priors on the parameters of the \rate--\mass--\redshift\ relation, the self-calibrated results from the modeling of the cluster abundance alone based on the full, low-\redshift\ and high-\redshift\ samples are shown by the blue solid, red dashed, and green dotted lines, respectively.
The $68\percent$ confidence level of the X-ray completeness from the joint modeling of the weak-lensing mass calibration and the cluster abundance assuming the single power-law scaling ($\Brate = \Bgrp$) of the count rate is presented by the filled interval.
Including the broken power-law feature into the joint modeling leads to the completeness shown by the brown dash-dotted line.
Note that the joint modeling (the filled interval and the brown dash-dotted line) is consistently obtained with the Gaussian priors applied to the parameters of the X-ray completeness function.
The richness-inferred measurements of the completeness (see Section~\ref{sec:completeness_measurements}) are marked by the black circles. 
Restricting the X-ray selection to the criterion of $\Lext>15$ leads to the completeness function shown by the purple dash-dotted line.
}
\label{fig:completeness_model}
\end{figure}
\begin{figure}
\centering
\resizebox{0.45\textwidth}{!}{
\includegraphics[scale=1]{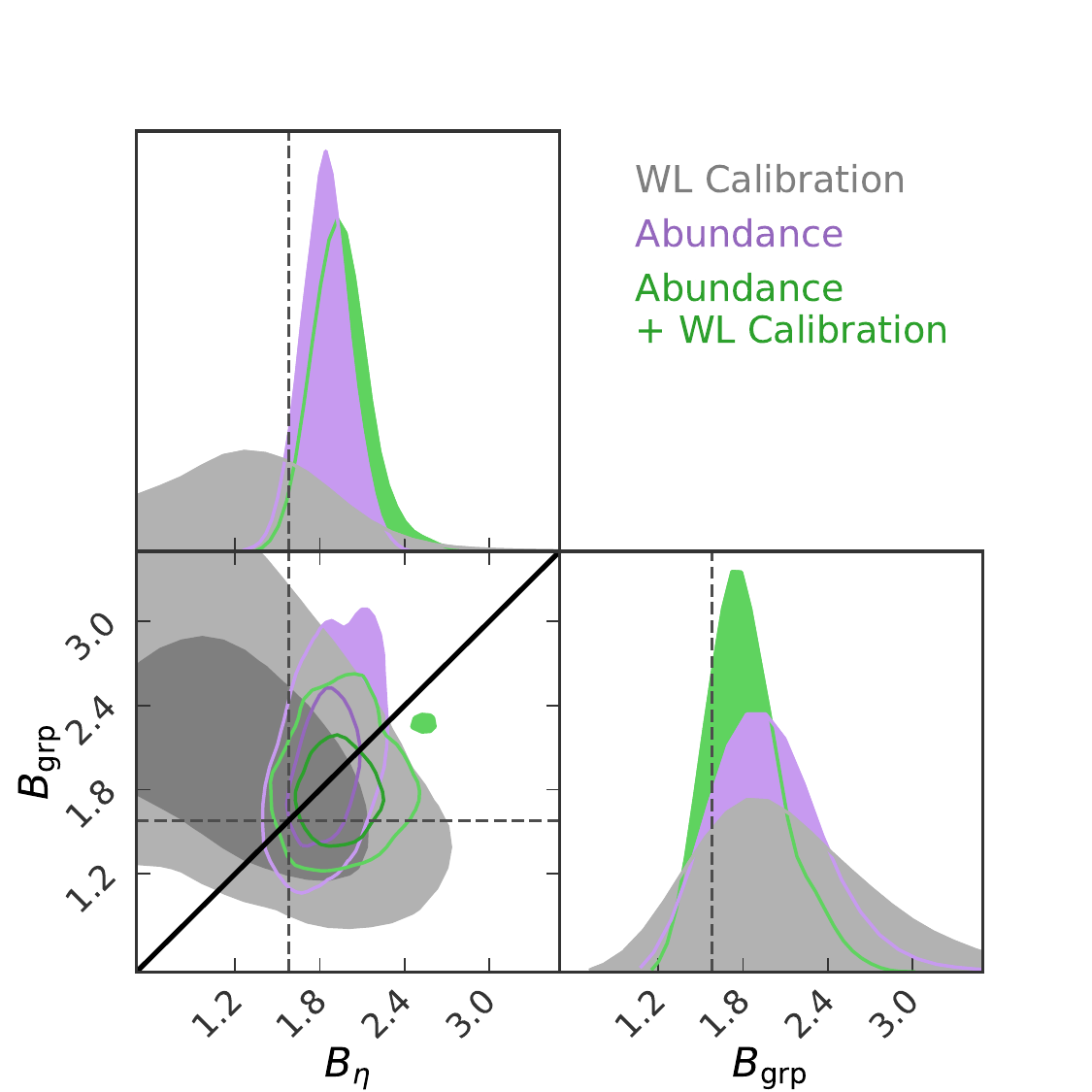}
}
\vspace{-0.2cm}
\caption{
The joint constraints on the mass scaling of \Brate\ and \Bgrp\ at the scales of cluster ($\mass\gtrsim\mpiv$) and groups ($\mass\lesssim\mpiv$), respectively.
The results from the weak-lensing mass calibration, the cluster abundance, and the joint modeling of them are in grey, purple, and green, respectively.
The black solid line represents the single power-law scaling ($\Brate = \Bgrp$) of the \rate--\mass--\redshift\ relation with the mass slope from \citet{chiu22} indicated by the dash lines.
The contours indicate the $68\percent$ and $95\percent$ confidence levels.
No sign of a broken power-law mass scaling of the count rate is revealed.
}
\label{fig:bpl}
\end{figure}
\begin{figure*}
\centering
\resizebox{0.9\textwidth}{!}{
\includegraphics[scale=1]{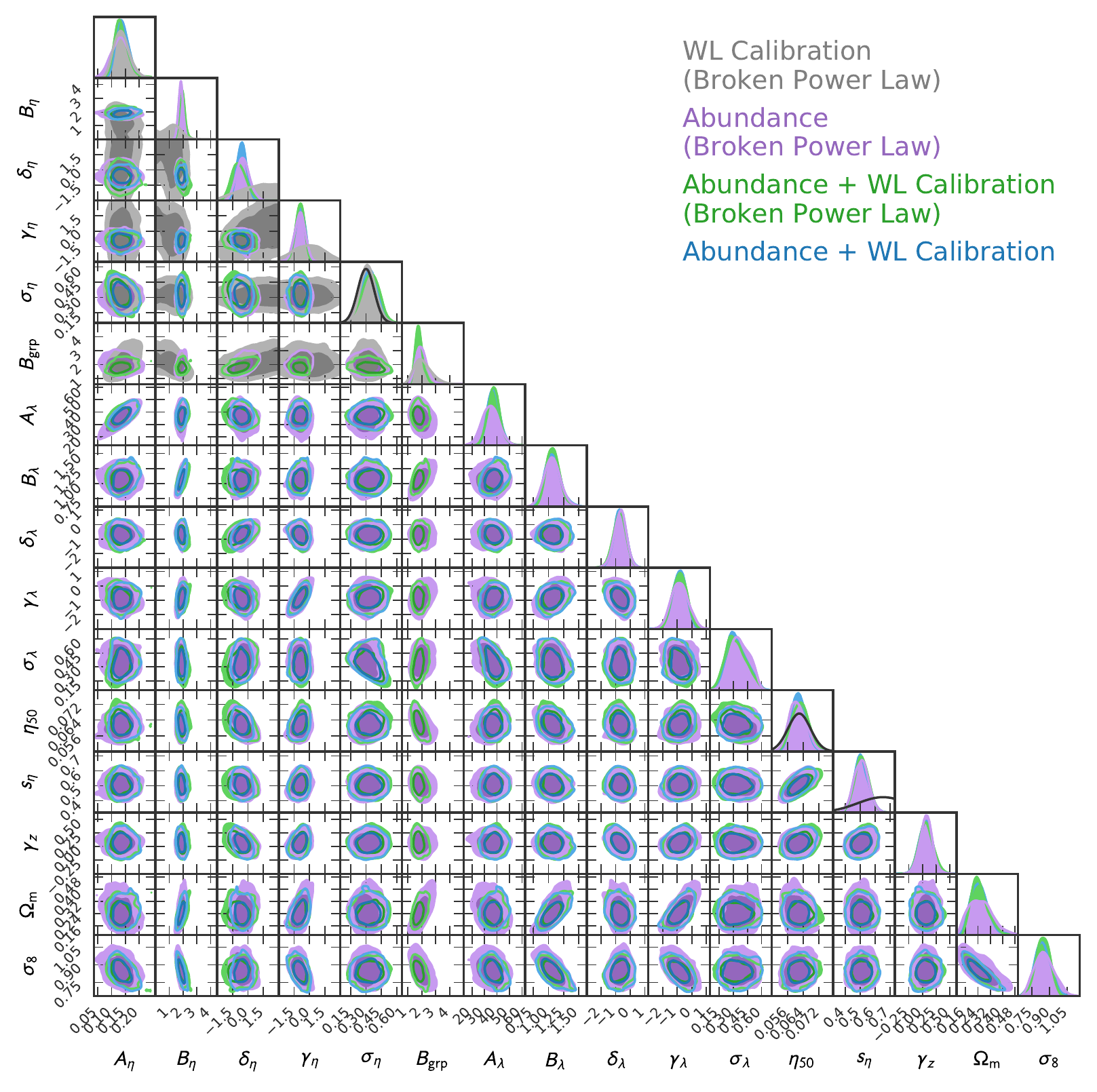}
}
\vspace{-0.3cm}
\caption{
The comparison of the constraints from the modeling of the weak-lensing mass calibration (grey contours), the cluster abundance (purple contours), and the joint modeling of them (green contours) with the broken-power law scaling of the \rate--\mass--\redshift\ relation.
The results from the joint modeling assuming the single power-law scaling of the count rate are shown in blue.
All results shown here are consistently obtained with the Gaussian priors applied to the parameters of the X-ray completeness when including the modeling of the cluster abundance.
For the modeling of the cluster abundance (purple contours), the informative priors are applied to the parameters of the \rate--\mass--\redshift\ relation (see Section~\ref{sec:statistics}).
The posteriors of and the covariance between the parameters are shown in the on-diagonal and off-diagonal plots, respectively.
The Gaussian priors applied to the parameters are shown as the black solid lines in the on-diagonal plots.
The contours indicate the $68\percent$ and $95\percent$ confidence levels.
}
\label{fig:bpl_sr_cosmo}
\end{figure*}

\subsubsection{The goodness of fit}
\label{sec:results_gof}

We quantify the goodness of fit of the best-fit model by comparing the cluster numbers between the model prediction and the observed data in the observable space of the count rate \rate, richness \rich, and the redshift \redshift.
Specifically, we bin each observable dimension (\rate, \rich, \redshift) spanned by the \eFEDS\ sample into $10$ logarithmic bins, resulting in a total $10^{3}$ cells in the three-dimensional space, and assess the consistency of the cluster numbers between the prediction and the data.
Given the nature of Poisson distributions, the modified \cite{cash79} statistic, which is attributed to Castor \citep{kaastra17,baker84}, is used to estimate the goodness of fit in terms of the test statistic $C_{\chi}$.
For a sufficiently large number of events, as in the case of $\approx450$ clusters in this work, the test statistic $C_{\chi}$ has a useful property that allows us to estimate the goodness of fit for a Poissonian model in a manner similar to the $\chi^2$ test for a Gaussian distribution.

Adopting the best-fit parameters obtained in the joint modeling 
with a single power-law \rate--\mass--\redshift\ relation, the test statistic expects a mean value $\left\langle C_{ \chi } \right\rangle$ and variance ${\Delta C_{\chi} }^2$ of $\left\langle C_{\chi} \right\rangle =  328.57$ and ${\Delta C_{ \chi } }^2 =  23.02^2$, respectively.
Meanwhile, the data statistic with respect to the best-fit model has a value $C_{\mathrm{data}} = 337.47$, which is in excellent agreement with the Gaussian distribution of $\mathcal{N}\left(\left\langle C_{\chi} \right\rangle, { \Delta C_{\chi} }^2\right)$.
Assuming the \wCDM\ cosmology or including the broken power-law feature in the \rate--\mass--\redshift\ relation both give goodness of fit in agreement with the expectations at a similar level.
Based on this test statistic, we conclude that the best-fit model provides an excellent description of the \eFEDS\ dataset.

\subsubsection{The X-ray completeness}
\label{sec:results_completeness}

The X-ray completeness \Comp(\rate,\redshift) of a count rate select sample is presented and measured in Section~\ref{sec:completeness}, and its parameters are also self-calibrated in the modeling of the cluster abundance 
using different samples.
We find that the self-calibrated completenesses obtained from the full, low-\redshift\ and high-\redshift\ samples are all in excellent agreement with the directly measured completeness that employs a method based on the cluster richness distributions.
No redshift dependence of the X-ray completeness is preferred by the data.
Moreover, the modeling results on the cosmology and scaling relations with and without applying the Gaussian priors, which are inferred using the richness-based direct completeness measurements, to the parameters of the completeness function are fully consistent.
This is strong evidence that our empirical modeling of the X-ray selection is robust.

We present the resulting completeness of the count rate in Figure~\ref{fig:completeness_model}.
Based on the joint modeling of the weak-lensing mass calibration and the cluster abundance with the Gaussian priors on $\left(\ratefiveo, \srate\right)$, we obtain
\begin{eqnarray}
\label{eq:results_completeness_constraints}
\frac{\ratefiveo}{ \mathrm{counts}/\mathrm{sec} }   &=  & \ansratefiveo  \, ,  \\
\srate         &=  & \anssrate       \, ,  \\
\gammaz   &=  & \ansgammaz  \, , 
\end{eqnarray}
suggesting that the $50\percent$ completeness at the pivotal redshift $\zpiv = 0.35$ occurs at the count rate of 
$\ratefiveo \approx 0.061~\mathrm{counts}/\mathrm{sec}$.
This result (the filled region) is consistent with those obtained by the self-calibration based on the full (the blue line), low-\redshift\ (the red dashed line), and high-\redshift\ (the green dotted line) samples.
Allowing the broken power-law feature (the brown dash-dotted line) in the mass scaling of the \rate--\mass--\redshift\ relation does not significantly change the result.
These constraints are all in agreement with the measurements (the black points).

Increasing the extent likelihood selection ($\Lext > 15$) in the X-ray selection results in the completeness model characterized by $\left(\ratefiveo, \srate, \gammaz\right) = \left( \ansratefiveoHghlext, \anssrateHghlext, \ansgammazHghlext\right)$, corresponding to a count rate where the $50\percent$ completeness occurs at $\ratefiveo \approx 0.094~\mathrm{counts}/\mathrm{sec}$.

\begin{figure}
\centering
\resizebox{0.45\textwidth}{!}{
\includegraphics[scale=1]{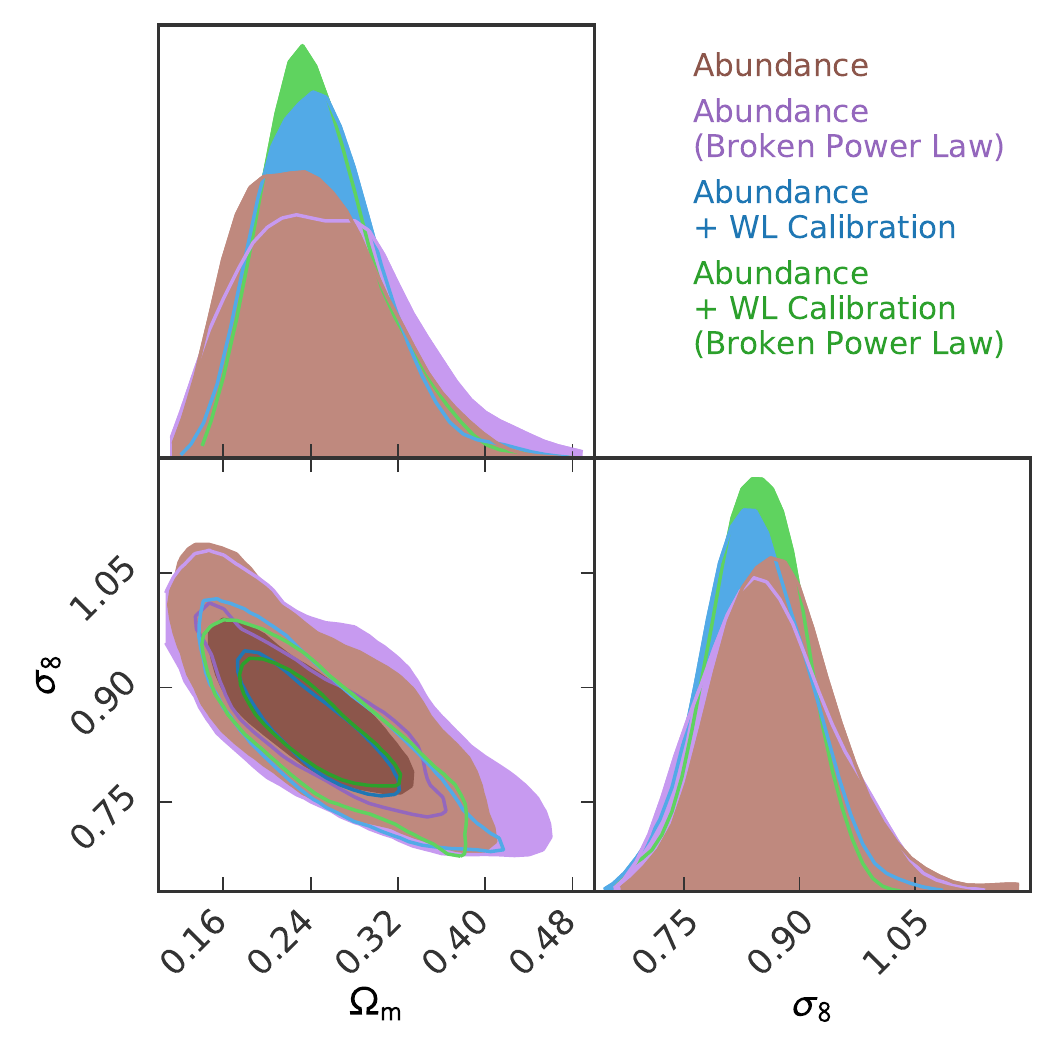}
}
\caption{
The constraints on \OmegaM\ and \sigmaeight\ obtained from the modeling of the cluster abundance and that jointly with the weak-lensing mass calibration.
The results based on the cluster abundance (the joint modeling) with and without the broken power-law scaling of the \rate--\mass--\redshift\ relation are in purple and brown (green and blue), respectively.
For the modeling of the cluster abundance (brown and purple contours), the informative priors are applied to the parameters of the \rate--\mass--\redshift\ relation (see Section~\ref{sec:statistics}).
The contours indicate the $68\percent$ and $95\percent$ confidence levels.
}
\label{fig:bpl_cosmo_om_sigma8}
\end{figure}

\subsubsection{The broken power-law feature of the \rate--\mass--\redshift\ relation}
\label{sec:results_bpl}

The broken power law of the \rate--\mass--\redshift\ relation with a steeper mass scaling at the low-mass end ($\Bgrp>\Brate$ at $\mass\lesssim\mgrp$) results in a deficit of the cluster number at the low-\Lext\ end, which could resolve the discrepancy between the data and the \eFEDS\ simulations (see discussions in Section~\ref{sec:brokenpowerlaw}).
In Figure~\ref{fig:bpl}, we present the results allowing the broken power-law feature in the modeling of the 
weak-lensing mass calibration (grey), the cluster abundance (purple), and the joint modeling (green).
As seen, they are in excellent agreement with each other and the solid line of $\Brate=\Bgrp$, providing no 
suggestions of the need for a broken power-law feature at the low-mass end.
This implies that the deficit of the \eFEDS\ clusters with respect to the simulations \citep{liuteng21} at the 
low-\Lext\ end cannot be resolved by the different mass scalings of the count rate on the group mass scale.

Allowing the broken power-law \rate--\mass--\redshift\ relation does not significantly change the overall modeling of the scaling relations and the X-ray completeness.
This is visualized in Figure~\ref{fig:bpl_sr_cosmo}, where the results from the modeling of the weak-lensing mass calibration (grey), cluster abundance (purple), and the joint modeling with (green) and without (blue) the broken power-law scaling of the count rate are in excellent agreement.
Moreover, as shown in Figure~\ref{fig:bpl_cosmo_om_sigma8}, the constraints on the cosmological parameters \OmegaM\ and \sigmaeight\ are insensitive to the broken power-law mass scaling of the count rate.
That is, the inclusion of the broken power-law mass scaling of the count rate has no significant impact on the cosmological constraints.

\begin{table*}
\centering
\caption{
The posteriors of the cosmological parameters, \OmegaM, \sigmaeight, \seight, and \w.
The parameter \seight\ is defined as $\seight\equiv \sigmaeight\left(\OmegaM/0.3\right)^{0.3}$ in this work.
The column (1) indicates the parameters.
The columns (2) and (3) contain the results solely based on the cluster abundance obtained with and without the Gaussian priors applied to the parameters of the completeness, $\left(\ratefiveo, \srate\right)$, respectively, assuming the single power-law mass scaling of the count rate.
The column (4) records the constraints obtained with the Gaussian priors and the broken power-law mass scaling of the count rate in the modeling of the cluster abundance alone.
The columns (2) to (4) are obtained with the weak-lensing informed priors on the parameters of the \rate--\mass--\redshift\ relation.
The columns (5) and (6) represent the results of the \LCDM\ cosmology from the joint modeling assuming the \rate--\mass--\redshift\ relation assuming the single and broken power-law mass scaling of the count rate, respectively.
The columns (7) and (8) contains the results obtained in the \wCDM\ model with the same configuration as in the columns (5) and (6), respectively.
The columns (5) and (7) are our fiducial results in the \LCDM\ and \wCDM\ models, respectively.
}
\label{tab:cosmology}
\resizebox{\textwidth}{!}{
\begin{tabular}{cccccccc}
\hline
\multirow{3}{*}{Parameters} 
&\multicolumn{3}{c}{Cluster Abundance}
&\multicolumn{4}{c}{Cluster Abundance + WL Calibration} \\
\cmidrule(lr){2-4}\cmidrule(lr){5-8}
&Without Comp. Prior &With Comp. Prior & With Comp. Prior
&\multicolumn{2}{c}{\LCDM} &\multicolumn{2}{c}{\wCDM} \\
&and $\Brate = \Bgrp$ & and $\Brate = \Bgrp$ & and $\Brate \neq \Bgrp$ &$\Brate = \Bgrp$ & $\Brate \neq \Bgrp$& $\Brate = \Bgrp$ & $\Brate \neq \Bgrp$ \\
(1) &(2) &(3) &(4) &(5) &(6) &(7) &(8) \\
\hline
		$\Omega_{\rm{m}}$ & $0.229^{+0.065}_{-0.062}$ & $0.230^{+0.063}_{-0.069}$ & $0.230^{+0.092}_{-0.064}$ & $0.245^{+0.048}_{-0.058}$ & $0.230^{+0.060}_{-0.043}$ & $0.234^{+0.048}_{-0.070}$ & $0.228^{+0.061}_{-0.051}$ \\ [3pt]
		$\sigma_8$ & $0.848^{+0.092}_{-0.072}$ & $0.867^{+0.073}_{-0.082}$ & $0.843^{+0.090}_{-0.081}$ & $0.833^{+0.075}_{-0.063}$ & $0.847\pm 0.061$ & $0.846^{+0.092}_{-0.066}$ & $0.838^{+0.075}_{-0.055}$ \\ [3pt]
		$S_8$ & $0.788^{+0.050}_{-0.035}$ & $0.792^{+0.049}_{-0.036}$ & $0.788^{+0.055}_{-0.039}$ & $0.791^{+0.028}_{-0.031}$ & $0.796^{+0.026}_{-0.032}$ & $0.784^{+0.034}_{-0.027}$ & $0.790^{+0.023}_{-0.032}$ \\ [3pt]
		$w$ & -- & -- & -- & -- & -- & $-1.25\pm 0.47$ & $-1.13^{+0.47}_{-0.41}$
 \\
\hline
\end{tabular}
}
\end{table*}
\begin{figure}
\centering
\resizebox{0.49\textwidth}{!}{
\includegraphics[scale=1]{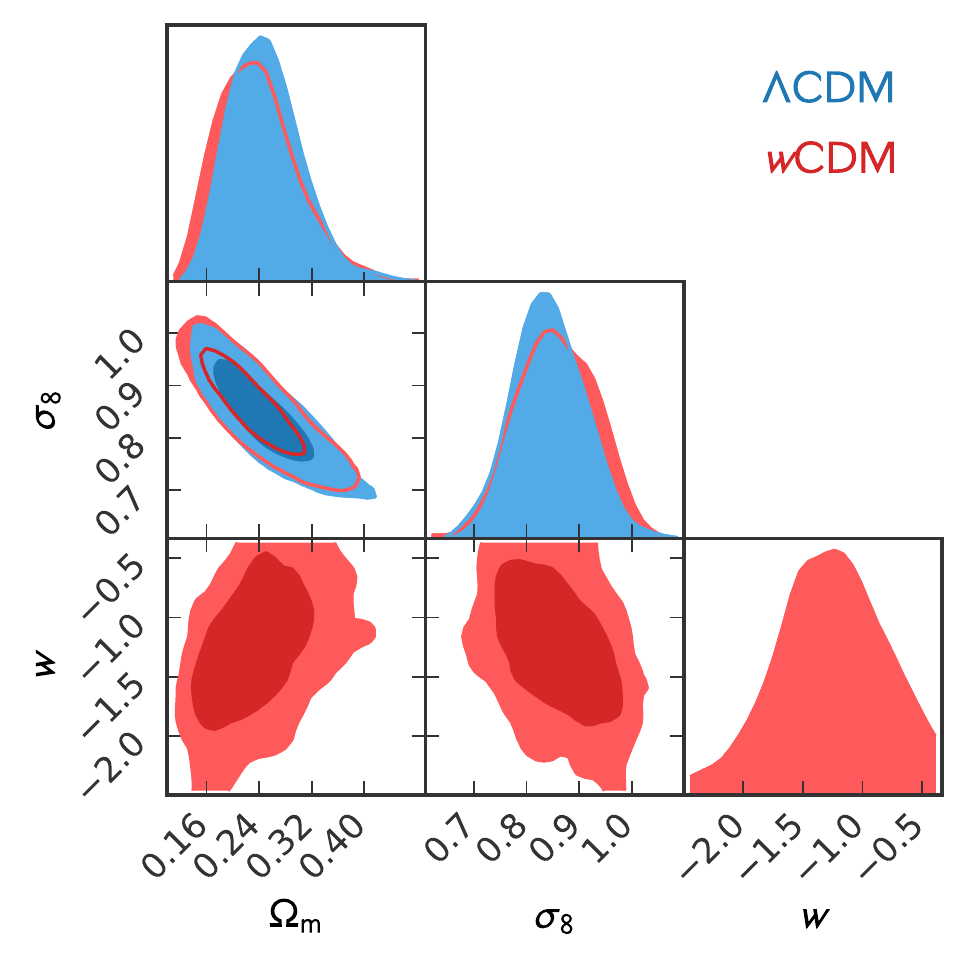}
}
\vspace{-0.5cm}
\caption{
The cosmological constraints from the \eFEDS\ clusters in the \LCDM\ (blue) and \wCDM\ (red) models.
These constraints are obtained in the joint modeling of the weak-lensing mass calibration and the cluster abundance with the single power-law mass scaling of the count rate and with the Gaussian priors applied to the parameters of the X-ray completeness.
The contours indicate the $68\percent$ and $95\percent$ confidence levels.
}
\label{fig:cosmos_efeds_only}
\end{figure}

\subsection{Cosmological constraints}
\label{sec:results_cosmos}

In Sections~\ref{sec:results_mcalib} to~\ref{sec:results_joint}, we present the constraints on the scaling relations assuming the \LCDM\ model.
In a blinding analysis, we also examine the impact raised from (1) the empirical modeling of the X-ray completeness and (2) the broken power-law mass scaling in the \rate--\mass--\redshift\ relation.
Neither of these elements affects the cosmological constraints.
Therefore, we use the modeling with the Gaussian priors applied to the parameters of the X-ray completeness, $\left(\ratefiveo, \srate\right)$, and with the single power-law \rate--\mass--\redshift\ relation as the baseline analysis in this work.

In what follows, we present the cosmological constraints with the extension to the \wCDM\ model.
Based on the sample of $455$ \eROSITA\ clusters selected in the \eFEDS\ survey, the cosmological constraints obtained in the joint modeling of the weak-lensing mass calibration and the cluster abundance are visualized in Figure~\ref{fig:cosmos_efeds_only}.
The constraints on the cosmological parameters of $\left(\OmegaM, \sigmaeight, \seight, \w\right)$ obtained in the different modeling are tabulated in Table~\ref{tab:cosmology}.
In this work, the quantity \seight\ for the \eFEDS\ survey is defined as 
\begin{equation}
\label{eq:s8_def}
\seight\equiv\sigmaeight\left(\frac{\OmegaM}{0.3}\right)^{\alpha} \, ~\mathrm{with}~\alpha \equiv 0.3 \, .
\end{equation}
\begin{figure*}
\centering
\resizebox{0.44\textwidth}{!}{
\includegraphics[scale=1]{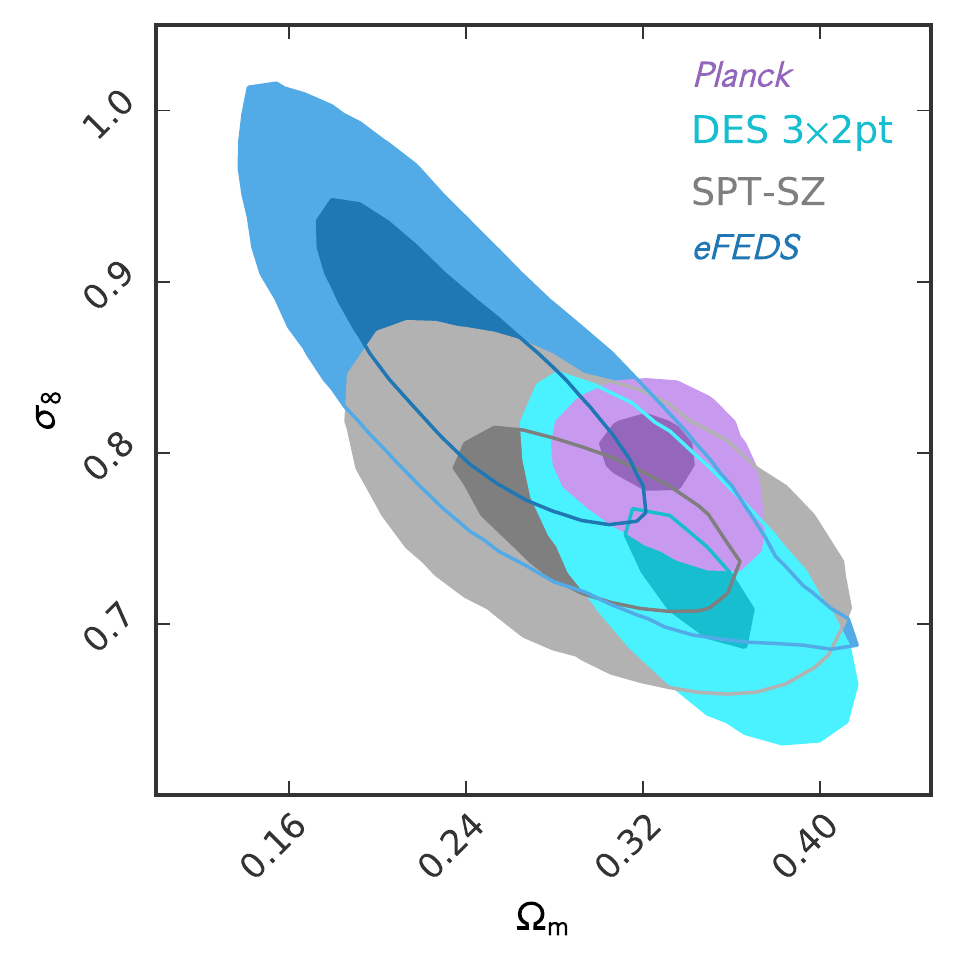}
}
\resizebox{0.46\textwidth}{!}{
\includegraphics[scale=1]{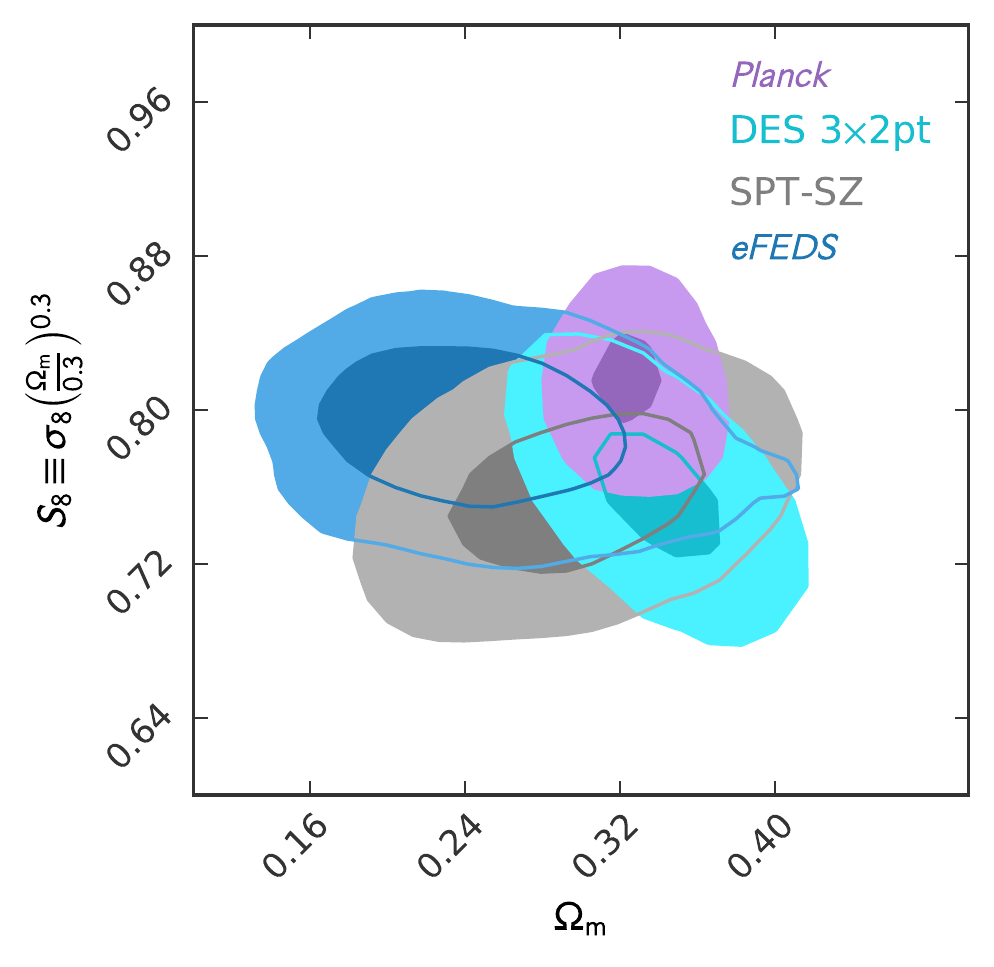}
}
\vspace{-0.15cm}
\caption{
The comparisons of the cosmological parameters assuming the \LCDM\ cosmology between the \eFEDS\ clusters (blue) and the external results, including 
the anisotropy and polarization ($\mathtt{TTTEEE}+\mathtt{lowE}$) of CMB temperatures from \PLANCK\ \citep[purple;][]{PlanckCollaboration20}, 
the 3$\times$2-point analysis from the Dark Energy Survey \citep[cyan;][]{desy3kp-3x2}, and
the clusters in the SPT-SZ survey \citep[grey;][]{bocquet19}.
In the left (right) panel, the constraints on \OmegaM\ and \sigmaeight\ ($S_{8}\equiv\sigmaeight\left(\OmegaM/0.3\right)^{0.3}$) are shown.
The contours indicate the $68\percent$ and $95\percent$ confidence levels.
The \eFEDS\ results are in agreement with the external constraints at a level of $\lesssim1.2\sigma$.
}
\label{fig:lcdm_with_externals}
\end{figure*}

\subsubsection{The \LCDM\ cosmology}

In the \LCDM\ model, we obtain 
\begin{eqnarray}
\label{eq:lcdm_constraints}
\OmegaM        &=      &\OmegamLCDM       \, ,  \\[3pt]
\sigmaeight     &=      &\sigmaeightLCDM    \, , \\[3pt]
\seight            &=      &\seightLCDM    \, ,
\end{eqnarray}
from the joint modeling of the weak-lensing mass calibration and the cluster abundance with the Gaussian priors applied to $\left(\ratefiveo, \srate\right)$ and with the single power-law \rate--\mass--\redshift\ relation.
Including the broken power-law \rate--\mass--\redshift\ relation leads to a fully consistent result with 
$\OmegaM = \OmegamLCDMBPL$, $\sigmaeight = \sigmaeightLCDMBPL$, and $\seight = \seightLCDMBPL$.
Excluding the weak-lensing mass calibration, the cluster abundance alone yields the constraints on $\left(\OmegaM, \sigmaeight, \seight\right)$ that are statistically consistent with the fiducial result, as seen in Table~\ref{tab:cosmology}.
This is not surprising, given that the constraints are dominated by the cluster abundance.

In Figure~\ref{fig:lcdm_with_externals}, we compare the constraints on \OmegaM, \sigmaeight, and \seight\ with those from other methods and surveys, including the anisotropy and polarization of the CMB temperature from the \PLANCK\ mission \citep[the \texttt{TTTEEE+lowE} constraints;][]{PlanckCollaboration20}, the galaxy-galaxy clustering and lensing (i.e., 3$\times$2-point analysis) from the Dark Energy Survey \citep[DES;][]{desy3kp-3x2}, and the state-of-the-art cluster constraints from the SPT-SZ survey \citep{bocquet19}.
The \eFEDS\ constraints are in good agreement\footnote{The (in)consistency is calculated using the chains of parameters by the code that is available in \url{https://github.com/SebastianBocquet/PosteriorAgreement} \citep[see also][for more details]{bocquet19}.} with all of these other results: the consistency with the results from \PLANCK, DES, and SPT are at levels of $1.2\sigma$, $1.0\sigma$, and $0.6\sigma$, respectively.
Our results based on the \eFEDS\ clusters do not provide evidence for the ``\sigmaeight\ tension'' with the CMB constraint from \PLANCK.

\begin{figure*}
\centering
\resizebox{0.46\textwidth}{!}{
\includegraphics[scale=1]{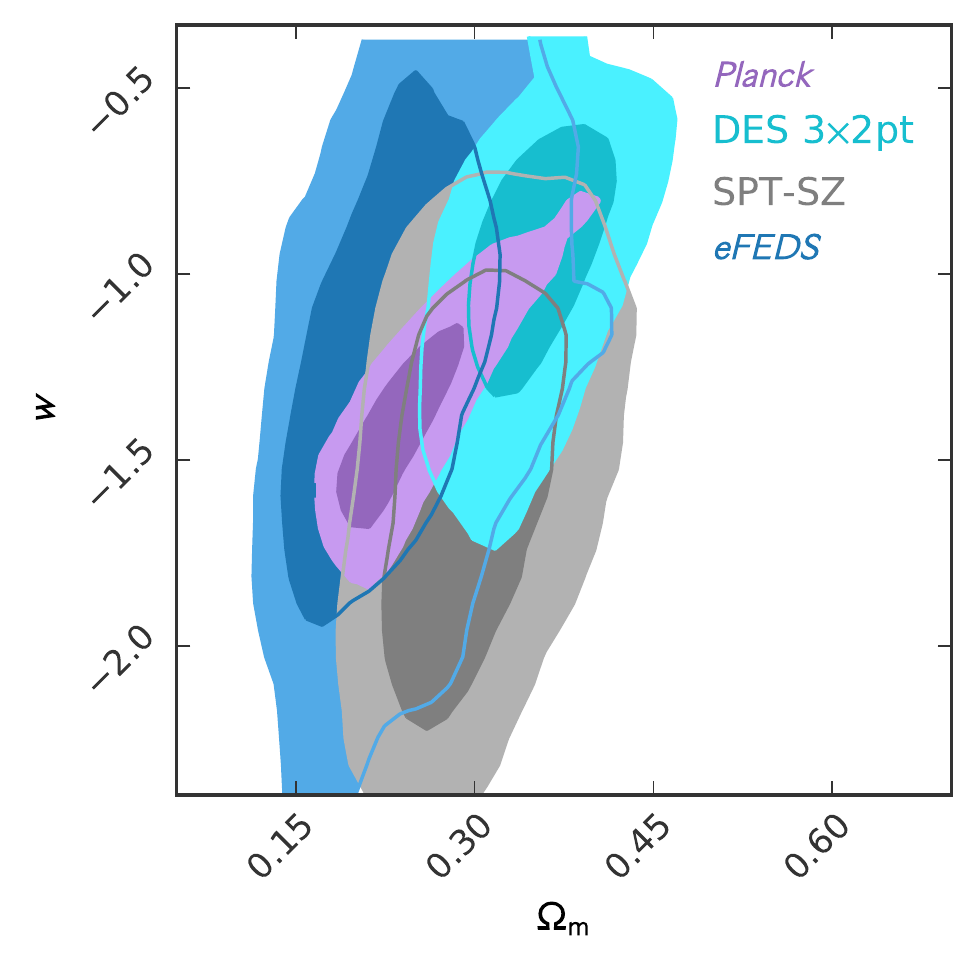}
}
\resizebox{0.46\textwidth}{!}{
\includegraphics[scale=1]{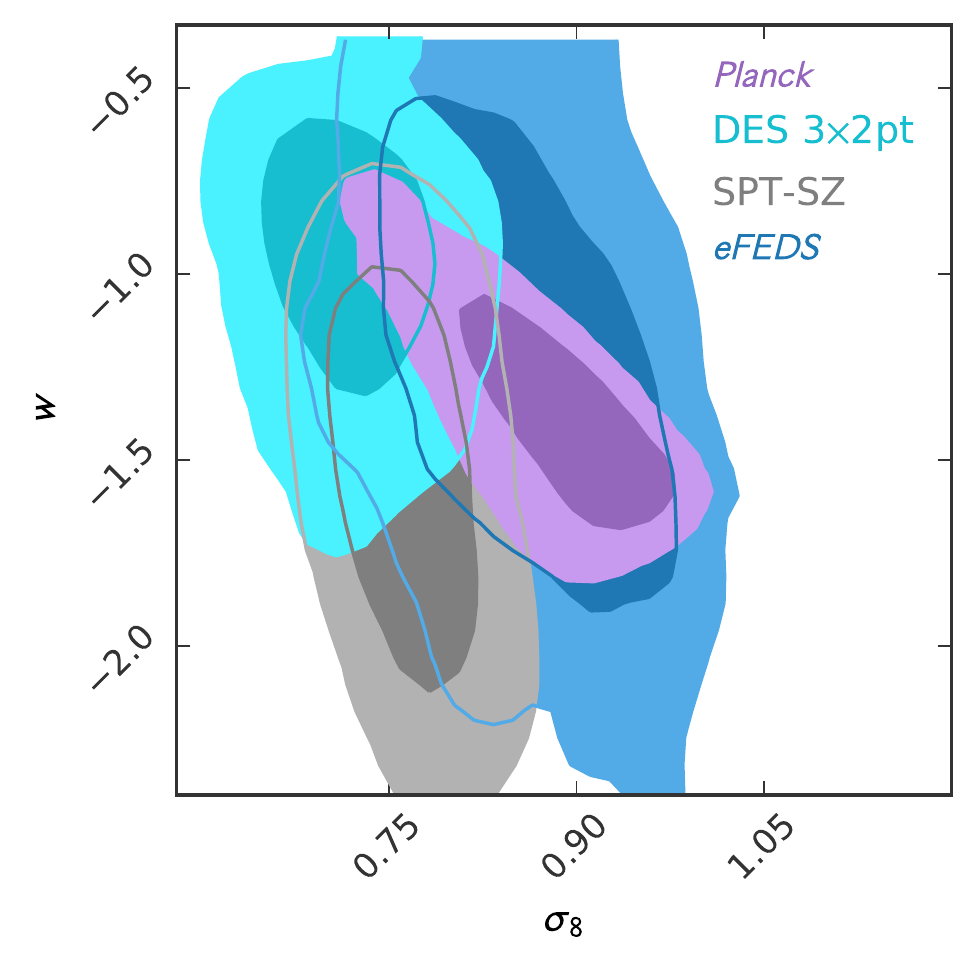}
}
\vspace{-0.15cm}
\caption{
The comparisons of the cosmological parameters \OmegaM, \sigmaeight, and \w\ in the \wCDM\ cosmology between the \eFEDS\ clusters (blue) and the external results with the same color codes as in Figure~\ref{fig:lcdm_with_externals}.
In the left (right) panel, the constraints in the parameter space spanned by \OmegaM\ (\sigmaeight) and \w\  is shown.
The contours indicate the $68\percent$ and $95\percent$ confidence levels.
The \eFEDS\ results are in agreement with the external constraints at a level of $\lesssim1\sigma$.
}
\label{fig:wcdm_with_externals}
\end{figure*}

\subsubsection{The \wCDM\ cosmology}

Now we turn to the constraints on the \wCDM\ model.
The results based on the joint modeling of the cluster abundance and the weak-lensing mass calibration in the \eFEDS\ survey yields
\begin{eqnarray}
\label{eq:wcdm_constraints}
\OmegaM        &=      &\OmegamWCDM       \, , \\[3pt]
\sigmaeight     &=     &\sigmaeightWCDM    \, , \\[3pt]
\seight            &=      &\seightWCDM            \, , \\[3pt]
\w                   &=      &\wWCDM                   \, ,
\end{eqnarray}
which are obtained with the Gaussian priors applied to the X-ray completeness parameters 
$\left(\ratefiveo, \srate\right)$ and with the single power-law \rate--\mass--\redshift\ relation.
Allowing the broken power-law mass scaling of the count rate or removing the Gaussian priors 
on $\left(\ratefiveo, \srate\right)$ all leads to a negligible difference in the parameter constraints, as seen in Table~\ref{tab:cosmology}.
It is worth mentioning that the constraints on \OmegaM\ and \sigmaeight\ are insensitive to the inclusion of \w, which is seen in Figure~\ref{fig:cosmos_efeds_only}.
Although the parameter uncertainty is large, the \eFEDS\ constraint on \w\ supports a model where the 
dark energy is a cosmological constant ($\w = -1$).

In Figure~\ref{fig:wcdm_with_externals}, we compare the \eFEDS\ \wCDM\ results with those from the external experiments.
It is clear that there is good consistency with the \PLANCK, DES, and SPT constraints at
 levels of $0.1\sigma$, $1.1\sigma$, and $0.8\sigma$, respectively.
Overall, there is excellent agreement between \eFEDS\ and these external results, clearly demonstrating the success of the empirical modeling in deriving cosmological constraints using a sample of X-ray selected clusters in a synergy with a wide-field weak-lensing survey, for the first time.

\begin{figure*}
\centering
\resizebox{0.9\textwidth}{!}{
\includegraphics[scale=1]{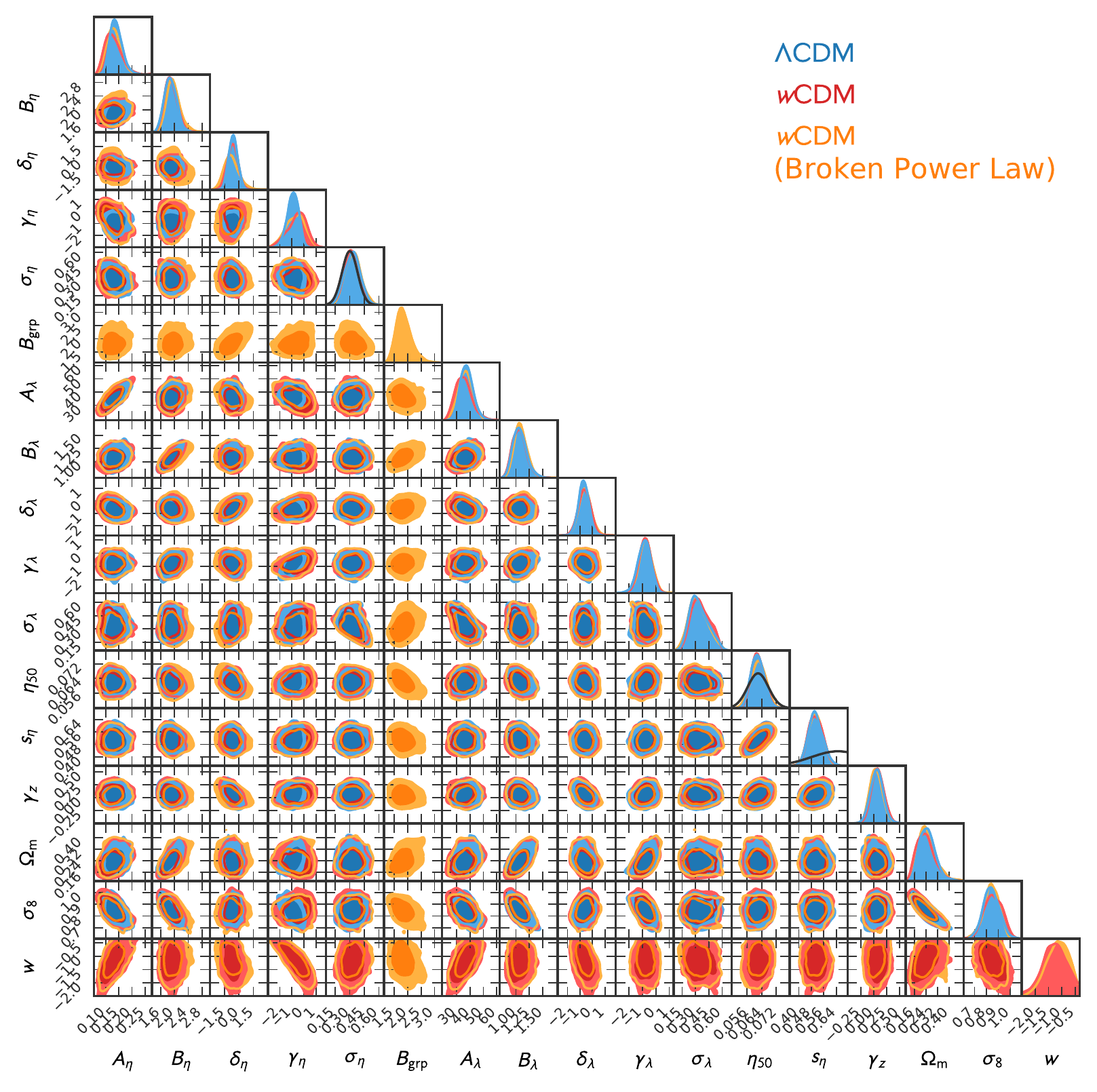}
}
\vspace{-0.5cm}
\caption{
The fully marginalized and joint posteriors of the highlighted parameters obtained the joint modeling of the weak-lensing mass calibration and the cluster abundance.
The results obtained in the \wCDM\ model assuming the single and broken power-law scaling of the \rate--\mass--\redshift\ relation are in red and orange, respectively.
The constraints obtained in the \LCDM\ model with the single power-law scaling of the count rate are in blue.
All these results are derived with the Gaussian priors applied to the parameters  of the completeness function, $\left(\ratefiveo, \srate\right)$.
The contours indicate the $68\percent$ and $95\percent$ confidence levels.
}
\label{fig:wcdm_contours}
\end{figure*}

We note that while the negative deviation from the self-similar redshift trend of the \rate--\mass--\redshift\ relation is observed in the \LCDM\ cosmology ($\gammarate = \ansgammarate$), the redshift scaling of the relation is constrained to be statistically consistent with no deviation ($\gammarate = -0.38^{+0.72}_{-0.82}$) in the \wCDM\ model.
This is due to the fact that there is strong degeneracy between the redshift scaling parameter \gammarate\ and the equation of state of dark energy \w.
This is visualized in Figure~\ref{fig:wcdm_contours}, where we find that \w\ has the strongest degeneracy with \gammarate.
This result suggests that a tight constraint on the redshift trend of the count rate is of critical importance to put a stringent constraint on \w\ \citep[see also][for the similar finding in SZE-selected clusters]{bocquet19}.

%

%
%

\section{Discussions}
\label{sec:discussions}

In this work, we obtain cosmological constraints through an analysis of the optically confirmed subset of the \eFEDS\ X-ray selected cluster sample, with support from weak lensing mass calibration carried out using the HSC Subaru Strategy Program S16A dataset.
Leveraging prior knowledge of structure formation from theory, simulations, and observations accumulated over the last few decades, we adopt flexible functional forms in the X-ray, optical and weak lensing observable-to-mass-and-redshift relations, and then empirically calibrate them within the cosmological analysis.  This empirical calibration ensures that the calibrated relations follow the mass and redshift variations of the actual cluster dataset, and is crucial for minimizing systematic errors in the parameter posteriors.

The X-ray selection is modeled using an empirically determined completeness \Comp(\rate,\redshift) of the X-ray selected sample as a function of count rate and redshift with minimal assumptions informed by the data and simulations.  This approach allows us to bypass the direct modeling of the complex X-ray selection on the quantities of \Lext, \Ldet, and \ext\ that is typically subject to biases due to inconsistencies between the clusters in the real sample and those in image simulations.
We calibrate the X-ray completeness function using 
(1) independent measurements inferred from the richness distribution of the extent selected and full cluster sample (see Section~\ref{sec:completeness_measurements}) and 
(2) a self-calibration approach where the parameters of $\Comp\left(\rate,\redshift\right)$ are allowed to float in the cosmological analysis.  
Both approaches are in good agreement, and the small differences do not significantly affect the cosmological constraints.
In addition, neither the weak-lensing mass calibration nor the cluster abundance implies a mass scaling of the X-ray count rate \rate\ that deviates from a single power law over the considerable mass range of the \eFEDS\ sample.
Validation tests indicate that the empirical modeling provides an accurate description of the data, allowing us to extract cosmological parameter posteriors.

Despite these successes, one issue demands further investigation in a future work.
The characterization of the mapping $P\left(\vect{\Xlabel} | \rate\right)$ between the observed count rate \rate\ and the X-ray observables $\vect{\Xlabel}\equiv\left\lbrace\Lext, \ext,, \Ldet\right\rbrace$.
The former is the X-ray mass proxy adopted in this work and the latter are the X-ray selection observables adopted in defining the initial \eFEDS\ cluster candidate list.
modeling the mapping $P\left(\vect{\Xlabel} | \rate\right)$ is challenging and is sensitive to the detailed pipeline configuration used in the cluster detection.
Given a fixed signal-to-noise ratio of the count rate, denoted as $\snr\equiv\rate/\delta_{\rate}$, we find that there exists an extremely tight power-law relation between \snr\ and \Ldet.
This is expected, because the detection likelihood describes the (log-)probability of a source being a real detection as opposed to a noise fluctuation.
Given the detection likelihood \Ldet, the \eROSITA\ pipeline additionally proceeds the estimation of the extent likelihood \Lext\ and the extent \ext.
There exhibits a strong and positive correlation between \Lext\ and \Ldet\ with a considerable amount of scatter in \Lext\ at a fixed \Ldet\ \citep[see Figure~16 in][]{klein22}.
Moreover, the scatter of \Lext\ at a fixed \Ldet\ is strongly driven by the extent \ext.
Specifically, at a fixed \Ldet, a cluster with higher extent \ext\ will also exhibit a higher extent likelihood \Lext.
This is again expected, because a higher \ext\ implies a more extended source and naturally a higher \Lext.

Therefore, it is possible to establish a relation between $\vect{\Xlabel}$ and \snr\ via a two-step empirical modeling, namely
$P\left(\vect{\Xlabel}| \snr \right) \equiv P\left( \Lext, \ext, \Ldet | \snr \right) \approx P\left( \Lext, \ext | \Ldet \right) P(\Ldet | \snr)$, in which $P\left( \Lext, \ext | \Ldet \right)$ characterizes the correlation between \Lext\ and \ext\ at a fixed \Ldet.
Note that  in the above equation we assume that the scatter of \Lext\ and \ext\ at a fixed \Ldet\ does not correlate with \snr.
That is, $P\left( \Lext, \ext | \Ldet, \snr \right) \approx P\left( \Lext, \ext | \Ldet \right)$.
This assumption is physically intuitive, meaning that the extendedness of a source does not depend on the signal-to-noise ratio of the detection, to first order.
Since we can empirically model the relation between \snr\ and \rate, in which they are related via the measurement uncertainty $\delta_{\rate}$, we can then model the mapping between $\vect{\Xlabel}$ and \rate\ as 
\begin{eqnarray}
P\left(\vect{\Xlabel} | \rate\right)  &\equiv &\int\dif \snr P\left( \Lext, \ext, \Ldet | \snr \right) P\left(\snr|\rate\right) \nonumber \\
&\approx &P\left( \Lext, \ext | \Ldet \right)  P\left(\Ldet|\rate\right) \, , \nonumber
\end{eqnarray}
where $P\left(\Ldet|\rate\right) \equiv \int\dif \snr  P(\Ldet | \snr) P\left(\snr|\rate\right)$.
This establishes the direct mapping between $\vect{\Xlabel}$ and \rate.
Moreover, given the well behaved relation between count rate \rate\ and halo mass \mass\ at each redshift \redshift, it is possible to construct a direct mapping between $\vect{\Xlabel}$ and the quantities of mass and redshift.

The modeling of $P(\Ldet | \snr)$ is straightforward, by design.
However, the characterization of $P\left( \Lext, \ext | \Ldet \right)$ is challenging due to the scatter of \Lext\ and \ext, depending on the configuration of the \eROSITA\ source detection.
We find that the bivariate distribution of $\left(\Lext, \ext\right)$ at a fixed \Ldet\ contains a mix of at least two components, each with different patterns of the scatter.
The first pattern represents the population of the clusters with $\ext$ smaller than the upper limit of the detection, i.e., $\ext < 60\arcsec$.
For such a population, the relation between \Lext\ and \ext, which causes the scatter of \Lext\ at a fixed \Ldet, does not follow a behavior that can be easily parameterized.
This is consistent with \citet[][see their Figure~9]{clerc18}, where they did not find a well characterized relation between \ext\ and \Lext.
The other pattern is attributed to the population with $\ext = 60\arcsec$ forced by the detection pipeline.
In this population, the shape of the ICM profile is essentially fixed in the source detection regardless of the actual angular size of the cluster.
This results in an upper limit of \Lext\ at a fixed \Ldet, denoted as ${\Lext}_{,\mathrm{upper}}(\Ldet)$, and the characterization of \Lext\ and \Ldet\ for the cluster is determined along the ${\Lext}_{,\mathrm{upper}}(\Ldet)$--\Ldet\ relation.
The upper limit of \ext\ alters the ``native'' relation between \ext\ and \Lext\ that could have been estimated with $\ext>60\arcsec$.
Moreover, a combination of these two populations ($\ext<60\arcsec$ and $\ext = 60\arcsec$) is challenging to model and hence requires the modeling based on either intensive simulations or an empirical method as done in this work.
As mentioned in Section~\ref{sec:brokenpowerlaw}, the simulation-based modeling 
\citep[as in][]{liu21,liuteng21,bulbul21} requires accurate and complete information on cluster physics 
and, inevitably, inherits systematic uncertainties due to the lack of realism in the simulations.
Any discrepancy seen between the data and simulations (as the case for the \eFEDS\ simulations which over-predict the low-\Lext\ clusters) would pose a concern that the associated cosmological posteriors would be biased.
In this work, we therefore choose the approach of empirical modeling and calibration informed by the data themselves.

Recall that the quantity \ext\ acts as a nuisance parameter to maximize \Lext\ in the source detection rather than an actual measurement of the core radius of the ICM profile.
Therefore, we do not expect \ext\ to contain much information about the cluster ICM distribution.
In fact, this is in line with the fact that no well behaved relation between \ext\ and other X-ray properties is seen for the \eFEDS\ clusters \cite[see also][for simulation validations]{clerc18,liuteng21}.
This is also supported by our results (in Section~\ref{sec:results_completeness}) that the X-ray completeness does not reveal a sign of redshift dependence, suggesting that (1) \ext\ is nearly insensitive to the actual angular size of the core radius, which is expected to be inversely proportional to the angular diameter distance, $\theta\propto {D_{\mathrm{A}}}^{-1}$, and that (2) the  detectability of a source mainly depends on the count rate.
Moreover, it is worth mentioning that the maximization of \Lext\ (through three free parameters, R.A., Dec., and \ext) in X-ray imaging is in analogy to that of the signal-to-noise ratio in SZE surveys \citep[][]{vanderlinde10}.
The former resides in a regime of Poisson noise, owing to the statistical property of X-ray sky images, while the latter has Gaussian noise that is straightforward to characterize \citep[see discussions in Appendix~A in][]{grandis19}.
It is then possible to study the noise properties of \Lext\ in a Poisson regime and derive a maximum likelihood estimator that can be directly used as a mass proxy, as in the case of SZE surveys.
This task is beyond the scope of this work and is indeed worth investigating in the future. 

\begin{figure}
\centering
\resizebox{0.45\textwidth}{!}{
\includegraphics[scale=1]{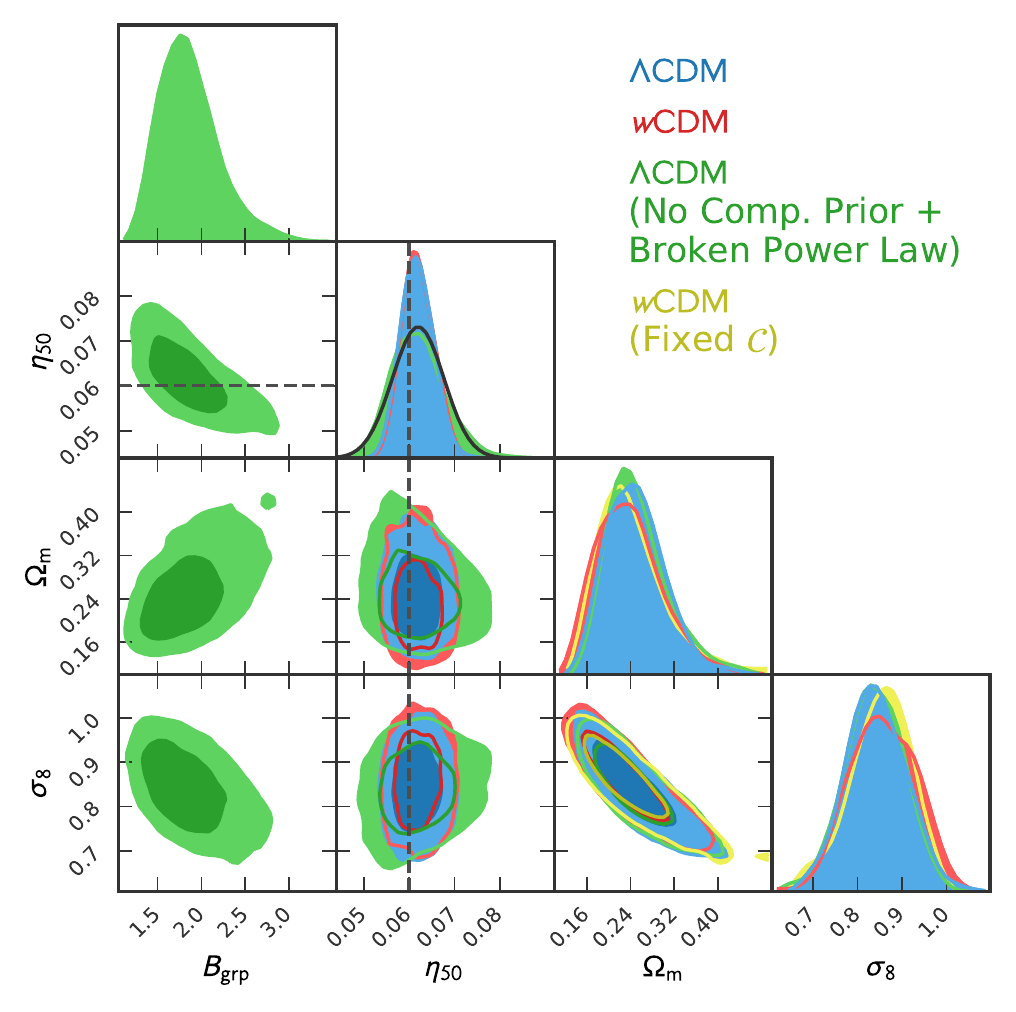}
}
\vspace{-0.15cm}
\caption{
The parameter constraints on \Bgrp, \ratefiveo, \OmegaM, and \sigmaeight.
The fiducial analyses assuming the \LCDM\ and \wCDM\ cosmology are in blue and red, respectively.
The results in the \LCDM\ cosmology without the Gaussian priors applied to the parameters  of the completeness function and with the broken power-law mass scaling of the count rate are in green.
The constraints assuming \wCDM\ cosmology with the parameters of \Comp\ fixed to the best-fit values are in yellow.
The dashed line indicates the fixed value of $\ratefiveo = 0.06~\mathrm{counts}/\mathrm{s}$.
The black line represents the constraint on the parameter \ratefiveo\ inferred from the richness distribution (see Section~\ref{sec:completeness_measurements}).
The contours indicate the $68\percent$ and $95\percent$ confidence levels.
}
\label{fig:sys_contours}
\end{figure}

The form of the X-ray selection in terms of the completeness function \Comp(\rate,\redshift) is such that the number of clusters with low-\rate\ are suppressed.
Meanwhile, the inclusion of the broken power-law mass scaling ($\Bgrp>\Brate$) of the count rate reduces the population of low-mass clusters at a fixed \rate\ given the scatter \sigmarate.
Therefore, we expect some degeneracy between \Bgrp\ and the parameters of \Comp(\rate,\redshift).
In our analysis, we find that \Bgrp\ is degenerate with \ratefiveo\ (but not strongly with \srate\ and \gammaz) when we remove the Gaussian priors on the X-ray completeness.
This is shown in the parameter space spanned by \Bgrp\ and \ratefiveo\ in Figure~\ref{fig:sys_contours}, where we compare the fiducial analyses (blue and red contours) with the \LCDM\ results (green contours) obtained without the Gaussian priors applied to the parameters $\left(\ratefiveo, \srate\right)$ and with the inclusion of \Bgrp.
Moreover, no significant difference in the cosmological parameters \OmegaM\ and \sigmaeight\ is seen.
In an extreme case where we fix the parameters of the completeness to the best-fit values (i.e., $\ratefiveo = 0.06~\mathrm{counts}/\mathrm{sec}$, $\srate = 0.5$, and $\gammaz=0$), the constraints on the cosmological parameters (yellow contours) do not show any significant changes.
This suggests that (1) the current cosmological constraints in \eFEDS\ are not limited by the precision of the completeness function, and that (2)
a more precise characterization on the completeness function in a future work will provide a better constraint on the mass scaling of clusters at the low mass end.

It is worth mentioning that fixing the X-ray completeness to the ``raw'' simulation-based estimate (without the corrections described in Section~\ref{sec:completeness_measurements}) does not significantly alter the cosmological constraints on \OmegaM\ and \sigmaeight\ but 
has a larger impact on the parameters of the count rate relation---both \Brate\ and \gammarate\ would be significantly biased high at a level of $\gtrsim3\sigma$ with respect to those inferred using the cluster abundance and the weak-lensing mass calibration.
Additionally, including the correction methods (either using the method  1 or 2; see Section~\ref{sec:completeness_measurements}) would alleviate the tension to $\approx2\sigma$.
This again affirms that the simulations from \cite{liuteng21} are not a perfect representation of the \eFEDS\ data.

%

%
%

\section{Conclusions}
\label{sec:conclusions}

We present the first cosmological constraints using a sample of $455$ \eROSITA\ clusters with a mass range of $2\times10^{13}\Msunh \lesssim\mass\lesssim 5\times10^{14}\Msunh$ at redshift $0.1<\redshift<1.2$ in the \eFEDS\ survey.
The sample is selected in X-rays and further confirmed in the optical imaging using the MCMF algorithm \citep{klein22}, resulting in a joint selection on both the X-ray and optical observables.
The galaxy richness \rich\ is used as the optical observable to facilitate the optical confirmation, for which the contamination due to the random line-of-sight superposition of X-ray and optical systems is required to be below $30\percent$ (i.e., $\fcont<0.3$).
With the optical confirmation to remove spurious sources, the contamination in the cluster sample is expected to be at a level of $6\percent$, which is significantly lower than that ($\approx20\percent$) of the catalog purely constructed in X-rays.

We make use of the public data of the HSC survey, which covers $177$ \eFEDS\ clusters, to perform the weak-lensing mass calibration that was thoroughly and carefully studied in \cite{chiu22}.
The weak-lensing analysis approach employed in this work is the same as in \cite{chiu22}, with two differences:
First, the public weak lensing data covers less of the eFEDS region, resulting a factor of $\approx30\percent$ reduction in the statistical power.
Second, an improved analysis allows us to update the uncertainty on the weak-lensing systematic error floor from $5.7\percent$ to $3\percent$, but this has a negligible impact on the final results.
The weak-lensing systematics are intensively calibrated against large cosmological simulations, in which the observed weak-lensing properties, as quantified in \cite{chiu22}, were added in \cite{grandis21}.
Importantly, the weak-lensing systematics associated with baryonic effects is also taken into account.

With the minimal assumptions, namely 
\begin{itemize}
\item the scaling relations between the observables (X-ray, optical, weak lensing) and the underlying halo mass \mass\ of clusters across a range of redshift \redshift\ can be described by a power law function depending on \mass\ and \redshift\ with log-normal intrinsic scatter,
\item the technique of weak lensing, which is nearly independent of any baryonic assumptions and hence is a direct probe to the full potential, can provide an unbiased estimate of the cluster halo mass when calibrated against simulations, and
\item the X-ray selection can be modeled using a completeness function \Comp(\rate,\redshift) in the X-ray mass proxy \rate\ whose form is well supported by the data and the dedicated simulations \citep{clerc18,liuteng21,bulbul21},
\end{itemize}
we perform the empirical modeling to simultaneously fit for the scaling relations, X-ray completeness, and cosmology.
The empirical modeling refers to the approach that allows the precise model to be determined by the data themselves.
This avoids the potential biases raised in a methodology that heavily relies on image and catalog simulations, where there are characteristically differences between real dataset and the simulations.
In fact, this is the case for the previously published \eFEDS\ simulations, which 
over-predict the clusters with low extent likelihood ($\Lext\lesssim10$) compared to the observations \citep{liuteng21}.

In this work, three scaling relations are modeled, including 
the count rate-to-mass-and-redshift (\rate--\mass--\redshift) relation (X-ray mass observable), 
the richness-to-mass-and-redshift (\rich--\mass--\redshift) relation (optical mass observable), and
the weak-lensing mass-to-mass-and-redshift (\Mwl--\mass--\redshift) relation.
The optical selection or cluster confirmation, which reduces the contamination relative to the X-ray-only selected sample by a factor of 3.3, is accounted for by a redshift-dependent cut on the cluster richness.
Meanwhile, the X-ray selection is modeled using a completeness function \Comp(\rate,\redshift) of the observed count rate with the dependence on the cluster redshift.
The weak-lensing systematics are taken into account through a marginalization over the priors applied to the parameters of the \Mwl--\mass--\redshift\ relation. 

Two novel aspects are introduced and studied in detail in this paper.
First, we generalize the \rate--\mass--\redshift\ relation to include a broken power-law scaling at the low-mass end, with the attempt to reconcile the discrepancy seen between the \eFEDS\ observations and simulations.  No compelling evidence for a broken power law is found.
Second, we perform an independent measurement of the X-ray completeness using the observed richness distribution in the optical.
The constraints from the richness-inferred completeness are applied as priors to the parameters of the completeness function in the cosmological analysis.

In this work, the $50\percent$ completeness in X-rays is constrained to occur at the count rate of $\rate \approx 0.06~\mathrm{counts}/\mathrm{s}$ with a scale factor of $\srate \approx 0.5$.
We find no clear dependence on the cluster redshift in the X-ray completeness.
There exhibits good agreement between the richness-inferred X-ray completeness and the self-calibrated completeness extracted using the cluster abundance.
Removing the priors to allow the self-calibration on the X-ray completeness has no significant impact on the cosmological constraints.

In a blind analysis, the cosmological constraints are derived in the joint modeling of the cluster abundance and the weak-lensing mass calibration.
The cluster abundance is modeled in the observable space of the count rate \rate, the richness \rich, and the redshift \redshift, accounting for the selection.
In the \LCDM\ cosmology, we obtain the constraints on \OmegaM, \sigmaeight\ and \seight$\left(\equiv\sigmaeight\left(\OmegaM/0.3\right)^{0.3}\right)$
as
\begin{eqnarray}
\OmegaM        &=      &\OmegamLCDM       \, ,  \nonumber \\
\sigmaeight     &=      &\sigmaeightLCDM    \, ,  \nonumber \\
\seight            &=      &\seightLCDM            \, .  \nonumber 
\end{eqnarray}
Extending the analysis to the \wCDM\ cosmology yields the fully consistent results of \OmegaM\ and \sigmaeight\ with the constraint on the equation of state of dark energy,
\begin{eqnarray}
\w                   &=      &\wWCDM                   \, . \nonumber
\end{eqnarray}
We assess the consistency between the best-fit model and the data, showing excellent agreement without bias in the distributions of the observables \rate, \rich, and \redshift.
These cosmological constraints are robust against
the priors applied to the parameters of the completeness function,
the assumed functional form (single or broken power law) for the mass scaling of the \rate--\mass--\redshift\ relation, and
the choice of different sample selections, including the subsamples at low redshift $0.1<\redshift<0.35$, at high redshift $0.35<\redshift<1.2$, and with high extent likelihood $\Lext>15$.

We compare the cosmological constraints with those from the \PLANCK\ mission (the $\mathtt{TTTEEE}+\mathtt{lowE}$ constraint), the third-year 3$\times$2-point analysis from DES, and the cluster sample from the SPT-SZ survey.
The \eFEDS\ results on \OmegaM, \sigmaeight, \seight, and \w\ are in good agreement with the external constraints at a level of $\lesssim1\sigma$.
Moreover, based on a sample of $455$ \eFEDS\ clusters, our results deliver constraints that are 
comparable to these independent methods, demonstrating the promise of the analysis based on the full \eROSITA\ survey in the future.

Apart from the cosmological constraints, we present the \rate--\mass--\redshift\ and \rich--\mass--\redshift\ relations.
The constraints on the scaling relations are 
significantly stronger than and also statistically consistent with the previous results inferred from the weak-lensing calibration alone \citep{chiu22}.
The \rich--\mass--\redshift\ relation shows a mass scaling that is consistent with the self-similar prediction ($\rich\propto \mass^{1}$) and a mildly decreasing redshift trend ($\deltarich=\ansdeltarich$) at a level of $\approx1.7\sigma$.
Meanwhile, we find that the \rate--\mass--\redshift\ prefers a mass trend that is steeper than, but still statistically consistent with, the slope inferred from the weak-lensing calibration only.
Hence, including the modeling of the cluster abundance is expected to increase the mass slope of other X-ray observables, leading to better agreement with scaling relation studies using the SPT sample \citep{chiu16a,chiu18a,bulbul19}.
In both the modeling of the weak-lensing mass calibration and the cluster abundance, there is no strong evidence supporting the broken power-law feature in the mass scaling of the count rate.
In Appendix~\ref{app:detailed_results}, we provide the updated and improved mass estimates of the individual clusters. 
We release the chains of the cosmological parameters online\footnote{\url{https://github.com/inonchiu/eFEDSproducts}}.

\subsection*{Outlook}
\label{sec:outlook}

These are the first fully self consistent constraints on cosmology using the abundance of X-ray selected clusters calibrated by weak lensing from a wide-field photometric survey.
Moreover, this paper demonstrates the success of empirical modeling in X-ray cluster cosmology, following the precursor studies using X-ray and SZE-selected clusters \citep{vikhlinin09a,mantz10a,benson13,deHaan16,bocquet19,costanzi21}.
It is also important to stress that the cosmological constraints are obtained using a large sample of clusters optically confirmed by the MCMF algorithm \citep{klein18} with the contamination rate significantly reduced from $\approx20\percent$ to $\approx6\percent$ without discarding a large number of real clusters. 
In what follows, we provide some aspects future improvements to pave a way forward.

It is of critical importance to understand the discrepancy between the \eFEDS\ data and the simulations to have a more accurate description of the clusters at the low-\Lext\ end (or with low signal-to-noise ratios).
This will support the construction and calibration of the X-ray completeness in a future work.

Meanwhile, according to our finding in the distribution of the X-ray observables $\left\lbrace \Lext, \Ldet, \ext\right\rbrace$, we advocate to remove the upper limit on \ext\ in the cluster detection to have a more well behaved scaling between \Lext\ and \Ldet.
As an analogy to the SZE experiments, moreover, it is possible and worth an investigation to quantify
the uncertainty of the maximum likelihood estimator of the X-ray cluster detection in a regime of Poisson noise.
Such a study will require intensive validations using accurate and end-to-end simulations, if available.

In this work, we only explore the mass-dependent slope in the \rate--\mass\ relation in terms of the broken power law.
With the all-sky sample from the \eROSITA\ survey, we will be in the position to constrain the mass and redshift dependence of the intrinsic scatter, whereas in the current analysis we assume the same value for the whole population of clusters in the \eFEDS\ survey. 

With the optical confirmation, the residual contamination to the \eFEDS\ sample is at a level of $6\percent$, for which we do not expect to have a significant impact on the final results.
In a future work with a larger sample, wherein the Poisson noise of cluster counts becomes significantly smaller, a proper treatment of the contamination will be required \citep[as in][]{bocquet19}.

Given the good agreement between the weak-lensing mass calibration and the cluster abundance of the \eFEDS\ sample, the next step forward is to combine the halo clustering of clusters to obtain a comprehensive constraint on cosmology \citep{chiu20b,marulli21}. 

As the sample size increases, we anticipate the increasing demand on computing resources in a future work.
Therefore, investigations into how to expedite the calculation are inevitably needed \citep[e.g.,][]{to22,boruah22}.

%
%

\section*{\scriptsize Acknowledgements}

\begin{scriptsize}

The authors thank the anonymous referee for providing constructive comments on this paper.
I-Non Chiu thanks Weikang Lin, Ying Zu, Hung-Hsu Chan, and Chia-Ying Lin for useful discussions that lead to improvements of this work.
The author acknowledges the great productivity stimulated by the lockdown in Shanghai in 2022 to significantly expedite this paper.
This work is supported by the National Science and Technology Council in Taiwan (Grant NSTC 111-2112-M-006-037-MY3).
This work is supported by the national science foundation of China (Nos. 11833005,  11890691, 11890692, 11621303), 111 project No. B20019 and Shanghai Natural Science Foundation, grant No. 19ZR1466800.
This work made use of the Gravity Supercomputer at the Department of Astronomy, Shanghai Jiao Tong University.
This work made use of the computing resources in the National Center for High-Performance Computing (NCHC) in Taiwan.

We also acknowledge the support of the Max Planck Society Faculty Fellowship program at MPE, the support of 
the DFG Cluster of Excellence ``Origin and Structure of the Universe'' and the Ludwig-Maximilians-Universit\"at.

This work is based on data from \eROSITA, the soft X-ray instrument aboard SRG, a joint Russian-German science mission supported by the Russian Space Agency (Roskosmos), in the interests of the Russian Academy of Sciences represented by its Space Research Institute (IKI), and the Deutsches Zentrum f{\"{u}}r Luft- und Raumfahrt (DLR). The SRG spacecraft was built by Lavochkin Association (NPOL) and its subcontractors, and is operated by NPOL with support from the Max Planck Institute for Extraterrestrial Physics (MPE).
The development and construction of the \eROSITA\ X-ray instrument was led by MPE, with contributions from the Dr. Karl Remeis Observatory Bamberg \& ECAP (FAU Erlangen-Nuernberg), the University of Hamburg Observatory, the Leibniz Institute for Astrophysics Potsdam (AIP), and the Institute for Astronomy and Astrophysics of the University of T{\"{u}}bingen, with the support of DLR and the Max Planck Society. The Argelander Institute for Astronomy of the University of Bonn and the Ludwig Maximilians Universit{\"{a}}t Munich also participated in the science preparation for \eROSITA.

The \eROSITA\ data shown here were processed using the \texttt{eSASS/NRTA} software system developed by the German \eROSITA\ consortium.

The Hyper Suprime-Cam (HSC) collaboration includes the astronomical communities of Japan and Taiwan, and Princeton University.  The HSC instrumentation and software were developed by the National Astronomical Observatory of Japan (NAOJ), the Kavli Institute for the Physics and Mathematics of the Universe (Kavli IPMU), the University of Tokyo, the High Energy Accelerator Research Organization (KEK), the Academia Sinica Institute for Astronomy and Astrophysics in Taiwan (ASIAA), and Princeton University.  Funding was contributed by the FIRST program from the Japanese Cabinet Office, the Ministry of Education, Culture, Sports, Science and Technology (MEXT), the Japan Society for the Promotion of Science (JSPS), Japan Science and Technology Agency  (JST), the Toray Science  Foundation, NAOJ, Kavli IPMU, KEK, ASIAA, and Princeton University.

This paper makes use of software developed for the Legacy Survey of Space and Time carried out by the Vera C. Rubin Observatory. We thank the LSST Project for making their code available as free software at \url{http://dm.lsst.org}.

This paper is based in part on data collected at the Subaru Telescope and retrieved from the HSC data archive system, which is operated by Subaru Telescope and Astronomy Data Center (ADC) at NAOJ. Data analysis was in part carried out with the cooperation of Center for Computational Astrophysics (CfCA), NAOJ.
We are honored and grateful for the opportunity of observing the Universe from Maunakea, which has the cultural, historical and natural significance in Hawaii.

The Pan-STARRS1 Surveys (PS1) and the PS1 public science archive have been made possible through contributions by the Institute for Astronomy, the University of Hawaii, the Pan-STARRS Project Office, the Max Planck Society and its participating institutes, the Max Planck Institute for Astronomy, Heidelberg, and the Max Planck Institute for Extraterrestrial Physics, Garching, The Johns Hopkins University, Durham University, the University of Edinburgh, the Queen's University Belfast, the Harvard-Smithsonian Center for Astrophysics, the Las Cumbres Observatory Global Telescope Network Incorporated, the National Central University of Taiwan, the Space Telescope Science Institute, the National Aeronautics and Space Administration under grant No. NNX08AR22G issued through the Planetary Science Division of the NASA Science Mission Directorate, the National Science Foundation grant No. AST-1238877, the University of Maryland, Eotvos Lorand University (ELTE), the Los Alamos National Laboratory, and the Gordon and Betty Moore Foundation.

This work is possible because of the efforts in the LSST \citep{juric17,ivezic19} and PS1 \citep{chambers16, schlafly12, tonry12, magnier13}, and in the HSC \citep{aihara18a} developments including the deep imaging of the COSMOS field \citep{tanaka17}, the on-site quality-assurance system \citep{furusawa18}, the Hyper Suprime-Cam \citep{miyazaki15, miyazaki18, komiyama18}, the design of the filters \citep{kawanomoto18},  the data pipeline \citep{bosch18}, the design of bright-star masks \citep{coupon18}, the characterization of the photometry by the code \texttt{Synpipe} \citep{huang18}, the photometric redshift estimation \citep{tanaka18}, the shear calibration \citep{mandelbaum18}, and the public data releases \citep{aihara18b, aihara19}.

This work made use of the IPython package \citep{ipython}, \texttt{SciPy} \citep{scipy}, \texttt{TOPCAT}, an interactive graphical viewer and editor for tabular data \citep{topcat1,topcat2}, \texttt{matplotlib}, a Python library for publication quality graphics \citep{matplotlib}, \texttt{Astropy}, a community-developed core Python package for Astronomy \citep{astropy}, \texttt{NumPy} \citep{van2011numpy}. 
This work made use of \texttt{Pathos} \citep{pathos} in parallel computing. 
This work made use of \citet{bocquet16b} and \cite{hinton2016} for producing the corner plots for the parameter constraints.
The code \texttt{Colossus} \citep{diemer18} is heavily used to calculate cosmology-dependent quantities in this work.

\end{scriptsize}

%
%

\section*{\scriptsize Data Availability}

\begin{scriptsize}

The \eFEDS\ cluster sample and their observed X-ray properties are publicly available in the \eROSITA\ Early Data Release via \url{https://erosita.mpe.mpg.de/edr}.
The weak-lensing data are publicly available in the second Public Data Release of the HSC survey via \url{https://hsc-release.mtk.nao.ac.jp/doc/index.php/s16a-shape-catalog-pdr2}.
The chains of the cosmological parameters and the mass estimates of individual \eFEDS\ clusters are publicly available via \url{https://github.com/inonchiu/eFEDSproducts}.
The other data products underlying this article will be shared upon a reasonable request to the corresponding author.

\end{scriptsize}

%
%

\bibliographystyle{mnras}
\bibliography{literature} 

\begin{thebibliography}{}
\makeatletter
\relax
\def\mn@urlcharsother{\let\do\@makeother \do\$\do\&\do\#\do\^\do\_\do\%\do\~}
\def\mn@doi{\begingroup\mn@urlcharsother \@ifnextchar [ {\mn@doi@}
  {\mn@doi@[]}}
\def\mn@doi@[#1]#2{\def\@tempa{#1}\ifx\@tempa\@empty \href
  {http://dx.doi.org/#2} {doi:#2}\else \href {http://dx.doi.org/#2} {#1}\fi
  \endgroup}
\def\mn@eprint#1#2{\mn@eprint@#1:#2::\@nil}
\def\mn@eprint@arXiv#1{\href {http://arxiv.org/abs/#1} {{\tt arXiv:#1}}}
\def\mn@eprint@dblp#1{\href {http://dblp.uni-trier.de/rec/bibtex/#1.xml}
  {dblp:#1}}
\def\mn@eprint@#1:#2:#3:#4\@nil{\def\@tempa {#1}\def\@tempb {#2}\def\@tempc
  {#3}\ifx \@tempc \@empty \let \@tempc \@tempb \let \@tempb \@tempa \fi \ifx
  \@tempb \@empty \def\@tempb {arXiv}\fi \@ifundefined
  {mn@eprint@\@tempb}{\@tempb:\@tempc}{\expandafter \expandafter \csname
  mn@eprint@\@tempb\endcsname \expandafter{\@tempc}}}

\bibitem[\protect\citeauthoryear{{Abbott} et~al.,}{{Abbott}
  et~al.}{2020}]{abbott20desy1clustercosmology}
{Abbott} T.~M.~C.,  et~al., 2020, \mn@doi [\prd] {10.1103/PhysRevD.102.023509},
  \href {https://ui.adsabs.harvard.edu/abs/2020PhRvD.102b3509A} {102, 023509}

\bibitem[\protect\citeauthoryear{{Abbott} et~al.,}{{Abbott}
  et~al.}{2022}]{desy3kp-3x2}
{Abbott} T.~M.~C.,  et~al., 2022, \mn@doi [\prd] {10.1103/PhysRevD.105.023520},
  \href {https://ui.adsabs.harvard.edu/abs/2022PhRvD.105b3520A} {105, 023520}

\bibitem[\protect\citeauthoryear{{Adami} et~al.,}{{Adami}
  et~al.}{2018}]{adami18}
{Adami} C.,  et~al., 2018, \mn@doi [\aap] {10.1051/0004-6361/201731606}, \href
  {https://ui.adsabs.harvard.edu/abs/2018A&A...620A...5A} {620, A5}

\bibitem[\protect\citeauthoryear{{Aihara} et~al.,}{{Aihara}
  et~al.}{2018a}]{aihara18a}
{Aihara} H.,  et~al., 2018a, \mn@doi [\pasj] {10.1093/pasj/psx066}, \href
  {http://adsabs.harvard.edu/abs/2018PASJ...70S...4A} {70, S4}

\bibitem[\protect\citeauthoryear{{Aihara} et~al.,}{{Aihara}
  et~al.}{2018b}]{aihara18b}
{Aihara} H.,  et~al., 2018b, \mn@doi [\pasj] {10.1093/pasj/psx081}, \href
  {http://adsabs.harvard.edu/abs/2018PASJ...70S...8A} {70, S8}

\bibitem[\protect\citeauthoryear{{Aihara} et~al.,}{{Aihara}
  et~al.}{2019}]{aihara19}
{Aihara} H.,  et~al., 2019, \mn@doi [\pasj] {10.1093/pasj/psz103}, \href
  {https://ui.adsabs.harvard.edu/abs/2019PASJ...71..114A} {71, 114}

\bibitem[\protect\citeauthoryear{Allen, Evrard  \& Mantz}{Allen
  et~al.}{2011}]{allen11}
Allen S.,  Evrard A.,   Mantz A.,  2011, \mn@doi [\araa]
  {10.1146/annurev-astro-081710-102514}, 49, 409

\bibitem[\protect\citeauthoryear{{Astropy Collaboration} et~al.,}{{Astropy
  Collaboration} et~al.}{2013}]{astropy}
{Astropy Collaboration} et~al., 2013, \mn@doi [\aap]
  {10.1051/0004-6361/201322068}, \href
  {http://adsabs.harvard.edu/abs/2013A%26A...558A..33A} {558, A33}

\bibitem[\protect\citeauthoryear{{Bahar} et~al.,}{{Bahar}
  et~al.}{2022}]{bahar21}
{Bahar} Y.~E.,  et~al., 2022, \mn@doi [\aap] {10.1051/0004-6361/202142462},
  \href {https://ui.adsabs.harvard.edu/abs/2022A&A...661A...7B} {661, A7}

\bibitem[\protect\citeauthoryear{{Baker} \& {Cousins}}{{Baker} \&
  {Cousins}}{1984}]{baker84}
{Baker} S.,  {Cousins} R.~D.,  1984, \mn@doi [Nuclear Instruments and Methods
  in Physics Research] {10.1016/0167-5087(84)90016-4}, \href
  {https://ui.adsabs.harvard.edu/abs/1984NIMPR.221..437B} {221, 437}

\bibitem[\protect\citeauthoryear{{Barnes} et~al.,}{{Barnes}
  et~al.}{2017}]{barnes17}
{Barnes} D.~J.,  et~al., 2017, \mn@doi [\mnras] {10.1093/mnras/stx1647}, \href
  {https://ui.adsabs.harvard.edu/abs/2017MNRAS.471.1088B} {471, 1088}

\bibitem[\protect\citeauthoryear{Bartelmann \& Schneider}{Bartelmann \&
  Schneider}{2001}]{bartelmann01}
Bartelmann M.,  Schneider P.,  2001, \mn@doi [\physrep]
  {10.1016/S0370-1573(00)00082-X}, 340, 291

\bibitem[\protect\citeauthoryear{Benson et~al.,}{Benson
  et~al.}{2013}]{benson13}
Benson B.,  et~al., 2013, \mn@doi [\apj] {10.1088/0004-637X/763/2/147}, 763,
  147

\bibitem[\protect\citeauthoryear{Bleem et~al.,}{Bleem et~al.}{2015}]{bleem15}
Bleem L.,  et~al., 2015, \mn@doi [\apjs] {10.1088/0067-0049/216/2/27}, 216, 27

\bibitem[\protect\citeauthoryear{{Bleem} et~al.,}{{Bleem}
  et~al.}{2020}]{bleem20}
{Bleem} L.~E.,  et~al., 2020, \mn@doi [\apjs] {10.3847/1538-4365/ab6993}, \href
  {https://ui.adsabs.harvard.edu/abs/2020ApJS..247...25B} {247, 25}

\bibitem[\protect\citeauthoryear{{Bocquet} \& {Carter}}{{Bocquet} \&
  {Carter}}{2016}]{bocquet16b}
{Bocquet} S.,  {Carter} F.~W.,  2016, \mn@doi [The Journal of Open Source
  Software] {10.21105/joss.00046}, \href
  {https://ui.adsabs.harvard.edu/abs/2016JOSS....1...46B} {1, 46}

\bibitem[\protect\citeauthoryear{Bocquet et~al.,}{Bocquet
  et~al.}{2015}]{bocquet15}
Bocquet S.,  et~al., 2015, \mn@doi [\apj] {10.1088/0004-637X/799/2/214}, 799,
  214

\bibitem[\protect\citeauthoryear{Bocquet, Saro, Dolag  \& Mohr}{Bocquet
  et~al.}{2016}]{bocquet16}
Bocquet S.,  Saro A.,  Dolag K.,   Mohr J.,  2016, \mn@doi [\mnras]
  {10.1093/mnras/stv2657}, 456, 2361

\bibitem[\protect\citeauthoryear{{Bocquet} et~al.,}{{Bocquet}
  et~al.}{2019}]{bocquet19}
{Bocquet} S.,  et~al., 2019, \mn@doi [\apj] {10.3847/1538-4357/ab1f10}, \href
  {https://ui.adsabs.harvard.edu/abs/2019ApJ...878...55B} {878, 55}

\bibitem[\protect\citeauthoryear{B{\"{o}}hringer et~al.,}{B{\"{o}}hringer
  et~al.}{2004}]{bohringer04}
B{\"{o}}hringer H.,  et~al., 2004, \mn@doi [\aap] {10.1051/0004-6361:20034484},
  425, 367

\bibitem[\protect\citeauthoryear{{B{\"o}hringer}, {Dolag}  \&
  {Chon}}{{B{\"o}hringer} et~al.}{2012}]{boehringer12}
{B{\"o}hringer} H.,  {Dolag} K.,   {Chon} G.,  2012, \mn@doi [\aap]
  {10.1051/0004-6361/201118000}, \href
  {https://ui.adsabs.harvard.edu/abs/2012A&A...539A.120B} {539, A120}

\bibitem[\protect\citeauthoryear{{Boller}, {Freyberg}, {Tr{\"u}mper}, {Haberl},
  {Voges}  \& {Nandra}}{{Boller} et~al.}{2016}]{boller16}
{Boller} T.,  {Freyberg} M.~J.,  {Tr{\"u}mper} J.,  {Haberl} F.,  {Voges} W.,
  {Nandra} K.,  2016, \mn@doi [\aap] {10.1051/0004-6361/201525648}, \href
  {https://ui.adsabs.harvard.edu/abs/2016A&A...588A.103B} {588, A103}

\bibitem[\protect\citeauthoryear{{Boruah}, {Eifler}, {Miranda}  \&
  {M}}{{Boruah} et~al.}{2022}]{boruah22}
{Boruah} S.~S.,  {Eifler} T.,  {Miranda} V.,   {M} S. K.~P.,  2022, arXiv
  e-prints, \href {https://ui.adsabs.harvard.edu/abs/2022arXiv220306124B} {p.
  arXiv:2203.06124}

\bibitem[\protect\citeauthoryear{{Bosch} et~al.,}{{Bosch}
  et~al.}{2018}]{bosch18}
{Bosch} J.,  et~al., 2018, \mn@doi [\pasj] {10.1093/pasj/psx080}, \href
  {http://adsabs.harvard.edu/abs/2018PASJ...70S...5B} {70, S5}

\bibitem[\protect\citeauthoryear{{Brunner} et~al.,}{{Brunner}
  et~al.}{2022}]{brunner21}
{Brunner} H.,  et~al., 2022, \mn@doi [\aap] {10.1051/0004-6361/202141266},
  \href {https://ui.adsabs.harvard.edu/abs/2022A&A...661A...1B} {661, A1}

\bibitem[\protect\citeauthoryear{{Bulbul} et~al.,}{{Bulbul}
  et~al.}{2019}]{bulbul19}
{Bulbul} E.,  et~al., 2019, \mn@doi [\apj] {10.3847/1538-4357/aaf230}, \href
  {https://ui.adsabs.harvard.edu/abs/2019ApJ...871...50B} {871, 50}

\bibitem[\protect\citeauthoryear{{Bulbul} et~al.,}{{Bulbul}
  et~al.}{2022}]{bulbul21}
{Bulbul} E.,  et~al., 2022, \mn@doi [\aap] {10.1051/0004-6361/202142460}, \href
  {https://ui.adsabs.harvard.edu/abs/2022A&A...661A..10B} {661, A10}

\bibitem[\protect\citeauthoryear{Cash}{Cash}{1979}]{cash79}
Cash W.,  1979, \mn@doi [\apj] {10.1086/156922}, 228, 939

\bibitem[\protect\citeauthoryear{{Chambers} et~al.,}{{Chambers}
  et~al.}{2016}]{chambers16}
{Chambers} K.~C.,  et~al., 2016, arXiv e-prints, \href
  {https://ui.adsabs.harvard.edu/abs/2016arXiv161205560C} {p. arXiv:1612.05560}

\bibitem[\protect\citeauthoryear{Chiu et~al.,}{Chiu et~al.}{2016a}]{chiu16a}
Chiu I.,  et~al., 2016a, \mn@doi [\mnras] {10.1093/mnras/stv2303}, 455, 258

\bibitem[\protect\citeauthoryear{Chiu et~al.,}{Chiu et~al.}{2016b}]{chiu16c}
Chiu I.,  et~al., 2016b, \mn@doi [\mnras] {10.1093/mnras/stw292}, 458, 379

\bibitem[\protect\citeauthoryear{{Chiu} et~al.,}{{Chiu} et~al.}{2018}]{chiu18a}
{Chiu} I.,  et~al., 2018, \mn@doi [\mnras] {10.1093/mnras/sty1284}, \href
  {http://adsabs.harvard.edu/abs/2018MNRAS.478.3072C} {478, 3072}

\bibitem[\protect\citeauthoryear{{Chiu}, {Okumura}, {Oguri}, {Agrawal},
  {Umetsu}  \& {Lin}}{{Chiu} et~al.}{2020}]{chiu20b}
{Chiu} I.~N.,  {Okumura} T.,  {Oguri} M.,  {Agrawal} A.,  {Umetsu} K.,   {Lin}
  Y.-T.,  2020, \mn@doi [\mnras] {10.1093/mnras/staa2440}, \href
  {https://ui.adsabs.harvard.edu/abs/2020MNRAS.498.2030C} {498, 2030}

\bibitem[\protect\citeauthoryear{{Chiu} et~al.,}{{Chiu} et~al.}{2022}]{chiu22}
{Chiu} I.~N.,  et~al., 2022, \mn@doi [\aap] {10.1051/0004-6361/202141755},
  \href {https://ui.adsabs.harvard.edu/abs/2022A&A...661A..11C} {661, A11}

\bibitem[\protect\citeauthoryear{{Clerc} et~al.,}{{Clerc}
  et~al.}{2018}]{clerc18}
{Clerc} N.,  et~al., 2018, \mn@doi [\aap] {10.1051/0004-6361/201732119}, \href
  {https://ui.adsabs.harvard.edu/abs/2018A&A...617A..92C} {617, A92}

\bibitem[\protect\citeauthoryear{{Comparat} et~al.,}{{Comparat}
  et~al.}{2019}]{comparat19}
{Comparat} J.,  et~al., 2019, \mn@doi [\mnras] {10.1093/mnras/stz1390}, \href
  {https://ui.adsabs.harvard.edu/abs/2019MNRAS.487.2005C} {487, 2005}

\bibitem[\protect\citeauthoryear{{Comparat} et~al.,}{{Comparat}
  et~al.}{2020}]{comparat20}
{Comparat} J.,  et~al., 2020, \mn@doi [The Open Journal of Astrophysics]
  {10.21105/astro.2008.08404}, \href
  {https://ui.adsabs.harvard.edu/abs/2020OJAp....3E..13C} {3, 13}

\bibitem[\protect\citeauthoryear{{Costanzi} et~al.,}{{Costanzi}
  et~al.}{2019}]{costanzi19}
{Costanzi} M.,  et~al., 2019, \mn@doi [\mnras] {10.1093/mnras/sty2665}, \href
  {https://ui.adsabs.harvard.edu/abs/2019MNRAS.482..490C} {482, 490}

\bibitem[\protect\citeauthoryear{{Costanzi} et~al.,}{{Costanzi}
  et~al.}{2021}]{costanzi21}
{Costanzi} M.,  et~al., 2021, \mn@doi [\prd] {10.1103/PhysRevD.103.043522},
  \href {https://ui.adsabs.harvard.edu/abs/2021PhRvD.103d3522C} {103, 043522}

\bibitem[\protect\citeauthoryear{{Coupon}, {Czakon}, {Bosch}, {Komiyama},
  {Medezinski}, {Miyazaki}  \& {Oguri}}{{Coupon} et~al.}{2018}]{coupon18}
{Coupon} J.,  {Czakon} N.,  {Bosch} J.,  {Komiyama} Y.,  {Medezinski} E.,
  {Miyazaki} S.,   {Oguri} M.,  2018, \mn@doi [\pasj] {10.1093/pasj/psx047},
  \href {http://adsabs.harvard.edu/abs/2018PASJ...70S...7C} {70, S7}

\bibitem[\protect\citeauthoryear{{Dey} et~al.,}{{Dey} et~al.}{2019}]{dey19}
{Dey} A.,  et~al., 2019, \mn@doi [\aj] {10.3847/1538-3881/ab089d}, \href
  {https://ui.adsabs.harvard.edu/abs/2019AJ....157..168D} {157, 168}

\bibitem[\protect\citeauthoryear{{Diemer}}{{Diemer}}{2018}]{diemer18}
{Diemer} B.,  2018, \mn@doi [\apjs] {10.3847/1538-4365/aaee8c}, \href
  {https://ui.adsabs.harvard.edu/abs/2018ApJS..239...35D} {239, 35}

\bibitem[\protect\citeauthoryear{{Diemer} \& {Kravtsov}}{{Diemer} \&
  {Kravtsov}}{2015}]{diemer15}
{Diemer} B.,  {Kravtsov} A.~V.,  2015, \mn@doi [\apj]
  {10.1088/0004-637X/799/1/108}, \href
  {http://adsabs.harvard.edu/abs/2015ApJ...799..108D} {799, 108}

\bibitem[\protect\citeauthoryear{{Dietrich} et~al.,}{{Dietrich}
  et~al.}{2019}]{dietrich19}
{Dietrich} J.~P.,  et~al., 2019, \mn@doi [\mnras] {10.1093/mnras/sty3088},
  \href {http://adsabs.harvard.edu/abs/2019MNRAS.483.2871D} {483, 2871}

\bibitem[\protect\citeauthoryear{{Finoguenov} et~al.,}{{Finoguenov}
  et~al.}{2020}]{finoguenov20}
{Finoguenov} A.,  et~al., 2020, \mn@doi [\aap] {10.1051/0004-6361/201937283},
  \href {https://ui.adsabs.harvard.edu/abs/2020A&A...638A.114F} {638, A114}

\bibitem[\protect\citeauthoryear{Fixsen et~al.,}{Fixsen
  et~al.}{2009}]{fixsen09}
Fixsen D.,  et~al., 2009, submitted to \apj

\bibitem[\protect\citeauthoryear{Foreman-Mackey, Hogg, Lang  \&
  Goodman}{Foreman-Mackey et~al.}{2013}]{foreman13}
Foreman-Mackey D.,  Hogg D.,  Lang D.,   Goodman J.,  2013, \mn@doi [\pasp]
  {10.1086/670067}, 125, 306

\bibitem[\protect\citeauthoryear{{Foreman-Mackey} et~al.,}{{Foreman-Mackey}
  et~al.}{2019}]{foreman19}
{Foreman-Mackey} D.,  et~al., 2019, \mn@doi [The Journal of Open Source
  Software] {10.21105/joss.01864}, \href
  {https://ui.adsabs.harvard.edu/abs/2019JOSS....4.1864F} {4, 1864}

\bibitem[\protect\citeauthoryear{{Furusawa} et~al.,}{{Furusawa}
  et~al.}{2018}]{furusawa18}
{Furusawa} H.,  et~al., 2018, \mn@doi [\pasj] {10.1093/pasj/psx079}, \href
  {https://ui.adsabs.harvard.edu/abs/2018PASJ...70S...3F} {70, S3}

\bibitem[\protect\citeauthoryear{{Garrel} et~al.,}{{Garrel}
  et~al.}{2021}]{garrel21}
{Garrel} C.,  et~al., 2021, arXiv e-prints, \href
  {https://ui.adsabs.harvard.edu/abs/2021arXiv210913171G} {p. arXiv:2109.13171}

\bibitem[\protect\citeauthoryear{{Ghirardini} et~al.,}{{Ghirardini}
  et~al.}{2021}]{ghirardini21}
{Ghirardini} V.,  et~al., 2021, \mn@doi [\aap] {10.1051/0004-6361/202039554},
  \href {https://ui.adsabs.harvard.edu/abs/2021A&A...647A...4G} {647, A4}

\bibitem[\protect\citeauthoryear{{Gladders} \& {Yee}}{{Gladders} \&
  {Yee}}{2000}]{gladders00}
{Gladders} M.~D.,  {Yee} H.~K.~C.,  2000, \mn@doi [\aj] {10.1086/301557}, \href
  {https://ui.adsabs.harvard.edu/abs/2000AJ....120.2148G} {120, 2148}

\bibitem[\protect\citeauthoryear{Gladders, Yee, Majumdar, Barrientos, Hoekstra,
  Hall  \& Infante}{Gladders et~al.}{2007}]{gladders07}
Gladders M.,  Yee H.,  Majumdar S.,  Barrientos L.,  Hoekstra H.,  Hall P.,
  Infante L.,  2007, \mn@doi [\apj] {10.1086/509909}, 655, 128

\bibitem[\protect\citeauthoryear{{Grandis}, {Mohr}, {Dietrich}, {Bocquet},
  {Saro}, {Klein}, {Paulus}  \& {Capasso}}{{Grandis} et~al.}{2019}]{grandis19}
{Grandis} S.,  {Mohr} J.~J.,  {Dietrich} J.~P.,  {Bocquet} S.,  {Saro} A.,
  {Klein} M.,  {Paulus} M.,   {Capasso} R.,  2019, \mn@doi [\mnras]
  {10.1093/mnras/stz1778}, \href
  {https://ui.adsabs.harvard.edu/abs/2019MNRAS.488.2041G} {488, 2041}

\bibitem[\protect\citeauthoryear{{Grandis} et~al.,}{{Grandis}
  et~al.}{2020}]{grandis20}
{Grandis} S.,  et~al., 2020, \mn@doi [\mnras] {10.1093/mnras/staa2333}, \href
  {https://ui.adsabs.harvard.edu/abs/2020MNRAS.498..771G} {498, 771}

\bibitem[\protect\citeauthoryear{{Grandis}, {Bocquet}, {Mohr}, {Klein}  \&
  {Dolag}}{{Grandis} et~al.}{2021}]{grandis21}
{Grandis} S.,  {Bocquet} S.,  {Mohr} J.~J.,  {Klein} M.,   {Dolag} K.,  2021,
  \mn@doi [\mnras] {10.1093/mnras/stab2414}, \href
  {https://ui.adsabs.harvard.edu/abs/2021MNRAS.tmp.2274G} {}

\bibitem[\protect\citeauthoryear{Haiman, Mohr  \& Holder}{Haiman
  et~al.}{2001}]{haiman01}
Haiman Z.,  Mohr J.,   Holder G.,  2001, \apj, 553, 545

\bibitem[\protect\citeauthoryear{{Hilton} et~al.,}{{Hilton}
  et~al.}{2021}]{hilton21}
{Hilton} M.,  et~al., 2021, \mn@doi [\apjs] {10.3847/1538-4365/abd023}, \href
  {https://ui.adsabs.harvard.edu/abs/2021ApJS..253....3H} {253, 3}

\bibitem[\protect\citeauthoryear{{Hinton}}{{Hinton}}{2016}]{hinton2016}
{Hinton} S.~R.,  2016, \mn@doi [The Journal of Open Source Software]
  {10.21105/joss.00045}, \href
  {http://adsabs.harvard.edu/abs/2016JOSS....1...45H} {1, 00045}

\bibitem[\protect\citeauthoryear{{Hoekstra}}{{Hoekstra}}{2003}]{hoekstra03}
{Hoekstra} H.,  2003, \mn@doi [\mnras] {10.1046/j.1365-8711.2003.06264.x},
  \href {https://ui.adsabs.harvard.edu/abs/2003MNRAS.339.1155H} {339, 1155}

\bibitem[\protect\citeauthoryear{Hoekstra, Herbonnet, Muzzin, Babul, Mahdavi,
  Viola  \& Cacciato}{Hoekstra et~al.}{2015}]{hoekstra15}
Hoekstra H.,  Herbonnet R.,  Muzzin A.,  Babul A.,  Mahdavi A.,  Viola M.,
  Cacciato M.,  2015, \mn@doi [\mnras] {10.1093/mnras/stv275}, 449, 685

\bibitem[\protect\citeauthoryear{{Hsieh} \& {Yee}}{{Hsieh} \&
  {Yee}}{2014}]{hsieh14}
{Hsieh} B.~C.,  {Yee} H.~K.~C.,  2014, \mn@doi [\apj]
  {10.1088/0004-637X/792/2/102}, \href
  {http://adsabs.harvard.edu/abs/2014ApJ...792..102H} {792, 102}

\bibitem[\protect\citeauthoryear{{Huang} et~al.,}{{Huang}
  et~al.}{2018}]{huang18}
{Huang} S.,  et~al., 2018, \mn@doi [\pasj] {10.1093/pasj/psx126}, \href
  {http://adsabs.harvard.edu/abs/2018PASJ...70S...6H} {70, S6}

\bibitem[\protect\citeauthoryear{{Huang} et~al.,}{{Huang}
  et~al.}{2020}]{huang20}
{Huang} N.,  et~al., 2020, \mn@doi [\aj] {10.3847/1538-3881/ab6a96}, \href
  {https://ui.adsabs.harvard.edu/abs/2020AJ....159..110H} {159, 110}

\bibitem[\protect\citeauthoryear{Hunter}{Hunter}{2007}]{matplotlib}
Hunter J.~D.,  2007, Computing In Science \& Engineering, 9, 90

\bibitem[\protect\citeauthoryear{{Huterer} et~al.,}{{Huterer}
  et~al.}{2015}]{huterer15}
{Huterer} D.,  et~al., 2015, \mn@doi [Astroparticle Physics]
  {10.1016/j.astropartphys.2014.07.004}, \href
  {https://ui.adsabs.harvard.edu/abs/2015APh....63...23H} {63, 23}

\bibitem[\protect\citeauthoryear{{Ider Chitham} et~al.,}{{Ider Chitham}
  et~al.}{2020}]{iderchitham20}
{Ider Chitham} J.,  et~al., 2020, \mn@doi [\mnras] {10.1093/mnras/staa3044},
  \href {https://ui.adsabs.harvard.edu/abs/2020MNRAS.499.4768I} {499, 4768}

\bibitem[\protect\citeauthoryear{{Ivezic} et~al.,}{{Ivezic}
  et~al.}{2019}]{ivezic19}
{Ivezic} {\v{Z}}.,  et~al., 2019, \mn@doi [\apj] {10.3847/1538-4357/ab042c},
  \href {https://ui.adsabs.harvard.edu/abs/2019ApJ...873..111I} {873, 111}

\bibitem[\protect\citeauthoryear{{Juric} et~al.,}{{Juric}
  et~al.}{2017}]{juric17}
{Juric} M.,  et~al., 2017, {The LSST Data Management System}.
p.~279

\bibitem[\protect\citeauthoryear{{Kaastra}}{{Kaastra}}{2017}]{kaastra17}
{Kaastra} J.~S.,  2017, \mn@doi [\aap] {10.1051/0004-6361/201629319}, \href
  {https://ui.adsabs.harvard.edu/abs/2017A&A...605A..51K} {605, A51}

\bibitem[\protect\citeauthoryear{Kaiser}{Kaiser}{1986}]{kaiser1986}
Kaiser N.,  1986, \mn@doi [\mnras] {10.1093/mnras/222.2.323}, 222, 323

\bibitem[\protect\citeauthoryear{{Kawanomoto} et~al.,}{{Kawanomoto}
  et~al.}{2018}]{kawanomoto18}
{Kawanomoto} S.,  et~al., 2018, \mn@doi [\pasj] {10.1093/pasj/psy056}, \href
  {https://ui.adsabs.harvard.edu/abs/2018PASJ...70...66K} {70, 66}

\bibitem[\protect\citeauthoryear{{Klein} et~al.,}{{Klein}
  et~al.}{2018}]{klein18}
{Klein} M.,  et~al., 2018, \mn@doi [\mnras] {10.1093/mnras/stx2929}, \href
  {http://adsabs.harvard.edu/abs/2018MNRAS.474.3324K} {474, 3324}

\bibitem[\protect\citeauthoryear{{Klein} et~al.,}{{Klein}
  et~al.}{2019}]{klein19}
{Klein} M.,  et~al., 2019, \mn@doi [\mnras] {10.1093/mnras/stz1463}, \href
  {https://ui.adsabs.harvard.edu/abs/2019MNRAS.488..739K} {488, 739}

\bibitem[\protect\citeauthoryear{{Klein} et~al.,}{{Klein}
  et~al.}{2022}]{klein22}
{Klein} M.,  et~al., 2022, \mn@doi [\aap] {10.1051/0004-6361/202141123}, \href
  {https://ui.adsabs.harvard.edu/abs/2022A&A...661A...4K} {661, A4}

\bibitem[\protect\citeauthoryear{{Komiyama} et~al.,}{{Komiyama}
  et~al.}{2018}]{komiyama18}
{Komiyama} Y.,  et~al., 2018, \mn@doi [\pasj] {10.1093/pasj/psx069}, \href
  {https://ui.adsabs.harvard.edu/abs/2018PASJ...70S...2K} {70, S2}

\bibitem[\protect\citeauthoryear{{Koulouridis} et~al.,}{{Koulouridis}
  et~al.}{2021}]{koulouridis21}
{Koulouridis} E.,  et~al., 2021, \mn@doi [\aap] {10.1051/0004-6361/202140566},
  \href {https://ui.adsabs.harvard.edu/abs/2021A&A...652A..12K} {652, A12}

\bibitem[\protect\citeauthoryear{Kravtsov \& Borgani}{Kravtsov \&
  Borgani}{2012}]{kravtsov12}
Kravtsov A.,  Borgani S.,  2012, \mn@doi [\araa]
  {10.1146/annurev-astro-081811-125502}, 50, 353

\bibitem[\protect\citeauthoryear{{Li} et~al.,}{{Li} et~al.}{2022}]{li21}
{Li} X.,  et~al., 2022, \mn@doi [\pasj] {10.1093/pasj/psac006}, \href
  {https://ui.adsabs.harvard.edu/abs/2022PASJ...74..421L} {74, 421}

\bibitem[\protect\citeauthoryear{Lin, Mohr  \& Stanford}{Lin
  et~al.}{2004}]{lin04a}
Lin Y.,  Mohr J.,   Stanford S.,  2004, \apj, 610, 745

\bibitem[\protect\citeauthoryear{{Liu} et~al.,}{{Liu} et~al.}{2015}]{liu15a}
{Liu} J.,  et~al., 2015, \mn@doi [\mnras] {10.1093/mnras/stv080}, \href
  {https://ui.adsabs.harvard.edu/abs/2015MNRAS.448.2085L} {448, 2085}

\bibitem[\protect\citeauthoryear{{Liu} et~al.,}{{Liu} et~al.}{2022a}]{liu21}
{Liu} A.,  et~al., 2022a, \mn@doi [\aap] {10.1051/0004-6361/202141120}, \href
  {https://ui.adsabs.harvard.edu/abs/2022A&A...661A...2L} {661, A2}

\bibitem[\protect\citeauthoryear{{Liu} et~al.,}{{Liu}
  et~al.}{2022b}]{liuteng21}
{Liu} T.,  et~al., 2022b, \mn@doi [\aap] {10.1051/0004-6361/202141178}, \href
  {https://ui.adsabs.harvard.edu/abs/2022A&A...661A..27L} {661, A27}

\bibitem[\protect\citeauthoryear{{Lovisari}, {Reiprich}  \&
  {Schellenberger}}{{Lovisari} et~al.}{2015}]{lovisari15}
{Lovisari} L.,  {Reiprich} T.~H.,   {Schellenberger} G.,  2015, \mn@doi [\aap]
  {10.1051/0004-6361/201423954}, \href
  {https://ui.adsabs.harvard.edu/abs/2015A&A...573A.118L} {573, A118}

\bibitem[\protect\citeauthoryear{{Magnier} et~al.,}{{Magnier}
  et~al.}{2013}]{magnier13}
{Magnier} E.~A.,  et~al., 2013, \mn@doi [\apjs] {10.1088/0067-0049/205/2/20},
  \href {https://ui.adsabs.harvard.edu/abs/2013ApJS..205...20M} {205, 20}

\bibitem[\protect\citeauthoryear{Majumdar \& Mohr}{Majumdar \&
  Mohr}{2004}]{majumdar04}
Majumdar S.,  Mohr J.,  2004, \mn@doi [\apj] {10.1086/422829}, 613, 41

\bibitem[\protect\citeauthoryear{{Mandelbaum} et~al.,}{{Mandelbaum}
  et~al.}{2018a}]{mandelbaum18a}
{Mandelbaum} R.,  et~al., 2018a, \mn@doi [\pasj] {10.1093/pasj/psx130}, \href
  {https://ui.adsabs.harvard.edu/abs/2018PASJ...70S..25M} {70, S25}

\bibitem[\protect\citeauthoryear{{Mandelbaum} et~al.,}{{Mandelbaum}
  et~al.}{2018b}]{mandelbaum18}
{Mandelbaum} R.,  et~al., 2018b, \mn@doi [\pasj] {10.1093/pasj/psx130}, \href
  {http://adsabs.harvard.edu/abs/2018PASJ...70S..25M} {70, S25}

\bibitem[\protect\citeauthoryear{{Mandelbaum} et~al.,}{{Mandelbaum}
  et~al.}{2018c}]{mandelbaum18b}
{Mandelbaum} R.,  et~al., 2018c, \mn@doi [\mnras] {10.1093/mnras/sty2420},
  \href {https://ui.adsabs.harvard.edu/abs/2018MNRAS.481.3170M} {481, 3170}

\bibitem[\protect\citeauthoryear{Mantz, Allen, Rapetti  \& Ebeling}{Mantz
  et~al.}{2010}]{mantz10a}
Mantz A.,  Allen S.,  Rapetti D.,   Ebeling H.,  2010, \mn@doi [\mnras]
  {10.1111/j.1365-2966.2010.16992.x}, 406, 1759

\bibitem[\protect\citeauthoryear{Mantz, Allen, Morris, Rapetti, Applegate,
  Kelly, von~der Linden  \& Schmidt}{Mantz et~al.}{2014}]{mantz14}
Mantz A.,  Allen S.,  Morris R.,  Rapetti D.,  Applegate D.,  Kelly P.,
  von~der Linden A.,   Schmidt R.,  2014, \mn@doi [\mnras]
  {10.1093/mnras/stu368}, 440, 2077

\bibitem[\protect\citeauthoryear{{Mantz} et~al.,}{{Mantz}
  et~al.}{2015}]{mantz15}
{Mantz} A.~B.,  et~al., 2015, \mn@doi [\mnras] {10.1093/mnras/stu2096}, \href
  {http://adsabs.harvard.edu/abs/2015MNRAS.446.2205M} {446, 2205}

\bibitem[\protect\citeauthoryear{Mantz et~al.,}{Mantz et~al.}{2016}]{mantz16b}
Mantz A.,  et~al., 2016, \mn@doi [\mnras] {10.1093/mnras/stw2250}, 463, 3582

\bibitem[\protect\citeauthoryear{{Marulli}, {Veropalumbo},
  {Garc{\'\i}a-Farieta}, {Moresco}, {Moscardini}  \& {Cimatti}}{{Marulli}
  et~al.}{2021}]{marulli21}
{Marulli} F.,  {Veropalumbo} A.,  {Garc{\'\i}a-Farieta} J.~E.,  {Moresco} M.,
  {Moscardini} L.,   {Cimatti} A.,  2021, \mn@doi [\apj]
  {10.3847/1538-4357/ac0e8c}, \href
  {https://ui.adsabs.harvard.edu/abs/2021ApJ...920...13M} {920, 13}

\bibitem[\protect\citeauthoryear{{Maturi}, {Bellagamba}, {Radovich},
  {Roncarelli}, {Sereno}, {Moscardini}, {Bardelli}  \& {Puddu}}{{Maturi}
  et~al.}{2019}]{maturi19}
{Maturi} M.,  {Bellagamba} F.,  {Radovich} M.,  {Roncarelli} M.,  {Sereno} M.,
  {Moscardini} L.,  {Bardelli} S.,   {Puddu} E.,  2019, \mn@doi [\mnras]
  {10.1093/mnras/stz294}, \href
  {https://ui.adsabs.harvard.edu/abs/2019MNRAS.485..498M} {485, 498}

\bibitem[\protect\citeauthoryear{{McKerns}, {Strand}, {Sullivan}, {Fang}  \&
  {Aivazis}}{{McKerns} et~al.}{2012}]{pathos}
{McKerns} M.~M.,  {Strand} L.,  {Sullivan} T.,  {Fang} A.,   {Aivazis} M.
  A.~G.,  2012, arXiv e-prints, \href
  {https://ui.adsabs.harvard.edu/abs/2012arXiv1202.1056M} {p. arXiv:1202.1056}

\bibitem[\protect\citeauthoryear{{Miyazaki}}{{Miyazaki}}{2015}]{miyazaki15}
{Miyazaki} S.,  2015, IAU General Assembly, \href
  {http://adsabs.harvard.edu/abs/2015IAUGA..2255916M} {22, 2255916}

\bibitem[\protect\citeauthoryear{{Miyazaki} et~al.,}{{Miyazaki}
  et~al.}{2018}]{miyazaki18}
{Miyazaki} S.,  et~al., 2018, \mn@doi [\pasj] {10.1093/pasj/psx063}, \href
  {https://ui.adsabs.harvard.edu/abs/2018PASJ...70S...1M} {70, S1}

\bibitem[\protect\citeauthoryear{Mohr \& Evrard}{Mohr \& Evrard}{1997}]{mohr97}
Mohr J.~J.,  Evrard A.~E.,  1997, \apj, 491, 38

\bibitem[\protect\citeauthoryear{Mohr, Mathiesen  \& Evrard}{Mohr
  et~al.}{1999}]{mohr99}
Mohr J.~J.,  Mathiesen B.,   Evrard A.~E.,  1999, \apj, 517, 627

\bibitem[\protect\citeauthoryear{{Murata}, {Sunayama}, {Oguri}, {More},
  {Nishizawa}, {Nishimichi}  \& {Osato}}{{Murata} et~al.}{2020}]{murata20}
{Murata} R.,  {Sunayama} T.,  {Oguri} M.,  {More} S.,  {Nishizawa} A.~J.,
  {Nishimichi} T.,   {Osato} K.,  2020, \mn@doi [\pasj] {10.1093/pasj/psaa041},
  \href {https://ui.adsabs.harvard.edu/abs/2020PASJ...72...64M} {72, 64}

\bibitem[\protect\citeauthoryear{{Oguri}}{{Oguri}}{2014}]{oguri14}
{Oguri} M.,  2014, \mn@doi [\mnras] {10.1093/mnras/stu1446}, \href
  {http://adsabs.harvard.edu/abs/2014MNRAS.444..147O} {444, 147}

\bibitem[\protect\citeauthoryear{{Oguri} et~al.,}{{Oguri}
  et~al.}{2018a}]{oguri18}
{Oguri} M.,  et~al., 2018a, \mn@doi [\pasj] {10.1093/pasj/psx042}, \href
  {http://adsabs.harvard.edu/abs/2018PASJ...70S..20O} {70, S20}

\bibitem[\protect\citeauthoryear{{Oguri} et~al.,}{{Oguri}
  et~al.}{2018b}]{oguri18b}
{Oguri} M.,  et~al., 2018b, \mn@doi [\pasj] {10.1093/pasj/psx070}, \href
  {https://ui.adsabs.harvard.edu/abs/2018PASJ...70S..26O} {70, S26}

\bibitem[\protect\citeauthoryear{{Oguri} et~al.,}{{Oguri}
  et~al.}{2021}]{oguri21}
{Oguri} M.,  et~al., 2021, \mn@doi [\pasj] {10.1093/pasj/psab047}, \href
  {https://ui.adsabs.harvard.edu/abs/2021PASJ...73..817O} {73, 817}

\bibitem[\protect\citeauthoryear{{Pacaud} et~al.,}{{Pacaud}
  et~al.}{2006}]{pacaud06}
{Pacaud} F.,  et~al., 2006, \mn@doi [\mnras]
  {10.1111/j.1365-2966.2006.10881.x}, \href
  {https://ui.adsabs.harvard.edu/abs/2006MNRAS.372..578P} {372, 578}

\bibitem[\protect\citeauthoryear{{Pacaud} et~al.,}{{Pacaud}
  et~al.}{2016}]{pacaud16}
{Pacaud} F.,  et~al., 2016, \mn@doi [\aap] {10.1051/0004-6361/201526891}, \href
  {https://ui.adsabs.harvard.edu/abs/2016A&A...592A...2P} {592, A2}

\bibitem[\protect\citeauthoryear{{Pacaud} et~al.,}{{Pacaud}
  et~al.}{2018}]{pacaud18}
{Pacaud} F.,  et~al., 2018, \mn@doi [\aap] {10.1051/0004-6361/201834022}, \href
  {https://ui.adsabs.harvard.edu/abs/2018A&A...620A..10P} {620, A10}

\bibitem[\protect\citeauthoryear{P\'erez \& Granger}{P\'erez \&
  Granger}{2007}]{ipython}
P\'erez F.,  Granger B.~E.,  2007, \mn@doi [Computing in Science and
  Engineering] {10.1109/MCSE.2007.53}, 9, 21

\bibitem[\protect\citeauthoryear{{Planck Collaboration} \& et al.}{{Planck
  Collaboration} \& et~al.}{2014}]{PlanckCollaboration2014}
{Planck Collaboration} et al. 2014, \mn@doi [\aap]
  {10.1051/0004-6361/201321521}, \href
  {https://ui.adsabs.harvard.edu/abs/2014A&A...571A..20P} {571, A20}

\bibitem[\protect\citeauthoryear{{Planck Collaboration} \& et al.}{{Planck
  Collaboration} \& et~al.}{2016}]{PlanckCollaboration2015b}
{Planck Collaboration} et al. 2016, \mn@doi [\aap]
  {10.1051/0004-6361/201525833}, \href
  {https://ui.adsabs.harvard.edu/abs/2016A&A...594A..24P} {594, A24}

\bibitem[\protect\citeauthoryear{{Planck Collaboration} et~al.,}{{Planck
  Collaboration} et~al.}{2015}]{PlanckCollaboration2015a}
{Planck Collaboration} P.,  et~al., 2015, \mn@doi [Astronomy {\&} Astrophysics,
  Volume 594, id.A27, 38 pp.] {10.1051/0004-6361/201525823}, 594

\bibitem[\protect\citeauthoryear{{Planck Collaboration} et~al.,}{{Planck
  Collaboration} et~al.}{2020}]{PlanckCollaboration20}
{Planck Collaboration} et~al., 2020, \mn@doi [\aap]
  {10.1051/0004-6361/201833910}, \href
  {https://ui.adsabs.harvard.edu/abs/2020A&A...641A...6P} {641, A6}

\bibitem[\protect\citeauthoryear{{Pop} et~al.,}{{Pop} et~al.}{2022}]{pop22}
{Pop} A.-R.,  et~al., 2022, arXiv e-prints, \href
  {https://ui.adsabs.harvard.edu/abs/2022arXiv220511528P} {p. arXiv:2205.11528}

\bibitem[\protect\citeauthoryear{{Predehl} et~al.,}{{Predehl}
  et~al.}{2021}]{predehl21}
{Predehl} P.,  et~al., 2021, \mn@doi [\aap] {10.1051/0004-6361/202039313},
  \href {https://ui.adsabs.harvard.edu/abs/2021A&A...647A...1P} {647, A1}

\bibitem[\protect\citeauthoryear{{Reiprich} \& {B{\"o}hringer}}{{Reiprich} \&
  {B{\"o}hringer}}{2002}]{reiprich02}
{Reiprich} T.~H.,  {B{\"o}hringer} H.,  2002, \mn@doi [\apj] {10.1086/338753},
  \href {https://ui.adsabs.harvard.edu/abs/2002ApJ...567..716R} {567, 716}

\bibitem[\protect\citeauthoryear{{Rykoff} et~al.,}{{Rykoff}
  et~al.}{2014}]{rykoff14}
{Rykoff} E.~S.,  et~al., 2014, \mn@doi [\apj] {10.1088/0004-637X/785/2/104},
  \href {https://ui.adsabs.harvard.edu/abs/2014ApJ...785..104R} {785, 104}

\bibitem[\protect\citeauthoryear{{Salvati} et~al.,}{{Salvati}
  et~al.}{2021}]{salvati21}
{Salvati} L.,  et~al., 2021, arXiv e-prints, \href
  {https://ui.adsabs.harvard.edu/abs/2021arXiv211203606S} {p. arXiv:2112.03606}

\bibitem[\protect\citeauthoryear{{Salvato} et~al.,}{{Salvato}
  et~al.}{2022}]{salvato21}
{Salvato} M.,  et~al., 2022, \mn@doi [\aap] {10.1051/0004-6361/202141631},
  \href {https://ui.adsabs.harvard.edu/abs/2022A&A...661A...3S} {661, A3}

\bibitem[\protect\citeauthoryear{{Schellenberger} \&
  {Reiprich}}{{Schellenberger} \& {Reiprich}}{2017}]{schellenberger17}
{Schellenberger} G.,  {Reiprich} T.~H.,  2017, \mn@doi [\mnras]
  {10.1093/mnras/stx1583}, \href
  {https://ui.adsabs.harvard.edu/abs/2017MNRAS.471.1370S} {471, 1370}

\bibitem[\protect\citeauthoryear{{Schlafly} et~al.,}{{Schlafly}
  et~al.}{2012}]{schlafly12}
{Schlafly} E.~F.,  et~al., 2012, \mn@doi [\apj] {10.1088/0004-637X/756/2/158},
  \href {https://ui.adsabs.harvard.edu/abs/2012ApJ...756..158S} {756, 158}

\bibitem[\protect\citeauthoryear{{Schrabback} et~al.,}{{Schrabback}
  et~al.}{2018}]{schrabback18}
{Schrabback} T.,  et~al., 2018, \mn@doi [\mnras] {10.1093/mnras/stx2666}, \href
  {http://adsabs.harvard.edu/abs/2018MNRAS.474.2635S} {474, 2635}

\bibitem[\protect\citeauthoryear{{Sommer}, {Schrabback}, {Applegate},
  {Hilbert}, {Ansarinejad}, {Floyd}  \& {Grandis}}{{Sommer}
  et~al.}{2022}]{sommer21}
{Sommer} M.~W.,  {Schrabback} T.,  {Applegate} D.~E.,  {Hilbert} S.,
  {Ansarinejad} B.,  {Floyd} B.,   {Grandis} S.,  2022, \mn@doi [\mnras]
  {10.1093/mnras/stab3052}, \href
  {https://ui.adsabs.harvard.edu/abs/2022MNRAS.509.1127S} {509, 1127}

\bibitem[\protect\citeauthoryear{{Song}, {Mohr}, {Barkhouse}, {Warren}, {Dolag}
   \& {Rude}}{{Song} et~al.}{2012}]{song12a}
{Song} J.,  {Mohr} J.~J.,  {Barkhouse} W.~A.,  {Warren} M.~S.,  {Dolag} K.,
  {Rude} C.,  2012, \mn@doi [\apj] {10.1088/0004-637X/747/1/58}, \href
  {https://ui.adsabs.harvard.edu/abs/2012ApJ...747...58S} {747, 58}

\bibitem[\protect\citeauthoryear{Staniszewski et~al.,}{Staniszewski
  et~al.}{2009}]{staniszewski09}
Staniszewski Z.,  et~al., 2009, \mn@doi [\apj] {10.1088/0004-637X/701/1/32},
  701, 32

\bibitem[\protect\citeauthoryear{{Sunayama}}{{Sunayama}}{2022}]{sunayama22}
{Sunayama} T.,  2022, arXiv e-prints, \href
  {https://ui.adsabs.harvard.edu/abs/2022arXiv220503233S} {p. arXiv:2205.03233}

\bibitem[\protect\citeauthoryear{{Sunayama} et~al.,}{{Sunayama}
  et~al.}{2020}]{sunayama20}
{Sunayama} T.,  et~al., 2020, \mn@doi [\mnras] {10.1093/mnras/staa1646}, \href
  {https://ui.adsabs.harvard.edu/abs/2020MNRAS.496.4468S} {496, 4468}

\bibitem[\protect\citeauthoryear{Sunyaev \& Zel'dovich}{Sunyaev \&
  Zel'dovich}{1972}]{sunyaev72}
Sunyaev R.,  Zel'dovich Y.,  1972, Comments on Astrophysics and Space Physics,
  4, 173

\bibitem[\protect\citeauthoryear{{Tanaka} et~al.,}{{Tanaka}
  et~al.}{2017}]{tanaka17}
{Tanaka} M.,  et~al., 2017, arXiv e-prints, \href
  {https://ui.adsabs.harvard.edu/abs/2017arXiv170600566T} {p. arXiv:1706.00566}

\bibitem[\protect\citeauthoryear{{Tanaka} et~al.,}{{Tanaka}
  et~al.}{2018}]{tanaka18}
{Tanaka} M.,  et~al., 2018, \mn@doi [\pasj] {10.1093/pasj/psx077}, \href
  {http://adsabs.harvard.edu/abs/2018PASJ...70S...9T} {70, S9}

\bibitem[\protect\citeauthoryear{{Taylor}}{{Taylor}}{2005}]{topcat1}
{Taylor} M.~B.,  2005, in {Shopbell} P.,  {Britton} M.,   {Ebert} R.,  eds,
  Astronomical Society of the Pacific Conference Series Vol. 347, Astronomical
  Data Analysis Software and Systems XIV. p.~29

\bibitem[\protect\citeauthoryear{{Taylor}}{{Taylor}}{2006}]{topcat2}
{Taylor} M.~B.,  2006, in {Gabriel} C.,  {Arviset} C.,  {Ponz} D.,   {Enrique}
  S.,  eds,  Astronomical Society of the Pacific Conference Series Vol. 351,
  Astronomical Data Analysis Software and Systems XV. p.~666

\bibitem[\protect\citeauthoryear{{The Dark Energy Survey Collaboration}}{{The
  Dark Energy Survey Collaboration}}{2005}]{des05}
{The Dark Energy Survey Collaboration} 2005, arXiv e-prints, \href
  {https://ui.adsabs.harvard.edu/abs/2005astro.ph.10346T} {p. astro:0510346}

\bibitem[\protect\citeauthoryear{{The Dark Energy Survey Collaboration}
  et~al.,}{{The Dark Energy Survey Collaboration} et~al.}{2016}]{des16}
{The Dark Energy Survey Collaboration} et~al., 2016, \mn@doi [\mnras]
  {10.1093/mnras/stw641}, \href
  {https://ui.adsabs.harvard.edu/abs/2016MNRAS.460.1270D} {460, 1270}

\bibitem[\protect\citeauthoryear{{To} et~al.,}{{To} et~al.}{2021}]{to21}
{To} C.,  et~al., 2021, \mn@doi [\prl] {10.1103/PhysRevLett.126.141301}, \href
  {https://ui.adsabs.harvard.edu/abs/2021PhRvL.126n1301T} {126, 141301}

\bibitem[\protect\citeauthoryear{{To}, {Rozo}, {Krause}, {Wu}, {Wechsler}  \&
  {Salcedo}}{{To} et~al.}{2022}]{to22}
{To} C.-H.,  {Rozo} E.,  {Krause} E.,  {Wu} H.-Y.,  {Wechsler} R.~H.,
  {Salcedo} A.~N.,  2022, arXiv e-prints, \href
  {https://ui.adsabs.harvard.edu/abs/2022arXiv220305583T} {p. arXiv:2203.05583}

\bibitem[\protect\citeauthoryear{{Tonry} et~al.,}{{Tonry}
  et~al.}{2012}]{tonry12}
{Tonry} J.~L.,  et~al., 2012, \mn@doi [\apj] {10.1088/0004-637X/750/2/99},
  \href {https://ui.adsabs.harvard.edu/abs/2012ApJ...750...99T} {750, 99}

\bibitem[\protect\citeauthoryear{{Umetsu}}{{Umetsu}}{2020}]{umetsu20b}
{Umetsu} K.,  2020, \mn@doi [\aapr] {10.1007/s00159-020-00129-w}, \href
  {https://ui.adsabs.harvard.edu/abs/2020A&ARv..28....7U} {28, 7}

\bibitem[\protect\citeauthoryear{{Umetsu} et~al.,}{{Umetsu}
  et~al.}{2020}]{umetsu20}
{Umetsu} K.,  et~al., 2020, \mn@doi [\apj] {10.3847/1538-4357/ab6bca}, \href
  {https://ui.adsabs.harvard.edu/abs/2020ApJ...890..148U} {890, 148}

\bibitem[\protect\citeauthoryear{Van Der~Walt, Colbert  \& Varoquaux}{Van
  Der~Walt et~al.}{2011}]{van2011numpy}
Van Der~Walt S.,  Colbert S.~C.,   Varoquaux G.,  2011, Computing in Science \&
  Engineering, 13, 22

\bibitem[\protect\citeauthoryear{Vanderlinde et~al.,}{Vanderlinde
  et~al.}{2010}]{vanderlinde10}
Vanderlinde K.,  et~al., 2010, \mn@doi [\apj] {10.1088/0004-637X/722/2/1180},
  722, 1180

\bibitem[\protect\citeauthoryear{Vikhlinin et~al.,}{Vikhlinin
  et~al.}{2009a}]{vikhlinin09a}
Vikhlinin A.,  et~al., 2009a, \mn@doi [\apj] {10.1088/0004-637X/692/2/1033},
  692, 1033

\bibitem[\protect\citeauthoryear{Vikhlinin et~al.,}{Vikhlinin
  et~al.}{2009b}]{vikhlinin09b}
Vikhlinin A.,  et~al., 2009b, \mn@doi [\apj] {10.1088/0004-637X/692/2/1060},
  692, 1060

\bibitem[\protect\citeauthoryear{{Virtanen} et~al.,}{{Virtanen}
  et~al.}{2020}]{scipy}
{Virtanen} P.,  et~al., 2020, \mn@doi [Nature Methods]
  {https://doi.org/10.1038/s41592-019-0686-2}, \href {https://rdcu.be/b08Wh}
  {17, 261}

\bibitem[\protect\citeauthoryear{{Weinberg}, {Mortonson}, {Eisenstein},
  {Hirata}, {Riess}  \& {Rozo}}{{Weinberg} et~al.}{2013}]{weinberg13}
{Weinberg} D.~H.,  {Mortonson} M.~J.,  {Eisenstein} D.~J.,  {Hirata} C.,
  {Riess} A.~G.,   {Rozo} E.,  2013, \mn@doi [\physrep]
  {10.1016/j.physrep.2013.05.001}, \href
  {https://ui.adsabs.harvard.edu/abs/2013PhR...530...87W} {530, 87}

\bibitem[\protect\citeauthoryear{{Wittman}, {Dell'Antonio}, {Hughes},
  {Margoniner}, {Tyson}, {Cohen}  \& {Norman}}{{Wittman}
  et~al.}{2006}]{wittman06}
{Wittman} D.,  {Dell'Antonio} I.~P.,  {Hughes} J.~P.,  {Margoniner} V.~E.,
  {Tyson} J.~A.,  {Cohen} J.~G.,   {Norman} D.,  2006, \mn@doi [\apj]
  {10.1086/502621}, \href
  {https://ui.adsabs.harvard.edu/abs/2006ApJ...643..128W} {643, 128}

\bibitem[\protect\citeauthoryear{{Wu} et~al.,}{{Wu} et~al.}{2022}]{wu22}
{Wu} H.-Y.,  et~al., 2022, arXiv e-prints, \href
  {https://ui.adsabs.harvard.edu/abs/2022arXiv220305416W} {p. arXiv:2203.05416}

\bibitem[\protect\citeauthoryear{{Yang}, {Mo}, {van den Bosch}, {Pasquali},
  {Li}  \& {Barden}}{{Yang} et~al.}{2007}]{yang07}
{Yang} X.,  {Mo} H.~J.,  {van den Bosch} F.~C.,  {Pasquali} A.,  {Li} C.,
  {Barden} M.,  2007, \mn@doi [\apj] {10.1086/522027}, \href
  {https://ui.adsabs.harvard.edu/abs/2007ApJ...671..153Y} {671, 153}

\bibitem[\protect\citeauthoryear{{Zhang} \& {Annis}}{{Zhang} \&
  {Annis}}{2022}]{zhang22}
{Zhang} Y.,  {Annis} J.,  2022, \mn@doi [\mnras] {10.1093/mnrasl/slac002},
  \href {https://ui.adsabs.harvard.edu/abs/2022MNRAS.511L..30Z} {511, L30}

\bibitem[\protect\citeauthoryear{{Zu}, {Weinberg}, {Rozo}, {Sheldon}, {Tinker}
  \& {Becker}}{{Zu} et~al.}{2014}]{zu14}
{Zu} Y.,  {Weinberg} D.~H.,  {Rozo} E.,  {Sheldon} E.~S.,  {Tinker} J.~L.,
  {Becker} M.~R.,  2014, \mn@doi [\mnras] {10.1093/mnras/stu033}, \href
  {https://ui.adsabs.harvard.edu/abs/2014MNRAS.439.1628Z} {439, 1628}

\bibitem[\protect\citeauthoryear{{Zu}, {Mandelbaum}, {Simet}, {Rozo}  \&
  {Rykoff}}{{Zu} et~al.}{2017}]{zu17}
{Zu} Y.,  {Mandelbaum} R.,  {Simet} M.,  {Rozo} E.,   {Rykoff} E.~S.,  2017,
  \mn@doi [\mnras] {10.1093/mnras/stx1264}, \href
  {https://ui.adsabs.harvard.edu/abs/2017MNRAS.470..551Z} {470, 551}

\bibitem[\protect\citeauthoryear{{Zu} et~al.,}{{Zu} et~al.}{2022}]{zu22}
{Zu} Y.,  et~al., 2022, \mn@doi [\mnras] {10.1093/mnras/stac125}, \href
  {https://ui.adsabs.harvard.edu/abs/2022MNRAS.511.1789Z} {511, 1789}

\bibitem[\protect\citeauthoryear{de Haan et~al.,}{de~Haan
  et~al.}{2016}]{deHaan16}
de Haan T.,  et~al., 2016, \mn@doi [\apj] {10.3847/0004-637X/832/1/95}, 832, 95

\makeatother
\end{thebibliography}

%
%

\onecolumn

\appendix

\section{The measurement uncertainty of observed count rates}
\label{app:meas_observed_rate}

To evaluate equation~(\ref{eq:measurement_uncertainty_rate}) in the number counts likelihood, we empirically derive the measurement uncertainty $\delta_{\rate}$ based on the observed data following a method introduced in \citet[][see their Appendix~A]{grandis20}.
Specifically, we model the measurement uncertainty $\delta_{\rate}$ as a log-normal distribution with intrinsic scatter around the mean value described as a power-law function of the observed count rate \rate, the exposure time \texp\, and the cluster redshift \redshift.
In the sample of the \eFEDS\ clusters selected in this work, we find that the logarithmic measurement uncertainty at a given $\left(\rate, \texp, \redshift\right)$ is described by the mean value,
\begin{multline}
\label{eq:measurement_uncertainty_rate_details}
\left\langle
\ln\left(
\frac{ 
\delta_{\rate}
}{~\mathrm{counts/sec} }
\Bigg|\rate,\redshift,\texp
\right)
\right\rangle
= 
\left( 0.0156 \pm 0.00016 \right) + 
\left( 0.565\pm 0.014 \right) \times \ln\left( \frac{\rate}{0.1~\mathrm{counts/sec}}\right) +
\left( -0.078\pm0.021 \right) \times \ln\left( \frac{\redshift}{0.35} \right)  \\
+
\left( -0.383\pm 0.034 \right) \times \ln \left( \frac{\texp}{1180~\mathrm{sec}} \right)
\, ,
\end{multline}
with the log-normal scatter of $0.211\pm0.007$.

In this work, we use the mean value of the measurement uncertainty predicted at the grid of the observed count rate \rate\ and the cluster redshift \redshift\ to evaluate equation~(\ref{eq:measurement_uncertainty_rate}) in the number counts likelihood.
Given that the exposure time is nearly uniform in the \eFEDS\ imaging and does not depend on the physical properties (e.g., \mass, \redshift, and \rate), we ignore the dependence on the exposure time when predicting $\delta_{\rate}$, i.e., \texp\ is fixed to the median exposure time $1180$~seconds in equation~(\ref{eq:measurement_uncertainty_rate_details}).
Because the scatter of the measurement uncertainty is a second-order perturbation to the count rate distribution of clusters at a given mass and redshift, we also choose to ignore the intrinsic scatter of $\delta_{\rate}$.
In addition, this scatter is expected to be absorbed into the intrinsic scatter of the intrinsic count rate at a fixed mass and redshift (see equation~(\ref{eq:richness_to_mass})), given that we empirically model the parameter \sigmarate\ in the likelihood.
Therefore, we don't expect that the approximations made in predicting $\delta_{\rate}$ would introduce significant bias to our results.

\begin{figure*}
\centering
\resizebox{0.33\textwidth}{!}{
\includegraphics[scale=1]{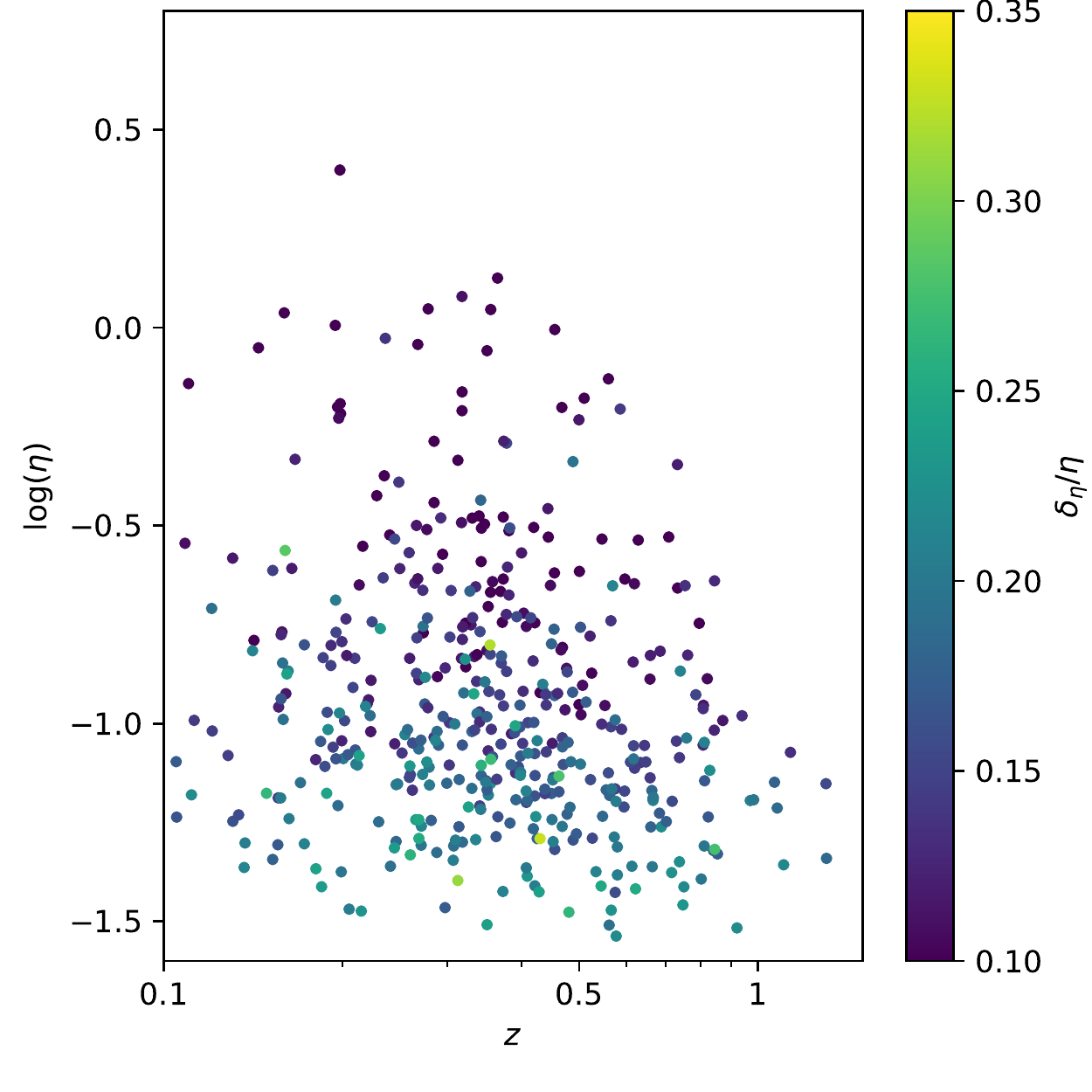}
}
\resizebox{0.33\textwidth}{!}{
\includegraphics[scale=1]{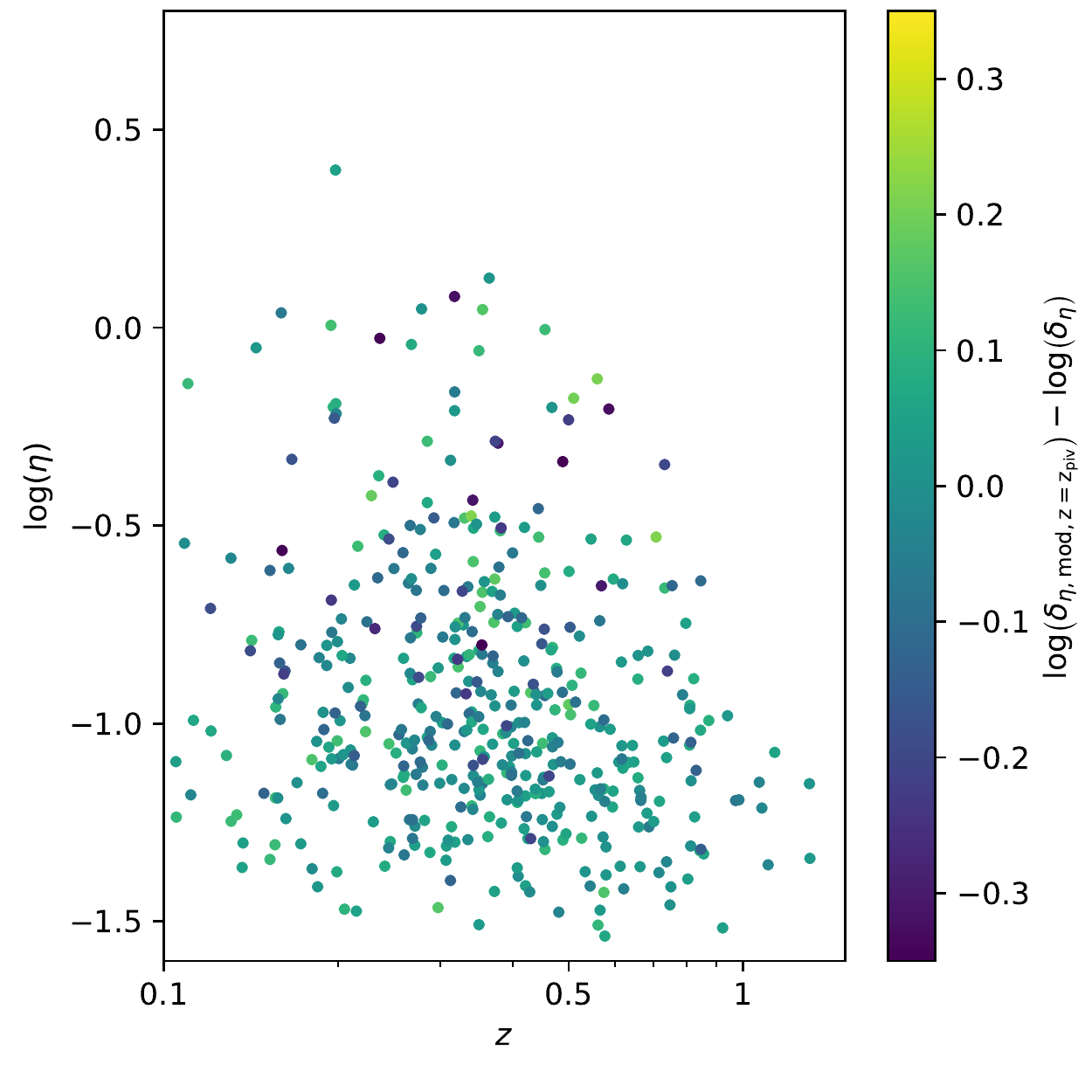}
}
\resizebox{0.33\textwidth}{!}{
\includegraphics[scale=1]{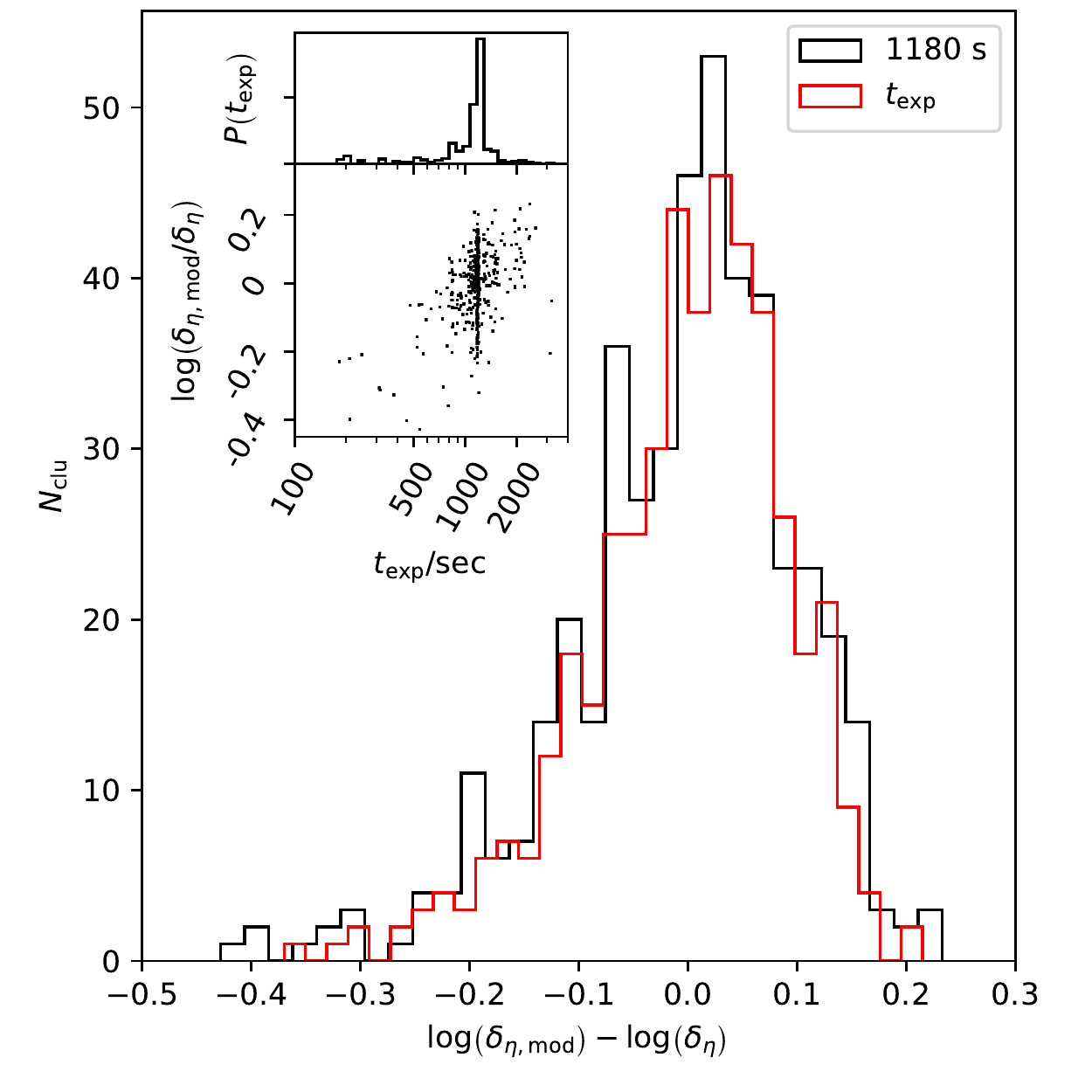}
}
\caption{
The empirical modeling of the observed count rate error $\delta_{\rate}$.
In the left panel, we show the observed fractional error $\delta_{\rate}/\rate$ of the \eFEDS\ clusters in the space of the count rate \rate\ and cluster redshift \redshift.
In the middle plot, the \eFEDS\ clusters in the same observable space of  (\rate, \redshift) are color-coded by the modeling residual, which is the difference between the predicted count rate error at the pivotal redshift ($\zpiv=0.35$) and the observed $\delta_{\rate}$.
Note that the predicted count rate error is evaluated at the median exposure time, $\texp=1180~\mathrm{sec}$, i.e., ignoring the \texp\ dependence in equation~(\ref{eq:measurement_uncertainty_rate_details}).
In the right panel, we show the distributions of the modeling residuals with and without the \texp\ dependence in red and black, respectively.
As seen, there is no significant difference between the predicted count rate errors 
and that of observed for individual clusters, except that the scatter of the residuals is marginally smaller when including the \texp\ dependence.
We additionally show the modeling residuals (with \texp\ fixed to $1180$~seconds) against the exposure time of the \eFEDS\ clusters in the embedded plot in the right panel, suggesting that the exposure time is nearly uniform and that the \texp\ dependence is mostly driven by the outliers in the exposure time.
}
\label{fig:rate_err}
\end{figure*}

In Figure~\ref{fig:rate_err}, we show the observed fractional error of the count rate, $\delta_{\rate}/\rate$, of the \eFEDS\ clusters in the left panel, while the modeling residual is presented in the middle panel.
As seen in the middle panel, no obvious trend of the modeling residual is revealed with respect to the count rate and redshift, suggesting that the modeling of $\delta_{\rate}$ indeed provides a good description of the data.
In the right panel, the overall distribution of the modeling residual is shown, suggesting no bias in the modeling with (red) and without (black) the dependence on the exposure time.
This validates our approach that we ignore the exposure-time dependence in predicting $\delta_{\rate}$.
The modeling residual as a function of the exposure time is shown in the embedded plot of the right panel, where we can see that (1) the exposure time is nearly uniform centering on the median of $1180$~seconds and that (2) the \texp\ trend is mostly driven by the outliers with extremely low ($\lesssim500$~seconds) and high ($\gtrsim2000$~seconds) exposure times.

It is worth mentioning that  the count rate error at a fixed count rate is expected to scale as $1/\sqrt{\texp}$ assuming a Poisson distribution of photon counts.
In the \eROSITA\ All-Sky Survey, the exposure time will vary across the sky and hence the inclusion of the \texp\ dependence will be needed.
We defer the improvements, the inclusion of \texp\ dependence and the scatter of $\delta_{\rate}$, to a future work with a much larger sample.

\section{Details of the analyses}
\label{app:detailed_results}

We present the mass estimates of individual clusters, the weak-lensing measurements,  and the constraints of the full lists of the parameters in this appendix.

\begin{figure}
\centering
\resizebox{0.33\textwidth}{!}{
\includegraphics[scale=1]{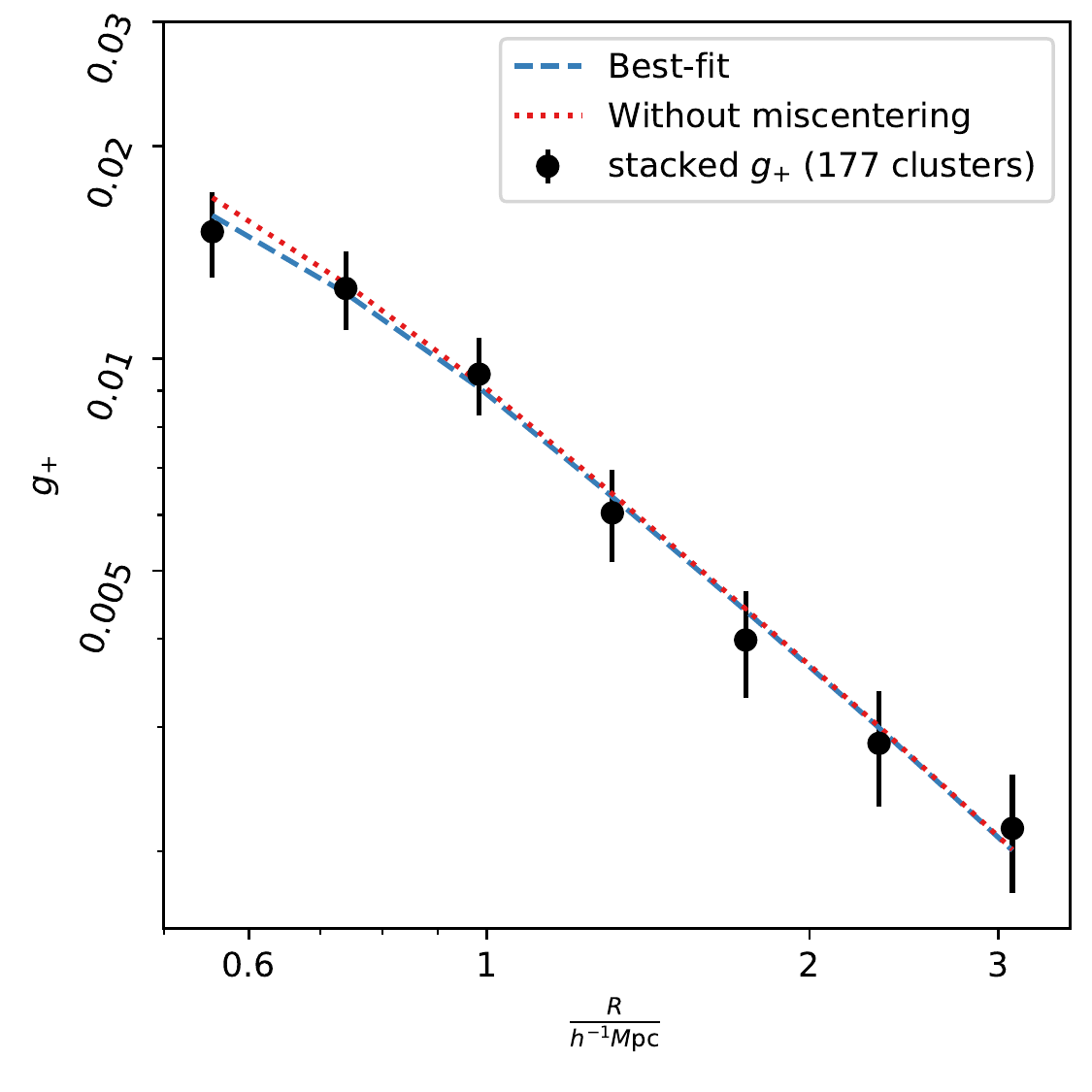}
}
\resizebox{0.33\textwidth}{!}{
\includegraphics[scale=1]{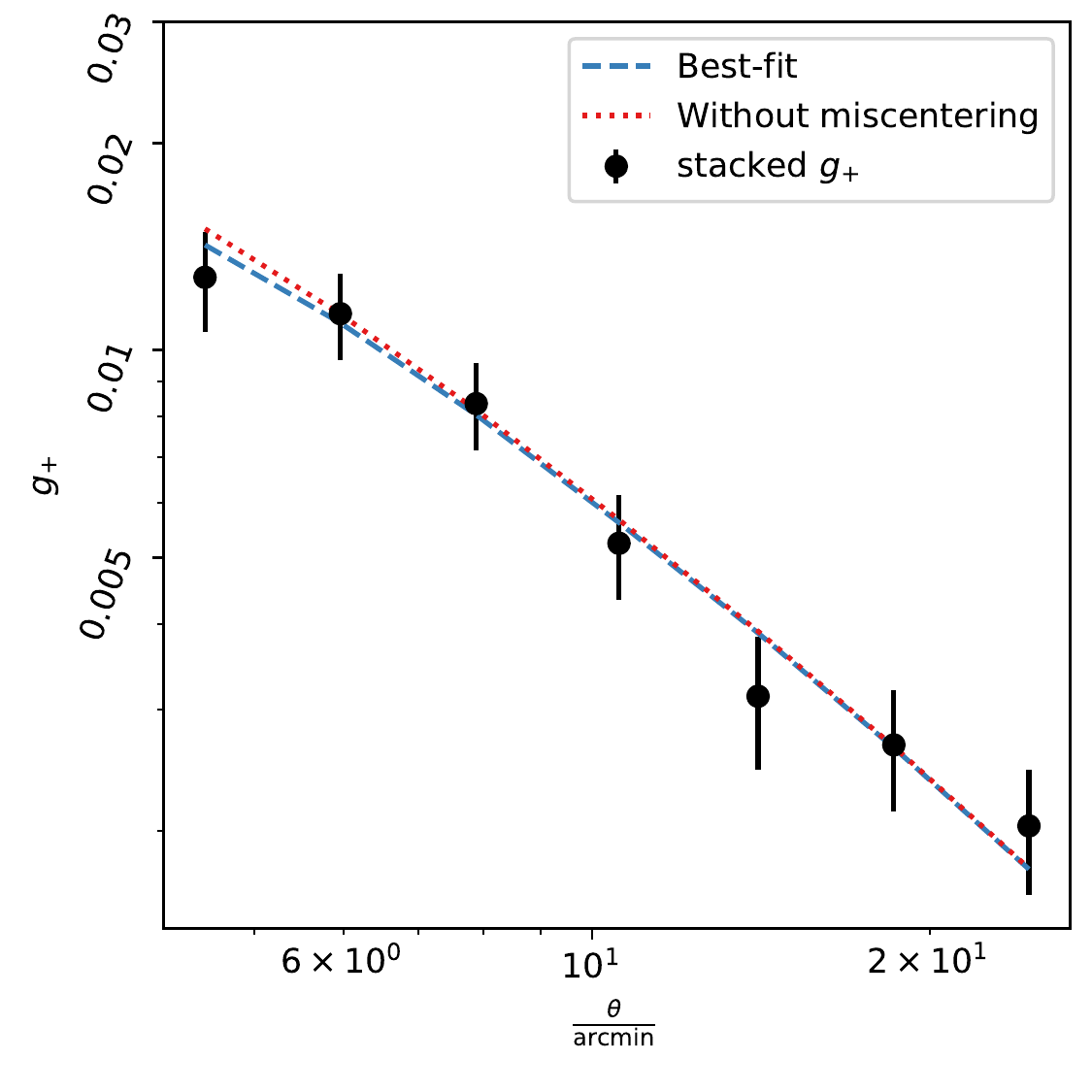}
}
\resizebox{0.33\textwidth}{!}{
\includegraphics[scale=1]{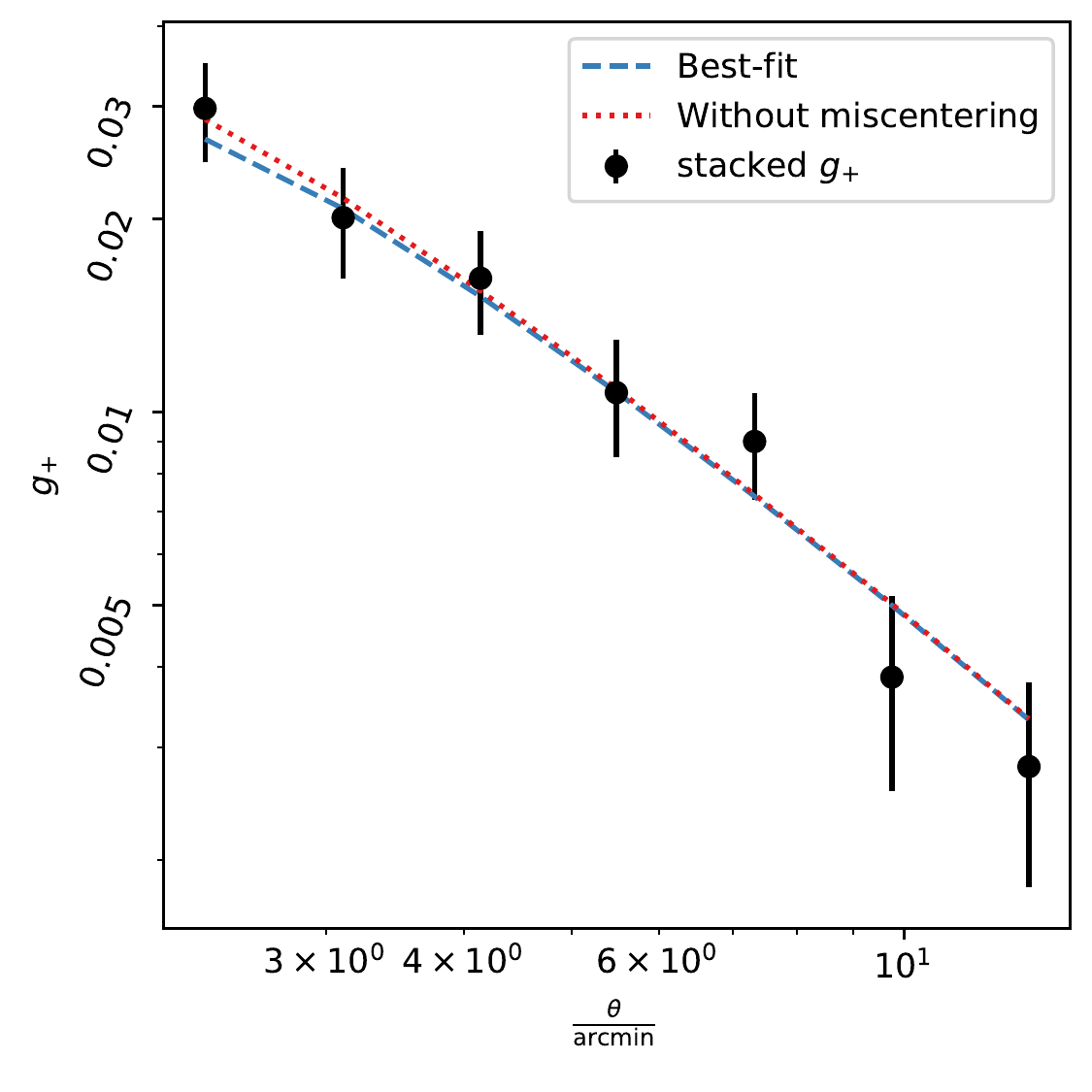}
}
\caption{
The observed shear profile \gshear\ of the \eFEDS\ clusters using the public S16A weak-lensing data from the HSC survey.
There are $177$ \eFEDS\ clusters with available weak-lensing observables in this work.
The stack profiles of the individual \gshear\ of the full, low-\redshift\ ($0.1<\redshift<0.35$) and high-\redshift\ ($0.35<\redshift<1.2$) samples are shown by the black points in the left, middle, and right panels, respectively.
The best-fit models with (without) the modeling of the miscentering of X-ray centers are shown by the blue dashed (red-dotted) lines.
For the full sample, the profiles are presented as a function of the physical radius.
For the low and high redshift samples, the profiles are presented with the radii in arcmin.
}
\label{fig:wlprofiles}
\end{figure}

The mass \Mfiveoo\ of the eFEDS clusters is presented in Table~\ref{tab:mass}.
The stacked weak-lensing profiles of the full, low-\redshift\ ($0.1<\redshift<0.35$), and high-\redshift\ ($0.35<\redshift<1.2$) samples are contained in Figure~\ref{fig:wlprofiles}.
As seen, the best-fit models well describe the data.

\begin{figure*}
\resizebox{\textwidth}{!}{
\includegraphics[scale=1]{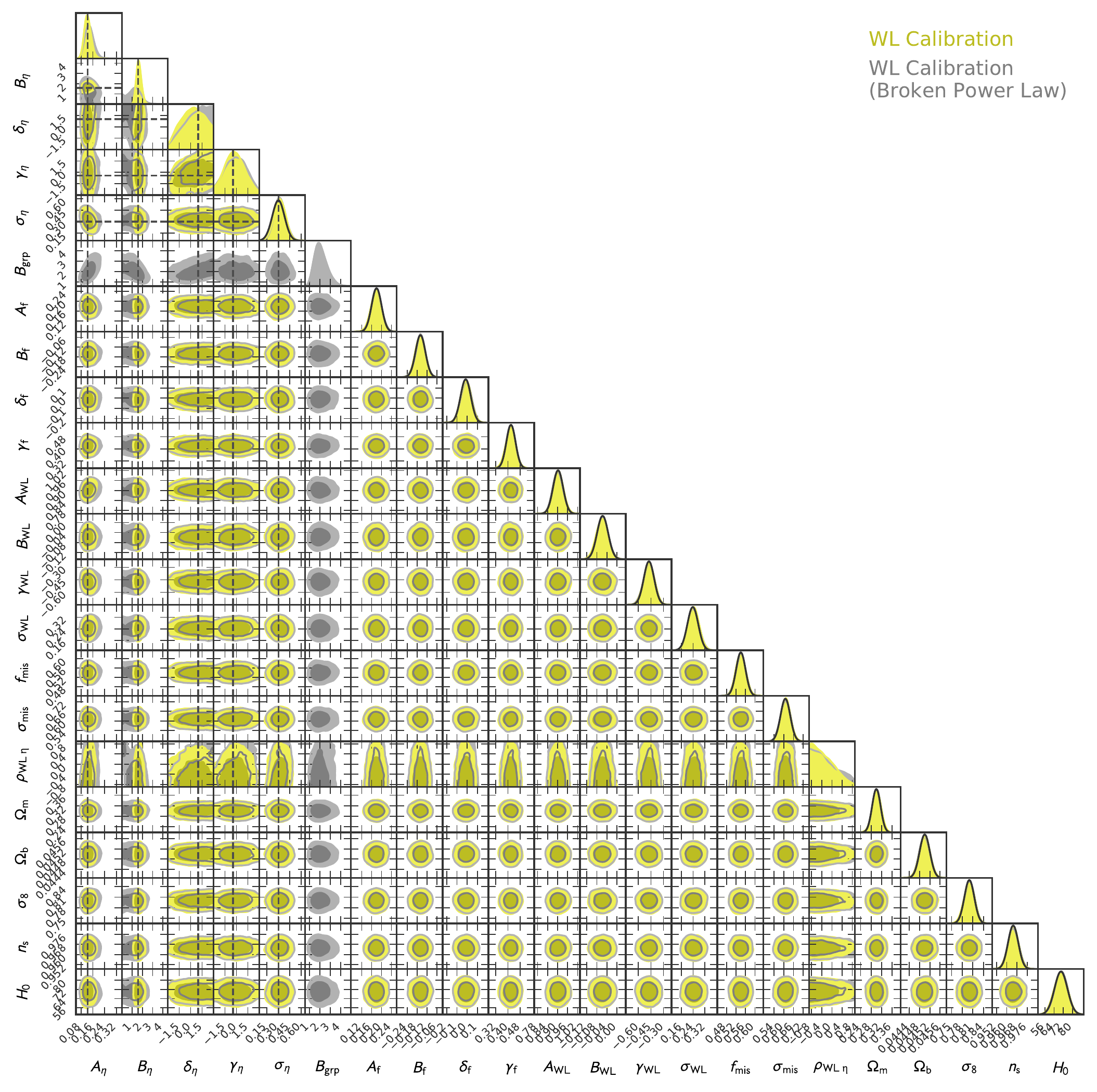}
}
\caption{
The fully marginalized and joint posteriors of the parameters in the modeling of the mass calibration alone.
The same plotting scheme as in Figure~\ref{fig:nbc_gtc_smf} is used here.
}
\label{fig:mcalib_gtc_full}
\end{figure*}

\begin{figure*}
\resizebox{\textwidth}{!}{
\includegraphics[scale=1]{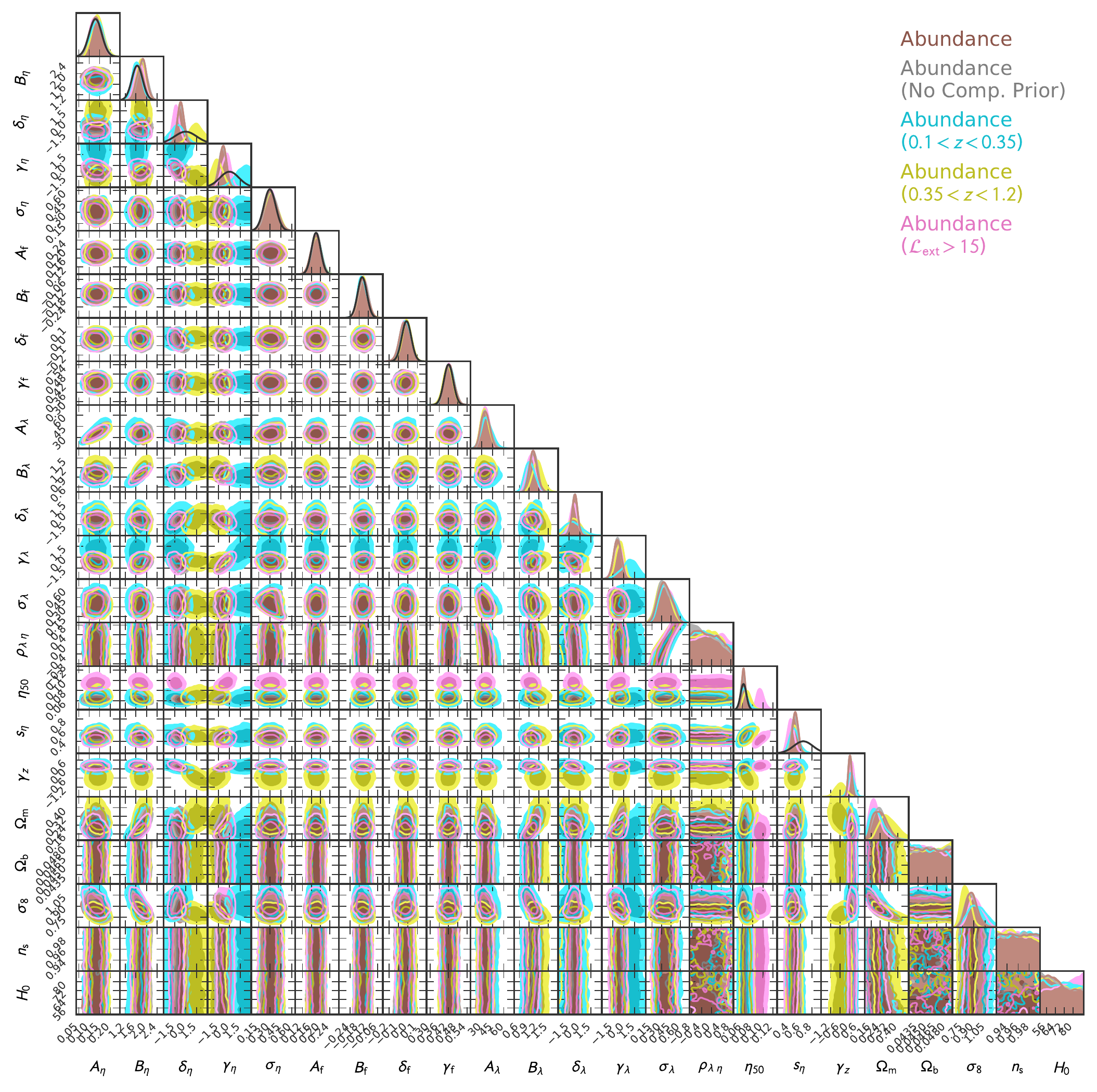}
}
\caption{
The fully marginalized and joint posteriors of 
the parameters in the modeling of the cluster abundance alone.
The same plotting scheme as in Figure~\ref{fig:nbc_gtc_smf} is used here.
}
\label{fig:nbc_gtc_full}
\end{figure*}

\begin{figure*}
\resizebox{\textwidth}{!}{
\includegraphics[scale=1]{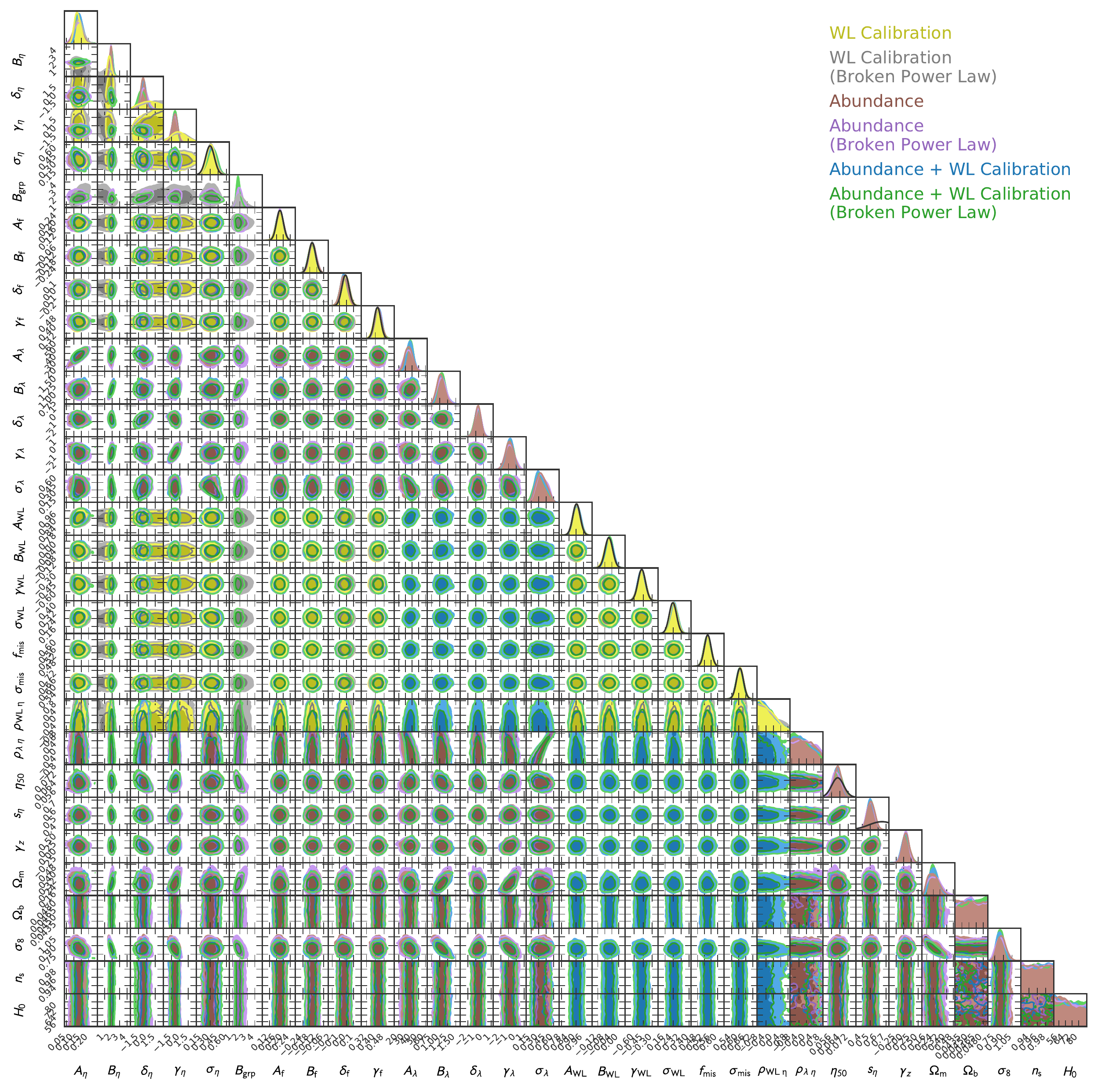}
}
\caption{
The comparisons between the modeling with and without the broken power-law mass scaling of the count rate.
The results of the mass calibration (the cluster abundance, the joint modeling) with the single and broken power-law \rate--\mass--\redshift\ relations are shown in yellow and grey (brown and purple, blue and green), respectively.
}
\label{fig:mcalib_nbc_gtc_full}
\end{figure*}

\begin{figure*}
\resizebox{\textwidth}{!}{
\includegraphics[scale=1]{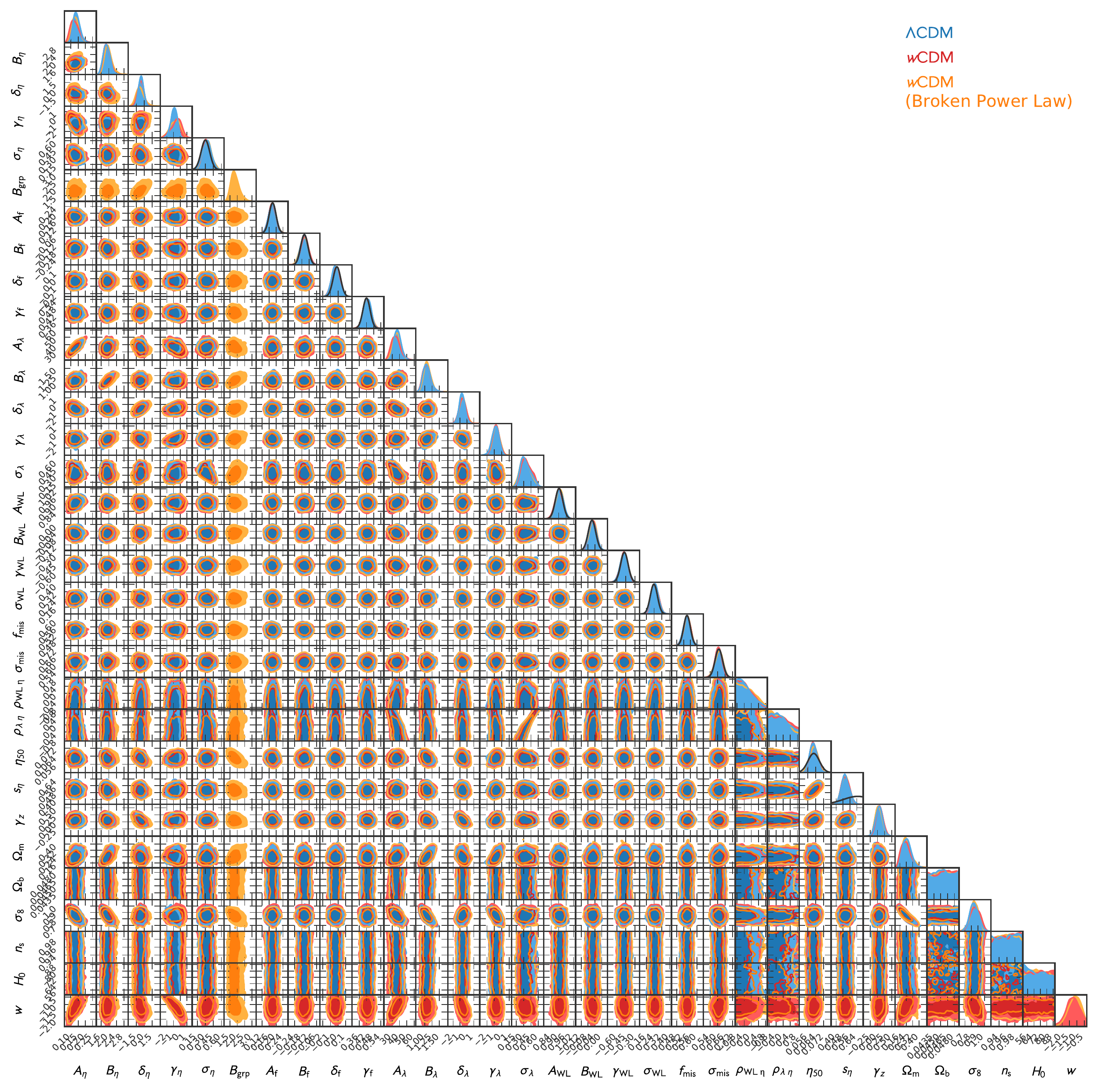}
}
\caption{
The comparisons between the modeling assuming the \LCDM\ and \wCDM\ models.
The same plotting scheme as in Figure~\ref{fig:cosmos_efeds_only} is used here.
}
\label{fig:lcdm_wcdm_full}
\end{figure*}

The constraints of the full parameter list in the modeling of the weak-lensing mass calibration and the cluster abundance are in Figures~\ref{fig:mcalib_gtc_full} and \ref{fig:nbc_gtc_full}, respectively.
The comparisons between the results with and without the broken power-law mass scaling in the \rate--\mass--\redshift\ relations are contained in Figure~\ref{fig:mcalib_nbc_gtc_full}.
The full comparisons between the \LCDM\ and \wCDM\ models are shown in Figure~\ref{fig:lcdm_wcdm_full}.

\begin{table*}
\centering
\caption{
The mass measurements of \eFEDS\ clusters.
Column (1): The cluster name.
Column (2): The cluster redshift (the quantity $\mathtt{Z\_BEST\_COMB}$ in the official \eFEDS\ catalog).
Column (3): The cluster radius \Rfiveoo\ in arcmin.
Column (4): The ensemble mass ${\Mfiveoo}_{,\mathrm{ens}}$ in the \LCDM\ cosmology.
Column (5): The median of the mass posterior $P(\mass|\rate,\redshift,\vect{p})$, where $\vect{p}$ are the best-fit parameters in the \LCDM\ cosmology.
Column (6): The same as in Column (4) but in the \wCDM\ cosmology.
Column (7): The same as in Column (5) but in the \wCDM\ cosmology.
The full catalog is publicly available via \url{https://github.com/inonchiu/eFEDSproducts}.
}
\label{tab:mass}
\begin{tabular}{ccccccc}
\hline\hline
& & \multicolumn{3}{c}{\LCDM} & \multicolumn{2}{c}{\wCDM} \\
\cmidrule(lr){3-5}\cmidrule(lr){6-7}
\multirow{1}{*}{Cluster name}  & 
\multirow{1}{*}{Redshift} & $\frac{\Rfiveoo}{\mathrm{arcmin}}$
& $\log\left( \frac{ {\Mfiveoo}_{,\mathrm{ens}} }{ \Msunh } \right)$ & $\log\left(\frac{ \Mfiveoo }{ \Msunh }\right)$ 
& $\log\left( \frac{ {\Mfiveoo}_{,\mathrm{ens}} }{ \Msunh } \right)$ & $\log\left(\frac{ \Mfiveoo }{ \Msunh }\right)$   \\
(1)  &(2) &(3) &(4) &(5) &(6) &(7) \\
\hline
J093712.9$+$031652   &  $  0.247  $  & $  2.584 $  & $  13.723 $  &  $  {    13.741    }^{+    0.099    }_{-    0.091    }  $  & $  13.861  $  &  $  {    13.761    }^{+    0.098    }_{-    0.086    }  $  \\    
J083811.9$-$015935   &  $  0.560  $  & $  2.875 $  & $  14.758 $  &  $  {    14.699    }^{+    0.082    }_{-    0.085    }  $  & $  14.771  $  &  $  {    14.713    }^{+    0.080    }_{-    0.083    }  $  \\    
J093521.0$+$023234   &  $  0.510  $  & $  2.908 $  & $  14.617 $  &  $  {    14.626    }^{+    0.086    }_{-    0.081    }  $  & $  14.633  $  &  $  {    14.641    }^{+    0.081    }_{-    0.078    }  $  \\    
J092121.2$+$031726   &  $  0.355  $  & $  3.686 $  & $  14.595 $  &  $  {    14.581    }^{+    0.080    }_{-    0.082    }  $  & $  14.608  $  &  $  {    14.595    }^{+    0.078    }_{-    0.079    }  $  \\    
J085751.7$+$031039   &  $  0.198  $  & $  5.605 $  & $  14.512 $  &  $  {    14.506    }^{+    0.073    }_{-    0.079    }  $  & $  14.517  $  &  $  {    14.511    }^{+    0.070    }_{-    0.076    }  $  \\    
J092647.5$+$050033   &  $  0.455  $  & $  3.270 $  & $  14.686 $  &  $  {    14.670    }^{+    0.085    }_{-    0.081    }  $  & $  14.700  $  &  $  {    14.685    }^{+    0.083    }_{-    0.079    }  $  \\    
J084528.7$+$032739   &  $  0.350  $  & $  3.549 $  & $  14.524 $  &  $  {    14.517    }^{+    0.079    }_{-    0.082    }  $  & $  14.538  $  &  $  {    14.531    }^{+    0.077    }_{-    0.079    }  $  \\    
J092002.2$+$010220   &  $  0.017  $  & $  21.144 $  & $  13.327 $  &  $  {    13.232    }^{+    0.068    }_{-    0.074    }  $  & $  13.330  $  &  $  {    13.238    }^{+    0.066    }_{-    0.071    }  $  \\    
J090131.2$+$030057   &  $  0.194  $  & $  4.827 $  & $  14.213 $  &  $  {    14.291    }^{+    0.078    }_{-    0.077    }  $  & $  14.224  $  &  $  {    14.299    }^{+    0.075    }_{-    0.073    }  $  \\    
J083651.3$+$030002   &  $  0.196  $  & $  4.410 $  & $  14.269 $  &  $  {    14.183    }^{+    0.080    }_{-    0.078    }  $  & $  14.275  $  &  $  {    14.193    }^{+    0.077    }_{-    0.075    }  $  \\  \\
\multicolumn{7}{c}{$\cdots$}\\
\hline\hline
\end{tabular}
\end{table*}
%


\bsp	
\label{lastpage}
\end{document}